%% file: main.tex
\documentclass[journal]{IEEEtran}

\usepackage{amsmath,amssymb,amsfonts}
\usepackage{lipsum}
\usepackage{nicefrac,multirow,multicol}
\usepackage{algorithmic,siunitx}
\usepackage{xcolor}
\usepackage{colortbl}
\usepackage{subfig} 
\usepackage{graphicx,tikz,pgfplots,graphics,booktabs}
\usepackage{stfloats} 
\usepackage{url,cite}
\usepackage{bm}
\usepackage{listings}
\usepackage{pgfplots}
\pgfplotsset{compat=1.8}
\pgfplotsset{
  set layers,
  every axis plot/.append style={
    mark layer=foreground,
  },
}
\def\BibTeX{{\rm B\kern-.05em{\sc i\kern-.025em b}\kern-.08em
    T\kern-.1667em\lower.7ex\hbox{E}\kern-.125emX}}
\usepackage{balance}

\begin{document}

\title{From Symmetry to Stability: Quantifying Converter–Grid Impedance Asymmetry as Indicator of Stability Margin}

\author{Chirag Ramgopal Shah,~\IEEEmembership{Graduate Student Member,~IEEE}, Marta Molinas,~\IEEEmembership{Fellow,~IEEE}, Sjur Føyen,~\IEEEmembership{Member,~IEEE}, Roy Nilsen
\thanks{

This manuscript has been submitted to the IEEE Transactions on Power Electronics on 09 April 2026 and has been rejected on 01 June 2026 on the grounds of lacking a theoretical framework to support the empirical observations.

 Chirag Ramgopal Shah (Corresponding author), Sjur Føyen and Roy Nilsen are with the Department of Electric Energy at the Norwegian University of Science and Technology, Norway. (email: chirag.shah@ntnu.no, foyen.sjur@ntnu.no, roy.nilsen@ntnu.no)

Marta Molinas is with the Department of Engineering Cybernetics at the Norwegian University of Science and Technology, Norway. (email: marta.molinas@ntnu.no)
}
}

\markboth{preprint}%
{How to Use the IEEEtran \LaTeX \ Templates}
\maketitle

\begin{abstract}

Although symmetricity in the converter controller is desirable for robust stability margins, a direct link between system-level asymmetricity and instability has yet to be clearly established. Converter control introduces three-phase asymmetricity through loops such as DC-link voltage control, a phase-locked loop , and a power synchronization loop. Furthermore, the inherently asymmetric topology of the two-level voltage-source converter, which converts a DC voltage into a three-phase balanced set, acts as the underlying origin of the asymmetries that propagate into the control structure. Consequently, establishing a direct relationship between system asymmetricity (rather than control asymmetricity alone) and the stability margin is essential for understanding the underlying instability mechanisms. In this work, asymmetricity is quantified using the Asymmetricity Quantification Index (\textit{AQI}), derived from the sequence-domain representation of the interconnected converter-grid impedance. Within this domain, symmetricity is identified through the definition of symmetrical matrices, which serve as the benchmark against which asymmetricity is measured. A robust and generalized analysis correlates \textit{AQI} with the stability margin, including both grid-following and grid-forming control structures connected to the power grid. It is found that instability arises from increased asymmetricity in the combined converter-grid system, which is dominated by asymmetric control loops and operating points. Thus, reducing asymmetricity without compromising controller functionality can improve stability margins. The analysis is validated in both control-hardware-in-the-loop and power-hardware-in-the-loop environments.
\end{abstract}


\section{Introduction}
\IEEEPARstart{I}{mpedance-based} small-signal stability has gained significant attention in power electronic-dominated systems due to its practical applicability \cite{wu_survey_2023}. Therefore, understanding impedance characteristics is crucial for analyzing interactions between power converters and the grid. Recent industry guidelines, such as the UNIFI report, have begun linking specific impedance characteristics to converter behavior, for example, associating high-frequency voltage-phasor stiffness with resistive-inductive behavior within $f_{1}\pm$\SI{40}{\hertz} at the point of common coupling \cite{UNIFI2026}. However, while such specifications describe desired impedance shapes, a rigorous understanding on how these characteristics fundamentally impact stability, irrespective on control objectives, remains an open question. This work addresses this question by providing generalized insights into the impedance structural characteristics and their direct link to the stability of the converter-grid, irrespective of the control structures employed. 

Previous research has implicitly and qualitatively suggested that increasing controller symmetry enhances the stability of converter-grid systems.  Asymmetricity in the controller arises from control loops such as a classical Phase Locked Loop (PLL), DC-link voltage control, and power controllers \cite{rygg_modified_2016}. Several studies have demonstrated that mitigating this asymmetry improves stability: eliminating the PLL and DC voltage controller asymmetry enhances stability in weak grids \cite{zhang_symmetrical_2020}, a symmetrical PLL design improves weak grid stability \cite{yang_symmetrical_2020}, and extending these principles to include power control loops further reinforces the trend \cite{sun_symmetrical_2025}. Compensating for $dq$ frame coupling introduced by AC voltage control and PLL has also been shown to improve stability \cite{wu_enhanced_2025}. Similarly, reducing sequence domain coupling in grid-forming converters enhances stability \cite{elshenawy_analysis_2024}, and reducing DC voltage controller asymmetry consistently improves overall system stability \cite{lin_filtered_2026}. Adaptive suppression of frequency coupling has been proposed to extend stability improvements across a wider operating range \cite{wang_adaptive_2025}.
Despite this body of work, two critical limitations persist: First, existing studies are largely confined to specific control structures, leaving the generality of the asymmetry-stability relationship unverified. Second, there has been no clear quantitative metric for asymmetricity that enables a rigorous analysis linking instability to asymmetry. This paper addresses this gap by establishing a direct, quantifiable link between the asymmetry of the converter-grid system and its instability. The presented work introduces the Asymmetry Quantification Index (AQI), derived from the sequence domain representation of the interconnected impedance, and demonstrates that this metric provides a generalized indicator of stability margins across both grid following and grid forming control structures. Analytical validation further substantiates the explicit relationship between system asymmetricity and instability. 

The availability of a quantified asymmetricity metric enables several novel analysis. By examining how the AQI varies with system stability margin, we directly observe how asymmetry evolves as stability increases or decreases. Parametric sweeps of asymmetric controller parameters under fixed grid conditions reveal a robust methodology for control parameter tuning based on asymmetricity. Furthermore, the dependence of system asymmetricity on the operating point provides critical insight into stability limitations: for a given set of control parameters, the converter's stable output range becomes evident through the lens of system asymmetry. Thus, analyzing the quantified asymmetricity together with the stability margin yields fundamentally new understanding of converter-grid instability mechanisms. 

In summary, the main contributions of this work are as follows.
\begin{enumerate}
    \item A systematic asymmetricity quantification of  a three-phase converter-grid system is proposed, enabling a direct examination of the relationship between asymmetricity and system stability margins.
    \item The proposed Asymmetricity Quantification Index (\textit{AQI}) is analyzed for both grid-forming and grid-following control structures, demonstrating its general applicability.
    \item A comprehensive analysis of the \textit{AQI} with respect to different control parameters and operating conditions provides in-depth insight into the explicit link between controller-induced asymmetry and system stability margins.
\end{enumerate}

The rest of the paper is structured as follows: Section \ref{Sec:System_Description_and_Impedance_Modelling} describes the system under consideration and presents impedance modeling for both Grid forming and Grid following converters. In Section \ref{Sec:Impedance-based_Stability_Analysis_and_Asymmetricity_Quantification}, an impedance-based stability analysis is introduced and the quantification of system asymmetricity via the \textit{AQI} is proposed. Section \ref{Sec:Case_Study}, analyzes the asymmetricity of the system in relation to the stability margin for two distinct control structures. Section \ref{Sec:Experiment_validation}, provides experimental validation of the analysis using both hardware in the loop and power hardware in the loop environments. Finally, conclusions are drawn in Section \ref{Sec:Conclusion}.

\section{System Description and Impedance Modeling}
\label{Sec:System_Description_and_Impedance_Modelling}
\begin{figure*}[tb]
    \centering
    \subfloat[]{\includegraphics[width=0.45\linewidth]{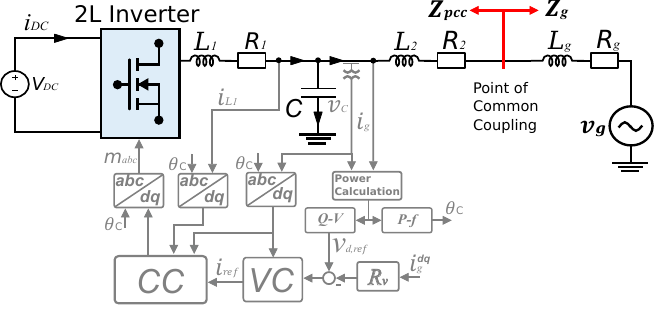}} \hspace{2mm}
    \subfloat[]{\includegraphics[width=0.47\linewidth]{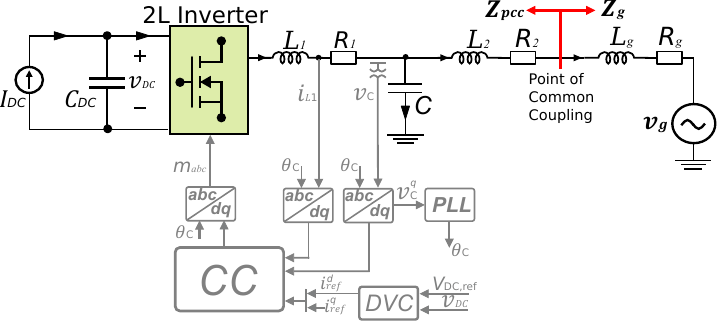}}
    \caption{(a) Grid Forming controlled converter-grid System. (b) Grid Following controlled converter-grid System. CC: PI-based Current Control with voltage feed-forward, VC: PI-based Voltage  control. , PLL: Conventional PI-based Phase Locked Loop, DVC: PI-based DC-link Voltage Control.}
    \label{fig:Sys_Description}
\end{figure*}

\begin{figure}[tb]
\centering
    \subfloat[]{\includegraphics[width=0.5\linewidth]{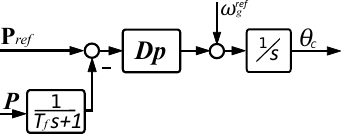}} \hspace{1mm}
    \subfloat[]{\includegraphics[width=0.45\linewidth]{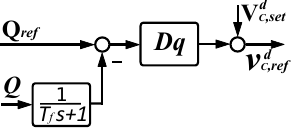}}
    \\
    \subfloat[]{\includegraphics[width=0.45\linewidth]{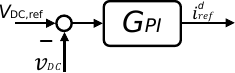}} \hspace{1mm}
    \subfloat[]{\includegraphics[width=0.45\linewidth]{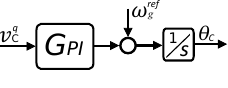}} 
    \\
    \subfloat[]{\scalebox{0.7}{\includegraphics[keepaspectratio]{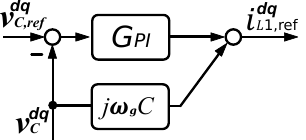}}} \hspace{1mm}
    \subfloat[]{\scalebox{0.7}{\includegraphics[keepaspectratio]{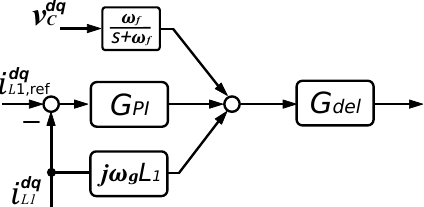}}}
\caption{Control Loops. (a) $P-f$ Droop. (b) $Q-V$ Droop. (c) DVC. (d) PLL. (e) Voltage Control (VC). (f) Current Control (CC).}
\label{fig:Control_loops}
\end{figure}
In this work, two types of converter control structures are considered. The $LCL$ filter is used on the output of the converter terminal. The power grid is considered as an inductive grid with an ideal voltage source. Detailed description of both types of converter controls is given below with their impedance modeling and validation.

\subsection{Grid Forming Control}

The schematic of a GFM controlled converter connected to an inductive grid is shown in Fig. \ref{fig:Sys_Description}(a) with its control structure. The primary purpose of the presented GFM control structure is to provide voltage and frequency support to the grid. The $P-f$ droop loop generates an angle for the controller frame, whereas the $Q-V$ droop provides the reference voltage for the $d-$ axis. When using a virtual resistor, the output of the virtual resistor loop is subtracted from the $Q-V$ droop output to generate a reference voltage for the voltage control loop. In the inner control, the cascaded voltage and current control loops are considered. The block diagram representations of the individual control loops of the GFM control are shown in Fig. \ref{fig:Control_loops}(a), Fig. \ref{fig:Control_loops}(b), Fig. \ref{fig:Control_loops}(e) and Fig. \ref{fig:Control_loops}(f).

The GFM controlled converter impedance equation is derived analytically and provided in (\ref{eq:GFM_zinv}). The derivation steps are provided in the appendix \ref{Append:GFM_Impedance_Derivation}. The analytical model is verified with single-tone frequency sweep in the Control-hardware-in the-loop (CHIL) environment. More information about the CHIL setup is provided in Section \ref{Sec:Case_Study}. As shown in Fig. \ref{fig:GFM_Impedance_verification_CHIL}, the analytical model matches very well with the measured impedance.
\begin{figure*}[!t]
\begin{equation}
\small
\begin{aligned}[b]
     Z_{inv}= \bigg(I-\bigg( G_{del} \bigg(G_{VC}^iG_A-G_{IC}^iG_B-&G_{CC}G_{VC}K_{qv}G_{QC}^i \bigg)- G_{MC}^i\frac{V_{DC}}{2}\bigg)  \bigg)^{-1} \\ & \left( I- \left( G_{del} \left(G_{VC}^vG_A-G_{IC}^vG_B-G_{CC}G_{VC}K_{qv}G_{QC}^v \right)-G_{MC}^v\frac{V_{DC}}{2}\right)\right)
    \end{aligned}
\label{eq:GFM_zinv}
\end{equation}
\end{figure*}
\begin{figure}[t]
	\centering
	\subfloat
	{
		\scalebox{0.5}{\input{updated_images/GFM_sweep_Z_dd}}
	}
	\subfloat
	{
		\scalebox{0.5}{\input{updated_images/GFM_sweep_Z_dq}}
	}
	\\
	\subfloat
	{
		\scalebox{0.5}{\input{updated_images/GFM_sweep_Z_qd}}
	}
	\subfloat 
	{
		\scalebox{0.5}{\input{updated_images/GFM_sweep_Z_qq}}
	}
	\caption{GFM controlled converter impedance verification through single-tone frequency sweep.}
	\label{fig:GFM_Impedance_verification_CHIL}
\end{figure}
\begin{figure}[t]
	\centering
	\subfloat
	{
		\scalebox{0.5}{\input{updated_images/GFL_sweep_Z_dd}}
	}
	\subfloat
	{
		\scalebox{0.5}{\input{updated_images/GFL_sweep_Z_dq}}
	}
	\\
	\subfloat
	{
		\scalebox{0.5}{\input{updated_images/GFL_sweep_Z_qd}}
	}
	\subfloat 
	{
		\scalebox{0.5}{\input{updated_images/GFL_sweep_Z_qq}}
	}
	\caption{GFL controlled converter impedance verification through single-tone frequency sweep.}
	\label{fig:GFL_Impedance_verification_CHIL}
\end{figure}
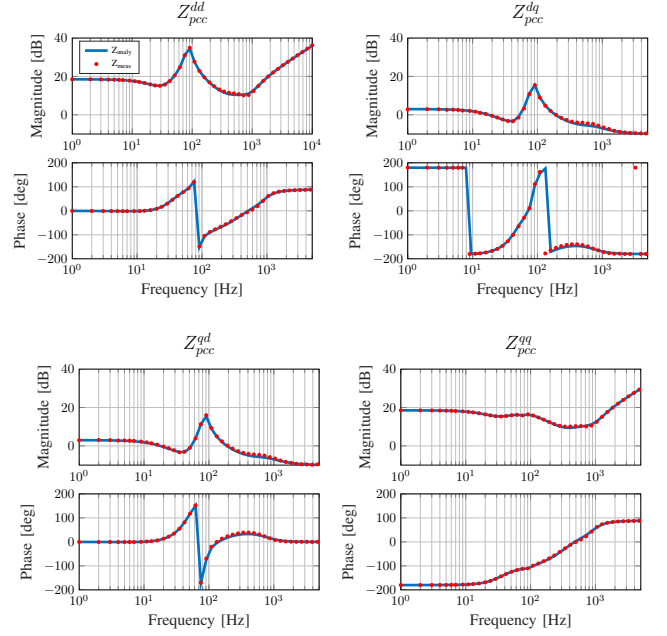

\subsection{Grid Following Control}

The schematic of a GFL controlled converter connected to an inductive grid is shown in Fig. \ref{fig:Sys_Description}(b) with its control structure. The primary purpose of the presented GFL control structure is to dispatch power in synchronization with the grid. The DC-link voltage is regulated by a PI-based controller. A classical phase-locked loop is used that aligns the capacitor voltage vector with the $d-$ axis. PI-based current control with decoupling term and voltage feedforward is implemented in the current controller, similar to the GFM control. The block diagrams of DVC and PLL are shown in Fig. \ref{fig:Control_loops}(c) and Fig. \ref{fig:Control_loops}(d), respectively. The current control loop block diagram is the same as shown in Fig. \ref{fig:Control_loops}(f).

The analytical impedance modeling of the GFL control is provided in \cite{yang_stability_2022} in the $dq$ frame. In this work, validation of the analytical model is provided in the CHIL environment. As shown in Fig. \ref{fig:GFL_Impedance_verification_CHIL}, the analytical model derived in \cite{yang_stability_2022} matches very well with the measurement results.

\section{Impedance-based Stability Analysis and Asymmetricity Quantification}
\label{Sec:Impedance-based_Stability_Analysis_and_Asymmetricity_Quantification}
\begin{figure}[tb]
    \centering
    \includegraphics[width=\linewidth]{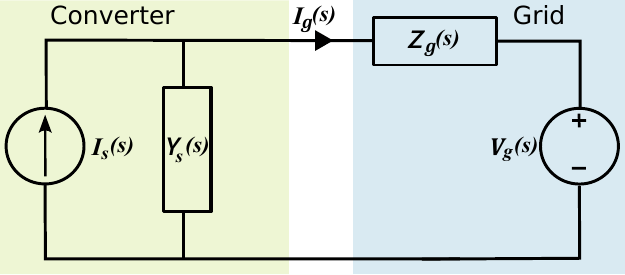 }
    \caption{Equivalent Circuit of a converter-grid System}
    \label{fig:Sys_equi_description}
\end{figure}

In a simplified converter-grid system, the system can be subdivided into two parts at the point of common coupling. The converter-side circuit is referred to as a source, and the grid-side circuit is referred to as a load. In the case of GFL controlled converter, since converter control is providing current to the power grid without any voltage/frequency support, converter-side can be modeled as a current source or a Norton equivalent. In contrast, in the case of GFM controlled converter, the converter is providing voltage and frequency support, therefore the converter-side can be modeled as a voltage source or Thevenin equivalent. The grid is assumed ideal with inductive impedance; therefore, the grid is modeled as a Thevenin equivalent. For stability analysis, the assumption is that the converter is stable when the grid impedance is zero and the grid is stable without the converter \cite{Sun_2011}. It is important to recognize that the Norton and Thevenin equivalence representations constitute mutually complementary characteristics of the same linear electrical circuit. The selection of Norton/Thevenin equivalence is done from the perception about the circuit's behavior only. The simplified converter-grid circuit is shown in Fig. \ref{fig:Sys_equi_description}. The grid current is given by (\ref{eq:control_equation_for_system}).

\begin{equation}
    I_g(s)=(1+Y_s(s)Z_g(s))^{-1}\left(I_s(s)-Y_s(s)V_g(s)\right)
	\label{eq:control_equation_for_system}
\end{equation}
Applying linear control theory to (\ref{eq:control_equation_for_system}), it is determined that the system is stable when the eigen-loci of $Y_s(s)Z_g(s)$ (also called a return-ratio, $L$) do not encircle the point $(-1,0)$ in the complex plane. The encirclement of eigen-loci is analyzed because the three-phase converter-grid system is a Multi-Input-Multi-Output system with a presence of frequency coupling due to asymmetrical control loop such as Droop control, DVC and PLL. If a balanced three-phase system is assumed (the zero sequence component is not excited), then the return-ratio can be described in a $2 \times 2$ matrix in the $dq$ frame as given by (\ref{eq:return_ratio_matrix}).
\begin{equation*}
	L(s)=Y_s(s)Z_g(s)
\end{equation*}
\begin{equation}
\small
	\begin{bmatrix}
		L_{dd}(s) & L_{dq}(s) \\ L_{qd}(s) & L_{qq}(s)
	\end{bmatrix}
	=
	\begin{bmatrix}
		Y_{s}^{dd}(s) & Y_s^{dq}(s) \\ Y_s^{qd}(s) & Y_s^{qq}(s)
	\end{bmatrix}
	\begin{bmatrix}
		Z_g^{dd}(s) & Z_g^{dq}(s) \\ Z_g^{qd}(s) & Z_g^{qq}(s)
	\end{bmatrix}
	\label{eq:return_ratio_matrix}
\end{equation}

\subsection{Impedance Symmetry}

The reason for analyzing the symmetry of the three-phase impedance is that the symmetrical system inherently provides a better control over the stability margin. However, the impedance of the converter introduces asymmetricity due to controllers such as DC-link voltage control, a classical phase-locked loop, and power synchronization control. Therefore, the impact of asymmetricity introduced by these control loops on the stability margin of the system is crucial to analyze the underlying instability mechanism. In order to analyze the asymmetricity of the impedance, it has to be quantified first. 

\begin{figure}[b]
    \centering
    \includegraphics[width=0.75\linewidth]{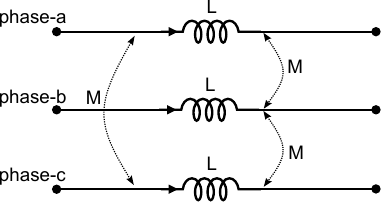}
    \caption{Symmetrical Three-phase Inductor}
    \label{fig:Symm_L}
\end{figure}

To understand the asymmetricity of the three-phase impedance, consider a simple example of an arbitrarily symmetrical three-phase inductor with equal mutual coupling as shown in Fig. \ref{fig:Symm_L}. In matrix form, the impedance is given by (\ref{eq:sym_L_eq}). Now, a symmetrical component transformation is applied to $Z_L^{abc}$ to obtain its sequence domain equivalence \cite{paap_symmetrical_2000}. The transformed impedance matrix in the sequence domain is given by (\ref{eq:sym_L_0pn}).

\begin{equation}
    Z_L^{abc}(s)=s\begin{bmatrix}
        L & M & M \\ M & L & M \\ M & M & L
    \end{bmatrix}
    \label{eq:sym_L_eq}
\end{equation}
\noindent Here, $'s'$ is a laplace operator.
\begin{equation}
    Z_L^{0pn}(s)= S^{-1}Z_L^{abc}(s)S=s\begin{bmatrix}
        L+2M & 0 & 0 \\ 0 & L-M & 0 \\ 0 & 0 & L-M
    \end{bmatrix}
    \label{eq:sym_L_0pn}
\end{equation}

where
\begin{equation}
    S = \frac{1}{\sqrt{3}}\begin{bmatrix}
        1 & 1 & 1 \\ 1 & a^{-1} & a^{-2} \\ 1 & a^{-2} & a^{-1}
    \end{bmatrix}
    ,
    a=e^{j\frac{2\pi}{3}}
\end{equation}

 For this work, the assumption is made that the considered three-phase system is balanced. Therefore, the zero sequence component is ignored. The $2\times 2$ sequence domain impedance is given by (\ref{eq:sym_L_pn}). An important aspect to observe here is that the off-diagonal elements are zero. If the three-phase inductors, shown in Fig. \ref{fig:Symm_L} are not symmetrical, then the off-diagonal in (\ref{eq:sym_L_pn}) will be non-zero. 
\begin{equation}
    Z_L^{pn}=s\begin{bmatrix}
        L-M & \textcolor{red}{0} \\ \textcolor{red}{0} & L-M
    \end{bmatrix}
    \label{eq:sym_L_pn}
\end{equation}

In the real closed-loop converter-grid system, the return-ratio matrix is observed for the stability analysis as discussed in the previous sub-section. Therefore, from a stability point of view, the system asymmetricity is apparent when the return-ratio matrix is analyzed in the sequence domain. That means that if the system is a three-phase symmetrical, the off-diagonal elements in the sequence domain return-ratio matrix will be zero. In other words, if there is a presence of off-diagonal elements in the sequence domain, then the system is three-phase asymmetrical. The return-ratio matrix, $L^{pn}(s)$, in the sequence domain is obtained by performing a matrix operation as shown in (\ref{eq:Lpn}) \cite{rygg_modified_2016}.
\begin{equation}
    L^{pn}(s)=A_zL^{dq}(s)A_z^{-1}=\begin{bmatrix}
        L^{pp}(s) & L^{pn}(s) \\ L^{np}(s) & L^{nn}(s)
    \end{bmatrix}
    \label{eq:Lpn}
\end{equation}
where
\begin{equation*}
    A_z=\frac{1}{\sqrt 2}\begin{bmatrix}
        1 & j \\ 1 & -j
    \end{bmatrix}
\end{equation*}

The eigenloci of the sequence domain return-ratio matrix are given by (\ref{eq:alpha12}). In the eigenloci equation, it is indicated that the last term $L^{pn}(s)L^{np}(s)$ is non-zero only when the system is three-phase asymmetrical. In other words, in the asymmetrical impedance case, eigenloci of the sequence domain return-ratio will differ from its diagonal element, whereas in the symmetrical case, eigenloci will be the same as diagonal elements of the sequence domain impedance.

\begin{equation}
	\begin{aligned}[b]
		\lambda^{p,n}(s)&=\frac{L^{pp}(s)+L^{nn}(s)}{2} \\ & \pm \sqrt{0.25(L^{pp}(s)-L^{nn}(s))^2+\underbrace{L^{pn}(s)L^{np}(s)}_{\text{non-zero when asymmetrical}}}
	\end{aligned}
	\label{eq:alpha12}
\end{equation}

Based on this observation, \textit{Asymmetry Quantification Index (AQI)} is defined as the absolute value of the multiplication of off-diagonal elements.
\begin{equation}
    AQI=\left| L^{pn}(s)\times L^{np}(s)\right|
\end{equation}
When the system is three-phase symmetrical, then $AQI$ will be zero. Any non-zero value of $AQI$ at a given frequency indicates the measure of asymmetricity. The significance of $AQI$ can be obtained when it is observed in relation to the closed-loop stability margin of the converter-grid system. To explore this relation, two case studies, corresponding to two converter controls structures, are presented in the next section.

\section{Case Studies}
\label{Sec:Case_Study}
\begin{figure}[t]
    \centering
    \includegraphics[width=\linewidth]{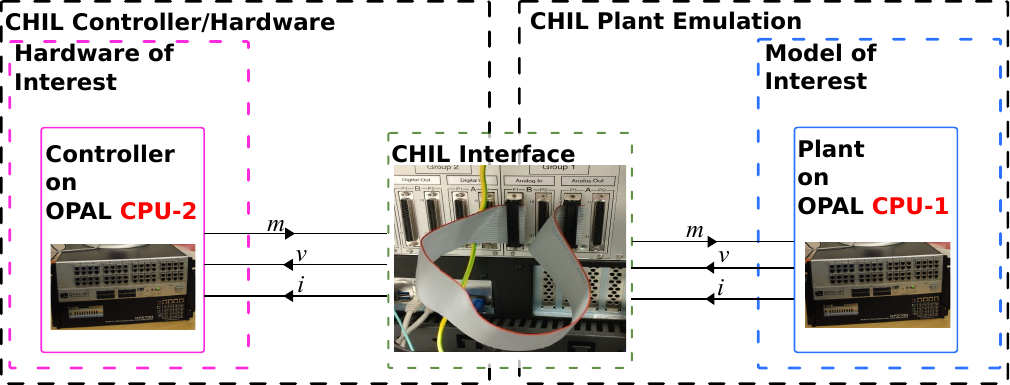}
    \caption{Control-Hardware-in-the-Loop Setup. The plant model is implemented on CPU-1, and the controller model is implemented on CPU-2.}
    \label{fig:CHIL_setup}
\end{figure}

In this section, $AQI$ of both GFM and GFL controlled converter-grid systems, shown in Fig. \ref{fig:Sys_Description}, are analyzed in relation to the system stability margin. The real-time simulator from OPAL-RT is used for the implementation of the circuit model. One of the CPU cores is used as a controller whereas the other CPU is used to emulate the plant model as shown in Fig. \ref{fig:CHIL_setup}. There is no direct communication between these two CPUs internally. The measured signals are scaled down before being taken out through analog-out ports. These signals are looped back-in as analog-in, rescaled, and conditioned before being used in the controller. Therefore, the setup can be considered as control-hardware-in-the loop (CHIL) setup \cite{noauthor_ieee_2025}. At first, instability is induced in the converter-grid system, and stability is analyzed utilizing corrected bode plots for the MIMO system \cite{shah_single_2025}. $AQI$ of both stable and unstable systems will be analyzed to observe the asymmetricity of the system. Moreover, through the parametric sweep, the stability margin of the system is analyzed against the system $AQI$, which will clearly indicate the direction of the phase margin when the asymmetry of the increases/decreases.

\subsection{Case Study 1: GFM controlled converter connected to power grid}

\begin{figure}[tb]
	\centering
    \subfloat[]{
		\scalebox{0.5}{\input{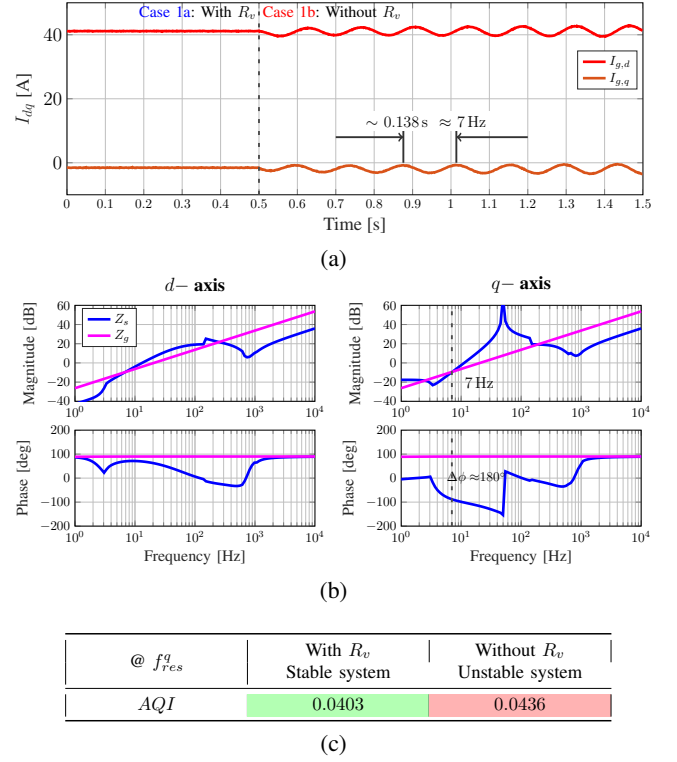}}
        }
        \\ [-0.1em]
    \subfloat[]{
    \scalebox{0.5}{\input{updated_images/Case_study1_dd}}
    \scalebox{0.5}{\input{updated_images/Case_study1_qq}}
	}
    \\
        \subfloat[]{
        \scalebox{0.7}{
            \begin{tabular}{|c|c|c|}
            \toprule
            \multirow{2}{3cm}{\centering @ $f_{res}^q$}& \multirow{2}{3cm}{\centering With $R_{v}$ \\ Stable system} & \multirow{2}{3cm}{\centering Without $R_{v}$ \\ Unstable system} \\
            \\
            \midrule
            $AQI$   &   \cellcolor{green!30}$0.0403$  &  \cellcolor{red!30}$0.0436$  \\ \bottomrule
            \end{tabular}
            }}
	\caption{GFM validation results. a)  Time-domain current in the $dq$ frame. At $t=$\SI{0.5}{\second}, virtual resistor is disabled. b) Single coordinate (corrected) bode plots without the virtual resistor, confirming the resonance frequency of \SI{7}{\hertz} in $q$ axis. c)$AQI$ at $f_{res}^q=$\SI{7}{\hertz}, indicating increase in asymmetricity without the virtual resistor.}
	\label{fig:CaseStudy1_results}
\end{figure}

 The system and control parameters of the GFM controlled converter connected to the power grid are mentioned in the appendix \ref{appx:System_and_Control_Parameters}. In Fig. \ref{fig:CaseStudy1_results}(a), the time domain waveform of the grid current is shown in the $dq$ frame. With $5\%$ frequency droop, at $t=\SI{0.5}{\second}$, when the virtual resistor is removed, an instability is observed with  resonances growing in amplitude. The resonance frequency is \SI{7}{\hertz} in the $dq$ frame. In \ref{fig:CaseStudy1_results}(b), the corrected Bode plots \cite{shah_single_2025} are shown without $R_v$. In the $q$ axis, it is clearly observed that the converter and the grid impedance are equal in magnitude at \SI{7}{\hertz} with a phase difference of \SI{180}{\degree}. Therefore, the corrected bode plots confirm the marginal instability. For both stable and unstable cases, the $AQI$s are calculated at the corresponding resonance frequencies of the $q-$ axis and shown in the table provided in Fig. \ref{fig:CaseStudy1_results}(c). It is clearly evident that $AQI$ of the converter-grid system increases when the virtual resistor loop is removed.

Now, it is more interesting to observe the trend in the phase margin with respect to the variation in $AQI$. To explore this, at first, frequency droop is varied. With variation in frequency droop, the potential resonance frequency of the $q-$ axis also changes, consequently, the phase margin changes in the $q$ axis. Please note that only the resonance frequency and the phase margin of the $q$ axis is observed here because the system is more prone to fulfill the instability criteria in the $q$ axis. In Fig. \ref{fig:Symmetricity_Stability_Margin_DP}, it can be seen that, when the frequency droop increases from $2\%$ to $5\%$, the resonance frequency varies slightly, but the phase margin reduces almost linearly, which eventually enters the unstable region just before $5\%$ of the droop. The color indicates that when the phase margin is reduced, the system $AQI$ also increases.

Similarly in Fig. \ref{fig:Symmetricity_Stability_Margin_Rv}, $q-$ axis phase margin and $AQI$ are analyzed with variation in virtual resistance. It was observed that without virtual resistance the system is unstable. The corresponding point in Fig. \ref{fig:Symmetricity_Stability_Margin_Rv} is shown with a red point. When $R_v$ gradually increases to \SI{0.5}{\ohm}, the phase margin of the system increases significantly. With $R_v=$\SI{0.5}{\ohm}, the stability phase margin of \SI{12}{\degree} is obtained. It is also observed that with an increase in the phase margin, the system $AQI$ also decreases, indicating a more symmetrical system. Therefore, with the GFM controlled converter-grid system, it is evidently observed that phase margin and $AQI$ are linked. When the phase margin is less, the system is more asymmetrical compared to the system with higher phase margin.
\begin{figure}[tb]
    \centering
    \scalebox{0.57}{\input{updated_images/Dp_Change}}
    \caption{Analyzing $q-$ axis system asymmetricity and stability margin with variation in frequency droop.}
    \label{fig:Symmetricity_Stability_Margin_DP}
\end{figure}
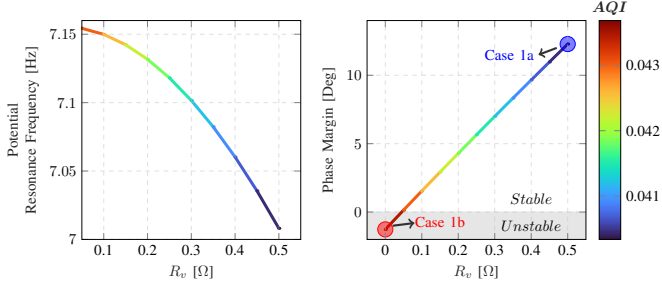
\begin{figure}[tb]
    \centering
    \scalebox{0.57}{\input{updated_images/Rv_Change}}
    \caption{Analyzing $q-$ axis system asymmetricity and stability margin with variation in virtual resistance.}
    \label{fig:Symmetricity_Stability_Margin_Rv}
\end{figure}

\subsection{Case Study 2: GFL controlled converter connected to power grid}

\begin{figure}[tb]
	\centering
    \subfloat[]{
		\scalebox{0.5}{\input{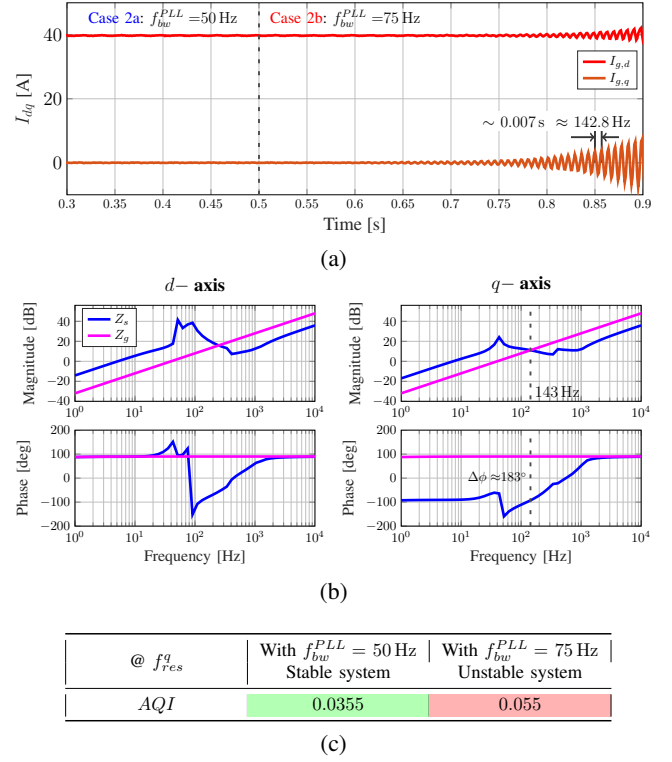}}
        }
        \\ [-0.1em]
    \subfloat[]{
    \scalebox{0.5}{\input{updated_images/Case_study2_dd}}
    \scalebox{0.5}{\input{updated_images/Case_study2_qq}}
	}
     \\
        \subfloat[]{
        \scalebox{0.7}{
            \begin{tabular}{|c|c|c|}
            \toprule
            \multirow{2}{3cm}{\centering @ $f_{res}^q$} & \multirow{2}{3cm}{\centering With $f_{bw}^{PLL}=$ \SI{50}{\hertz} \\ Stable system} & \multirow{2}{3cm}{\centering With $f_{bw}^{PLL}=$ \SI{75}{\hertz} \\ Unstable system} \\
            \\
            \midrule
            $AQI$   &   \cellcolor{green!30}$0.0355$  &  \cellcolor{red!30}$0.055$  \\ \bottomrule
            \end{tabular}
            }}
	\caption{GFL validation results. a)  Time-domain current in $dq$ frame. At $t=$\SI{0.5}{\second}, the PLL bandwidth is increased to \SI{75}{\hertz}. b) Single coordinate (corrected) bode plots with PLL bandwidth of \SI{75}{\hertz}, confirming \SI{143}{\hertz} resonance frequency. c) $AQI$ at $f_{res}^q$, indicating increase in asymmetricity with higher PLL bandwidth.}
	\label{fig:CaseStudy2_results}
\end{figure}

 The system and control parameters of the GFL controlled converter connected to the power grid are mentioned in the appendix \ref{appx:System_and_Control_Parameters}. In Fig. \ref{fig:CaseStudy2_results}(a), the time domain waveform of the grid current is shown in the $dq$ frame. At $t=\SI{0.5}{\second}$, when the PLL bandwidth ($f_{bw}^{PLL}$) is increased to \SI{75}{\hertz}, an instability is observed with  resonances growing in amplitude. The resonance frequency is $\approx$\SI{143}{\hertz} in the $dq$ frame. In \ref{fig:CaseStudy2_results}(b), the corrected Bode plots are shown with $f_{bw}^{PLL}=$ \SI{75}{\hertz}. In the $q$ axis, it is clearly observed that the converter and the grid impedance are equal in magnitude at \SI{143}{\hertz} with a phase difference of \SI{183}{\degree}. Therefore, the corrected bode plots confirm the instability. For both stable and unstable cases, the $AQI$s are calculated at the corresponding resonance frequencies of the $q-$ axis and shown in the table provided in Fig. \ref{fig:CaseStudy2_results}(c). It is clearly evident that $AQI$ of the converter-grid system increases when the PLL bandwidth is increased. 
 
In Fig. \ref{fig:Symmetricity_Stability_Margin_PLLbw_CaseStudy2}, the resonance frequency of the $q-$ axis and the phase margin are analyzed relative to $AQI$ when the PLL bandwidth is increased. It is clearly evident that when the PLL bandwidth is increased, the phase margin is decreased, which is in-line with current research work. However, more significantly, by observing the color of a line, it is evident that the $AQI$ is also increasing when phase margin is moving toward the unstable region. The stable and unstable cases shown in Fig. \ref{fig:CaseStudy2_results}(a) are highlighted with blue and red dots in Fig. \ref{fig:Symmetricity_Stability_Margin_PLLbw_CaseStudy2}, respectively.

The impedance of the converter also varies with the operating points. Therefore, the reference current $I_d$ is changed with a PLL bandwidth of \SI{75}{\hertz}. The corresponding resonance frequency and the phase margin of $q-$ axis is shown in Fig. \ref{fig:Symmetricity_Stability_Margin_Id_CaseStudy2}. It is observed that the converter-grid system can achieve stability, having a small phase margin at lower current with a PLL bandwidth of \SI{75}{\hertz}; however, the phase margin goes into the unstable region as the current increases. Moreover, it is also observed that $AQI$ increases when the phase margin moves from stable region to unstable region.

\begin{figure}[tb]
    \centering
    \scalebox{0.57}{\input{updated_images/PLLbw_Change}}
    \caption{Analyzing symmetry and stability margin with variation in PLL Bandwidth.}
    \label{fig:Symmetricity_Stability_Margin_PLLbw_CaseStudy2}
\end{figure}
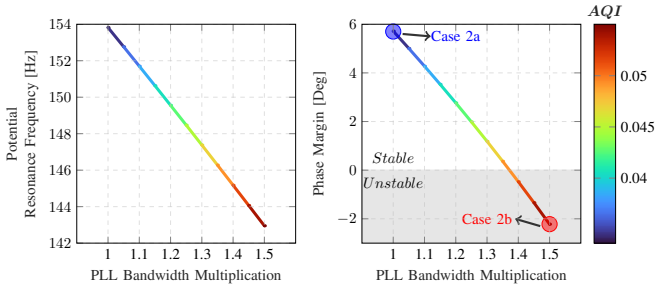
\begin{figure}[tb]
    \centering
    \scalebox{0.57}{\input{updated_images/Id_Change}}
    \caption{Analyzing symmetry and stability margin with variation in $I_d$ current.}
    \label{fig:Symmetricity_Stability_Margin_Id_CaseStudy2}
\end{figure}
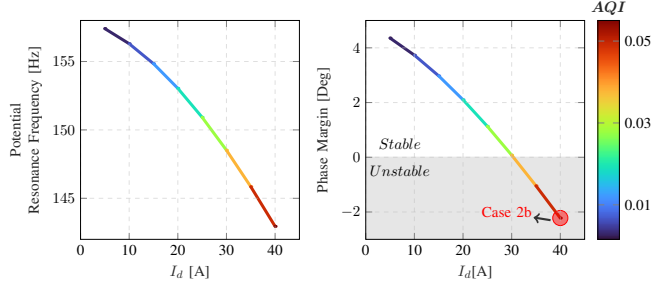

Analysis of both case studies has clearly shown and validated the relationship between the phase margin and the asymmetricity of the system. When the phase margin decreases, the system asymmetricity increases and eventually goes into the unstable range. In the next section, the analysis is expanded with experimental results and analysis.

\section{Experiment validation}
\label{Sec:Experiment_validation}

\begin{figure}[t]
	\begin{minipage}{\linewidth}
    \centering
		\scalebox{0.9}{
		    \includegraphics[width=1\linewidth]{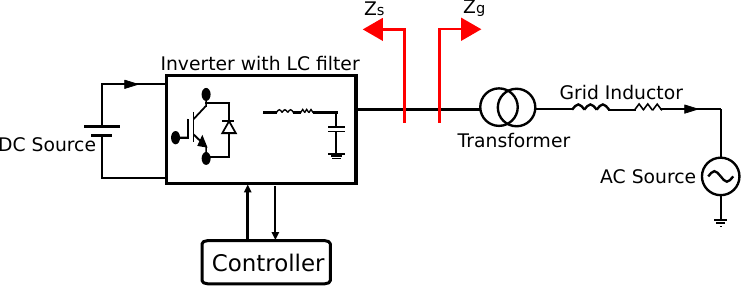} }
	\end{minipage}
	\begin{minipage}{\linewidth}
    \centering
	\scalebox{0.4}{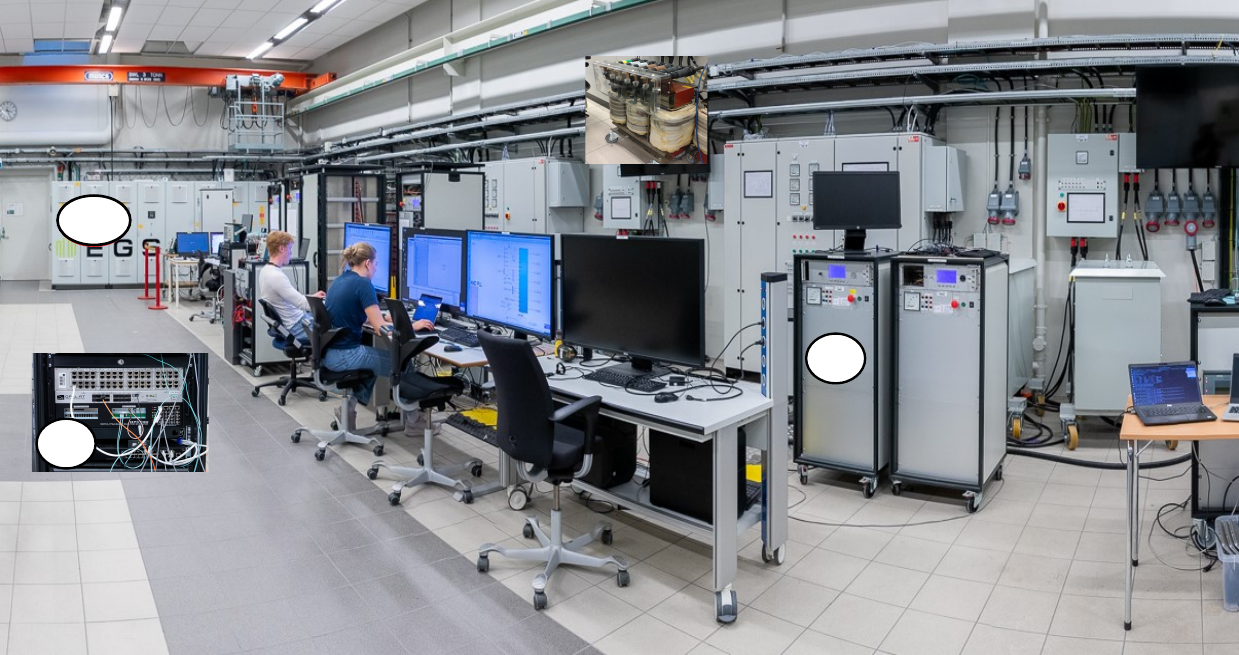}
	\end{minipage}
	\vfill
	\vspace{2mm}
	\begin{minipage}{\linewidth}
		\begin{tabular}{ll|ll}
			1. & DC source 					& 4. & Grid inductor\\
			2. & Converter with LC filter 	& 5. & Controller in OPAL-RT\\
			3. & Transformer				& 6. & AC Source
		\end{tabular}
		\caption{Experimental set-up at the Norwegian Smart Grid Laboratory. 2-level converter with $LC$ filter is connected to the power grid, which is emulated by a power amplifier.}
		\label{Fig:Experimental set-up at the National Smart Grid Laboratory}
	\end{minipage}
\end{figure}

The experiments are carried out in the Norwegian Smart Grid Laboratory. The schematic of the experiment set-up is shown in Fig. \ref{Fig:Experimental set-up at the National Smart Grid Laboratory}. A constant DC source is assumed, which is emulated by the power amplifier.  A two-level converter with an LC filter is used. The converter and power amplifiers are controlled by OPAL-RT OP5700. The Gate signals are generated by the field programmable gate array of the Zynq 7030 System on Chip, which is included in the local control board mounted on the converter. The local control board also performs signal conditioning and A/D measurements, which are transmitted to the OPAL-RT platform via fiber-optic cables. For the experiments presented, the control loops are implemented in OPAL-RT with a sampling time of \SI{100}{\micro\second}. 

\begin{figure}[t]
	\centering
	\subfloat
	{
		\scalebox{0.5}{\input{updated_images/Bode_freq_sweep_Z_dd}}
	}
	\subfloat
	{
		\scalebox{0.5}{\input{updated_images/Bode_freq_sweep_Z_dq}}
	}
	\\
	\subfloat
	{
		\scalebox{0.5}{\input{updated_images/Bode_freq_sweep_Z_qd}}
	}
	\subfloat 
	{
		\scalebox{0.5}{\input{updated_images/Bode_freq_sweep_Z_qq}}
	}
	\caption{Converter impedance verification through single-tone frequency sweep.}
	\label{fig:GFL_Impedance_verification_Lab}
\end{figure}
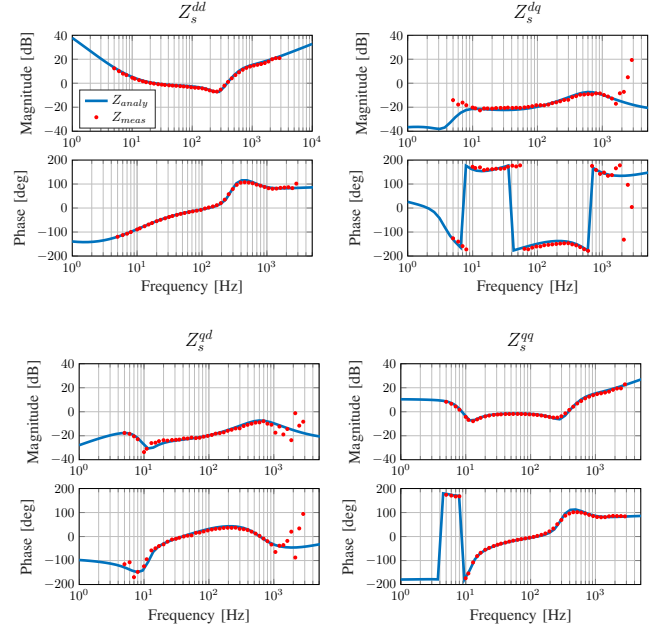

For the converter control, current control with a classical PLL is used. Active damping based on voltage measurement \cite{suul_synchronous_2011} and voltage feedforward are also utilized in current control. The analytical modeling of this controller is verified with the impedance measurement through a single-tone frequency sweep. The results, shown in Fig. \ref{fig:GFL_Impedance_verification_Lab}, show that the analytical model aligns very well with the measurement data.


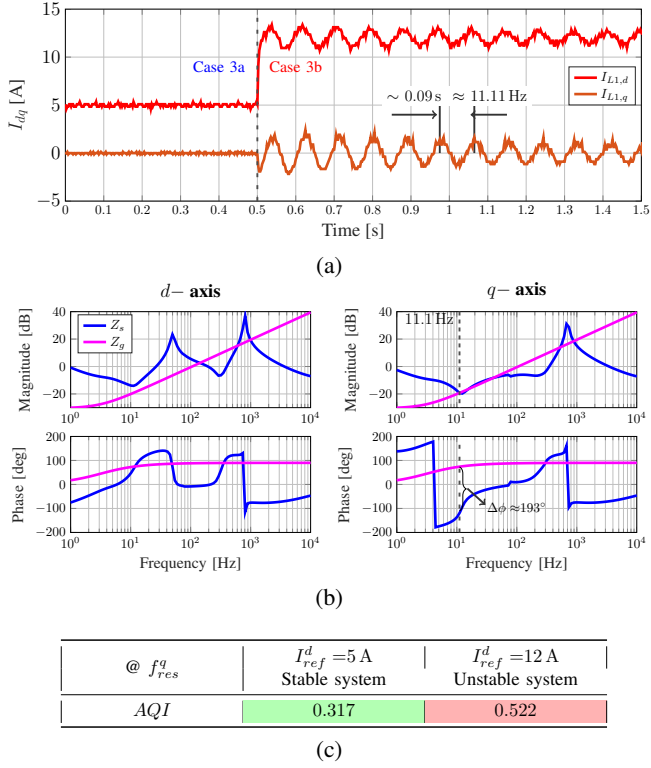
\begin{figure}[t!]
	\centering
    \subfloat[]{
		\scalebox{0.5}{\input{updated_images/test1_IL1dq}}
        }
        \\ [-0.1em]
    \subfloat[]{
    \scalebox{0.5}{\input{updated_images/test1_Zdd}}
    \scalebox{0.5}{\input{updated_images/test1_Zqq}}
	}
    \\
        \subfloat[]{
        \scalebox{0.7}{
            \begin{tabular}{|c|c|c|}
            \toprule
            \multirow{2}{3cm}{\centering @ $f_{res}^q$}& \multirow{2}{3cm}{\centering $I_{ref}^d=$\SI{5}{\ampere} \\ Stable system} & \multirow{2}{3cm}{\centering $I_{ref}^d=$\SI{12}{\ampere} \\ Unstable system} \\
            \\
            \midrule
            $AQI$   &   \cellcolor{green!30}$0.317$  &  \cellcolor{red!30}$0.522$  \\ \bottomrule
            \end{tabular}
            }}
	\caption{Experiment validation results. a) Time-domain current in $dq$ frame. At $t=$\SI{0.5}{\second}, the $d-$ axis current reference is changed from \SI{5}{\ampere} to \SI{12}{\ampere}. b) Single coordinate (corrected) bode plots with $I_{ref}^d=$\SI{12}{\ampere} confirms the resonance frequency of \SI{11.1}{\hertz}. c)$AQI$ at $f_{res}^q$, indicating increase in asymmetricity at higher current.}
	\label{fig:Case1_results}
\end{figure}

The system and control parameters for the experimental case studies are provided in the appendix \ref{appx:System_and_Control_Parameters}. Under these parameters, when the current reference ($I_{ref}^d$) is increased from \SI{5}{\ampere} to \SI{12}{\ampere}, resonances are observed as shown in Fig. \ref{fig:Case1_results}(a). The resonance frequency is \SI{11.1}{\hertz} in the $dq$ frame. The corrected bode plots in Fig. \ref{fig:Case1_results}(b) also show the potential resonance at \SI{11.1}{\hertz}. Now, when $AQI$ is also observed for these two operating points as shown in Fig. \ref{fig:Case1_results}(c), it is evident that the system has become more asymmetrical with \SI{12}{\ampere} of $d-$ axis current. 

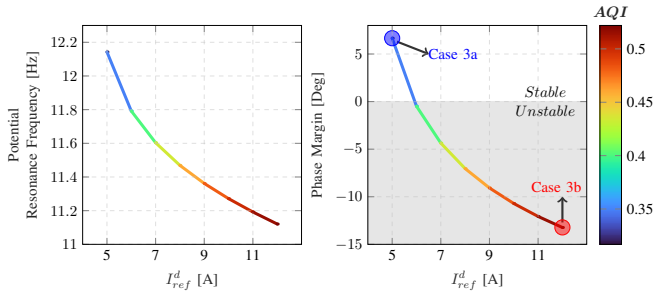
\begin{figure}[t]
    \centering
    \scalebox{0.57}{\input{updated_images/test1_Id_Change}}
    \caption{Analyzing $q-$ axis system asymmetry and stability margin with variation in frequency droop}
    \label{fig:Symmetricity_Stability_Margin_Id}
\end{figure}

Now, to validate the connection between asymmetricity and the system stability, the resonance frequency and phase margin of the $q$ axis are plotted with an increase in the reference current of the $d$ axis in conjunction with $AQI$ in Fig. \ref{fig:Symmetricity_Stability_Margin_Id}. The plots clearly show that as the current of the $d-$ axis increases, the phase margin of the system gradually decreases and moves into the unstable region. Moreover, by observing the color of the line, it is also evident that the system asymmetricity also increases gradually with a decrease in the stability margin. Therefore, the experimental case also validates the previously presented analysis.

\section{Conclusion}
\label{Sec:Conclusion}
In this work, a systematic framework is developed to relate asymmetricity in the converter-grid impedance to the stability margin of the overall system, building on the general premise that higher degrees of symmetry are typically associated with more stable dynamical behavior. Converter control inherently introduces asymmetry through elements such as the DC-link voltage controller, the classical phase-locked loop, and power synchronization control. The resulting system-level asymmetry is quantified in the frequency domain through an Asymmetry Quantification Index ($AQI$). The $AQI$ is analyzed for both grid-following and grid-forming controlled converter-grid system in relation to variations in system stability margins. The results consistently show that as the phase margin decreases, the $AQI$ increases, indicating a progressive decrease in the system symmetry as the system approaches instability. For grid-forming control, increasing the $P-f$ droop or reducing virtual resistance values leads to a degradation of the phase margin under certain grid conditions, accompanied by a corresponding increase in asymmetry. Similarly, in grid-following configurations, reducing the PLL bandwidth and increasing the $d-$-axis current result in lower phase margins and higher $AQI$ values. Experimental validation further confirms this trend, demonstrating that increased converter output current leads to reduced phase margin and increased system asymmetry. Overall, the analysis establishes a clear and consistent link between converter-grid impedance asymmetry and reduced stability margins. It further highlights that improved stability is not solely a consequence of symmetric control structures, but rather of the symmetry of the combined converter-control-grid system as a whole. When such symmetry cannot be fully achieved in practice, the $AQI$ provides a valuable guideline for tuning control parameters and defining safe operating regions to ensure stable converter-grid interaction.

\appendices
\section{System and Control Parameters}
\label{appx:System_and_Control_Parameters}
\begin{table}[h!]
	\centering      
	\caption{System and control parameters ($f_{g}=$\SI{50}{\hertz}) - CHIL Cases}
	\label{Tab: System and control parameters}
	\begin{tabular} {cc|cc}   \toprule
		\textbf{Parameter}		&\textbf{Value}			&\textbf{Parameter}		&\textbf{Value} 	\\    \midrule
		$V_{g,rms}^{ll}$	&\SI{400}{\volt}			& $L_{1},R_1$			&\SI{2}{\milli\henry},\SI{0.001}{\ohm}	  	\\
		$V_{dc}$			&\SI{700}{\volt}		    &$C,R_d$ 				&\SI{50}{\micro\farad},\SI{1}{\ohm}\\
		$P_{rated}$				&\SI{20}{\kilo\watt}			&$L_{2},R_2$			&\SI{1}{\milli\henry},\SI{0.001}{\ohm} \\
                            &                           & $L_{g},R_g$			&\SI{4}{\milli\henry},\SI{0.01}{\ohm}	\\ \midrule
        \multicolumn{2}{c|}{GFL controller} & \multicolumn{2}{c}{GFM controller}                 \\ \midrule   
		$K_p^{CC}/K_i^{CC}$			& $8.5 / 10000$		&$K_p^{CC}/K_i^{CC}$			& $8.5 / 10000$  \\ [0.3em]
		$K_{p}^{DVC}/K_{i}^{DVC}$		& $-0.38 / -68.63$		& $K_{p}^{VC}/K_{i}^{VC}$		& $0.07 / 18.27$\\ [0.3em]
        $K_{p}^{PLL}/K_i^{PLL}$		& $0.83/151$		&$Q-V$ droop		& $2 \%$ \\ [0.3em]
        $\omega_{VFF}$		&\SI{250}{\radian / \second}		&$P-f$ droop		& $5\%$ \\ [0.3em] 
       		&		&$R_v$ 		& \SI{0.5}{\Omega} \\
		\bottomrule
	\end{tabular}  
\end{table}

\begin{table}[h!]
	\centering      
	\caption{System and control parameters ($f_{g}=$\SI{50}{\hertz}) - Experiment Case}
	\label{Tab: System and control parameters exp}
	\begin{tabular} {cc|cc}   \toprule
		\textbf{Parameter}		&\textbf{Value}			&\textbf{Parameter}		&\textbf{Value} 	\\    \midrule
		$V_{g,ph}^{peak}$	&\SI{50}{\volt}			& $L_{1},R_1$			&\SI{0.75}{\milli\henry},\SI{0.065}{\ohm}	  	\\
		$V_{dc}$			&\SI{125}{\volt}		    &$C,R_d$ 				&\SI{50}{\micro\farad},\SI{0.3}{\ohm}\\
		$P_{rated}$				&\SI{2.5}{\kilo\watt}			& $L_{g},R_g$			&\SI{1.5}{\milli\henry},\SI{0.03}{\ohm}	\\ \midrule
		$K_p^{CC}/K_i^{CC}$			& $0.94/40$		&$K_{p}^{PLL}/K_i^{PLL}$		& $1.52/70$  \\ [0.3em]
		  \multirow{2}{2cm}{\centering $K_{AD}$  \\ (active damping)} & 0.07 & $\omega_{AD}$  & \SI{100}{\radian / \second} \\ [0.3em]
         		& & 		$\omega_{VFF}$		&\SI{40}{\radian / \second} \\
		\bottomrule
	\end{tabular}  
\end{table}

\section{Derivation of GFM impedance equation given by (\ref{eq:GFM_zinv})}
\label{Append:GFM_Impedance_Derivation}
The capital letters $I$, $V$, and $M$ denote steady state values of the current, voltage, and modulation index, respectively, while the small letters ($i,v,m$) denote current, voltage, and modulation index as variables. The entire frequency domain impedance modeling is performed in the synchronous reference frame $dq$. The subscript '$s$' denotes the system frame variables and the subscript $c$ denotes the control frame variables.

\subsection{Inner cascaded control loop}
\noindent The PI-based current control loop transfer function
\begin{equation*}
    G_{CC} = \begin{bmatrix}
        K_p^{CC}+\nicefrac{K_i^{CC}}{s} & 0 \\ 0 &  K_p^{CC}+\nicefrac{K_i^{CC}}{s}
    \end{bmatrix}
\end{equation*}
\noindent The PI-based voltage control loop transfer function
\begin{equation*}
    G_{VC} = \begin{bmatrix}
        K_p^{VC}+\nicefrac{K_i^{VC}}{s} & 0 \\ 0 &  K_p^{VC}+\nicefrac{K_i^{VC}}{s}
    \end{bmatrix}
\end{equation*}
\noindent The decoupling terms in both current and voltage control loops
\begin{equation*}
    G_{i,dec} = \begin{bmatrix}
        0 & \omega_g L_1 \\ -\omega_g L_1 & 0
    \end{bmatrix},
    G_{v,dec} = \begin{bmatrix}
        0 & \omega_g C \\ -\omega_g C & 0
    \end{bmatrix}
\end{equation*}
\noindent The voltage-feedforward with low-pass filter
\begin{equation*}
    G_{VFF}=\begin{bmatrix}
        \frac{\omega_{VFF}}{s+\omega_{VFF}} & 0 \\ 0 & \frac{\omega_{VFF}}{s+\omega_{VFF}}
    \end{bmatrix}
\end{equation*}
\noindent The filter inductor $L_1$ and virtual impedance
\begin{equation*}
    Z_{L1}=\begin{bmatrix}
        sL_1 & -\omega_g L_1 \\ \omega_g L_1 & sL_1
    \end{bmatrix},
    Z_{virt}=\begin{bmatrix}
        R_v & 0 \\ 0 & R_v
    \end{bmatrix}
\end{equation*}

\noindent Based on above defined transfer function, the converter output under the small-signal perturbation is given by,
\begin{equation}
\begin{aligned}
\begin{bmatrix} \hat{m}^{d,c} \\ \hat{m}^{q,c}\end{bmatrix} & \frac{V_{dc}}{2}  =  G_{del} \bigg(G_{CC} \bigg(G_{VC} \bigg(\begin{bmatrix} \hat{v}_{C,ref}^{d,c} \\ \hat{v}_{C,ref}^{q,c}\end{bmatrix}-Z_{virt} \begin{bmatrix} \hat{i}_{L2}^{d,c} \\ \hat{i}_{L2}^{q,c}\end{bmatrix}\bigg) \\ & -(G_{VC}+G_{v,dec} )\begin{bmatrix} \hat{v}_{C}^{d,c} \\ \hat{v}_{C}^{q,c}\end{bmatrix}\bigg)-(G_{CC}+G_{i,dec} )\begin{bmatrix} \hat{i}_{L1}^{d,c} \\ \hat{i}_{L1}^{q,c}\end{bmatrix} \\ & +G_{VFF} \begin{bmatrix} \hat{v}_{C}^{d,c} \\ \hat{v}_{C}^{q,c}\end{bmatrix}\bigg)
\end{aligned}
\label{eq:mvdc_lin}
\end{equation}
The ' $\hat{}$ ' above variables indicate the small-signal deviation.
\subsection{Active Power Loop}

\noindent Active power is given by,
\begin{equation*}
    P=1.5(i_{L2}^{d,s}v_C^{d,s}+i_{L2}^{q,s}v_C^{q,s})
\end{equation*}

\noindent Applying small-signal perturbation,
\begin{equation*}
    \hat{P}=1.5\bigg( \begin{bmatrix}
        I_{L2}^{d,s}  & I_{L2}^{q,s}
    \end{bmatrix} \begin{bmatrix} \hat{v}_{C}^{d,s} \\ \hat{v}_{C}^{q,s}\end{bmatrix}
    + \begin{bmatrix} V_{C}^{d,s} & V_{C}^{q,s}\end{bmatrix}\begin{bmatrix}
        \hat{i}_{L2}^{d,s}  \\  \hat{i}_{L2}^{q,s}
    \end{bmatrix} \bigg)
\end{equation*}
with
\begin{equation}
    \begin{bmatrix}
        \hat{i}_{L2}^{d,s}  \\  \hat{i}_{L2}^{q,s}
    \end{bmatrix} = \begin{bmatrix}
        \hat{i}_{L1}^{d,s}  \\  \hat{i}_{L1}^{q,s}
    \end{bmatrix} - \begin{bmatrix}
        sC & -\omega C \\ \omega C & sC
    \end{bmatrix} \begin{bmatrix} \hat{v}_{C}^{d,s} \\ \hat{v}_{C}^{q,s}\end{bmatrix}
    \label{eq:iL2_vc}
\end{equation}
Perturbed power can be written as,
\begin{equation}
\small
    \hat{P}=1.5 \bigg( \begin{bmatrix}
       G_{PS}^d &  G_{PS}^q
    \end{bmatrix}\begin{bmatrix} \hat{v}_{C}^{d,s} \\ \hat{v}_{C}^{q,s}\end{bmatrix}+\begin{bmatrix} V_{C}^{d,s} & V_{C}^{q,s}\end{bmatrix}\begin{bmatrix}
        \hat{i}_{L1}^{d,s}  \\  \hat{i}_{L1}^{q,s}
    \end{bmatrix} \bigg)
\end{equation}
where, $G_{PS}^d=I_{L2}^{d,s}-(V_{C}^{d,s}sC+V_{C}^{q,s}\omega C)$ and $G_{PS}^q=I_{L2}^{q,s}+(V_{C}^{d,s}sC-V_{C}^{q,s}\omega C)$.

\noindent The small-signal angle for the Clarke and Park transformation is given by ,
\begin{equation}
\begin{aligned}
    &\hat{\theta}=-\frac{1}{1+T_f s}K_{pf}\hat{P} \\ & \small
    =\underbrace{\frac{-1.5K_{pf}}{1+T_f s}}_{G_{PS}}\bigg(\begin{bmatrix}
       G_{PS}^d &  G_{PS}^q
    \end{bmatrix}\begin{bmatrix} \hat{v}_{C}^{d,s} \\ \hat{v}_{C}^{q,s}\end{bmatrix}+\begin{bmatrix} V_{C}^{d,s} & V_{C}^{q,s}\end{bmatrix}\begin{bmatrix}
        \hat{i}_{L1}^{d,s}  \\  \hat{i}_{L1}^{q,s}
    \end{bmatrix}\bigg)
\end{aligned}
\label{eq:theta}
\end{equation}
\noindent where, $T_f$ is filter time constant and $K_{pf}$ is a power-frequency droop co-efficient.

\noindent Under the small signal perturbation, the system frame to control frame voltage transformation is given by \cite{wen_analysis_2016},

\begin{equation*}
    \begin{bmatrix}
        \hat{v}_{C}^{d,c} \\ \hat{v}_{C}^{q,c}
    \end{bmatrix}
    \approx \begin{bmatrix}
        \hat{v}_{C}^{d,s}+V_C^{q,s}\hat{\theta} \\ \hat{v}_{C}^{q,s}-V_C^{d,s}\hat{\theta}
    \end{bmatrix}
\end{equation*}
\noindent Substituting $\hat{\theta}$ from (\ref{eq:theta}),
\begin{equation}
\small
\begin{aligned}[b]
    \begin{bmatrix}
        \hat{v}_{C}^{d,c} \\ \hat{v}_{C}^{q,c}
    \end{bmatrix}
     & \approx \underbrace{\begin{bmatrix}
	1+V_C^{q,s}G_{PS}G_{PS}^d & V_C^{q,s}G_{PS}G_{PS}^q \\ -V_C^{d,s}G_{PS}G_{PS}^d & 1-V_C^{d,s}G_{PS}G_{PS}^q
	\end{bmatrix}}_{G_{VC}^v} \begin{bmatrix}
        \hat{v}_{C}^{d,s} \\ \hat{v}_{C}^{q,s}
    \end{bmatrix} \\ & +\underbrace{\begin{bmatrix}
	V_C^{q,s}G_{PS}V_C^{d,s} & 	V_C^{q,s}G_{PS}V_C^{q,s} \\ -V_C^{d,s}G_{PS}V_C^{d,s} & -V_C^{d,s}G_{PS}V_C^{q,s}
	\end{bmatrix}}_{G_{VC}^i}\begin{bmatrix}
        \hat{i}_{L1}^{d,s}  \\  \hat{i}_{L1}^{q,s}
    \end{bmatrix}
    \end{aligned}
    \label{eq:vcs_to_vcc}
\end{equation}

\noindent Similarly, the current transformation is given by,

\begin{equation*}
    \begin{bmatrix}
        \hat{i}_{L1}^{d,c} \\ \hat{i}_{L1}^{q,c}
    \end{bmatrix}
    \approx \begin{bmatrix}
        \hat{i}_{L1}^{d,s}+I_{L1}^{q,s}\hat{\theta} \\ \hat{i}_{L1}^{q,s}-I_{L1}^{d,s}\hat{\theta}
    \end{bmatrix}
\end{equation*}
Substituting $\hat{\theta}$ from (\ref{eq:theta}),
\begin{equation}
\small
\begin{aligned}[b]
    \begin{bmatrix}
        \hat{i}_{L1}^{d,c} \\ \hat{i}_{L1}^{q,c}
    \end{bmatrix}
     & \approx \underbrace{\begin{bmatrix}
	I_{L1}^{q,s}G_{PS}G_{PS}^d & I_{L1}^{q,s}G_{PS}G_{PS}^q \\ -I_{L1}^{d,s}G_{PS}G_{PS}^d & -I_{L1}^{d,s}G_{PS}G_{PS}^q
	\end{bmatrix}}_{G_{IC}^v} \begin{bmatrix}
        \hat{v}_{C}^{d,s} \\ \hat{v}_{C}^{q,s}
    \end{bmatrix} \\ & +\underbrace{\begin{bmatrix}
	1+I_{L1}^{q,s}G_{PS}V_C^{d,s} & 	I_{L1}^{q,s}G_{PS}V_C^{q,s} \\ -I_{L1}^{d,s}G_{PS}V_C^{d,s} & 1-I_{L1}^{d,s}G_{PS}V_C^{q,s}
	\end{bmatrix}}_{G_{IC}^i}\begin{bmatrix}
        \hat{i}_{L1}^{d,s}  \\  \hat{i}_{L1}^{q,s}
    \end{bmatrix}
    \end{aligned}
    \label{eq:iL1s_to_iL1c}
\end{equation}

\noindent Also, the modulating signal transformation is given by,

\begin{equation*}
    \begin{bmatrix}
        \hat{m}^{d,c} \\ \hat{m}^{q,c}
    \end{bmatrix}
    \approx \begin{bmatrix}
        \hat{m}^{d,s}+M^{q,s}\hat{\theta} \\ \hat{m}^{q,s}-M^{d,s}\hat{\theta}
    \end{bmatrix}
\end{equation*}
Substituting $\hat{\theta}$ from (\ref{eq:theta}),
\begin{equation}
\small
\begin{aligned}[b]
    \begin{bmatrix}
        \hat{m}^{d,c} \\ \hat{m}^{q,c}
    \end{bmatrix}
     & \approx \begin{bmatrix}
        \hat{m}^{d,s} \\ \hat{m}^{q,s}
    \end{bmatrix} + \underbrace{\begin{bmatrix}
	M^{q,s}G_{PS}G_{PS}^d & M^{q,s}G_{PS}G_{PS}^q \\ -M^{d,s}G_{PS}G_{PS}^d & -M^{d,s}G_{PS}G_{PS}^q
	\end{bmatrix}}_{G_{MC}^v} \begin{bmatrix}
        \hat{v}_{C}^{d,s} \\ \hat{v}_{C}^{q,s}
    \end{bmatrix} \\ & +\underbrace{\begin{bmatrix}
	M^{q,s}G_{PS}V_C^{d,s} & 	M^{q,s}G_{PS}V_C^{q,s} \\ -M^{d,s}G_{PS}V_C^{d,s} & -M^{d,s}G_{PS}V_C^{q,s}
	\end{bmatrix}}_{G_{MC}^i}\begin{bmatrix}
        \hat{i}_{L1}^{d,s}  \\  \hat{i}_{L1}^{q,s}
    \end{bmatrix}
    \end{aligned}
    \label{eq:ms_to_mc}
\end{equation}

\subsection{Reactive Power Loop}

\noindent Reactive power is given by,

\begin{equation}
    Q=1.5(i_{L2}^{d,s}v_C^{q,s}-i_{L2}^{q,s}v_C^{d,s})
\end{equation}

\noindent Perturbed reactive power is given by,
\begin{equation}
\small
    \hat{Q}=1.5\bigg( \begin{bmatrix}
        -I_{L2}^{q,s}  & I_{L2}^{d,s}
    \end{bmatrix} \begin{bmatrix} \hat{v}_{C}^{d,s} \\ \hat{v}_{C}^{q,s}\end{bmatrix}
    + \begin{bmatrix} V_{C}^{q,s} & -V_{C}^{q,s}\end{bmatrix}\begin{bmatrix}
        \hat{i}_{L2}^{d,s}  \\  \hat{i}_{L2}^{q,s}
    \end{bmatrix} \bigg)
\end{equation}

\noindent Substituting $i_{L2}^{dq}$ from (\ref{eq:iL2_vc}),

\begin{equation}
\footnotesize
\begin{aligned}[b]
    \begin{bmatrix}
        \hat{Q} \\ 0
    \end{bmatrix}= &
    \underbrace{1.5\begin{bmatrix}
        -I_{L1}^{q,s}-V_C^{q,s}sC+V_C^{d,s}\omega C &  I_{L1}^{d,s}+V_C^{d,s}sC+V_C^{q,s}\omega C 
    \end{bmatrix}}_{G_{QC}^v} \\ & \begin{bmatrix} \hat{v}_{C}^{d,s} \\ \hat{v}_{C}^{q,s}\end{bmatrix}  +\underbrace{1.5\begin{bmatrix}
        V_C^{q,s} & -V_C^{d,s} 
    \end{bmatrix}}_{G_{QC}^i} \begin{bmatrix}
        \hat{i}_{L1}^{d,s}  \\  \hat{i}_{L1}^{q,s}
    \end{bmatrix}
\end{aligned}
\label{eq:Q_lin}
\end{equation}

\noindent the reference voltage is given by,
\begin{equation}
    \begin{bmatrix} \hat{v}_{C,ref}^{d,c} \\ \hat{v}_{C,ref}^{q,c}\end{bmatrix} = -K_{qv}\begin{bmatrix}
        \hat{Q} \\ 0
    \end{bmatrix}
    \label{eq:vcref_lin}
\end{equation}
\noindent where, $K_{qv}$ is reactive power-voltage droop coefficent.
\subsection{Impedance equation}
\noindent Combining (\ref{eq:mvdc_lin}) with (\ref{eq:vcs_to_vcc}),(\ref{eq:iL1s_to_iL1c}),(\ref{eq:ms_to_mc}),(\ref{eq:Q_lin}) and (\ref{eq:vcref_lin}),
\begin{equation}
\small
    \begin{aligned}
        \begin{bmatrix} \hat{m}^{d,s} \\ \hat{m}^{q,s}\end{bmatrix} & \frac{V_{dc}}{2}=\bigg(G_{del}(G_{VC}^vG_A-G_{IC}^vG_B-G_{CC}G_{VC}K_{qv}G_{QC}^v)- \\ &G_{MC}^v\frac{V_{dc}}{2}\bigg)\begin{bmatrix} \hat{v}_{C}^{d,s} \\ \hat{v}_{C}^{q,s}\end{bmatrix}+  \bigg(G_{del}(G_{VC}^iG_A-G_{IC}^iG_B-\\ &G_{CC}G_{VC}K_{qv}G_{QC}^i)-G_{MC}^i\frac{V_{dc}}{2}  \bigg)\begin{bmatrix}
        \hat{i}_{L1}^{d,s}  \\  \hat{i}_{L1}^{q,s}
    \end{bmatrix}
    \end{aligned}
    \label{eq:mvdc_final}
\end{equation}
where,
\begin{equation*}
    G_A=(-G_{CC}(G_{VC}+G_{v,dec}-G_{VC}Z_{virt}Y_C)+G_{VFF})
\end{equation*}
\begin{equation*}
    G_B=G_{CC}+G_{i,dec}+G_{CC}G_{VC}Z_{virt}
\end{equation*}

\noindent The capacitor voltage under small-signal perturbation is given by
\begin{equation}
    \begin{bmatrix} \hat{v}_{C}^{d,s} \\ \hat{v}_{C}^{q,s}\end{bmatrix}=\begin{bmatrix} \hat{m}^{d,s} \\ \hat{m}^{q,s}\end{bmatrix} \frac{V_{dc}}{2}-Z_{L1}\begin{bmatrix}
        \hat{i}_{L1}^{d,s}  \\  \hat{i}_{L1}^{q,s}
    \end{bmatrix}
    \label{eq:vc_filter}
\end{equation}

\noindent Combining (\ref{eq:mvdc_final}) with (\ref{eq:vc_filter}),
\begin{equation}
    \begin{bmatrix} \hat{v}_{C}^{d,s} \\ \hat{v}_{C}^{q,s}\end{bmatrix}=Z_{inv}\begin{bmatrix}
        \hat{i}_{L1}^{d,s}  \\  \hat{i}_{L1}^{q,s}
    \end{bmatrix}
\end{equation}
where, $Z_{inv}$ is shown in (\ref{eq:GFM_zinv}).



\ifCLASSOPTIONcaptionsoff
  \newpage
\fi

\bibliographystyle{IEEEtran}
\bibliography{mybibfile}

%








\end{document}

%% file: updated_images/GFM_sweep_Z_dd.tex
%
%
\definecolor{mycolor1}{rgb}{0.00000,0.44700,0.74100}%
\begin{tikzpicture}

\begin{axis}[%
width=2.5in,
height=1in,
at={(0.758in,2.0in)},
scale only axis,
xmode=log,
xmin=1,
xmax=10000,
xminorticks=true,
xlabel style={font=\color{white!15!black}},
ymin=-25,
ymax=40,
ylabel style={font=\large\color{white!15!black}},
ylabel={Magnitude [dB]},
axis background/.style={fill=white},
title style={font=\Large\bfseries},
title={$Z_{pcc}^{dd}$},
xmajorgrids,
xminorgrids,
ymajorgrids,
legend pos=north west,
legend style={legend cell align=left, align=left,font=\scriptsize, draw=white!15!black}
]
\addplot [color=mycolor1, line width=2.0pt]
  table[row sep=crcr]{%
1	0.592630469041993\\
1.20679264063934	0.826124120487911\\
1.45634847750124	1.14541530919289\\
1.7575106248548	1.5720445271788\\
2.12095088792018	2.12649026545998\\
2.55954792269953	2.82441447909459\\
3.0888435964775	3.67301956088751\\
3.72759372031492	4.66903018578328\\
4.49843266896945	5.79925353875278\\
5.4286754393239	7.04336935742144\\
6.5512855685955	8.3775660731965\\
7.90604321090773	9.77758775078574\\
9.54095476349988	11.2204428331673\\
16.76832936811	15.5960520462012\\
20.2358964772516	17.0003987571041\\
24.4205309454863	18.3385534747476\\
29.4705170255181	19.5828879584684\\
35.5648030622315	20.7034459163328\\
42.9193426012876	21.6698955330415\\
51.7947467923122	22.454419432314\\
62.5055192527392	23.0344867198747\\
75.4312006335461	23.3942431024462\\
91.0298177991526	23.5240493135347\\
109.854114198755	23.4189944139654\\
132.571136559011	23.0776103747158\\
159.985871960607	22.5007162724442\\
193.069772888325	21.6884685826453\\
232.995181051538	20.6336555000461\\
281.176869797421	19.3116096887127\\
339.322177189533	17.668716520039\\
409.491506238045	15.6095595260611\\
494.171336132382	12.9873712883278\\
596.362331659466	9.70903047002155\\
719.685673001147	6.90701675545942\\
868.511373751352	8.28655102274277\\
1048.11313415469	11.8697199965898\\
1264.85521685529	15.0538542971677\\
1526.41796717524	17.6983647994541\\
1842.06996932673	19.9820210614157\\
2222.99648252619	22.0385703421563\\
2682.69579527974	23.9512070269498\\
3237.45754281763	25.7710060244795\\
3906.93993705462	27.5298554444943\\
4714.86636345743	29.2481237268747\\
6866.48845004303	32.6115825350527\\
10000	35.9228118347935\\
};
\addlegendentry{$\text{Z}_{\text{analy}}$}
\addplot [color=red, line width=1.0pt, only marks, mark size=1pt, mark=*, mark options={solid, red}]
  table[row sep=crcr]{%
1	0.622912842285437\\
2	1.97260439017737\\
3	3.574162358846\\
4	5.13202810294658\\
5	6.5490151215398\\
6	7.80715926456281\\
7	8.93424409666439\\
9.00000000000001	10.8588058324785\\
10.9999999999999	12.4277926932325\\
12.9999999999999	13.7299282713396\\
16	15.3361154623853\\
20	17.0169727847178\\
24	18.3349152731794\\
28.9999999999999	19.5885411407691\\
35	20.7209173767126\\
42	21.6677940999618\\
51.0000000000002	22.4811614912541\\
61.9999999999998	23.06209252274\\
75	23.3859364898177\\
90.9999999999993	23.486948427112\\
109	23.3521344153278\\
131.999999999999	22.9966107255053\\
159	22.4509349694825\\
192.999999999999	21.7235291971945\\
231.999999999999	20.8734256656872\\
280.999999999999	19.786379129424\\
339.000000000002	18.4463926941967\\
409	16.661624036\\
494	14.1774873187438\\
595.999999999999	10.7636984845977\\
719	6.83145636945821\\
867.999999999998	7.34401790532608\\
1047.99999999999	11.3953849865981\\
1263.99999999999	14.8559797601169\\
1525.99999999999	17.614613769163\\
1842	19.9643973533649\\
2222	22.0453829421586\\
2682	23.9911616499477\\
3236.99999999999	25.8407832035605\\
3905.99999999999	27.6314896735399\\
4714.00000000002	29.3975576750215\\
5689	31.1554097589283\\
6866.00000000003	32.927864058312\\
8286.00000000004	34.7346095720274\\
10000	36.6069587312498\\
};
\addlegendentry{$\text{Z}_{\text{meas}}$}
\end{axis}

\begin{axis}[%
width=2.5in,
height=1in,
at={(0.758in,0.7in)},
scale only axis,
xmode=log,
xmin=1,
xmax=5000,
xminorticks=true,
xlabel style={font=\large\color{white!15!black}},
xlabel={Frequency [Hz]},
ymin=-200,
ymax=200,
ylabel style={font=\large\color{white!15!black}},
ylabel={Phase [deg]},
axis background/.style={fill=white},
xmajorgrids,
xminorgrids,
ymajorgrids,
]
\addplot [color=mycolor1, line width=2.0pt]
  table[row sep=crcr]{%
1	19.5548945339118\\
1.20679264063932	22.9629618986066\\
1.45634847750126	26.8263337935542\\
1.7575106248548	31.0899031025202\\
2.12095088792018	35.6456923126754\\
2.55954792269958	40.3327860991333\\
3.0888435964775	44.9516811411162\\
3.72759372031492	49.2909900769739\\
4.49843266896938	53.1571045604859\\
5.4286754393239	56.3954664466953\\
6.5512855685955	58.8973860458969\\
7.9060432109076	60.5940814154987\\
9.54095476350004	61.4440925144388\\
11.5139539932645	61.4202026381384\\
13.8949549437312	60.4997568686242\\
16.7683293681103	58.6601384896269\\
20.2358964772516	55.8799662481472\\
24.4205309454863	52.1459581067228\\
29.4705170255186	47.46459752306\\
35.5648030622315	41.8761194284743\\
42.9193426012876	35.4661090168533\\
51.7947467923114	28.3688615522245\\
62.5055192527403	20.7591469019351\\
91.0298177991511	4.80430287804188\\
109.854114198757	-3.12871292100114\\
132.571136559011	-10.7597976372192\\
159.985871960604	-17.891091206611\\
193.069772888328	-24.3379910677406\\
232.995181051538	-29.9240270253319\\
281.176869797421	-34.4349075676649\\
339.322177189539	-37.5032853324765\\
409.491506238045	-38.3650281477061\\
494.171336132382	-35.2348006058713\\
596.362331659456	-23.3859343843144\\
719.685673001159	7.36555996230213\\
868.511373751352	46.9191671448069\\
1048.11313415467	67.6263035834421\\
1264.85521685531	76.5714245809581\\
1526.41796717524	80.9802917287637\\
1842.0699693267	83.4766586488884\\
2222.99648252623	85.0547683911175\\
2682.69579527974	86.1374808447114\\
3237.45754281763	86.9245171452376\\
3906.93993705456	87.5198921115324\\
4714.86636345743	87.9828434517902\\
5689.86602901828	88.3497901094618\\
6866.48845004291	88.644575244648\\
8286.42772854693	88.8836338656176\\
10000	89.0787855292682\\
};

\addplot [color=red, line width=1.0pt, only marks, mark size=1pt, mark=*, mark options={solid, red}]
  table[row sep=crcr]{%
1	18.8227430978759\\
2	33.9865517890067\\
3	44.1083116529247\\
4	50.749545344165\\
5	55.0445501023459\\
6	57.7774915583077\\
6.99999999999989	59.4708437456339\\
9.00000000000001	61.2127899790905\\
11.0000000000001	61.4426436099237\\
12.9999999999999	60.9211970141246\\
16	59.1295210195289\\
20	55.9899564151837\\
24	52.4090363807057\\
29.0000000000003	47.7393774269336\\
35.0000000000006	42.194137054749\\
41.9999999999993	35.8678106612811\\
50.9999999999994	28.7511984623239\\
62.0000000000008	20.7966643680371\\
75	12.7808859279187\\
91.0000000000008	4.86966478726117\\
109.000000000001	-2.38129122568552\\
132.000000000001	-9.61146488261581\\
159	-15.8246959874662\\
192.999999999999	-21.5250057075017\\
232.000000000003	-26.3792651755948\\
281.000000000003	-30.7739620132512\\
338.999999999996	-34.6406781342451\\
409	-37.408291902321\\
494.000000000008	-37.5240771168283\\
596.000000000009	-30.0695154874045\\
719	-3.64786999849136\\
867.999999999998	43.4180863340505\\
1048.00000000001	67.8400737564483\\
1264.00000000001	77.1461573474692\\
1525.99999999999	81.4447996858675\\
1842	83.8145057058828\\
2222.00000000003	85.2616141457801\\
2682.00000000004	86.2733366089414\\
3237.00000000004	87.0153380411086\\
3905.99999999999	87.5823178747722\\
4714.00000000002	88.0281557822601\\
5689	88.3956255663361\\
6866.00000000003	88.6904227972052\\
8285.99999999991	88.9389897150253\\
10000	89.1428643277165\\
};

\end{axis}
\end{tikzpicture}%

%% file: updated_images/GFM_sweep_Z_dq.tex
%
%
\definecolor{mycolor1}{rgb}{0.00000,0.44700,0.74100}%
\begin{tikzpicture}

\begin{axis}[%
width=2.5in,
height=1in,
at={(0.758in,2.0in)},
scale only axis,
xmode=log,
xmin=1,
xmax=5000,
xminorticks=true,
xlabel style={font=\color{white!15!black}},
ymin=-25,
ymax=40,
ylabel style={font=\large\color{white!15!black}},
ylabel={Magnitude [dB]},
axis background/.style={fill=white},
title style={font=\Large\bfseries},
title={$Z_{pcc}^{dq}$},
xmajorgrids,
xminorgrids,
ymajorgrids,
legend pos=north west,
legend style={legend cell align=left, align=left,font=\normalsize, draw=white!15!black}
]
\addplot [color=mycolor1, line width=2.0pt, forget plot]
  table[row sep=crcr]{%
1	-8.53860740231125\\
1.7575106248548	-8.53532564024517\\
2.55954792269953	-8.53010071093077\\
3.08884359647747	-8.52565226113778\\
3.72759372031495	-8.51925438586397\\
4.49843266896945	-8.51009598364859\\
5.42867543932386	-8.49708185120512\\
6.5512855685955	-8.47878881408942\\
7.90604321090767	-8.45347966444195\\
9.54095476349996	-8.41925497869393\\
11.5139539932645	-8.37446568678198\\
13.8949549437314	-8.3185080398355\\
16.76832936811	-8.25293300101745\\
24.4205309454865	-8.11117638432611\\
29.4705170255181	-8.03622338081862\\
35.5648030622312	-7.92145741431534\\
42.919342601288	-7.66901643675932\\
51.7947467923122	-7.12823935043898\\
62.5055192527398	-6.21101977134982\\
91.0298177991519	-3.83978105707778\\
109.854114198756	-2.90144288385253\\
132.571136559011	-2.3887031200373\\
159.985871960606	-2.41501545505795\\
193.069772888325	-3.07014330513163\\
232.995181051538	-4.46453231419987\\
281.176869797424	-6.79925792926588\\
339.322177189533	-10.5435420039068\\
409.491506238042	-17.1684996669696\\
494.171336132382	-26.1181030757097\\
596.362331659466	-16.5771515219857\\
719.685673001153	-12.4638798786573\\
868.511373751352	-10.4828908551195\\
1048.11313415468	-9.5354389955132\\
1264.8552168553	-9.17658955113709\\
1526.41796717524	-9.14751976342794\\
1842.06996932672	-9.27381874225951\\
2222.99648252619	-9.44585667447518\\
2682.69579527972	-9.60770998328557\\
3237.45754281765	-9.7392632505379\\
3906.93993705462	-9.83862908977967\\
4714.86636345739	-9.91076678644336\\
5689.86602901828	-9.96198252554587\\
6866.48845004303	-9.9978717416097\\
8286.42772854686	-10.0228154645721\\
10000	-10.0400498693692\\
};
\addplot [color=red, line width=1.0pt, only marks, mark size=1pt, mark=*, mark options={solid, red}, forget plot]
  table[row sep=crcr]{%
1	-8.52251209996083\\
2	-8.48376397671693\\
3	-8.50446232065612\\
4	-8.46040364429147\\
5	-8.42946108490677\\
6	-8.46766064128215\\
7	-8.49390664237808\\
9.00000000000001	-8.36946818129098\\
11	-8.30041496934277\\
13	-8.24593319806992\\
16	-8.23098940880494\\
20	-7.92639795002775\\
24	-7.80685207571084\\
29.0000000000001	-7.59258132412133\\
35	-7.3030152976487\\
42	-7.30320939678717\\
50.9999999999998	-6.45782616043967\\
61.9999999999998	-6.6866311128894\\
75	-6.78901236614706\\
91.0000000000001	-5.5529449041116\\
109	-5.20904015845536\\
132	-5.48164813179598\\
159	-5.88905192747136\\
193	-6.73077686687101\\
232.000000000001	-8.61257717869852\\
281.000000000001	-11.459231894176\\
338.999999999999	-18.2394852824086\\
409	-25.4563178911011\\
494	-15.5274447855426\\
595.999999999999	-11.4883318342055\\
719	-9.67256003252569\\
867.999999999998	-9.2935030100792\\
1048	-8.97052735020336\\
1264	-8.83633691005369\\
1525.99999999999	-9.08951324759008\\
1842	-9.51757642592373\\
2222	-9.54721618206484\\
2682	-9.51327310328836\\
3237.00000000001	-9.67995406690775\\
3905.99999999999	-9.57250999196053\\
4714.00000000002	-9.51018510820863\\
5689	-9.3728294886754\\
6866.00000000003	-9.0690348702496\\
8285.99999999997	-8.72693343660315\\
10000	-7.97681833141322\\
};
\end{axis}

\begin{axis}[%
width=2.5in,
height=1in,
at={(0.758in,0.7in)},
scale only axis,
xmode=log,
xmin=1,
xmax=5000,
xminorticks=true,
xlabel style={font=\large\color{white!15!black}},
xlabel={Frequency [Hz]},
ymin=-200,
ymax=200,
ylabel style={font=\large\color{white!15!black}},
ylabel={Phase [deg]},
axis background/.style={fill=white},
xmajorgrids,
xminorgrids,
ymajorgrids,
]
\addplot [color=mycolor1, line width=2.0pt, forget plot]
  table[row sep=crcr]{%
1	-179.704812894309\\
1.20679264063936	-179.623422299509\\
1.45634847750121	-179.528879110098\\
1.7575106248548	-179.41798223994\\
2.12095088792024	-179.287060258244\\
2.55954792269949	-179.131922394927\\
3.0888435964775	-178.947848332282\\
3.72759372031504	-178.729654096208\\
4.49843266896938	-178.471893929002\\
5.4286754393239	-178.169285648683\\
6.55128556859571	-177.817467045336\\
7.9060432109076	-177.414162922266\\
9.54095476350004	-176.960665383298\\
11.5139539932641	-176.462996554391\\
13.8949549437312	-175.930909844443\\
16.7683293681103	-175.370759170963\\
20.235896477251	-174.765924250118\\
24.4205309454863	-174.039701652376\\
29.4705170255186	-173.011084476381\\
35.5648030622303	-171.403856767206\\
42.9193426012876	-169.056764741377\\
51.7947467923131	-166.442647542971\\
62.5055192527382	-165.033009487787\\
75.4312006335461	-166.593126508547\\
91.0298177991541	-171.988203385182\\
109.854114198753	178.998723502896\\
132.571136559011	167.095129744272\\
159.98587196061	153.084104142318\\
193.069772888321	137.614068375465\\
232.995181051538	121.119110351293\\
281.176869797431	103.707922074467\\
339.322177189527	84.6769851287856\\
409.491506238045	58.7415967477703\\
494.171336132398	-47.5657468678613\\
596.362331659456	-112.053884607126\\
719.685673001159	-131.524215577642\\
868.51137375138	-144.956118329496\\
1048.11313415467	-155.396138799918\\
1264.85521685531	-163.412183803991\\
1526.41796717519	-169.240244635447\\
1842.0699693267	-173.204214303964\\
2222.99648252623	-175.746389922978\\
2682.69579527965	-177.314168590873\\
3237.45754281763	-178.264801589896\\
3906.93993705468	-178.842280316469\\
4714.86636345727	-179.198543180454\\
5689.86602901828	-179.423791722266\\
6866.48845004314	-179.57052735989\\
8286.42772854666	-179.669205487996\\
10000	-179.737602845837\\
};
\addplot [color=red, line width=1.0pt, only marks, mark size=1pt, mark=*, mark options={solid, red}, forget plot]
  table[row sep=crcr]{%
1	-179.496584618081\\
2	-178.984664633763\\
3	-178.633245744225\\
4	-178.39743043264\\
5	-178.013422441812\\
6	-178.07344468898\\
6.99999999999989	-178.108426195567\\
9.00000000000001	-177.819280563604\\
11.0000000000001	-177.249832719404\\
13.0000000000003	-176.362176847196\\
16	-176.284900073495\\
20	-175.680529976577\\
24	-176.535173969594\\
29.0000000000003	-175.187908976377\\
34.9999999999994	-176.682915169317\\
41.9999999999993	-177.367159474517\\
50.9999999999994	-179.005537499173\\
62.0000000000008	178.491316073891\\
75	177.772994046789\\
91.0000000000008	170.922024023191\\
109.000000000001	162.170706926554\\
132.000000000001	153.130462889648\\
159.000000000005	143.793552649175\\
192.999999999999	130.789046793895\\
232.000000000003	117.823203497302\\
281.000000000003	100.498345404476\\
339.000000000007	72.8920791833029\\
409.000000000013	-41.8647317427636\\
494.000000000008	-106.624077158448\\
596.000000000009	-126.774905809244\\
719.000000000023	-145.556620664667\\
867.999999999998	-153.216196532706\\
1048.00000000001	-164.631236849502\\
1264.00000000001	-170.059162510556\\
1525.99999999999	-173.003684317384\\
1841.99999999994	-175.561086312594\\
2221.99999999996	-176.525246378283\\
2682.00000000004	-177.78960234406\\
3236.99999999993	-178.337952052592\\
3905.99999999999	-179.242074698665\\
4714.00000000002	-179.130167765631\\
5689	-179.568384875634\\
6865.9999999998	-179.277431469567\\
8286.00000000018	-179.972078775828\\
10000	179.671943572346\\
};
\end{axis}
\end{tikzpicture}%

%% file: updated_images/GFM_sweep_Z_qd.tex
%
%
\definecolor{mycolor1}{rgb}{0.00000,0.44700,0.74100}%
\begin{tikzpicture}

\begin{axis}[%
width=2.5in,
height=1in,
at={(0.758in,2.0in)},
scale only axis,
xmode=log,
xmin=1,
xmax=5000,
xminorticks=true,
xlabel style={font=\color{white!15!black}},
ymin=-25,
ymax=40,
ylabel style={font=\large\color{white!15!black}},
ylabel={Magnitude [dB]},
axis background/.style={fill=white},
title style={font=\Large\bfseries},
title={$Z_{pcc}^{qd}$},
xmajorgrids,
xminorgrids,
ymajorgrids,
legend pos=north west,
legend style={legend cell align=left, align=left,font=\normalsize, draw=white!15!black}
]
\addplot [color=mycolor1, line width=2.0pt, forget plot]
  table[row sep=crcr]{%
1	2.51530880869006\\
1.20679264063933	0.987657942169264\\
1.45634847750124	-0.497070441116357\\
1.7575106248548	-1.92349756008286\\
2.12095088792019	-3.27305082103814\\
2.55954792269953	-4.52503004230158\\
3.08884359647747	-5.6589728272658\\
3.72759372031495	-6.6582568707126\\
4.49843266896945	-7.51412496625119\\
5.42867543932386	-8.22869892886861\\
6.5512855685955	-8.81574912081861\\
7.90604321090767	-9.29913338763562\\
9.54095476349996	-9.71008937289646\\
13.8949549437314	-10.4645671320933\\
16.76832936811	-10.8945168089148\\
20.2358964772516	-11.4251521150213\\
24.4205309454865	-12.102196613191\\
29.4705170255181	-12.9216624283176\\
35.5648030622312	-13.677888248134\\
42.919342601288	-13.7149228032786\\
51.7947467923122	-12.3577952262066\\
75.4312006335461	-7.73870407218329\\
91.0298177991519	-5.83911860464844\\
109.854114198756	-4.50972305015397\\
132.571136559011	-3.80059608635739\\
159.985871960606	-3.74839975586403\\
193.069772888325	-4.40932141367638\\
232.995181051538	-5.89530503545816\\
281.176869797424	-8.45789994722144\\
339.322177189533	-12.7642313418646\\
409.491506238042	-21.366502660732\\
494.171336132382	-22.9234189370991\\
596.362331659466	-14.9979365212354\\
719.685673001153	-11.6600741794122\\
868.511373751352	-9.99462284910487\\
1048.11313415468	-9.21023793409032\\
1264.8552168553	-8.94891903799752\\
1526.41796717524	-8.98424122526427\\
1842.06996932672	-9.15531394510828\\
2222.99648252619	-9.35871184397644\\
2682.69579527972	-9.54215874227156\\
3237.45754281765	-9.68828923347661\\
3906.93993705462	-9.79737909204409\\
4714.86636345739	-9.87598538881276\\
5689.86602901828	-9.93152140903774\\
6866.48845004303	-9.97032121671925\\
8286.42772854686	-9.9972520773539\\
10000	-10.0158717639747\\
};
\addplot [color=red, line width=1.0pt, only marks, mark size=1pt, mark=*, mark options={solid, red}, forget plot]
  table[row sep=crcr]{%
1	2.53607107514597\\
2	-2.84154500501574\\
3	-5.48134244802252\\
4	-6.97288663405402\\
5	-7.88779456862819\\
6	-8.50727236643353\\
7	-8.95338224647147\\
9.00000000000001	-9.51297744022201\\
11	-9.84790723402494\\
13	-10.0151628994537\\
16	-10.5235971685055\\
20	-10.7596722046476\\
24	-10.6124151123051\\
29.0000000000001	-11.1918491238568\\
35	-10.5237641553186\\
42	-9.64396367389939\\
50.9999999999998	-10.0072989960457\\
61.9999999999998	-8.1151862384911\\
75	-7.81288843554101\\
91.0000000000001	-6.76597085390312\\
109	-6.28077960845194\\
132	-6.3284146842345\\
159	-6.46631783440976\\
193	-7.93279049150203\\
232.000000000001	-9.73117264582958\\
281.000000000001	-13.128389210531\\
338.999999999999	-20.2730419879049\\
409	-20.7832834482988\\
494	-13.3670374745737\\
595.999999999999	-10.5210294454426\\
719	-9.27874656843892\\
867.999999999998	-8.84850168816649\\
1048	-8.60078548145786\\
1264	-8.85989272247638\\
1525.99999999999	-8.92294786510476\\
1842	-9.24016684668428\\
2222	-9.35352869526863\\
2682	-9.43621800320963\\
3237.00000000001	-9.48082920748598\\
3905.99999999999	-9.48831794266088\\
4714.00000000002	-9.39729324345973\\
5689	-9.33812665234383\\
6866.00000000003	-9.10766384965243\\
8285.99999999997	-8.62178827896761\\
10000	-8.02434029182827\\
};
\end{axis}

\begin{axis}[%
width=2.5in,
height=1in,
at={(0.758in,0.7in)},
scale only axis,
xmode=log,
xmin=1,
xmax=5000,
xminorticks=true,
xlabel style={font=\large\color{white!15!black}},
xlabel={Frequency [Hz]},
ymin=-200,
ymax=200,
ylabel style={font=\large\color{white!15!black}},
ylabel={Phase [deg]},
axis background/.style={fill=white},
xmajorgrids,
xminorgrids,
ymajorgrids,
]
\addplot [color=mycolor1, line width=2.0pt, forget plot]
  table[row sep=crcr]{%
1	-78.3381505665206\\
1.20679264063932	-75.5802241748765\\
1.45634847750126	-72.4262288710332\\
1.7575106248548	-68.8389287429938\\
2.12095088792018	-64.8063373826461\\
2.55954792269958	-60.3549887350599\\
3.0888435964775	-55.5614379540947\\
4.49843266896938	-45.512365579387\\
5.4286754393239	-40.6240849001436\\
6.5512855685955	-36.0716768218202\\
7.9060432109076	-31.9942862344295\\
9.54095476350004	-28.4714004399906\\
11.5139539932645	-25.5136718957064\\
13.8949549437312	-23.0525798962551\\
16.7683293681103	-20.9119206342994\\
20.2358964772516	-18.7350683832522\\
24.4205309454863	-15.8275311643519\\
29.4705170255186	-10.8886316253912\\
35.5648030622315	-1.93413883339771\\
42.9193426012876	11.7893932464555\\
51.7947467923114	25.1924671921784\\
62.5055192527403	31.3083457776626\\
75.4312006335461	29.3695896587173\\
91.0298177991511	21.6546210446764\\
109.854114198757	10.1294518900132\\
132.571136559011	-3.91810768582921\\
159.985871960604	-19.6331137008694\\
193.069772888328	-36.4358020297386\\
232.995181051538	-53.9900004482021\\
281.176869797421	-72.3230819826272\\
339.322177189539	-92.5840352587368\\
409.491506238045	-125.421525066592\\
494.171336132382	98.1334014394075\\
596.362331659456	64.2922086540275\\
719.685673001159	47.552971939548\\
868.511373751352	34.7564617373644\\
1048.11313415467	24.4925510654466\\
1264.85521685531	16.5095893320172\\
1526.41796717524	10.6739565580361\\
1842.0699693267	6.70134206878386\\
2222.99648252623	4.1604089195698\\
2682.69579527974	2.60172375137348\\
3237.45754281763	1.66310087136388\\
3906.93993705456	1.09709964315246\\
4714.86636345743	0.750299686283739\\
5689.86602901828	0.532208143400084\\
6866.48845004291	0.390518021566663\\
8286.42772854693	0.295077545853687\\
10000	0.228375586859144\\
};
\addplot [color=red, line width=1.0pt, only marks, mark size=1pt, mark=*, mark options={solid, red}, forget plot]
  table[row sep=crcr]{%
1	-78.1965515178433\\
2	-65.9901648083737\\
3	-56.1543455075375\\
4	-48.4655358780479\\
5	-42.412249621045\\
6	-37.7459822943553\\
6.99999999999989	-33.8243641514175\\
9.00000000000001	-28.414736619457\\
11.0000000000001	-24.859999960926\\
13.0000000000003	-20.7630984157699\\
16	-16.5334566065678\\
20	-12.2427762785913\\
24	-8.44803158534299\\
29.0000000000003	-2.41404256507613\\
34.9999999999994	-0.953121850961338\\
41.9999999999993	9.9523865491424\\
50.9999999999994	5.49346105771261\\
62.0000000000008	5.08046764252001\\
75	3.8585738638952\\
91.0000000000008	-2.23963547498184\\
109.000000000001	-10.5557239213827\\
132.000000000001	-25.7496791413455\\
159.000000000005	-37.8215754846523\\
192.999999999999	-50.1602139253803\\
232.000000000003	-70.8719545280499\\
281.000000000003	-91.6288481229999\\
339.000000000007	-137.478709764881\\
409.000000000013	117.378245258057\\
494.000000000008	78.3015103928998\\
596.000000000009	50.8233915854002\\
719.000000000023	37.6797656029999\\
867.999999999998	23.2231445101674\\
1048.00000000001	15.4802950567001\\
1264.00000000001	10.7421042552793\\
1525.99999999999	6.30821348663048\\
1841.99999999994	3.87100985398874\\
2221.99999999996	3.12577384739549\\
2682.00000000004	1.16753469787258\\
3236.99999999993	1.38197211921656\\
3905.99999999999	0.971720257115379\\
4714.00000000002	0.397529674365359\\
5689	0.797425500860896\\
6865.9999999998	0.842569667518944\\
8286.00000000018	0.191604923832244\\
10000	0.391145943191987\\
};
\end{axis}
\end{tikzpicture}%

%% file: updated_images/GFM_sweep_Z_qq.tex
%
%
\definecolor{mycolor1}{rgb}{0.00000,0.44700,0.74100}%
\begin{tikzpicture}

\begin{axis}[%
width=2.5in,
height=1in,
at={(0.758in,2.0in)},
scale only axis,
xmode=log,
xmin=1,
xmax=5000,
xminorticks=true,
xlabel style={font=\color{white!15!black}},
ymin=-25,
ymax=40,
ylabel style={font=\large\color{white!15!black}},
ylabel={Magnitude [dB]},
axis background/.style={fill=white},
title style={font=\Large\bfseries},
title={$Z_{pcc}^{qq}$},
xmajorgrids,
xminorgrids,
ymajorgrids,
legend pos=north west,
legend style={legend cell align=left, align=left,font=\normalsize, draw=white!15!black}
]
\addplot [color=mycolor1, line width=2.0pt, forget plot]
  table[row sep=crcr]{%
1	0.550434746562303\\
1.20679264063934	0.786566113099632\\
1.45634847750124	1.1092456100582\\
1.7575106248548	1.54003020241601\\
2.12095088792018	2.09929765388862\\
2.55954792269953	2.80247299870145\\
3.0888435964775	3.65641738254988\\
3.72759372031492	4.65749119357957\\
4.49843266896945	5.79220749562447\\
5.4286754393239	7.04008005496797\\
6.5512855685955	8.37726262490403\\
7.90604321090773	9.7795597188174\\
9.54095476349988	11.2240884951621\\
16.76832936811	15.6022352961644\\
20.2358964772516	17.0068887649757\\
24.4205309454863	18.3451738635837\\
29.4705170255181	19.5894951663317\\
35.5648030622315	20.7099218631885\\
42.9193426012876	21.6761456969535\\
51.7947467923122	22.4603739464712\\
62.5055192527392	23.0401007449908\\
75.4312006335461	23.3994933511502\\
91.0298177991526	23.5289268757247\\
109.854114198755	23.4234983758064\\
132.571136559011	23.0817493097918\\
159.985871960607	22.5045227843881\\
193.069772888325	21.6920230239396\\
232.995181051538	20.6370985386769\\
281.176869797421	19.3151079391874\\
339.322177189533	17.6723447150817\\
409.491506238045	15.6130643337066\\
494.171336132382	12.9896874552855\\
596.362331659466	9.70704984615914\\
719.685673001147	6.89649339766206\\
868.511373751352	8.27913564804469\\
1048.11313415469	11.8680533757577\\
1264.85521685529	15.0539327703561\\
1526.41796717524	17.6987266022421\\
1842.06996932673	19.9823057466283\\
2222.99648252619	22.0387440931187\\
2682.69579527974	23.9513044263172\\
3237.45754281763	25.7710596439909\\
3906.93993705462	27.5298854051359\\
4714.86636345743	29.2481409748465\\
6866.48845004303	32.6115888510814\\
10000	35.9228144073519\\
};
\addplot [color=red, line width=1.0pt, only marks, mark size=1pt, mark=*, mark options={solid, red}, forget plot]
  table[row sep=crcr]{%
1	0.51161491417556\\
2	1.89279399644752\\
3	3.49912577910538\\
4	5.06211843902724\\
5	6.48632781492637\\
6	7.74692945682134\\
7	8.86782823861123\\
9.00000000000001	10.7846604417962\\
10.9999999999999	12.3449694871175\\
12.9999999999999	13.6635932799353\\
16	15.2544021274804\\
20	16.9259613648502\\
24	18.2228831959937\\
28.9999999999999	19.4785419570051\\
35	20.6136950800986\\
42	21.5452326996232\\
51.0000000000002	22.3381350782298\\
61.9999999999998	22.9516065220661\\
75	23.275542370276\\
90.9999999999993	23.3624955850175\\
109	23.2606042577917\\
131.999999999999	22.9095873253174\\
159	22.3983353535709\\
192.999999999999	21.6775742934902\\
231.999999999999	20.8568836818553\\
280.999999999999	19.8062280787089\\
339.000000000002	18.4768574936376\\
409	16.7267592496039\\
494	14.2600600711058\\
595.999999999999	10.8119896903177\\
719	6.88283863660565\\
867.999999999998	7.31010956636324\\
1047.99999999999	11.3772649164933\\
1263.99999999999	14.8247404017397\\
1525.99999999999	17.6132414413451\\
1842	19.9617020680225\\
2222	22.0497737959289\\
2682	23.9904871877797\\
3236.99999999999	25.8386326457883\\
3905.99999999999	27.6317006030546\\
4714.00000000002	29.3970306472026\\
5689	31.1563392258568\\
6866.00000000003	32.9280158592661\\
8286.00000000004	34.7345141488982\\
10000	36.6070310178726\\
};
\end{axis}

\begin{axis}[%
width=2.5in,
height=1in,
at={(0.758in,0.7in)},
scale only axis,
xmode=log,
xmin=1,
xmax=5000,
xminorticks=true,
xlabel style={font=\large\color{white!15!black}},
xlabel={Frequency [Hz]},
ymin=-200,
ymax=200,
ylabel style={font=\large\color{white!15!black}},
ylabel={Phase [deg]},
axis background/.style={fill=white},
xmajorgrids,
xminorgrids,
ymajorgrids,
]
\addplot [color=mycolor1, line width=2.0pt, forget plot]
  table[row sep=crcr]{%
1	16.146624429733\\
1.20679264063932	20.1770942189379\\
1.45634847750126	24.5586733348782\\
1.7575106248548	29.2522761034303\\
3.0888435964775	43.9931122763237\\
3.72759372031492	48.5200611252971\\
4.49843266896938	52.5351927226929\\
5.4286754393239	55.8914748782352\\
6.5512855685955	58.4867640990835\\
7.9060432109076	60.2576600976622\\
9.54095476350004	61.1669346675735\\
11.5139539932645	61.1906089432176\\
13.8949549437312	60.3084800145742\\
16.7683293681103	58.4997963721685\\
20.2358964772516	55.7446189455479\\
24.4205309454863	52.0308092629557\\
29.4705170255186	47.3657845203188\\
35.5648030622315	41.790560351325\\
42.9193426012876	35.3913768775637\\
51.7947467923114	28.3030623168312\\
62.5055192527403	20.700792973908\\
91.0298177991511	4.75744725460767\\
109.854114198757	-3.17113484659913\\
132.571136559011	-10.7984205667183\\
159.985871960604	-17.9263821435038\\
193.069772888328	-24.3703949955751\\
232.995181051538	-29.9543357084167\\
281.176869797421	-34.4648312481326\\
339.322177189539	-37.5360320471253\\
409.491506238045	-38.4056081870479\\
494.171336132382	-35.2895681557128\\
596.362331659456	-23.4567589755538\\
719.685673001159	7.32192892698812\\
868.511373751352	46.9426224877401\\
1048.11313415467	67.6499976107212\\
1264.85521685531	76.5838008899319\\
1526.41796717524	80.9857316939159\\
1842.0699693267	83.4788323997724\\
2222.99648252623	85.0555886449327\\
2682.69579527974	86.13778193126\\
3237.45754281763	86.9246267851652\\
3906.93993705456	87.5199318812321\\
4714.86636345743	87.9828575130756\\
5689.86602901828	88.3497945718386\\
6866.48845004291	88.6445761006369\\
8286.42772854693	88.8836334027791\\
10000	89.0787846413573\\
};
\addplot [color=red, line width=1.0pt, only marks, mark size=1pt, mark=*, mark options={solid, red}, forget plot]
  table[row sep=crcr]{%
1	17.4001268623593\\
2	33.2861868504303\\
3	43.7692564048455\\
4	50.486907862223\\
5	54.8525961200195\\
6	57.6465422941387\\
6.99999999999989	59.4634089806133\\
9.00000000000001	61.137415280192\\
11.0000000000001	61.3504514279508\\
12.9999999999999	60.8124488362284\\
16	59.1200640937467\\
20	55.9514794169788\\
24	52.3778334125997\\
29.0000000000003	47.7762515471587\\
35.0000000000006	42.2005388049426\\
41.9999999999993	36.1666762269128\\
50.9999999999994	28.848262674006\\
62.0000000000008	20.9514708092539\\
75	13.1968990888591\\
91.0000000000008	5.25091207083231\\
109.000000000001	-1.94552733216968\\
132.000000000001	-9.01416491672687\\
159	-15.0635346673313\\
192.999999999999	-20.8805355860422\\
232.000000000003	-25.7119411780679\\
281.000000000003	-30.1800031909556\\
338.999999999996	-34.0526022566599\\
409	-37.0116460762388\\
494.000000000008	-37.2911651688287\\
596.000000000009	-30.4094577202503\\
719	-3.76396819760811\\
867.999999999998	42.9723722964441\\
1048.00000000001	67.904625510713\\
1264.00000000001	77.228029461323\\
1525.99999999999	81.4138257995141\\
1842	83.8046509196302\\
2222.00000000003	85.2482289603001\\
2682.00000000004	86.2695807268548\\
3237.00000000004	87.0118678487062\\
3905.99999999999	87.5846180106113\\
4714.00000000002	88.0349387468081\\
5689	88.3943997839829\\
6866.00000000003	88.6880567003635\\
8285.99999999991	88.9343363167785\\
10000	89.1394732903322\\
};
\end{axis}
\end{tikzpicture}%

%% file: updated_images/GFL_sweep_Z_dd.tex
%
%
\definecolor{mycolor1}{rgb}{0.00000,0.44700,0.74100}%
\begin{tikzpicture}

\begin{axis}[%
width=2.5in,
height=1in,
at={(0.758in,2.0in)},
scale only axis,
xmode=log,
xmin=1,
xmax=10000,
xminorticks=true,
xlabel style={font=\color{white!15!black}},
ymin=-10,
ymax=40,
ylabel style={font=\large\color{white!15!black}},
ylabel={Magnitude [dB]},
axis background/.style={fill=white},
title style={font=\Large\bfseries},
title={$Z_{pcc}^{dd}$},
xmajorgrids,
xminorgrids,
ymajorgrids,
legend pos=north west,
legend style={legend cell align=left, align=left,font=\scriptsize, draw=white!15!black}
]
\addplot [color=mycolor1, line width=2.0pt]
  table[row sep=crcr]{%
1	18.4816178121379\\
1.20679264063934	18.4775130027794\\
1.45634847750124	18.4715389818262\\
1.7575106248548	18.4628472392571\\
2.12095088792018	18.4502071049119\\
2.55954792269953	18.43183703824\\
3.0888435964775	18.4051653452467\\
3.72759372031492	18.3664953971708\\
4.49843266896945	18.3105473868889\\
5.4286754393239	18.2298539287246\\
6.5512855685955	18.1140157260742\\
7.90604321090773	17.9489126581873\\
9.54095476349988	17.7162005567318\\
11.5139539932645	17.3940029875278\\
13.8949549437315	16.9610811837669\\
16.76832936811	16.4098718057704\\
20.2358964772516	15.7797923950287\\
24.4205309454863	15.2265487126827\\
29.4705170255181	15.1044761023944\\
35.5648030622315	15.881810677984\\
42.9193426012876	17.775447103273\\
51.7947467923122	20.7040536772296\\
62.5055192527392	24.8218281554384\\
75.4312006335461	31.1420367567624\\
91.0298177991526	34.5004321181284\\
109.854114198755	27.3826785128272\\
132.571136559011	22.7245524365237\\
159.985871960607	19.4335706816374\\
193.069772888325	16.8265256501718\\
232.995181051538	14.6559618861976\\
281.176869797421	12.8723440919669\\
339.322177189533	11.5409017608026\\
409.491506238045	10.7438270202487\\
494.171336132382	10.4339165377966\\
596.362331659466	10.393425330678\\
719.685673001147	10.4677523001705\\
868.511373751352	11.0112877722445\\
1048.11313415469	12.669065415275\\
1264.85521685529	15.1196309761425\\
1526.41796717524	17.6217988103637\\
1842.06996932673	19.9095665177222\\
2222.99648252619	21.9899495222273\\
2682.69579527974	23.9212778367008\\
3237.45754281763	25.7529333465463\\
3906.93993705462	27.5188694664076\\
4714.86636345743	29.2413296907611\\
6866.48845004303	32.6088241227577\\
10000	35.9216161183962\\
};
\addlegendentry{$\text{Z}_{\text{analy}}$}
\addplot [color=red, line width=1.0pt, only marks, mark size=1pt, mark=*, mark options={solid, red}]
  table[row sep=crcr]{%
1	18.4670188132903\\
2	18.4671900101622\\
3	18.4228672245295\\
4	18.3337403638912\\
5	18.282303844316\\
6	18.1873175880232\\
7	18.0630875835703\\
9.00000000000001	17.8115746030383\\
10.9999999999999	17.4975717680591\\
12.9999999999999	17.1302910436533\\
16	16.5752604476147\\
20	15.8392715115222\\
24	15.2919736073813\\
28.9999999999999	15.109460750668\\
35	15.7810037092825\\
42	17.5326985556237\\
51.0000000000002	20.4828933895837\\
61.9999999999998	24.6547929797128\\
75	31.0281159528351\\
90.9999999999993	34.8882976228353\\
109	27.672337455948\\
131.999999999999	22.8273323068103\\
159	19.5556587778513\\
192.999999999999	16.9090323812711\\
231.999999999999	14.8751819286049\\
280.999999999999	13.2201611445443\\
339.000000000002	12.1328606981012\\
409	11.5021927552599\\
494	11.1527422741898\\
595.999999999999	10.7512334790183\\
719	10.1600562117521\\
867.999999999998	10.3408166246137\\
1047.99999999999	12.33655767215\\
1263.99999999999	15.0538233614161\\
1525.99999999999	17.639785851643\\
1842	19.9344060301925\\
2222	22.0138650922617\\
2682	23.9462702228185\\
3236.99999999999	25.7879862928754\\
3905.99999999999	27.564755045257\\
4714.00000000002	29.3053238899904\\
5689	31.0290514283684\\
6866.00000000003	32.747014713271\\
8286.00000000004	34.4714200292988\\
10000	36.216969325483\\
};
\addlegendentry{$\text{Z}_{\text{meas}}$}
\end{axis}

\begin{axis}[%
width=2.5in,
height=1in,
at={(0.758in,0.7in)},
scale only axis,
xmode=log,
xmin=1,
xmax=5000,
xminorticks=true,
xlabel style={font=\large\color{white!15!black}},
xlabel={Frequency [Hz]},
ymin=-200,
ymax=200,
ylabel style={font=\large\color{white!15!black}},
ylabel={Phase [deg]},
axis background/.style={fill=white},
xmajorgrids,
xminorgrids,
ymajorgrids,
]
\addplot [color=mycolor1, line width=2.0pt]
  table[row sep=crcr]{%
1	-0.319012070028549\\
1.20679264063936	-0.383882706385634\\
1.45634847750121	-0.461336041184552\\
1.7575106248548	-0.553344116825173\\
2.12095088792024	-0.661810102532371\\
2.55954792269949	-0.788193804853535\\
3.0888435964775	-0.932788197041333\\
3.72759372031504	-1.09337132280649\\
4.49843266896938	-1.26275054032902\\
5.4286754393239	-1.42435850247466\\
6.55128556859571	-1.54445260468572\\
7.9060432109076	-1.55846407395688\\
9.54095476350004	-1.34746741896359\\
11.5139539932641	-0.698583323483291\\
13.8949549437312	0.758617596469918\\
16.7683293681103	3.6434726599347\\
20.235896477251	8.91379401932846\\
24.4205309454863	17.7871095090094\\
29.4705170255186	31.0228227913979\\
42.9193426012876	63.982917872222\\
51.7947467923131	79.0956370992794\\
62.5055192527382	94.7868883860177\\
75.4312006335461	122.181803609818\\
91.0298177991541	-150.821241364276\\
109.854114198753	-105.546305676197\\
132.571136559011	-89.7022112155052\\
159.98587196061	-78.8389169819315\\
193.069772888321	-68.7818598480096\\
232.995181051538	-58.0487266954996\\
281.176869797431	-45.8745647694521\\
339.322177189527	-32.1056259695268\\
409.491506238045	-17.4444942152211\\
494.171336132398	-3.12249151780375\\
596.362331659456	10.4506191622362\\
719.685673001159	24.9062429866082\\
1048.11313415467	60.1637115918118\\
1264.85521685531	72.3966757596728\\
1526.41796717519	79.0972392539913\\
1842.0699693267	82.6652425353896\\
2222.99648252623	84.7032528664135\\
2682.69579527965	85.9816445336626\\
3237.45754281763	86.8532700607186\\
3906.93993705468	87.4861810711397\\
4714.86636345727	87.9663294007322\\
5689.86602901828	88.3414338999015\\
6866.48845004314	88.6402259347073\\
8286.42772854666	88.8813170088924\\
10000	89.0775287069701\\
};

\addplot [color=red, line width=1.0pt, only marks, mark size=1pt, mark=*, mark options={solid, red}]
  table[row sep=crcr]{%
1	-0.313526302558017\\
2	-0.622254747706506\\
3	-0.909633439623974\\
4	-1.16554210807294\\
5	-1.35060342952252\\
6	-1.48763005976556\\
6.99999999999989	-1.55953221349731\\
9.00000000000001	-1.43713416580343\\
11.0000000000001	-0.895728829448188\\
13.0000000000003	0.133261108462619\\
16	2.75520945500824\\
20	8.51359021821901\\
24	16.780277105839\\
29.0000000000003	29.6998585603206\\
34.9999999999994	45.8810781078423\\
41.9999999999993	61.964687123024\\
50.9999999999994	77.6449057305221\\
62.0000000000008	93.4334078518348\\
75	119.43754007477\\
91.0000000000008	-148.749039648811\\
109.000000000001	-104.426158283412\\
132.000000000001	-88.2279266838492\\
159.000000000005	-77.2734019446\\
192.999999999999	-66.6243272336727\\
232.000000000003	-55.8675513175947\\
281.000000000003	-43.4091619638203\\
339.000000000007	-30.3487111163775\\
409.000000000013	-17.5856576964804\\
494.000000000008	-6.13540376200828\\
596.000000000009	4.803899039409\\
719.000000000023	19.5727069607581\\
867.999999999998	40.8255923191151\\
1048.00000000001	61.3795418895887\\
1264.00000000001	73.3972781675721\\
1525.99999999999	79.6167684129469\\
1841.99999999994	82.890138425211\\
2221.99999999996	84.7854606421639\\
2682.00000000004	86.0226598763436\\
3236.99999999993	86.8674105471553\\
3905.99999999999	87.492215582787\\
4714.00000000002	87.9785540704255\\
5689	88.3598823611747\\
6865.9999999998	88.6589467285764\\
8286.00000000018	88.9047071053064\\
10000	89.1060285875928\\
};

\end{axis}
\end{tikzpicture}%

%% file: updated_images/GFL_sweep_Z_dq.tex
%
%
\definecolor{mycolor1}{rgb}{0.00000,0.44700,0.74100}%
\begin{tikzpicture}

\begin{axis}[%
width=2.5in,
height=1in,
at={(0.758in,2.0in)},
scale only axis,
xmode=log,
xmin=1,
xmax=5000,
xminorticks=true,
xlabel style={font=\color{white!15!black}},
ymin=-10,
ymax=40,
ylabel style={font=\large\color{white!15!black}},
ylabel={Magnitude [dB]},
axis background/.style={fill=white},
title style={font=\Large\bfseries},
title={$Z_{pcc}^{dq}$},
xmajorgrids,
xminorgrids,
ymajorgrids,
legend pos=north west,
legend style={legend cell align=left, align=left,font=\normalsize, draw=white!15!black}
]
\addplot [color=mycolor1, line width=2.0pt, forget plot]
  table[row sep=crcr]{%
1	2.96405342236446\\
1.20679264063933	2.95839399810938\\
1.45634847750124	2.95016045280098\\
1.75751062485479	2.93818765055862\\
2.12095088792019	2.92078941787075\\
2.55954792269953	2.89553263327162\\
3.08884359647748	2.85892132171228\\
3.72759372031494	2.80596395019638\\
4.49843266896945	2.72960001865178\\
5.42867543932386	2.61998144792356\\
6.55128556859552	2.46366114766192\\
7.9060432109077	2.24287390765875\\
9.54095476349992	1.93536224420662\\
11.5139539932645	1.51565589388857\\
13.8949549437314	0.959276432150398\\
16.7683293681101	0.251401783450589\\
20.2358964772516	-0.600570937698123\\
24.4205309454865	-1.55609035397241\\
29.4705170255181	-2.5305049221251\\
35.5648030622314	-3.30389335469353\\
42.9193426012878	-3.14592747365205\\
51.794746792312	-0.785341588731043\\
62.5055192527398	4.00804686678733\\
75.4312006335461	11.263764270138\\
91.0298177991523	15.4006172462326\\
109.854114198756	8.89339291990236\\
132.571136559011	4.75018550651237\\
159.985871960606	1.94214365494112\\
193.069772888325	-0.175246180107079\\
232.995181051537	-1.84444297844506\\
281.176869797424	-3.1595180464623\\
339.322177189533	-4.1548859057791\\
409.491506238042	-4.85124810575313\\
494.171336132384	-5.29553554964305\\
596.362331659464	-5.60005129945033\\
719.685673001153	-5.94665947844546\\
868.511373751352	-6.48540985324193\\
1264.8552168553	-7.87477952039931\\
1526.41796717524	-8.45615278877608\\
1842.06996932672	-8.90795324617151\\
2222.9964825262	-9.24514154077049\\
2682.69579527973	-9.48980887684322\\
3237.45754281764	-9.66368718357346\\
3906.93993705462	-9.78549938959894\\
4714.86636345739	-9.87006829141837\\
5689.86602901831	-9.92847245163903\\
6866.488450043	-9.96869127895916\\
8286.42772854686	-9.99634596966889\\
10000	-10.0153474108747\\
};
\addplot [color=red, line width=1.0pt, only marks, mark size=1pt, mark=*, mark options={solid, red}, forget plot]
  table[row sep=crcr]{%
1	2.79316771611997\\
2	2.93582808859932\\
3	2.87214613712686\\
4	2.60567164362747\\
5	2.67865946235413\\
6	2.55545952508442\\
7	2.22769848815154\\
9.00000000000001	2.05037617612877\\
11	1.64025635529593\\
13	1.0008983701392\\
16	0.456456588699883\\
20	-0.526992949080341\\
24	-1.40486448000786\\
29	-2.39736058252297\\
35	-3.20407702680068\\
42	-3.34338940169517\\
51	-1.46057964480929\\
62.0000000000001	3.2466679375369\\
75	10.7129298430731\\
91.0000000000001	15.4608001681953\\
109	8.93739855377316\\
132	4.76898406248716\\
159	2.0655434948303\\
193	0.00206764880795873\\
232	-1.54453616693544\\
281	-2.66216407263262\\
339	-3.5680577410832\\
409	-4.01464278877229\\
494	-4.11107872251825\\
596.000000000001	-4.50401026149557\\
719	-4.76477328767434\\
868.000000000001	-5.63477996991535\\
1048	-6.59936854127589\\
1264	-7.38249710768522\\
1526	-8.31260910549551\\
1842	-8.90114477214798\\
2222	-9.18648732812804\\
2682	-9.39625787395169\\
3237	-9.54669581554073\\
3906.00000000001	-9.68950491769641\\
4714	-9.70816695697086\\
5689	-9.56117242444895\\
6866	-9.61514224426328\\
8286.00000000001	-9.44954954244141\\
10000	-9.11988629164179\\
};
\end{axis}

\begin{axis}[%
width=2.5in,
height=1in,
at={(0.758in,0.7in)},
scale only axis,
xmode=log,
xmin=1,
xmax=5000,
xminorticks=true,
xlabel style={font=\large\color{white!15!black}},
xlabel={Frequency [Hz]},
ymin=-200,
ymax=200,
ylabel style={font=\large\color{white!15!black}},
ylabel={Phase [deg]},
axis background/.style={fill=white},
xmajorgrids,
xminorgrids,
ymajorgrids,
]
\addplot [color=mycolor1, line width=2.0pt, forget plot]
  table[row sep=crcr]{%
1	179.807937239943\\
1.20679264063936	179.769720523181\\
1.45634847750121	179.724735384301\\
1.7575106248548	179.672438311703\\
2.12095088792024	179.612816873317\\
2.55954792269949	179.546979110534\\
3.0888435964775	179.478215921468\\
3.72759372031504	179.413881491244\\
4.49843266896938	179.36866699426\\
5.4286754393239	179.370198376728\\
6.55128556859571	179.468393881864\\
7.9060432109076	179.750617278499\\
9.54095476350004	-179.634939426534\\
11.5139539932641	-178.44590132806\\
13.8949549437312	-176.304319592821\\
16.7683293681103	-172.657965498056\\
20.235896477251	-166.763708828507\\
24.4205309454863	-157.683815044837\\
29.4705170255186	-144.144646446689\\
35.5648030622303	-124.040593268918\\
42.9193426012876	-95.0143958442408\\
51.7947467923131	-60.5024285841124\\
62.5055192527382	-28.1843086782821\\
75.4312006335461	10.4739117531408\\
91.0298177991541	105.415964269491\\
109.854114198753	156.589441351164\\
132.571136559011	176.492878885972\\
159.98587196061	-170.721617849285\\
193.069772888321	-161.36545537255\\
232.995181051538	-154.490172428911\\
281.176869797431	-149.772996205817\\
339.322177189527	-147.120834286403\\
409.491506238045	-146.589036388003\\
494.171336132398	-148.338123346228\\
596.362331659456	-152.456682441515\\
719.685673001159	-158.507319596441\\
868.51137375138	-165.111425951857\\
1048.11313415467	-170.59706141929\\
1264.85521685531	-174.330560858039\\
1526.41796717519	-176.660813189134\\
1842.0699693267	-178.07409849473\\
2222.99648252623	-178.90996562003\\
2682.69579527965	-179.390984252753\\
3237.45754281763	-179.661802845071\\
3906.93993705468	-179.812312998315\\
4714.86636345727	-179.895563922663\\
5689.86602901828	-179.941653587583\\
6866.48845004314	-179.967268023253\\
8286.42772854666	-179.981571113657\\
10000	-179.989594048497\\
};
\addplot [color=red, line width=1.0pt, only marks, mark size=1pt, mark=*, mark options={solid, red}, forget plot]
  table[row sep=crcr]{%
1	179.786248049354\\
2	179.654954133504\\
3	179.477517337454\\
4	179.328262436582\\
5	179.351873738317\\
6	179.373983268567\\
6.99999999999989	179.573856226581\\
9.00000000000001	-179.900548832254\\
11.0000000000001	-178.760262338498\\
13.0000000000003	-177.138543899192\\
16	-173.692562155114\\
20	-167.202443955321\\
24	-158.800109181349\\
29.0000000000003	-145.921795912624\\
34.9999999999994	-126.866051496494\\
41.9999999999993	-100.308188252322\\
50.9999999999994	-64.6084723521013\\
62.0000000000008	-29.3327787004484\\
75	10.2331654879218\\
91.0000000000008	111.057903496368\\
109.000000000001	161.665992649668\\
132.000000000001	-177.655627377382\\
159.000000000005	-164.52476579134\\
192.999999999999	-154.320511310747\\
232.000000000003	-147.19725984566\\
281.000000000003	-142.191079670955\\
339.000000000007	-139.448503505512\\
409.000000000013	-139.646579930595\\
494.000000000008	-142.766838203699\\
596.000000000009	-148.381085001996\\
719.000000000023	-157.319450422658\\
867.999999999998	-164.772672381774\\
1048.00000000001	-172.157348801486\\
1264.00000000001	-176.53510440629\\
1525.99999999999	-178.909479096052\\
1841.99999999994	-178.460535527952\\
2221.99999999996	-179.663376073557\\
2682.00000000004	-179.633406013878\\
3236.99999999993	179.72718188188\\
3905.99999999999	-179.582433171907\\
4714.00000000002	-179.836905119363\\
5689	-179.574407503819\\
6865.9999999998	-179.906625986888\\
8286.00000000018	179.651588871052\\
10000	-179.843863514572\\
};
\end{axis}
\end{tikzpicture}%

%% file: updated_images/GFL_sweep_Z_qd.tex
%
%
\definecolor{mycolor1}{rgb}{0.00000,0.44700,0.74100}%
\begin{tikzpicture}

\begin{axis}[%
width=2.5in,
height=1in,
at={(0.758in,2.0in)},
scale only axis,
xmode=log,
xmin=1,
xmax=5000,
xminorticks=true,
xlabel style={font=\color{white!15!black}},
ymin=-10,
ymax=40,
ylabel style={font=\large\color{white!15!black}},
ylabel={Magnitude [dB]},
axis background/.style={fill=white},
title style={font=\Large\bfseries},
title={$Z_{pcc}^{qd}$},
xmajorgrids,
xminorgrids,
ymajorgrids,
legend pos=north west,
legend style={legend cell align=left, align=left,font=\normalsize, draw=white!15!black}
]
\addplot [color=mycolor1, line width=2.0pt, forget plot]
  table[row sep=crcr]{%
1	2.96406372933073\\
1.20679264063933	2.95840901255889\\
1.45634847750124	2.9501823273687\\
1.75751062485479	2.93821952513052\\
2.12095088792019	2.92083587557905\\
2.55954792269953	2.89560037065204\\
3.08884359647748	2.85902013744035\\
3.72759372031494	2.80610821171485\\
4.49843266896945	2.72981085354449\\
5.42867543932386	2.62029004939563\\
6.55128556859552	2.464113819444\\
7.9060432109077	2.24353986705893\\
9.54095476349992	1.93634584885124\\
11.5139539932645	1.51711596554121\\
13.8949549437314	0.961456726217639\\
16.7683293681101	0.25467800837597\\
20.2358964772516	-0.595621961614047\\
24.4205309454865	-1.54859880808232\\
29.4705170255181	-2.51924756310341\\
35.5648030622314	-3.28778332122659\\
42.9193426012878	-3.1270894501614\\
51.794746792312	-0.770922447359881\\
62.5055192527398	4.0166472521467\\
75.4312006335461	11.2696006944099\\
91.0298177991523	15.4047599594962\\
109.854114198756	8.89459443549024\\
132.571136559011	4.74542838762814\\
159.985871960606	1.92691337094997\\
193.069772888325	-0.206786533949437\\
232.995181051537	-1.89912534073583\\
281.176869797424	-3.2443003409371\\
339.322177189533	-4.27484303433601\\
409.491506238042	-5.00576797436234\\
494.171336132384	-5.47273517333663\\
719.685673001153	-6.0733607513208\\
868.511373751352	-6.5382030871993\\
1048.11313415469	-7.17266568210562\\
1264.8552168553	-7.84082754546536\\
1526.41796717524	-8.42573660854482\\
1842.06996932672	-8.8881980244718\\
2222.9964825262	-9.2339300142592\\
2682.69579527973	-9.48375593701096\\
3237.45754281764	-9.66042347186307\\
3906.93993705462	-9.78369183602583\\
4714.86636345739	-9.86902695567856\\
5689.86602901831	-9.92784757770896\\
6866.488450043	-9.96830270901258\\
8286.42772854686	-9.99609742678678\\
10000	-10.0151850523542\\
};
\addplot [color=red, line width=1.0pt, only marks, mark size=1pt, mark=*, mark options={solid, red}, forget plot]
  table[row sep=crcr]{%
1	2.97318030686789\\
2	3.08120648054978\\
3	3.03817825616199\\
4	2.807008434735\\
5	2.84233032892539\\
6	2.73844644394873\\
7	2.61037099591302\\
9.00000000000001	2.24563477911416\\
11	1.83505246499024\\
13	1.36741315055449\\
16	0.640195086913828\\
20	-0.342892077758881\\
24	-1.19947711052343\\
29	-2.25768486172621\\
35	-3.27706326242635\\
42	-3.01439042003372\\
51	-0.990397245076078\\
62.0000000000001	3.86788774884358\\
75	11.2618809830297\\
91.0000000000001	15.9687771305401\\
109	9.35675603505625\\
132	5.05685822750527\\
159	2.3530032915169\\
193	0.173353135391928\\
232	-1.40767461821575\\
281	-2.73665617921981\\
339	-3.64295864988582\\
409	-4.17033516174652\\
494	-4.45610884372955\\
596.000000000001	-4.65050414655441\\
719	-5.06377490464572\\
868.000000000001	-5.85623723125322\\
1048	-6.58184089060023\\
1264	-7.57025126288651\\
1526	-8.31986970498507\\
1842	-8.84721927243687\\
2222	-9.31174361459145\\
2682	-9.56444618846835\\
3237	-9.6570685769328\\
3906.00000000001	-9.84062095252952\\
4714	-9.62407321685118\\
5689	-9.67076233953194\\
6866	-9.53065199061956\\
8286.00000000001	-9.29910072708283\\
10000	-9.25463008307099\\
};
\end{axis}

\begin{axis}[%
width=2.5in,
height=1in,
at={(0.758in,0.7in)},
scale only axis,
xmode=log,
xmin=1,
xmax=5000,
xminorticks=true,
xlabel style={font=\large\color{white!15!black}},
xlabel={Frequency [Hz]},
ymin=-200,
ymax=200,
ylabel style={font=\large\color{white!15!black}},
ylabel={Phase [deg]},
axis background/.style={fill=white},
xmajorgrids,
xminorgrids,
ymajorgrids,
]
\addplot [color=mycolor1, line width=2.0pt, forget plot]
  table[row sep=crcr]{%
1	-0.192062880795646\\
1.20679264063936	-0.230279690081602\\
1.45634847750121	-0.275264993231787\\
1.7575106248548	-0.327562358779261\\
2.12095088792024	-0.387184322876124\\
2.55954792269949	-0.453023037350391\\
3.0888435964775	-0.52178796995554\\
3.72759372031504	-0.586125645667693\\
4.49843266896938	-0.631346309065407\\
5.4286754393239	-0.629826942503115\\
6.55128556859571	-0.531655550854992\\
7.9060432109076	-0.249482118242497\\
9.54095476350004	0.364854285968562\\
11.5139539932641	1.5536570480063\\
13.8949549437312	3.69470910221293\\
16.7683293681103	7.33985580977802\\
20.235896477251	13.2313511380227\\
24.4205309454863	22.3049206987642\\
29.4705170255186	35.8293817418257\\
35.5648030622303	55.8988364734163\\
42.9193426012876	84.8564681332039\\
51.7947467923131	119.296759573348\\
62.5055192527382	151.578461847268\\
75.4312006335461	-169.795133119547\\
91.0298177991541	-74.9021379183192\\
109.854114198753	-23.7979126769574\\
132.571136559011	-3.9802838255799\\
159.98587196061	8.71028261982843\\
193.069772888321	17.9742830901058\\
232.995181051538	24.7784917842598\\
281.176869797431	29.4726249412246\\
339.322177189527	32.1857083484516\\
409.491506238045	32.9011182449199\\
494.171336132398	31.4793618313871\\
596.362331659456	27.7939035246319\\
719.685673001159	22.139126698127\\
868.51137375138	15.6814835148029\\
1048.11313415467	10.0253696033762\\
1264.85521685531	6.00306183594418\\
1526.41796717519	3.46586293080992\\
1842.0699693267	1.95619620017504\\
2222.99648252623	1.08802733587186\\
2682.69579527965	0.600480348327068\\
3237.45754281763	0.33080537139071\\
3906.93993705468	0.182679360708846\\
4714.86636345727	0.10134837907745\\
5689.86602901828	0.0565221056301652\\
6866.48845004314	0.0316757146613611\\
8286.42772854666	0.0178231683267143\\
10000	0.0100601565792715\\
};
\addplot [color=red, line width=1.0pt, only marks, mark size=1pt, mark=*, mark options={solid, red}, forget plot]
  table[row sep=crcr]{%
1	-0.212797230472432\\
2	-0.441661960930617\\
3	-0.376624033458995\\
4	-0.655466547908986\\
5	-0.499547306056826\\
6	-0.349954297510578\\
6.99999999999989	-0.292383684163553\\
9.00000000000001	0.285888670338579\\
11.0000000000001	1.48759694155947\\
13.0000000000003	2.90841931721025\\
16	6.35037781366583\\
20	13.355345756243\\
24	21.936671055053\\
29.0000000000003	34.6743356581076\\
34.9999999999994	54.8345964750509\\
41.9999999999993	81.4662820741206\\
50.9999999999994	117.091226907201\\
62.0000000000008	151.831624295873\\
75	-169.990166837766\\
91.0000000000008	-69.7144319012862\\
109.000000000001	-20.3169028125244\\
132.000000000001	-0.00453355155377722\\
159.000000000005	13.6325823079465\\
192.999999999999	23.3344556095787\\
232.000000000003	30.3464458903821\\
281.000000000003	35.4915977682156\\
339.000000000007	38.6728929000049\\
409.000000000013	38.9547594148161\\
494.000000000008	36.5928680414931\\
596.000000000009	31.0678025153759\\
719.000000000023	23.2534643869475\\
867.999999999998	14.9428488862532\\
1048.00000000001	7.69711125802343\\
1264.00000000001	4.31626823765856\\
1525.99999999999	1.88213130841285\\
1841.99999999994	1.05602828324226\\
2221.99999999996	0.351477244584771\\
2682.00000000004	0.565496246161132\\
3236.99999999993	0.427891766328173\\
3905.99999999999	-0.357594486822535\\
4714.00000000002	0.471066539168845\\
5689	0.0763060549557792\\
6865.9999999998	0.113436981612921\\
8286.00000000018	-0.0573309358434813\\
10000	0.563870850466969\\
};
\end{axis}
\end{tikzpicture}%

%% file: updated_images/GFL_sweep_Z_qq.tex
%
%
\definecolor{mycolor1}{rgb}{0.00000,0.44700,0.74100}%
\begin{tikzpicture}

\begin{axis}[%
width=2.5in,
height=1in,
at={(0.758in,2.0in)},
scale only axis,
xmode=log,
xmin=1,
xmax=5000,
xminorticks=true,
xlabel style={font=\color{white!15!black}},
ymin=-10,
ymax=40,
ylabel style={font=\large\color{white!15!black}},
ylabel={Magnitude [dB]},
axis background/.style={fill=white},
title style={font=\Large\bfseries},
title={$Z_{pcc}^{qq}$},
xmajorgrids,
xminorgrids,
ymajorgrids,
legend pos=north west,
legend style={legend cell align=left, align=left,font=\normalsize, draw=white!15!black}
]
\addplot [color=mycolor1, line width=2.0pt, forget plot]
  table[row sep=crcr]{%
1	18.5146432000681\\
1.20679264063934	18.5113214896325\\
1.45634847750124	18.5064884853857\\
1.7575106248548	18.4994596122894\\
2.12095088792018	18.4892435799764\\
2.55954792269953	18.4744087310807\\
3.0888435964775	18.4528953562829\\
3.72759372031492	18.4217568702396\\
4.49843266896945	18.3768129147444\\
5.4286754393239	18.3122057322305\\
6.5512855685955	18.2198786286761\\
7.90604321090773	18.0890640522405\\
9.54095476349988	17.9060206570768\\
11.5139539932645	17.65457215319\\
13.8949549437315	17.3186239553769\\
16.76832936811	16.8890232440684\\
20.2358964772516	16.3789955321531\\
24.4205309454863	15.8524463669729\\
29.4705170255181	15.453025773582\\
35.5648030622315	15.3645878690665\\
42.9193426012876	15.6306073477462\\
51.7947467923122	16.0265376178741\\
62.5055192527392	16.2149360651448\\
75.4312006335461	15.8841723284034\\
91.0298177991526	16.3282815630737\\
109.854114198755	15.8137094649965\\
132.571136559011	14.8026422227527\\
159.985871960607	13.6480259550568\\
193.069772888325	12.4097475718777\\
232.995181051538	11.1926941262708\\
281.176869797421	10.1638000821973\\
339.322177189533	9.52660057235605\\
409.491506238045	9.39809458084608\\
494.171336132382	9.6729348291587\\
596.362331659466	10.0818490250352\\
719.685673001147	10.4637193080833\\
868.511373751352	11.1477139212711\\
1048.11313415469	12.7814019933908\\
1264.85521685529	15.1719160392877\\
1526.41796717524	17.642421777986\\
1842.06996932673	19.9176125271489\\
2222.99648252619	21.9931698452038\\
2682.69579527974	23.9226038803022\\
3237.45754281763	25.7534935359351\\
3906.93993705462	27.519111552938\\
4714.86636345743	29.2414364043611\\
6866.48845004303	32.6088458529736\\
10000	35.9216207329276\\
};
\addplot [color=red, line width=1.0pt, only marks, mark size=1pt, mark=*, mark options={solid, red}, forget plot]
  table[row sep=crcr]{%
1	18.5171945693244\\
2	18.4947144682113\\
3	18.4560747029554\\
4	18.4054543677097\\
5	18.3401929396166\\
6	18.2712456905356\\
7	18.1820716883474\\
9.00000000000001	17.9742186520253\\
10.9999999999999	17.7217243144516\\
12.9999999999999	17.444928072447\\
16	16.9994189515494\\
20	16.4072865994064\\
24	15.895229876912\\
28.9999999999999	15.4590203231201\\
35	15.3307191390048\\
42	15.5555235445431\\
51.0000000000002	15.9663257095583\\
61.9999999999998	16.1765137435917\\
75	15.8684486595661\\
90.9999999999993	16.4049970210643\\
109	15.82666383013\\
131.999999999999	14.7945374940865\\
159	13.6786310498626\\
192.999999999999	12.4508436876146\\
231.999999999999	11.3645609140933\\
280.999999999999	10.4946367196468\\
339.000000000002	10.0620012728785\\
409	10.0894820697033\\
494	10.2792233078093\\
595.999999999999	10.3482736456909\\
719	10.1728952356916\\
867.999999999998	10.5993382457488\\
1047.99999999999	12.5101736076217\\
1263.99999999999	15.1311544053698\\
1525.99999999999	17.6591937372558\\
1842	19.9447699644341\\
2222	22.0098519753295\\
2682	23.9489342669628\\
3236.99999999999	25.7863818383564\\
3905.99999999999	27.5659377156503\\
4714.00000000002	29.3065465714878\\
5689	31.0291507857014\\
6866.00000000003	32.7461393169589\\
8286.00000000004	34.4715347635131\\
10000	36.2163661373959\\
};
\end{axis}

\begin{axis}[%
width=2.5in,
height=1in,
at={(0.758in,0.7in)},
scale only axis,
xmode=log,
xmin=1,
xmax=5000,
xminorticks=true,
xlabel style={font=\large\color{white!15!black}},
xlabel={Frequency [Hz]},
ymin=-200,
ymax=200,
ylabel style={font=\large\color{white!15!black}},
ylabel={Phase [deg]},
axis background/.style={fill=white},
xmajorgrids,
xminorgrids,
ymajorgrids,
]
\addplot [color=mycolor1, line width=2.0pt, forget plot]
  table[row sep=crcr]{%
1	-179.896842643428\\
1.20679264063936	-179.874692003525\\
1.45634847750121	-179.847341854088\\
1.7575106248548	-179.813249799238\\
2.12095088792024	-179.770203173979\\
2.55954792269949	-179.714918290187\\
3.0888435964775	-179.642364314541\\
3.72759372031504	-179.54461819808\\
4.49843266896938	-179.408927064962\\
5.4286754393239	-179.214448510386\\
6.55128556859571	-178.926829609177\\
7.9060432109076	-178.489358877233\\
9.54095476350004	-177.808931339856\\
11.5139539932641	-176.734745607787\\
13.8949549437312	-175.02832472199\\
16.7683293681103	-172.327910962879\\
20.235896477251	-168.127970922515\\
24.4205309454863	-161.850386980197\\
29.4705170255186	-153.188213468898\\
42.9193426012876	-132.480472401432\\
51.7947467923131	-123.835402676274\\
62.5055192527382	-117.020082314476\\
75.4312006335461	-112.306365537031\\
91.0298177991541	-110.984281100587\\
109.854114198753	-99.8034110621729\\
132.571136559011	-90.9682304475654\\
159.98587196061	-81.584710008104\\
193.069772888321	-70.9113501094386\\
232.995181051538	-58.488979257462\\
281.176869797431	-44.1018499925121\\
409.491506238045	-12.2859097320175\\
494.171336132398	2.17861464124479\\
596.362331659456	15.0187547728794\\
719.685673001159	28.1716642076423\\
868.51137375138	44.092524928629\\
1048.11313415467	60.6314154094204\\
1264.85521685531	72.4629928308732\\
1526.41796717519	79.0923112143351\\
1842.0699693267	82.6563102762813\\
2222.99648252623	84.6980238300662\\
2682.69579527965	85.9791192968887\\
3237.45754281763	86.8521409861623\\
3906.93993705468	87.485696563267\\
4714.86636345727	87.9661264729247\\
5689.86602901828	88.3413500672202\\
6866.48845004314	88.640191504127\\
8286.42772854666	88.8813028564297\\
10000	89.0775228487913\\
};
\addplot [color=red, line width=1.0pt, only marks, mark size=1pt, mark=*, mark options={solid, red}, forget plot]
  table[row sep=crcr]{%
1	-179.923807866818\\
2	-179.780218853878\\
3	-179.628489100368\\
4	-179.521389940344\\
5	-179.310572397238\\
6	-179.094193577298\\
6.99999999999989	-178.80065567046\\
9.00000000000001	-178.054486375107\\
11.0000000000001	-177.025272843703\\
13.0000000000003	-175.733693527161\\
16	-173.125477418837\\
20	-168.450727021366\\
24	-162.482345779645\\
29.0000000000003	-153.978201206095\\
34.9999999999994	-143.638279665356\\
41.9999999999993	-133.475978105774\\
50.9999999999994	-124.338991819137\\
62.0000000000008	-117.05826946893\\
75	-112.171633163852\\
91.0000000000008	-110.24877327095\\
109.000000000001	-99.0699468212185\\
132.000000000001	-89.9180632733433\\
159.000000000005	-80.2902248925381\\
192.999999999999	-68.9537583637558\\
232.000000000003	-56.4607823644548\\
281.000000000003	-41.8127922198449\\
339.000000000007	-26.8310130983707\\
409.000000000013	-12.7219176311133\\
494.000000000008	-0.697960378809256\\
596.000000000009	10.2521195483966\\
719.000000000023	23.3362206291352\\
867.999999999998	42.9116659464304\\
1048.00000000001	61.816011507508\\
1264.00000000001	73.4575240787416\\
1525.99999999999	79.5607539242462\\
1841.99999999994	82.813360481425\\
2221.99999999996	84.7930888338707\\
2682.00000000004	86.0126575716383\\
3236.99999999993	86.8640994310835\\
3905.99999999999	87.496987168055\\
4714.00000000002	87.9865234392815\\
5689	88.3578396329309\\
6865.9999999998	88.6625123424409\\
8286.00000000018	88.9100657049152\\
10000	89.1086515381997\\
};
\end{axis}
\end{tikzpicture}%

%% file: updated_images/Case_study1_dd.tex
%
%
\definecolor{mycolor1}{rgb}{1.00000,0.00000,1.00000}%
\begin{tikzpicture}

\begin{axis}[%
width=2.5in,
height=1in,
at={(0.758in,2.0in)},
scale only axis,
xmode=log,
xmin=1,
xmax=10000,
xminorticks=true,
xlabel style={font=\color{white!15!black}},
title style={font=\Large\bfseries},
title={$d-$ axis},
ymin=-40,
ymax=60,
ylabel style={font=\large\color{white!15!black}},
ylabel={Magnitude [dB]},
axis background/.style={fill=white},
xmajorgrids,
xminorgrids,
ymajorgrids,
legend pos=north west,
legend style={legend cell align=left, align=left,font=\normalsize, draw=white!15!black}
]
\addplot [color=blue, line width=2.0pt]
  table[row sep=crcr]{%
1	-42.7315733898626\\
1.45082877849593	-39.5284083780336\\
1.59228279334109	-38.7024497256656\\
1.74752840000768	-37.8386416057606\\
1.91791026167249	-36.8998488930382\\
2.10490414451203	-35.8193156950682\\
2.31012970008318	-34.4841344904586\\
2.5353644939701	-32.7183702584185\\
2.78255940220711	-30.2537545941294\\
3.05385550883341	-26.3920999954556\\
3.35160265093885	-21.3879838436107\\
3.67837977182865	-19.4872339519779\\
4.03701725859658	-17.8563036983196\\
4.43062145758385	-16.3267052835565\\
5.3366992312063	-13.3999112595374\\
7.05480231071869	-9.14558026972008\\
9.32603346883218	-4.97572967585681\\
12.3284673944207	-0.899188180932271\\
14.8496826225446	1.74392199501941\\
17.8864952905744	4.30430479821947\\
21.544346900319	6.75726642966713\\
23.6448941264543	7.93457688760577\\
25.9502421139972	9.07387292854612\\
28.4803586843579	10.1709765618665\\
31.2571584968824	11.2215768709465\\
34.3046928631493	12.2213137361635\\
37.6493580679249	13.1658845984158\\
41.3201240011537	14.0511716152298\\
45.3487850812855	14.8733833940523\\
49.7702356433209	15.6292026411884\\
54.6227721768434	16.3159293036343\\
59.9484250318942	16.9316087200798\\
65.7933224657571	17.475136093396\\
72.2080901838551	17.9463317530679\\
79.2482898353912	18.345985103057\\
86.974900261778	18.6758675070738\\
95.4548456661833	18.9387145291596\\
104.761575278967	19.1381753854565\\
114.975699539774	19.2787222748841\\
126.185688306603	19.3655049283159\\
138.488637139386	19.4041267574761\\
151.991108295293	25.2308266768278\\
183.073828029537	23.8801031178161\\
220.513073990306	22.5991913848815\\
291.505306282517	20.7438437770082\\
319.926713779738	20.0999416936072\\
351.119173421514	19.4077312425375\\
385.352859371055	18.6352614234324\\
422.924287438953	17.7447462948769\\
464.158883361275	16.6950232815978\\
509.413801481636	15.4447435460786\\
559.081018251222	13.957012235609\\
613.590727341319	12.2127548637999\\
673.415065775086	7.19890343317218\\
739.072203352584	5.87842826105234\\
811.130830789682	6.35278264492545\\
890.215085445036	8.14973955424826\\
977.009957299225	10.2141534408827\\
1072.26722201033	11.5764667750816\\
1176.811952435	13.3680955118914\\
1291.54966501489	14.9760670677143\\
1417.4741629268	16.4183922723271\\
1555.67614393047	17.7273027793698\\
1707.35264747069	18.9320172381526\\
1873.81742286039	20.0558086929464\\
2056.51230834866	21.1165520249247\\
2257.01971963394	22.1278713168838\\
2477.07635599169	23.1001808069495\\
2983.64724028333	24.9580352501163\\
3593.81366380464	26.7352155958482\\
4328.76128108303	28.4596495147576\\
5722.36765935022	30.9844856746273\\
8302.1756813197	34.286090134564\\
10000	35.9215019748263\\
};
\addlegendentry{${Z}_{s}$}
\addplot [color=mycolor1, line width=2.0pt]
  table[row sep=crcr]{%
1	-26.3047331018141\\
1417.4741629268	36.7237149018384\\
10000	53.6934118706305\\
};
\addlegendentry{${Z}_{g}$}
\end{axis}

\begin{axis}[%
width=2.5in,
height=1in,
at={(0.758in,0.7in)},
scale only axis,
xmode=log,
xmin=1,
xmax=10000,
xminorticks=true,
xlabel style={font=\large\color{white!15!black}},
xlabel={Frequency [Hz]},
ymin=-200,
ymax=200,
ylabel style={font=\large\color{white!15!black}},
ylabel={Phase [deg]},
axis background/.style={fill=white},
xmajorgrids,
xminorgrids,
ymajorgrids,
legend pos=north east,
legend style={legend cell align=left, align=left,font=\scriptsize, draw=white!15!black}
]
\addplot [color=blue, line width=2.0pt]
  table[row sep=crcr]{%
1	86.414868431879\\
1.09749876549306	85.6595575201867\\
1.20450354025879	84.6046608031883\\
1.32194114846604	83.1532706776088\\
1.45082877849596	81.1753748792707\\
1.59228279334111	78.4981380278193\\
1.74752840000771	74.8969345018737\\
1.91791026167246	70.0945603932047\\
2.104904144512	63.7870316812414\\
2.31012970008314	55.728832211168\\
2.5353644939701	45.8991229479936\\
2.78255940220711	34.6527892639989\\
3.05385550883341	22.531749006557\\
3.35160265093885	36.3450999170092\\
3.67837977182865	47.2158971667956\\
4.03701725859658	53.9996920709843\\
4.43062145758392	58.8730046576312\\
4.86260158006541	62.5459177292419\\
5.33669923120639	65.3564658925613\\
5.85702081805658	67.4983187246115\\
6.42807311728424	69.095786253988\\
7.05480231071857	70.2348816457715\\
7.74263682681121	70.9781630102463\\
8.4975343590864	71.3725860588687\\
9.32603346883218	71.454012823804\\
10.2353102189903	71.2499968557462\\
11.2332403297803	70.7816305108343\\
12.3284673944207	70.0648616264897\\
13.5304777457982	69.1115030019814\\
14.8496826225448	67.9300638824279\\
16.2975083462067	66.5264822127409\\
17.8864952905746	64.9048080672213\\
19.6304065004024	63.0678713543384\\
21.5443469003186	61.0179546783314\\
23.6448941264539	58.7574817075\\
25.9502421139972	56.2897206310934\\
28.4803586843579	53.6194904388459\\
31.2571584968824	50.7538450032491\\
34.3046928631493	47.7026975266396\\
37.6493580679249	44.4793380879531\\
41.3201240011537	41.1007926776042\\
45.3487850812863	37.58797593236\\
54.6227721768443	30.2618477086452\\
79.2482898353912	15.2793416003292\\
86.974900261778	11.6610230599041\\
95.4548456661833	8.15810361928023\\
104.761575278967	4.79770537751213\\
114.975699539774	1.6017441954787\\
126.185688306603	-1.4147257676488\\
138.488637139389	-4.24533557223522\\
151.991108295295	-13.12094432166\\
166.810053720008	-15.6526984930672\\
183.073828029534	-17.9143318531909\\
200.923300256502	-19.9255782358581\\
220.513073990302	-21.7206375633144\\
242.012826479436	-23.3469993488829\\
265.608778294667	-24.8615425282211\\
385.352859371055	-30.6680110324405\\
422.924287438953	-31.937684375391\\
464.158883361283	-32.8043111204909\\
509.413801481645	-32.86770289679\\
559.081018251231	-31.5007946437017\\
613.590727341309	-27.7012482185941\\
673.415065775075	-11.9480215818114\\
739.072203352571	10.0558201545601\\
811.130830789682	34.7550453599015\\
890.215085445036	53.0894718798532\\
977.009957299225	64.2913823779996\\
1072.26722201033	65.1996785360997\\
1176.811952435	71.7107736413594\\
1291.54966501489	75.9190481407524\\
1417.47416292682	78.7739020162993\\
1555.67614393049	80.800147483335\\
1707.35264747072	82.2962733649834\\
1873.81742286036	83.4390764802555\\
2056.51230834863	84.3375599933207\\
2257.0197196339	85.0614355926579\\
2477.07635599169	85.6567753748973\\
2718.58824273293	86.1549409344071\\
2983.64724028333	86.577860049593\\
3274.54916287773	86.9412454827499\\
3593.81366380464	87.2566149198371\\
3944.20605943768	87.5325911421961\\
4328.7612810831	87.7757582219207\\
4750.81016210285	87.9912371157108\\
5214.00828799976	88.1830799750641\\
5722.36765935031	88.3545449806024\\
6280.29144183417	88.5082909924935\\
6892.61210434962	88.646517485118\\
7564.63327554623	88.7710665699344\\
8302.1756813197	88.883498376356\\
9111.62756115487	88.9851474635764\\
10000	89.0771655608905\\
};

\addplot [color=mycolor1, line width=2.0pt]
  table[row sep=crcr]{%
1	88.8158950330041\\
1.09749876549306	88.9210615767232\\
1.20450354025879	89.0168917639324\\
1.32194114846604	89.1042134650887\\
1.45082877849596	89.1837813811422\\
1.59228279334111	89.2562834521707\\
1.74752840000771	89.3223467189517\\
1.91791026167246	89.382542680663\\
2.104904144512	89.4373921893748\\
2.31012970008314	89.4873699193824\\
2.5353644939701	89.5329084468019\\
2.78255940220711	89.5744019722728\\
3.05385550883341	89.6122097171298\\
3.35160265093885	89.6466590210324\\
3.67837977182865	89.6780481668042\\
4.03701725859658	89.7066489561331\\
4.43062145758392	89.7327090578236\\
4.86260158006541	89.7564541484745\\
5.33669923120639	89.7780898637683\\
5.85702081805658	89.7978035770076\\
6.42807311728424	89.8157660200995\\
7.05480231071857	89.8321327608767\\
7.74263682681121	89.8470455494336\\
8.4975343590864	89.8606335450525\\
10.2353102189903	89.8842954497863\\
12.3284673944207	89.9039400091235\\
14.8496826225448	89.9202492847605\\
17.8864952905746	89.9337895418501\\
21.5443469003186	89.9450309068466\\
25.9502421139972	89.9543636888089\\
34.3046928631493	89.9654778010562\\
45.3487850812863	89.9738852211606\\
59.9484250318932	89.9802451269156\\
86.974900261778	89.9863837319827\\
138.488637139389	89.9914485867526\\
242.012826479436	89.9951065669147\\
559.081018251231	89.9978817496287\\
2056.51230834863	89.9994241349443\\
10000	89.9998815726425\\
};

\end{axis}
\end{tikzpicture}%

%% file: updated_images/Case_study1_qq.tex
%
%
\definecolor{mycolor1}{rgb}{1.00000,0.00000,1.00000}%
\begin{tikzpicture}

\begin{axis}[%
width=2.5in,
height=1in,
at={(0.758in,2.0in)},
scale only axis,
xmode=log,
xmin=1,
xmax=10000,
xminorticks=true,
xlabel style={font=\color{white!15!black}},
title style={font=\Large\bfseries},
title={$q-$ axis},
ymin=-40,
ymax=60,
ylabel style={font=\large\color{white!15!black}},
ylabel={Magnitude [dB]},
axis background/.style={fill=white},
xmajorgrids,
xminorgrids,
ymajorgrids,
legend pos=north west,
legend style={legend cell align=left, align=left,font=\normalsize, draw=white!15!black}
]
\addplot [color=blue, line width=2.0pt, forget plot]
  table[row sep=crcr]{%
1	-17.621202383967\\
1.09749876549306	-17.6175975083759\\
1.20450354025879	-17.6191548361158\\
1.32194114846604	-17.6276706252622\\
1.45082877849596	-17.6460232916212\\
1.59228279334111	-17.6787358158248\\
1.74752840000771	-17.7329145452313\\
1.91791026167246	-17.8199047538999\\
2.104904144512	-17.9584513500683\\
2.31012970008314	-18.1814648738714\\
2.5353644939701	-18.553225272387\\
2.78255940220711	-19.2278671048155\\
3.05385550883341	-20.8458182229281\\
3.35160265093885	-23.1908428658978\\
3.67837977182865	-22.1252252574687\\
4.03701725859658	-20.5999330856092\\
4.43062145758392	-18.8725119470636\\
5.33669923120639	-15.1791507802438\\
10.2353102189903	-1.91462908854727\\
13.5304777457982	3.8541048975265\\
16.2975083462067	7.79801954120073\\
19.6304065004024	11.8825182584772\\
21.5443469003186	14.0041974251538\\
23.6448941264539	16.2009884179334\\
25.9502421139972	18.4984678116501\\
28.4803586843579	20.9349543796966\\
31.2571584968824	23.5710834087788\\
34.3046928631493	26.5100506305211\\
37.6493580679249	29.947034927776\\
41.3201240011537	34.3198205918671\\
45.3487850812863	40.9981290682112\\
49.7702356433217	68.2949359615386\\
54.6227721768443	43.4069603732174\\
59.9484250318932	37.8402614757939\\
65.793322465756	34.854846914966\\
72.208090183854	32.8603798613879\\
79.2482898353912	31.3716604218111\\
86.974900261778	30.17573616228\\
95.4548456661833	29.1609615700661\\
104.761575278967	28.2625950391853\\
114.975699539774	27.4409334067264\\
126.185688306603	26.6710656035125\\
138.488637139389	25.9374681125259\\
151.991108295295	19.4003087884573\\
166.810053720008	19.359396808177\\
183.073828029534	19.2856579889604\\
200.923300256502	19.181311439111\\
220.513073990302	19.0452549776576\\
242.012826479436	18.8715096453334\\
265.608778294667	18.6475335938325\\
291.505306282517	18.3527781294702\\
319.926713779738	17.9581288524417\\
351.119173421514	17.4270004503572\\
385.352859371055	16.718460435534\\
422.924287438953	15.7915957197091\\
464.158883361283	14.609110388766\\
509.413801481645	13.1390652907234\\
559.081018251231	11.3604215767323\\
613.590727341309	9.30061177506748\\
673.415065775075	10.2624864167623\\
739.072203352571	8.39608678412657\\
811.130830789682	7.39884670909453\\
890.215085445036	7.99142243804279\\
977.009957299225	9.66623918948271\\
1072.26722201033	12.1239115633923\\
1176.811952435	13.8169954875067\\
1291.54966501489	15.3222394325178\\
1417.47416292682	16.6795122302664\\
1555.67614393049	17.9226783408984\\
1707.35264747072	19.0778218004236\\
1873.81742286036	20.1645787103253\\
2056.51230834863	21.1977324603887\\
2257.0197196339	22.1885024316548\\
2477.07635599169	23.145492836574\\
2983.64724028333	24.9833678013648\\
3593.81366380464	26.7493672312363\\
4328.7612810831	28.4675246909834\\
5722.36765935031	30.9877026475195\\
8302.1756813197	34.2870034263389\\
10000	35.9219611561418\\
};
\addplot [color=mycolor1, line width=2.0pt, forget plot]
  table[row sep=crcr]{%
1	-26.3047331018141\\
1417.4741629268	36.7237149018384\\
10000	53.6934118706305\\
};
\draw[line width=0.5mm,draw=black!70,loosely dashed] (axis cs:7,-50)--(axis cs:7,70);
\node[color=black, fill=none,font=\large] at (axis cs: 20,-20) {\SI{7}{\hertz}};
\end{axis}

\begin{axis}[%
width=2.5in,
height=1in,
at={(0.758in,0.7in)},
scale only axis,
xmode=log,
xmin=1,
xmax=10000,
xminorticks=true,
xlabel style={font=\large\color{white!15!black}},
xlabel={Frequency [Hz]},
ymin=-200,
ymax=200,
ylabel style={font=\large\color{white!15!black}},
ylabel={Phase [deg]},
axis background/.style={fill=white},
xmajorgrids,
xminorgrids,
ymajorgrids,
legend pos=north east,
legend style={legend cell align=left, align=left,font=\scriptsize, draw=white!15!black}
]
\addplot [color=blue, line width=2.0pt, forget plot]
  table[row sep=crcr]{%
1	-4.85007313819946\\
1.09749876549306	-3.96103260991904\\
1.20450354025879	-3.11286350701329\\
1.32194114846604	-2.30003015015956\\
1.45082877849596	-1.51712288560336\\
1.74752840000771	-0.0172765794085876\\
2.10490414451206	1.45542891083394\\
2.31012970008321	2.22924217551022\\
2.53536449397018	3.09418744601572\\
2.78255940220721	4.19102312544834\\
3.05385550883351	5.91836732882163\\
3.35160265093874	-17.5333159773886\\
3.67837977182853	-36.6717442822464\\
4.03701725859645	-50.2230849252523\\
4.43062145758378	-60.5409189830104\\
4.86260158006525	-68.6184166663471\\
5.33669923120621	-75.0710309359528\\
5.85702081805658	-80.324795113828\\
6.42807311728424	-84.6867945845157\\
7.05480231071857	-88.3836539528602\\
7.74263682681121	-91.5859460128898\\
8.4975343590864	-94.4244733467592\\
9.32603346883218	-97.0013688847467\\
10.2353102189903	-99.39778196072\\
12.3284673944207	-103.899656802756\\
14.8496826225448	-108.329277770926\\
16.2975083462067	-110.608432944635\\
17.8864952905746	-112.968684101584\\
19.6304065004031	-115.433492888307\\
21.5443469003193	-118.022664844422\\
23.6448941264547	-120.75250468599\\
25.9502421139981	-123.635813343357\\
28.4803586843589	-126.681793138172\\
31.2571584968834	-129.895968302924\\
34.3046928631481	-133.280353952817\\
37.6493580679236	-136.834587842885\\
41.3201240011523	-140.561311814283\\
45.3487850812848	-144.503718232496\\
49.7702356433201	-153.332583133882\\
54.6227721768425	28.0262563031531\\
65.793322465756	19.6024301934519\\
79.2482898353912	11.332491821985\\
86.974900261778	7.30414764084713\\
95.4548456661833	3.39637158845821\\
104.761575278967	-0.352913650252589\\
114.975699539774	-3.90781506274556\\
126.185688306603	-7.236496066922\\
138.488637139389	-10.3132990601686\\
151.991108295295	-6.89433971486037\\
166.810053720008	-9.37851392046755\\
183.07382802954	-11.7286863350366\\
242.012826479444	-18.4909631100628\\
265.608778294676	-20.862297998566\\
291.505306282527	-23.3819662005846\\
385.352859371042	-31.4356941253248\\
422.92428743894	-33.6642079932287\\
464.158883361268	-35.0276940254467\\
509.413801481628	-34.8795991821139\\
559.081018251213	-32.2387648787272\\
613.590727341309	-25.4665317489669\\
673.415065775075	-19.8305098230907\\
739.072203352571	-5.59653357961747\\
811.130830789682	15.646993524965\\
890.215085445036	38.076472580311\\
1072.26722201033	71.0736944824877\\
1176.811952435	75.3918503339892\\
1291.54966501489	78.3020607172405\\
1417.47416292682	80.3655790862761\\
1555.67614393049	81.8929825635948\\
1707.35264747072	83.0648081436345\\
1873.81742286042	83.9909882763855\\
2056.51230834869	84.7412927666877\\
2257.01971963397	85.3616515482601\\
2477.07635599177	85.8832983162603\\
2718.58824273302	86.328099013579\\
2983.64724028343	86.7117681363465\\
3274.54916287762	87.0458717752202\\
3593.81366380452	87.3391104047833\\
3944.20605943755	87.5981612947153\\
4328.76128108296	87.8282444144967\\
4750.8101621027	88.0335104955434\\
5214.00828799959	88.2173121763132\\
5722.36765935013	88.3823967314708\\
6280.29144183417	88.5310452367819\\
6892.61210434962	88.6651745272827\\
7564.63327554623	88.7864129109871\\
8302.1756813197	88.8961571078311\\
9111.62756115487	88.9956155828255\\
10000	89.0858419023833\\
};
\addplot [color=mycolor1, line width=2.0pt, forget plot]
  table[row sep=crcr]{%
1	88.8158950330041\\
1.09749876549306	88.9210615767232\\
1.20450354025879	89.0168917639324\\
1.32194114846604	89.1042134650887\\
1.45082877849596	89.1837813811422\\
1.59228279334111	89.2562834521707\\
1.74752840000771	89.3223467189517\\
1.91791026167246	89.382542680663\\
2.104904144512	89.4373921893748\\
2.31012970008314	89.4873699193824\\
2.5353644939701	89.5329084468019\\
2.78255940220711	89.5744019722728\\
3.05385550883341	89.6122097171298\\
3.35160265093885	89.6466590210324\\
3.67837977182865	89.6780481668042\\
4.03701725859658	89.7066489561331\\
4.43062145758392	89.7327090578236\\
4.86260158006541	89.7564541484745\\
5.33669923120639	89.7780898637683\\
5.85702081805658	89.7978035770076\\
6.42807311728424	89.8157660200995\\
7.05480231071857	89.8321327608767\\
7.74263682681121	89.8470455494336\\
8.4975343590864	89.8606335450525\\
10.2353102189903	89.8842954497863\\
12.3284673944207	89.9039400091235\\
14.8496826225448	89.9202492847605\\
17.8864952905746	89.9337895418501\\
21.5443469003186	89.9450309068466\\
25.9502421139972	89.9543636888089\\
34.3046928631493	89.9654778010562\\
45.3487850812863	89.9738852211606\\
59.9484250318932	89.9802451269156\\
86.974900261778	89.9863837319827\\
138.488637139389	89.9914485867526\\
242.012826479436	89.9951065669147\\
559.081018251231	89.9978817496287\\
2056.51230834863	89.9994241349443\\
10000	89.9998815726425\\
};
\draw[line width=0.5mm,draw=black!70,loosely dashed] (axis cs:7,-200)--(axis cs:7,200);
\node[color=black, fill=none, font=\normalsize] at (axis cs: 18,0) {$\Delta\phi\approx$\SI{180}{\degree}};
\end{axis}
\end{tikzpicture}%

%% file: updated_images/Dp_Change.tex
%
%
\begin{tikzpicture}

\begin{axis}[%
width=2in,
height=2in,
at={(0.494in,0.431in)},
scale only axis,
point meta min=0.0078938906122036,
point meta max=0.0436830670101369,
xmin=0.015,
xmax=0.055,
xlabel style={font=\color{white!15!black}},
xlabel={Frequency Droop [\%]},
xtick = {0.02,0.03,0.04,0.05},
xticklabels={2,3,4,5},
scaled ticks=false,
ymin=7.04,
ymax=7.16,
ylabel style={align=center,font=\color{white!15!black}},
ylabel={Potential \\ Resonance Frequency [Hz]},
axis background/.style={fill=white},
xmajorgrids,
ymajorgrids,
grid style={dashed, opacity=0.5}
]

\addplot[%
mesh,shader=flat,
    mark=none,
    line join=round,
    line cap=round,
    line width=2pt,
    point meta=explicit, colormap={mymap}{[1pt] rgb(0pt)=(0.18995,0.07176,0.23217); rgb(1pt)=(0.19483,0.08339,0.26149); rgb(2pt)=(0.19956,0.09498,0.29024); rgb(3pt)=(0.20415,0.10652,0.31844); rgb(4pt)=(0.2086,0.11802,0.34607); rgb(5pt)=(0.21291,0.12947,0.37314); rgb(6pt)=(0.21708,0.14087,0.39964); rgb(7pt)=(0.22111,0.15223,0.42558); rgb(8pt)=(0.225,0.16354,0.45096); rgb(9pt)=(0.22875,0.17481,0.47578); rgb(10pt)=(0.23236,0.18603,0.50004); rgb(11pt)=(0.23582,0.1972,0.52373); rgb(12pt)=(0.23915,0.20833,0.54686); rgb(13pt)=(0.24234,0.21941,0.56942); rgb(14pt)=(0.24539,0.23044,0.59142); rgb(15pt)=(0.2483,0.24143,0.61286); rgb(16pt)=(0.25107,0.25237,0.63374); rgb(17pt)=(0.25369,0.26327,0.65406); rgb(18pt)=(0.25618,0.27412,0.67381); rgb(19pt)=(0.25853,0.28492,0.693); rgb(20pt)=(0.26074,0.29568,0.71162); rgb(21pt)=(0.2628,0.30639,0.72968); rgb(22pt)=(0.26473,0.31706,0.74718); rgb(23pt)=(0.26652,0.32768,0.76412); rgb(24pt)=(0.26816,0.33825,0.7805); rgb(25pt)=(0.26967,0.34878,0.79631); rgb(26pt)=(0.27103,0.35926,0.81156); rgb(27pt)=(0.27226,0.3697,0.82624); rgb(28pt)=(0.27334,0.38008,0.84037); rgb(29pt)=(0.27429,0.39043,0.85393); rgb(30pt)=(0.27509,0.40072,0.86692); rgb(31pt)=(0.27576,0.41097,0.87936); rgb(32pt)=(0.27628,0.42118,0.89123); rgb(33pt)=(0.27667,0.43134,0.90254); rgb(34pt)=(0.27691,0.44145,0.91328); rgb(35pt)=(0.27701,0.45152,0.92347); rgb(36pt)=(0.27698,0.46153,0.93309); rgb(37pt)=(0.2768,0.47151,0.94214); rgb(38pt)=(0.27648,0.48144,0.95064); rgb(39pt)=(0.27603,0.49132,0.95857); rgb(40pt)=(0.27543,0.50115,0.96594); rgb(41pt)=(0.27469,0.51094,0.97275); rgb(42pt)=(0.27381,0.52069,0.97899); rgb(43pt)=(0.27273,0.5304,0.98461); rgb(44pt)=(0.27106,0.54015,0.9893); rgb(45pt)=(0.26878,0.54995,0.99303); rgb(46pt)=(0.26592,0.55979,0.99583); rgb(47pt)=(0.26252,0.56967,0.99773); rgb(48pt)=(0.25862,0.57958,0.99876); rgb(49pt)=(0.25425,0.5895,0.99896); rgb(50pt)=(0.24946,0.59943,0.99835); rgb(51pt)=(0.24427,0.60937,0.99697); rgb(52pt)=(0.23874,0.61931,0.99485); rgb(53pt)=(0.23288,0.62923,0.99202); rgb(54pt)=(0.22676,0.63913,0.98851); rgb(55pt)=(0.22039,0.64901,0.98436); rgb(56pt)=(0.21382,0.65886,0.97959); rgb(57pt)=(0.20708,0.66866,0.97423); rgb(58pt)=(0.20021,0.67842,0.96833); rgb(59pt)=(0.19326,0.68812,0.9619); rgb(60pt)=(0.18625,0.69775,0.95498); rgb(61pt)=(0.17923,0.70732,0.94761); rgb(62pt)=(0.17223,0.7168,0.93981); rgb(63pt)=(0.16529,0.7262,0.93161); rgb(64pt)=(0.15844,0.73551,0.92305); rgb(65pt)=(0.15173,0.74472,0.91416); rgb(66pt)=(0.14519,0.75381,0.90496); rgb(67pt)=(0.13886,0.76279,0.8955); rgb(68pt)=(0.13278,0.77165,0.8858); rgb(69pt)=(0.12698,0.78037,0.8759); rgb(70pt)=(0.12151,0.78896,0.86581); rgb(71pt)=(0.11639,0.7974,0.85559); rgb(72pt)=(0.11167,0.80569,0.84525); rgb(73pt)=(0.10738,0.81381,0.83484); rgb(74pt)=(0.10357,0.82177,0.82437); rgb(75pt)=(0.10026,0.82955,0.81389); rgb(76pt)=(0.0975,0.83714,0.80342); rgb(77pt)=(0.09532,0.84455,0.79299); rgb(78pt)=(0.09377,0.85175,0.78264); rgb(79pt)=(0.09287,0.85875,0.7724); rgb(80pt)=(0.09267,0.86554,0.7623); rgb(81pt)=(0.0932,0.87211,0.75237); rgb(82pt)=(0.09451,0.87844,0.74265); rgb(83pt)=(0.09662,0.88454,0.73316); rgb(84pt)=(0.09958,0.8904,0.72393); rgb(85pt)=(0.10342,0.896,0.715); rgb(86pt)=(0.10815,0.90142,0.70599); rgb(87pt)=(0.11374,0.90673,0.69651); rgb(88pt)=(0.12014,0.91193,0.6866); rgb(89pt)=(0.12733,0.91701,0.67627); rgb(90pt)=(0.13526,0.92197,0.66556); rgb(91pt)=(0.14391,0.9268,0.65448); rgb(92pt)=(0.15323,0.93151,0.64308); rgb(93pt)=(0.16319,0.93609,0.63137); rgb(94pt)=(0.17377,0.94053,0.61938); rgb(95pt)=(0.18491,0.94484,0.60713); rgb(96pt)=(0.19659,0.94901,0.59466); rgb(97pt)=(0.20877,0.95304,0.58199); rgb(98pt)=(0.22142,0.95692,0.56914); rgb(99pt)=(0.23449,0.96065,0.55614); rgb(100pt)=(0.24797,0.96423,0.54303); rgb(101pt)=(0.2618,0.96765,0.52981); rgb(102pt)=(0.27597,0.97092,0.51653); rgb(103pt)=(0.29042,0.97403,0.50321); rgb(104pt)=(0.30513,0.97697,0.48987); rgb(105pt)=(0.32006,0.97974,0.47654); rgb(106pt)=(0.33517,0.98234,0.46325); rgb(107pt)=(0.35043,0.98477,0.45002); rgb(108pt)=(0.36581,0.98702,0.43688); rgb(109pt)=(0.38127,0.98909,0.42386); rgb(110pt)=(0.39678,0.99098,0.41098); rgb(111pt)=(0.41229,0.99268,0.39826); rgb(112pt)=(0.42778,0.99419,0.38575); rgb(113pt)=(0.44321,0.99551,0.37345); rgb(114pt)=(0.45854,0.99663,0.3614); rgb(115pt)=(0.47375,0.99755,0.34963); rgb(116pt)=(0.48879,0.99828,0.33816); rgb(117pt)=(0.50362,0.99879,0.32701); rgb(118pt)=(0.51822,0.9991,0.31622); rgb(119pt)=(0.53255,0.99919,0.30581); rgb(120pt)=(0.54658,0.99907,0.29581); rgb(121pt)=(0.56026,0.99873,0.28623); rgb(122pt)=(0.57357,0.99817,0.27712); rgb(123pt)=(0.58646,0.99739,0.26849); rgb(124pt)=(0.59891,0.99638,0.26038); rgb(125pt)=(0.61088,0.99514,0.2528); rgb(126pt)=(0.62233,0.99366,0.24579); rgb(127pt)=(0.63323,0.99195,0.23937); rgb(128pt)=(0.64362,0.98999,0.23356); rgb(129pt)=(0.65394,0.98775,0.22835); rgb(130pt)=(0.66428,0.98524,0.2237); rgb(131pt)=(0.67462,0.98246,0.2196); rgb(132pt)=(0.68494,0.97941,0.21602); rgb(133pt)=(0.69525,0.9761,0.21294); rgb(134pt)=(0.70553,0.97255,0.21032); rgb(135pt)=(0.71577,0.96875,0.20815); rgb(136pt)=(0.72596,0.9647,0.2064); rgb(137pt)=(0.7361,0.96043,0.20504); rgb(138pt)=(0.74617,0.95593,0.20406); rgb(139pt)=(0.75617,0.95121,0.20343); rgb(140pt)=(0.76608,0.94627,0.20311); rgb(141pt)=(0.77591,0.94113,0.2031); rgb(142pt)=(0.78563,0.93579,0.20336); rgb(143pt)=(0.79524,0.93025,0.20386); rgb(144pt)=(0.80473,0.92452,0.20459); rgb(145pt)=(0.8141,0.91861,0.20552); rgb(146pt)=(0.82333,0.91253,0.20663); rgb(147pt)=(0.83241,0.90627,0.20788); rgb(148pt)=(0.84133,0.89986,0.20926); rgb(149pt)=(0.8501,0.89328,0.21074); rgb(150pt)=(0.85868,0.88655,0.2123); rgb(151pt)=(0.86709,0.87968,0.21391); rgb(152pt)=(0.8753,0.87267,0.21555); rgb(153pt)=(0.88331,0.86553,0.21719); rgb(154pt)=(0.89112,0.85826,0.2188); rgb(155pt)=(0.8987,0.85087,0.22038); rgb(156pt)=(0.90605,0.84337,0.22188); rgb(157pt)=(0.91317,0.83576,0.22328); rgb(158pt)=(0.92004,0.82806,0.22456); rgb(159pt)=(0.92666,0.82025,0.2257); rgb(160pt)=(0.93301,0.81236,0.22667); rgb(161pt)=(0.93909,0.80439,0.22744); rgb(162pt)=(0.94489,0.79634,0.228); rgb(163pt)=(0.95039,0.78823,0.22831); rgb(164pt)=(0.9556,0.78005,0.22836); rgb(165pt)=(0.96049,0.77181,0.22811); rgb(166pt)=(0.96507,0.76352,0.22754); rgb(167pt)=(0.96931,0.75519,0.22663); rgb(168pt)=(0.97323,0.74682,0.22536); rgb(169pt)=(0.97679,0.73842,0.22369); rgb(170pt)=(0.98,0.73,0.22161); rgb(171pt)=(0.98289,0.7214,0.21918); rgb(172pt)=(0.98549,0.7125,0.2165); rgb(173pt)=(0.98781,0.7033,0.21358); rgb(174pt)=(0.98986,0.69382,0.21043); rgb(175pt)=(0.99163,0.68408,0.20706); rgb(176pt)=(0.99314,0.67408,0.20348); rgb(177pt)=(0.99438,0.66386,0.19971); rgb(178pt)=(0.99535,0.65341,0.19577); rgb(179pt)=(0.99607,0.64277,0.19165); rgb(180pt)=(0.99654,0.63193,0.18738); rgb(181pt)=(0.99675,0.62093,0.18297); rgb(182pt)=(0.99672,0.60977,0.17842); rgb(183pt)=(0.99644,0.59846,0.17376); rgb(184pt)=(0.99593,0.58703,0.16899); rgb(185pt)=(0.99517,0.57549,0.16412); rgb(186pt)=(0.99419,0.56386,0.15918); rgb(187pt)=(0.99297,0.55214,0.15417); rgb(188pt)=(0.99153,0.54036,0.1491); rgb(189pt)=(0.98987,0.52854,0.14398); rgb(190pt)=(0.98799,0.51667,0.13883); rgb(191pt)=(0.9859,0.50479,0.13367); rgb(192pt)=(0.9836,0.49291,0.12849); rgb(193pt)=(0.98108,0.48104,0.12332); rgb(194pt)=(0.97837,0.4692,0.11817); rgb(195pt)=(0.97545,0.4574,0.11305); rgb(196pt)=(0.97234,0.44565,0.10797); rgb(197pt)=(0.96904,0.43399,0.10294); rgb(198pt)=(0.96555,0.42241,0.09798); rgb(199pt)=(0.96187,0.41093,0.0931); rgb(200pt)=(0.95801,0.39958,0.08831); rgb(201pt)=(0.95398,0.38836,0.08362); rgb(202pt)=(0.94977,0.37729,0.07905); rgb(203pt)=(0.94538,0.36638,0.07461); rgb(204pt)=(0.94084,0.35566,0.07031); rgb(205pt)=(0.93612,0.34513,0.06616); rgb(206pt)=(0.93125,0.33482,0.06218); rgb(207pt)=(0.92623,0.32473,0.05837); rgb(208pt)=(0.92105,0.31489,0.05475); rgb(209pt)=(0.91572,0.3053,0.05134); rgb(210pt)=(0.91024,0.29599,0.04814); rgb(211pt)=(0.90463,0.28696,0.04516); rgb(212pt)=(0.89888,0.27824,0.04243); rgb(213pt)=(0.89298,0.26981,0.03993); rgb(214pt)=(0.88691,0.26152,0.03753); rgb(215pt)=(0.88066,0.25334,0.03521); rgb(216pt)=(0.87422,0.24526,0.03297); rgb(217pt)=(0.8676,0.2373,0.03082); rgb(218pt)=(0.86079,0.22945,0.02875); rgb(219pt)=(0.8538,0.2217,0.02677); rgb(220pt)=(0.84662,0.21407,0.02487); rgb(221pt)=(0.83926,0.20654,0.02305); rgb(222pt)=(0.83172,0.19912,0.02131); rgb(223pt)=(0.82399,0.19182,0.01966); rgb(224pt)=(0.81608,0.18462,0.01809); rgb(225pt)=(0.80799,0.17753,0.0166); rgb(226pt)=(0.79971,0.17055,0.0152); rgb(227pt)=(0.79125,0.16368,0.01387); rgb(228pt)=(0.7826,0.15693,0.01264); rgb(229pt)=(0.77377,0.15028,0.01148); rgb(230pt)=(0.76476,0.14374,0.01041); rgb(231pt)=(0.75556,0.13731,0.00942); rgb(232pt)=(0.74617,0.13098,0.00851); rgb(233pt)=(0.73661,0.12477,0.00769); rgb(234pt)=(0.72686,0.11867,0.00695); rgb(235pt)=(0.71692,0.11268,0.00629); rgb(236pt)=(0.7068,0.1068,0.00571); rgb(237pt)=(0.6965,0.10102,0.00522); rgb(238pt)=(0.68602,0.09536,0.00481); rgb(239pt)=(0.67535,0.0898,0.00449); rgb(240pt)=(0.66449,0.08436,0.00424); rgb(241pt)=(0.65345,0.07902,0.00408); rgb(242pt)=(0.64223,0.0738,0.00401); rgb(243pt)=(0.63082,0.06868,0.00401); rgb(244pt)=(0.61923,0.06367,0.0041); rgb(245pt)=(0.60746,0.05878,0.00427); rgb(246pt)=(0.5955,0.05399,0.00453); rgb(247pt)=(0.58336,0.04931,0.00486); rgb(248pt)=(0.57103,0.04474,0.00529); rgb(249pt)=(0.55852,0.04028,0.00579); rgb(250pt)=(0.54583,0.03593,0.00638); rgb(251pt)=(0.53295,0.03169,0.00705); rgb(252pt)=(0.51989,0.02756,0.0078); rgb(253pt)=(0.50664,0.02354,0.00863); rgb(254pt)=(0.49321,0.01963,0.00955); rgb(255pt)=(0.4796,0.01583,0.01055)}, mesh/rows=7]
table[row sep=crcr, point meta=\thisrow{c}] {%
x	y	c\\
0.02	7.05217025010271	0.0078938906122036\\
0.02	7.05217025010271	0.0078938906122036\\
0.025	7.07105916053855	0.0119339686361623\\
0.025	7.07105916053855	0.0119339686361623\\
0.03	7.08924892091013	0.0167670525041117\\
0.03	7.08924892091013	0.0167670525041117\\
0.035	7.10676713774943	0.0223739899301914\\
0.035	7.10676713774943	0.0223739899301914\\
0.04	7.1236383689321	0.0287381539702339\\
0.04	7.1236383689321	0.0287381539702339\\
0.045	7.13988440621521	0.0358452345575888\\
0.045	7.13988440621521	0.0358452345575888\\
0.05	7.15552451014913	0.0436830670101369\\
0.05	7.15552451014913	0.0436830670101369\\
};
\end{axis}

\begin{axis}[%
width=2in,
height=2in,
at={(3.1in,0.431in)},
scale only axis,
point meta min=0.0078938906122036,
point meta max=0.0436830670101369,
xmin=0.015,
xmax=0.055,
xlabel style={font=\color{white!15!black}},
xlabel={Frequency Droop [\%]},
xtick = {0.02,0.03,0.04,0.05},
xticklabels={2,3,4,5},
scaled ticks=false,
ymin=-2,
ymax=7,
ylabel style={font=\color{white!15!black}},
ylabel={Phase Margin [Deg]},
axis background/.style={fill=white},
xmajorgrids,
ymajorgrids,
grid style={dashed, opacity=0.5},
colormap={mymap}{[1pt] rgb(0pt)=(0.18995,0.07176,0.23217); rgb(1pt)=(0.19483,0.08339,0.26149); rgb(2pt)=(0.19956,0.09498,0.29024); rgb(3pt)=(0.20415,0.10652,0.31844); rgb(4pt)=(0.2086,0.11802,0.34607); rgb(5pt)=(0.21291,0.12947,0.37314); rgb(6pt)=(0.21708,0.14087,0.39964); rgb(7pt)=(0.22111,0.15223,0.42558); rgb(8pt)=(0.225,0.16354,0.45096); rgb(9pt)=(0.22875,0.17481,0.47578); rgb(10pt)=(0.23236,0.18603,0.50004); rgb(11pt)=(0.23582,0.1972,0.52373); rgb(12pt)=(0.23915,0.20833,0.54686); rgb(13pt)=(0.24234,0.21941,0.56942); rgb(14pt)=(0.24539,0.23044,0.59142); rgb(15pt)=(0.2483,0.24143,0.61286); rgb(16pt)=(0.25107,0.25237,0.63374); rgb(17pt)=(0.25369,0.26327,0.65406); rgb(18pt)=(0.25618,0.27412,0.67381); rgb(19pt)=(0.25853,0.28492,0.693); rgb(20pt)=(0.26074,0.29568,0.71162); rgb(21pt)=(0.2628,0.30639,0.72968); rgb(22pt)=(0.26473,0.31706,0.74718); rgb(23pt)=(0.26652,0.32768,0.76412); rgb(24pt)=(0.26816,0.33825,0.7805); rgb(25pt)=(0.26967,0.34878,0.79631); rgb(26pt)=(0.27103,0.35926,0.81156); rgb(27pt)=(0.27226,0.3697,0.82624); rgb(28pt)=(0.27334,0.38008,0.84037); rgb(29pt)=(0.27429,0.39043,0.85393); rgb(30pt)=(0.27509,0.40072,0.86692); rgb(31pt)=(0.27576,0.41097,0.87936); rgb(32pt)=(0.27628,0.42118,0.89123); rgb(33pt)=(0.27667,0.43134,0.90254); rgb(34pt)=(0.27691,0.44145,0.91328); rgb(35pt)=(0.27701,0.45152,0.92347); rgb(36pt)=(0.27698,0.46153,0.93309); rgb(37pt)=(0.2768,0.47151,0.94214); rgb(38pt)=(0.27648,0.48144,0.95064); rgb(39pt)=(0.27603,0.49132,0.95857); rgb(40pt)=(0.27543,0.50115,0.96594); rgb(41pt)=(0.27469,0.51094,0.97275); rgb(42pt)=(0.27381,0.52069,0.97899); rgb(43pt)=(0.27273,0.5304,0.98461); rgb(44pt)=(0.27106,0.54015,0.9893); rgb(45pt)=(0.26878,0.54995,0.99303); rgb(46pt)=(0.26592,0.55979,0.99583); rgb(47pt)=(0.26252,0.56967,0.99773); rgb(48pt)=(0.25862,0.57958,0.99876); rgb(49pt)=(0.25425,0.5895,0.99896); rgb(50pt)=(0.24946,0.59943,0.99835); rgb(51pt)=(0.24427,0.60937,0.99697); rgb(52pt)=(0.23874,0.61931,0.99485); rgb(53pt)=(0.23288,0.62923,0.99202); rgb(54pt)=(0.22676,0.63913,0.98851); rgb(55pt)=(0.22039,0.64901,0.98436); rgb(56pt)=(0.21382,0.65886,0.97959); rgb(57pt)=(0.20708,0.66866,0.97423); rgb(58pt)=(0.20021,0.67842,0.96833); rgb(59pt)=(0.19326,0.68812,0.9619); rgb(60pt)=(0.18625,0.69775,0.95498); rgb(61pt)=(0.17923,0.70732,0.94761); rgb(62pt)=(0.17223,0.7168,0.93981); rgb(63pt)=(0.16529,0.7262,0.93161); rgb(64pt)=(0.15844,0.73551,0.92305); rgb(65pt)=(0.15173,0.74472,0.91416); rgb(66pt)=(0.14519,0.75381,0.90496); rgb(67pt)=(0.13886,0.76279,0.8955); rgb(68pt)=(0.13278,0.77165,0.8858); rgb(69pt)=(0.12698,0.78037,0.8759); rgb(70pt)=(0.12151,0.78896,0.86581); rgb(71pt)=(0.11639,0.7974,0.85559); rgb(72pt)=(0.11167,0.80569,0.84525); rgb(73pt)=(0.10738,0.81381,0.83484); rgb(74pt)=(0.10357,0.82177,0.82437); rgb(75pt)=(0.10026,0.82955,0.81389); rgb(76pt)=(0.0975,0.83714,0.80342); rgb(77pt)=(0.09532,0.84455,0.79299); rgb(78pt)=(0.09377,0.85175,0.78264); rgb(79pt)=(0.09287,0.85875,0.7724); rgb(80pt)=(0.09267,0.86554,0.7623); rgb(81pt)=(0.0932,0.87211,0.75237); rgb(82pt)=(0.09451,0.87844,0.74265); rgb(83pt)=(0.09662,0.88454,0.73316); rgb(84pt)=(0.09958,0.8904,0.72393); rgb(85pt)=(0.10342,0.896,0.715); rgb(86pt)=(0.10815,0.90142,0.70599); rgb(87pt)=(0.11374,0.90673,0.69651); rgb(88pt)=(0.12014,0.91193,0.6866); rgb(89pt)=(0.12733,0.91701,0.67627); rgb(90pt)=(0.13526,0.92197,0.66556); rgb(91pt)=(0.14391,0.9268,0.65448); rgb(92pt)=(0.15323,0.93151,0.64308); rgb(93pt)=(0.16319,0.93609,0.63137); rgb(94pt)=(0.17377,0.94053,0.61938); rgb(95pt)=(0.18491,0.94484,0.60713); rgb(96pt)=(0.19659,0.94901,0.59466); rgb(97pt)=(0.20877,0.95304,0.58199); rgb(98pt)=(0.22142,0.95692,0.56914); rgb(99pt)=(0.23449,0.96065,0.55614); rgb(100pt)=(0.24797,0.96423,0.54303); rgb(101pt)=(0.2618,0.96765,0.52981); rgb(102pt)=(0.27597,0.97092,0.51653); rgb(103pt)=(0.29042,0.97403,0.50321); rgb(104pt)=(0.30513,0.97697,0.48987); rgb(105pt)=(0.32006,0.97974,0.47654); rgb(106pt)=(0.33517,0.98234,0.46325); rgb(107pt)=(0.35043,0.98477,0.45002); rgb(108pt)=(0.36581,0.98702,0.43688); rgb(109pt)=(0.38127,0.98909,0.42386); rgb(110pt)=(0.39678,0.99098,0.41098); rgb(111pt)=(0.41229,0.99268,0.39826); rgb(112pt)=(0.42778,0.99419,0.38575); rgb(113pt)=(0.44321,0.99551,0.37345); rgb(114pt)=(0.45854,0.99663,0.3614); rgb(115pt)=(0.47375,0.99755,0.34963); rgb(116pt)=(0.48879,0.99828,0.33816); rgb(117pt)=(0.50362,0.99879,0.32701); rgb(118pt)=(0.51822,0.9991,0.31622); rgb(119pt)=(0.53255,0.99919,0.30581); rgb(120pt)=(0.54658,0.99907,0.29581); rgb(121pt)=(0.56026,0.99873,0.28623); rgb(122pt)=(0.57357,0.99817,0.27712); rgb(123pt)=(0.58646,0.99739,0.26849); rgb(124pt)=(0.59891,0.99638,0.26038); rgb(125pt)=(0.61088,0.99514,0.2528); rgb(126pt)=(0.62233,0.99366,0.24579); rgb(127pt)=(0.63323,0.99195,0.23937); rgb(128pt)=(0.64362,0.98999,0.23356); rgb(129pt)=(0.65394,0.98775,0.22835); rgb(130pt)=(0.66428,0.98524,0.2237); rgb(131pt)=(0.67462,0.98246,0.2196); rgb(132pt)=(0.68494,0.97941,0.21602); rgb(133pt)=(0.69525,0.9761,0.21294); rgb(134pt)=(0.70553,0.97255,0.21032); rgb(135pt)=(0.71577,0.96875,0.20815); rgb(136pt)=(0.72596,0.9647,0.2064); rgb(137pt)=(0.7361,0.96043,0.20504); rgb(138pt)=(0.74617,0.95593,0.20406); rgb(139pt)=(0.75617,0.95121,0.20343); rgb(140pt)=(0.76608,0.94627,0.20311); rgb(141pt)=(0.77591,0.94113,0.2031); rgb(142pt)=(0.78563,0.93579,0.20336); rgb(143pt)=(0.79524,0.93025,0.20386); rgb(144pt)=(0.80473,0.92452,0.20459); rgb(145pt)=(0.8141,0.91861,0.20552); rgb(146pt)=(0.82333,0.91253,0.20663); rgb(147pt)=(0.83241,0.90627,0.20788); rgb(148pt)=(0.84133,0.89986,0.20926); rgb(149pt)=(0.8501,0.89328,0.21074); rgb(150pt)=(0.85868,0.88655,0.2123); rgb(151pt)=(0.86709,0.87968,0.21391); rgb(152pt)=(0.8753,0.87267,0.21555); rgb(153pt)=(0.88331,0.86553,0.21719); rgb(154pt)=(0.89112,0.85826,0.2188); rgb(155pt)=(0.8987,0.85087,0.22038); rgb(156pt)=(0.90605,0.84337,0.22188); rgb(157pt)=(0.91317,0.83576,0.22328); rgb(158pt)=(0.92004,0.82806,0.22456); rgb(159pt)=(0.92666,0.82025,0.2257); rgb(160pt)=(0.93301,0.81236,0.22667); rgb(161pt)=(0.93909,0.80439,0.22744); rgb(162pt)=(0.94489,0.79634,0.228); rgb(163pt)=(0.95039,0.78823,0.22831); rgb(164pt)=(0.9556,0.78005,0.22836); rgb(165pt)=(0.96049,0.77181,0.22811); rgb(166pt)=(0.96507,0.76352,0.22754); rgb(167pt)=(0.96931,0.75519,0.22663); rgb(168pt)=(0.97323,0.74682,0.22536); rgb(169pt)=(0.97679,0.73842,0.22369); rgb(170pt)=(0.98,0.73,0.22161); rgb(171pt)=(0.98289,0.7214,0.21918); rgb(172pt)=(0.98549,0.7125,0.2165); rgb(173pt)=(0.98781,0.7033,0.21358); rgb(174pt)=(0.98986,0.69382,0.21043); rgb(175pt)=(0.99163,0.68408,0.20706); rgb(176pt)=(0.99314,0.67408,0.20348); rgb(177pt)=(0.99438,0.66386,0.19971); rgb(178pt)=(0.99535,0.65341,0.19577); rgb(179pt)=(0.99607,0.64277,0.19165); rgb(180pt)=(0.99654,0.63193,0.18738); rgb(181pt)=(0.99675,0.62093,0.18297); rgb(182pt)=(0.99672,0.60977,0.17842); rgb(183pt)=(0.99644,0.59846,0.17376); rgb(184pt)=(0.99593,0.58703,0.16899); rgb(185pt)=(0.99517,0.57549,0.16412); rgb(186pt)=(0.99419,0.56386,0.15918); rgb(187pt)=(0.99297,0.55214,0.15417); rgb(188pt)=(0.99153,0.54036,0.1491); rgb(189pt)=(0.98987,0.52854,0.14398); rgb(190pt)=(0.98799,0.51667,0.13883); rgb(191pt)=(0.9859,0.50479,0.13367); rgb(192pt)=(0.9836,0.49291,0.12849); rgb(193pt)=(0.98108,0.48104,0.12332); rgb(194pt)=(0.97837,0.4692,0.11817); rgb(195pt)=(0.97545,0.4574,0.11305); rgb(196pt)=(0.97234,0.44565,0.10797); rgb(197pt)=(0.96904,0.43399,0.10294); rgb(198pt)=(0.96555,0.42241,0.09798); rgb(199pt)=(0.96187,0.41093,0.0931); rgb(200pt)=(0.95801,0.39958,0.08831); rgb(201pt)=(0.95398,0.38836,0.08362); rgb(202pt)=(0.94977,0.37729,0.07905); rgb(203pt)=(0.94538,0.36638,0.07461); rgb(204pt)=(0.94084,0.35566,0.07031); rgb(205pt)=(0.93612,0.34513,0.06616); rgb(206pt)=(0.93125,0.33482,0.06218); rgb(207pt)=(0.92623,0.32473,0.05837); rgb(208pt)=(0.92105,0.31489,0.05475); rgb(209pt)=(0.91572,0.3053,0.05134); rgb(210pt)=(0.91024,0.29599,0.04814); rgb(211pt)=(0.90463,0.28696,0.04516); rgb(212pt)=(0.89888,0.27824,0.04243); rgb(213pt)=(0.89298,0.26981,0.03993); rgb(214pt)=(0.88691,0.26152,0.03753); rgb(215pt)=(0.88066,0.25334,0.03521); rgb(216pt)=(0.87422,0.24526,0.03297); rgb(217pt)=(0.8676,0.2373,0.03082); rgb(218pt)=(0.86079,0.22945,0.02875); rgb(219pt)=(0.8538,0.2217,0.02677); rgb(220pt)=(0.84662,0.21407,0.02487); rgb(221pt)=(0.83926,0.20654,0.02305); rgb(222pt)=(0.83172,0.19912,0.02131); rgb(223pt)=(0.82399,0.19182,0.01966); rgb(224pt)=(0.81608,0.18462,0.01809); rgb(225pt)=(0.80799,0.17753,0.0166); rgb(226pt)=(0.79971,0.17055,0.0152); rgb(227pt)=(0.79125,0.16368,0.01387); rgb(228pt)=(0.7826,0.15693,0.01264); rgb(229pt)=(0.77377,0.15028,0.01148); rgb(230pt)=(0.76476,0.14374,0.01041); rgb(231pt)=(0.75556,0.13731,0.00942); rgb(232pt)=(0.74617,0.13098,0.00851); rgb(233pt)=(0.73661,0.12477,0.00769); rgb(234pt)=(0.72686,0.11867,0.00695); rgb(235pt)=(0.71692,0.11268,0.00629); rgb(236pt)=(0.7068,0.1068,0.00571); rgb(237pt)=(0.6965,0.10102,0.00522); rgb(238pt)=(0.68602,0.09536,0.00481); rgb(239pt)=(0.67535,0.0898,0.00449); rgb(240pt)=(0.66449,0.08436,0.00424); rgb(241pt)=(0.65345,0.07902,0.00408); rgb(242pt)=(0.64223,0.0738,0.00401); rgb(243pt)=(0.63082,0.06868,0.00401); rgb(244pt)=(0.61923,0.06367,0.0041); rgb(245pt)=(0.60746,0.05878,0.00427); rgb(246pt)=(0.5955,0.05399,0.00453); rgb(247pt)=(0.58336,0.04931,0.00486); rgb(248pt)=(0.57103,0.04474,0.00529); rgb(249pt)=(0.55852,0.04028,0.00579); rgb(250pt)=(0.54583,0.03593,0.00638); rgb(251pt)=(0.53295,0.03169,0.00705); rgb(252pt)=(0.51989,0.02756,0.0078); rgb(253pt)=(0.50664,0.02354,0.00863); rgb(254pt)=(0.49321,0.01963,0.00955); rgb(255pt)=(0.4796,0.01583,0.01055)},
colorbar,
colorbar style={ylabel style={rotate=-90,font=\color{white!15!black},at={(0.5,1)}, anchor=south,},scaled ticks=false,ytick={0.015,0.025,0.035},yticklabels={0.015,0.025,0.035}, ylabel={$\bm{AQI}$}}
]
\draw[fill=black, thick,opacity=0.1] (axis cs: 0,-2) -- (axis cs: 0.055,-2) -- (axis cs: 0.055,0) -- (axis cs: 0,0) -- cycle;
\node[color=black, fill=none] at (axis cs: 0.025,-0.5) {$Unstable$};
\node[color=black, fill=none] at (axis cs: 0.025,0.5) {$Stable$};
\addplot[%
mesh,
shader=flat,
    mark=none,
    line join=round,
    line cap=round,
    line width=2pt,
    point meta=explicit, colormap={mymap}{[1pt] rgb(0pt)=(0.18995,0.07176,0.23217); rgb(1pt)=(0.19483,0.08339,0.26149); rgb(2pt)=(0.19956,0.09498,0.29024); rgb(3pt)=(0.20415,0.10652,0.31844); rgb(4pt)=(0.2086,0.11802,0.34607); rgb(5pt)=(0.21291,0.12947,0.37314); rgb(6pt)=(0.21708,0.14087,0.39964); rgb(7pt)=(0.22111,0.15223,0.42558); rgb(8pt)=(0.225,0.16354,0.45096); rgb(9pt)=(0.22875,0.17481,0.47578); rgb(10pt)=(0.23236,0.18603,0.50004); rgb(11pt)=(0.23582,0.1972,0.52373); rgb(12pt)=(0.23915,0.20833,0.54686); rgb(13pt)=(0.24234,0.21941,0.56942); rgb(14pt)=(0.24539,0.23044,0.59142); rgb(15pt)=(0.2483,0.24143,0.61286); rgb(16pt)=(0.25107,0.25237,0.63374); rgb(17pt)=(0.25369,0.26327,0.65406); rgb(18pt)=(0.25618,0.27412,0.67381); rgb(19pt)=(0.25853,0.28492,0.693); rgb(20pt)=(0.26074,0.29568,0.71162); rgb(21pt)=(0.2628,0.30639,0.72968); rgb(22pt)=(0.26473,0.31706,0.74718); rgb(23pt)=(0.26652,0.32768,0.76412); rgb(24pt)=(0.26816,0.33825,0.7805); rgb(25pt)=(0.26967,0.34878,0.79631); rgb(26pt)=(0.27103,0.35926,0.81156); rgb(27pt)=(0.27226,0.3697,0.82624); rgb(28pt)=(0.27334,0.38008,0.84037); rgb(29pt)=(0.27429,0.39043,0.85393); rgb(30pt)=(0.27509,0.40072,0.86692); rgb(31pt)=(0.27576,0.41097,0.87936); rgb(32pt)=(0.27628,0.42118,0.89123); rgb(33pt)=(0.27667,0.43134,0.90254); rgb(34pt)=(0.27691,0.44145,0.91328); rgb(35pt)=(0.27701,0.45152,0.92347); rgb(36pt)=(0.27698,0.46153,0.93309); rgb(37pt)=(0.2768,0.47151,0.94214); rgb(38pt)=(0.27648,0.48144,0.95064); rgb(39pt)=(0.27603,0.49132,0.95857); rgb(40pt)=(0.27543,0.50115,0.96594); rgb(41pt)=(0.27469,0.51094,0.97275); rgb(42pt)=(0.27381,0.52069,0.97899); rgb(43pt)=(0.27273,0.5304,0.98461); rgb(44pt)=(0.27106,0.54015,0.9893); rgb(45pt)=(0.26878,0.54995,0.99303); rgb(46pt)=(0.26592,0.55979,0.99583); rgb(47pt)=(0.26252,0.56967,0.99773); rgb(48pt)=(0.25862,0.57958,0.99876); rgb(49pt)=(0.25425,0.5895,0.99896); rgb(50pt)=(0.24946,0.59943,0.99835); rgb(51pt)=(0.24427,0.60937,0.99697); rgb(52pt)=(0.23874,0.61931,0.99485); rgb(53pt)=(0.23288,0.62923,0.99202); rgb(54pt)=(0.22676,0.63913,0.98851); rgb(55pt)=(0.22039,0.64901,0.98436); rgb(56pt)=(0.21382,0.65886,0.97959); rgb(57pt)=(0.20708,0.66866,0.97423); rgb(58pt)=(0.20021,0.67842,0.96833); rgb(59pt)=(0.19326,0.68812,0.9619); rgb(60pt)=(0.18625,0.69775,0.95498); rgb(61pt)=(0.17923,0.70732,0.94761); rgb(62pt)=(0.17223,0.7168,0.93981); rgb(63pt)=(0.16529,0.7262,0.93161); rgb(64pt)=(0.15844,0.73551,0.92305); rgb(65pt)=(0.15173,0.74472,0.91416); rgb(66pt)=(0.14519,0.75381,0.90496); rgb(67pt)=(0.13886,0.76279,0.8955); rgb(68pt)=(0.13278,0.77165,0.8858); rgb(69pt)=(0.12698,0.78037,0.8759); rgb(70pt)=(0.12151,0.78896,0.86581); rgb(71pt)=(0.11639,0.7974,0.85559); rgb(72pt)=(0.11167,0.80569,0.84525); rgb(73pt)=(0.10738,0.81381,0.83484); rgb(74pt)=(0.10357,0.82177,0.82437); rgb(75pt)=(0.10026,0.82955,0.81389); rgb(76pt)=(0.0975,0.83714,0.80342); rgb(77pt)=(0.09532,0.84455,0.79299); rgb(78pt)=(0.09377,0.85175,0.78264); rgb(79pt)=(0.09287,0.85875,0.7724); rgb(80pt)=(0.09267,0.86554,0.7623); rgb(81pt)=(0.0932,0.87211,0.75237); rgb(82pt)=(0.09451,0.87844,0.74265); rgb(83pt)=(0.09662,0.88454,0.73316); rgb(84pt)=(0.09958,0.8904,0.72393); rgb(85pt)=(0.10342,0.896,0.715); rgb(86pt)=(0.10815,0.90142,0.70599); rgb(87pt)=(0.11374,0.90673,0.69651); rgb(88pt)=(0.12014,0.91193,0.6866); rgb(89pt)=(0.12733,0.91701,0.67627); rgb(90pt)=(0.13526,0.92197,0.66556); rgb(91pt)=(0.14391,0.9268,0.65448); rgb(92pt)=(0.15323,0.93151,0.64308); rgb(93pt)=(0.16319,0.93609,0.63137); rgb(94pt)=(0.17377,0.94053,0.61938); rgb(95pt)=(0.18491,0.94484,0.60713); rgb(96pt)=(0.19659,0.94901,0.59466); rgb(97pt)=(0.20877,0.95304,0.58199); rgb(98pt)=(0.22142,0.95692,0.56914); rgb(99pt)=(0.23449,0.96065,0.55614); rgb(100pt)=(0.24797,0.96423,0.54303); rgb(101pt)=(0.2618,0.96765,0.52981); rgb(102pt)=(0.27597,0.97092,0.51653); rgb(103pt)=(0.29042,0.97403,0.50321); rgb(104pt)=(0.30513,0.97697,0.48987); rgb(105pt)=(0.32006,0.97974,0.47654); rgb(106pt)=(0.33517,0.98234,0.46325); rgb(107pt)=(0.35043,0.98477,0.45002); rgb(108pt)=(0.36581,0.98702,0.43688); rgb(109pt)=(0.38127,0.98909,0.42386); rgb(110pt)=(0.39678,0.99098,0.41098); rgb(111pt)=(0.41229,0.99268,0.39826); rgb(112pt)=(0.42778,0.99419,0.38575); rgb(113pt)=(0.44321,0.99551,0.37345); rgb(114pt)=(0.45854,0.99663,0.3614); rgb(115pt)=(0.47375,0.99755,0.34963); rgb(116pt)=(0.48879,0.99828,0.33816); rgb(117pt)=(0.50362,0.99879,0.32701); rgb(118pt)=(0.51822,0.9991,0.31622); rgb(119pt)=(0.53255,0.99919,0.30581); rgb(120pt)=(0.54658,0.99907,0.29581); rgb(121pt)=(0.56026,0.99873,0.28623); rgb(122pt)=(0.57357,0.99817,0.27712); rgb(123pt)=(0.58646,0.99739,0.26849); rgb(124pt)=(0.59891,0.99638,0.26038); rgb(125pt)=(0.61088,0.99514,0.2528); rgb(126pt)=(0.62233,0.99366,0.24579); rgb(127pt)=(0.63323,0.99195,0.23937); rgb(128pt)=(0.64362,0.98999,0.23356); rgb(129pt)=(0.65394,0.98775,0.22835); rgb(130pt)=(0.66428,0.98524,0.2237); rgb(131pt)=(0.67462,0.98246,0.2196); rgb(132pt)=(0.68494,0.97941,0.21602); rgb(133pt)=(0.69525,0.9761,0.21294); rgb(134pt)=(0.70553,0.97255,0.21032); rgb(135pt)=(0.71577,0.96875,0.20815); rgb(136pt)=(0.72596,0.9647,0.2064); rgb(137pt)=(0.7361,0.96043,0.20504); rgb(138pt)=(0.74617,0.95593,0.20406); rgb(139pt)=(0.75617,0.95121,0.20343); rgb(140pt)=(0.76608,0.94627,0.20311); rgb(141pt)=(0.77591,0.94113,0.2031); rgb(142pt)=(0.78563,0.93579,0.20336); rgb(143pt)=(0.79524,0.93025,0.20386); rgb(144pt)=(0.80473,0.92452,0.20459); rgb(145pt)=(0.8141,0.91861,0.20552); rgb(146pt)=(0.82333,0.91253,0.20663); rgb(147pt)=(0.83241,0.90627,0.20788); rgb(148pt)=(0.84133,0.89986,0.20926); rgb(149pt)=(0.8501,0.89328,0.21074); rgb(150pt)=(0.85868,0.88655,0.2123); rgb(151pt)=(0.86709,0.87968,0.21391); rgb(152pt)=(0.8753,0.87267,0.21555); rgb(153pt)=(0.88331,0.86553,0.21719); rgb(154pt)=(0.89112,0.85826,0.2188); rgb(155pt)=(0.8987,0.85087,0.22038); rgb(156pt)=(0.90605,0.84337,0.22188); rgb(157pt)=(0.91317,0.83576,0.22328); rgb(158pt)=(0.92004,0.82806,0.22456); rgb(159pt)=(0.92666,0.82025,0.2257); rgb(160pt)=(0.93301,0.81236,0.22667); rgb(161pt)=(0.93909,0.80439,0.22744); rgb(162pt)=(0.94489,0.79634,0.228); rgb(163pt)=(0.95039,0.78823,0.22831); rgb(164pt)=(0.9556,0.78005,0.22836); rgb(165pt)=(0.96049,0.77181,0.22811); rgb(166pt)=(0.96507,0.76352,0.22754); rgb(167pt)=(0.96931,0.75519,0.22663); rgb(168pt)=(0.97323,0.74682,0.22536); rgb(169pt)=(0.97679,0.73842,0.22369); rgb(170pt)=(0.98,0.73,0.22161); rgb(171pt)=(0.98289,0.7214,0.21918); rgb(172pt)=(0.98549,0.7125,0.2165); rgb(173pt)=(0.98781,0.7033,0.21358); rgb(174pt)=(0.98986,0.69382,0.21043); rgb(175pt)=(0.99163,0.68408,0.20706); rgb(176pt)=(0.99314,0.67408,0.20348); rgb(177pt)=(0.99438,0.66386,0.19971); rgb(178pt)=(0.99535,0.65341,0.19577); rgb(179pt)=(0.99607,0.64277,0.19165); rgb(180pt)=(0.99654,0.63193,0.18738); rgb(181pt)=(0.99675,0.62093,0.18297); rgb(182pt)=(0.99672,0.60977,0.17842); rgb(183pt)=(0.99644,0.59846,0.17376); rgb(184pt)=(0.99593,0.58703,0.16899); rgb(185pt)=(0.99517,0.57549,0.16412); rgb(186pt)=(0.99419,0.56386,0.15918); rgb(187pt)=(0.99297,0.55214,0.15417); rgb(188pt)=(0.99153,0.54036,0.1491); rgb(189pt)=(0.98987,0.52854,0.14398); rgb(190pt)=(0.98799,0.51667,0.13883); rgb(191pt)=(0.9859,0.50479,0.13367); rgb(192pt)=(0.9836,0.49291,0.12849); rgb(193pt)=(0.98108,0.48104,0.12332); rgb(194pt)=(0.97837,0.4692,0.11817); rgb(195pt)=(0.97545,0.4574,0.11305); rgb(196pt)=(0.97234,0.44565,0.10797); rgb(197pt)=(0.96904,0.43399,0.10294); rgb(198pt)=(0.96555,0.42241,0.09798); rgb(199pt)=(0.96187,0.41093,0.0931); rgb(200pt)=(0.95801,0.39958,0.08831); rgb(201pt)=(0.95398,0.38836,0.08362); rgb(202pt)=(0.94977,0.37729,0.07905); rgb(203pt)=(0.94538,0.36638,0.07461); rgb(204pt)=(0.94084,0.35566,0.07031); rgb(205pt)=(0.93612,0.34513,0.06616); rgb(206pt)=(0.93125,0.33482,0.06218); rgb(207pt)=(0.92623,0.32473,0.05837); rgb(208pt)=(0.92105,0.31489,0.05475); rgb(209pt)=(0.91572,0.3053,0.05134); rgb(210pt)=(0.91024,0.29599,0.04814); rgb(211pt)=(0.90463,0.28696,0.04516); rgb(212pt)=(0.89888,0.27824,0.04243); rgb(213pt)=(0.89298,0.26981,0.03993); rgb(214pt)=(0.88691,0.26152,0.03753); rgb(215pt)=(0.88066,0.25334,0.03521); rgb(216pt)=(0.87422,0.24526,0.03297); rgb(217pt)=(0.8676,0.2373,0.03082); rgb(218pt)=(0.86079,0.22945,0.02875); rgb(219pt)=(0.8538,0.2217,0.02677); rgb(220pt)=(0.84662,0.21407,0.02487); rgb(221pt)=(0.83926,0.20654,0.02305); rgb(222pt)=(0.83172,0.19912,0.02131); rgb(223pt)=(0.82399,0.19182,0.01966); rgb(224pt)=(0.81608,0.18462,0.01809); rgb(225pt)=(0.80799,0.17753,0.0166); rgb(226pt)=(0.79971,0.17055,0.0152); rgb(227pt)=(0.79125,0.16368,0.01387); rgb(228pt)=(0.7826,0.15693,0.01264); rgb(229pt)=(0.77377,0.15028,0.01148); rgb(230pt)=(0.76476,0.14374,0.01041); rgb(231pt)=(0.75556,0.13731,0.00942); rgb(232pt)=(0.74617,0.13098,0.00851); rgb(233pt)=(0.73661,0.12477,0.00769); rgb(234pt)=(0.72686,0.11867,0.00695); rgb(235pt)=(0.71692,0.11268,0.00629); rgb(236pt)=(0.7068,0.1068,0.00571); rgb(237pt)=(0.6965,0.10102,0.00522); rgb(238pt)=(0.68602,0.09536,0.00481); rgb(239pt)=(0.67535,0.0898,0.00449); rgb(240pt)=(0.66449,0.08436,0.00424); rgb(241pt)=(0.65345,0.07902,0.00408); rgb(242pt)=(0.64223,0.0738,0.00401); rgb(243pt)=(0.63082,0.06868,0.00401); rgb(244pt)=(0.61923,0.06367,0.0041); rgb(245pt)=(0.60746,0.05878,0.00427); rgb(246pt)=(0.5955,0.05399,0.00453); rgb(247pt)=(0.58336,0.04931,0.00486); rgb(248pt)=(0.57103,0.04474,0.00529); rgb(249pt)=(0.55852,0.04028,0.00579); rgb(250pt)=(0.54583,0.03593,0.00638); rgb(251pt)=(0.53295,0.03169,0.00705); rgb(252pt)=(0.51989,0.02756,0.0078); rgb(253pt)=(0.50664,0.02354,0.00863); rgb(254pt)=(0.49321,0.01963,0.00955); rgb(255pt)=(0.4796,0.01583,0.01055)}, mesh/rows=7]
table[row sep=crcr, point meta=\thisrow{c}] {%
x	y	c\\
0.02	6.1503105999989	0.0078938906122036\\
0.02	6.1503105999989	0.0078938906122036\\
0.025	4.9026203095784	0.0119339686361623\\
0.025	4.9026203095784	0.0119339686361623\\
0.03	3.66090028687159	0.0167670525041117\\
0.03	3.66090028687159	0.0167670525041117\\
0.035	2.4240050026109	0.0223739899301914\\
0.035	2.4240050026109	0.0223739899301914\\
0.04	1.19083021127599	0.0287381539702339\\
0.04	1.19083021127599	0.0287381539702339\\
0.045	-0.0396964441864611	0.0358452345575888\\
0.045	-0.0396964441864611	0.0358452345575888\\
0.05	-1.26862410477304	0.0436830670101369\\
0.05	-1.26862410477304	0.0436830670101369\\
};
\addplot[
    only marks,
    mark=*,
    mark size=5pt,
    color=red, 
    mark options={
        fill=red,
        fill opacity=0.5, 
        draw opacity=1     
    },
]
coordinates {(0.05,-1.268)};
\node[color=black, fill=none] at (axis cs: 0.04,-0.8) {\textcolor{red}{Case 1b}};
\draw[line width=0.5mm,draw=black!80,->] (axis cs:0.048,-1.4)--(axis cs:0.044,-1.2);
\end{axis}
\end{tikzpicture}%

%% file: updated_images/Rv_Change.tex
%
%
\begin{tikzpicture}

\begin{axis}[%
width=2in,
height=2in,
at={(0.494in,0.431in)},
scale only axis,
point meta min=0.0403390495559979,
point meta max=0.0436830670101369,
xmin=0.05,
xmax=0.55,
xtick = {0,0.1,0.2,0.3,0.4,0.5},
xticklabels={0,0.1,0.2,0.3,0.4,0.5},
xlabel style={font=\color{white!15!black}},
xlabel={$R_v$ [$\Omega$]},
ymin=7,
ymax=7.16,
ylabel style={align=center,font=\color{white!15!black}},
ylabel={Potential \\ Resonance Frequency [Hz]},
axis background/.style={fill=white},
xmajorgrids,
ymajorgrids,
grid style={dashed, opacity=0.5}
]

\addplot[%
mesh,
shader=flat,
    mark=none,
    line join=round,
    line cap=round,
    line width=2pt,
    point meta=explicit, colormap={mymap}{[1pt] rgb(0pt)=(0.18995,0.07176,0.23217); rgb(1pt)=(0.19483,0.08339,0.26149); rgb(2pt)=(0.19956,0.09498,0.29024); rgb(3pt)=(0.20415,0.10652,0.31844); rgb(4pt)=(0.2086,0.11802,0.34607); rgb(5pt)=(0.21291,0.12947,0.37314); rgb(6pt)=(0.21708,0.14087,0.39964); rgb(7pt)=(0.22111,0.15223,0.42558); rgb(8pt)=(0.225,0.16354,0.45096); rgb(9pt)=(0.22875,0.17481,0.47578); rgb(10pt)=(0.23236,0.18603,0.50004); rgb(11pt)=(0.23582,0.1972,0.52373); rgb(12pt)=(0.23915,0.20833,0.54686); rgb(13pt)=(0.24234,0.21941,0.56942); rgb(14pt)=(0.24539,0.23044,0.59142); rgb(15pt)=(0.2483,0.24143,0.61286); rgb(16pt)=(0.25107,0.25237,0.63374); rgb(17pt)=(0.25369,0.26327,0.65406); rgb(18pt)=(0.25618,0.27412,0.67381); rgb(19pt)=(0.25853,0.28492,0.693); rgb(20pt)=(0.26074,0.29568,0.71162); rgb(21pt)=(0.2628,0.30639,0.72968); rgb(22pt)=(0.26473,0.31706,0.74718); rgb(23pt)=(0.26652,0.32768,0.76412); rgb(24pt)=(0.26816,0.33825,0.7805); rgb(25pt)=(0.26967,0.34878,0.79631); rgb(26pt)=(0.27103,0.35926,0.81156); rgb(27pt)=(0.27226,0.3697,0.82624); rgb(28pt)=(0.27334,0.38008,0.84037); rgb(29pt)=(0.27429,0.39043,0.85393); rgb(30pt)=(0.27509,0.40072,0.86692); rgb(31pt)=(0.27576,0.41097,0.87936); rgb(32pt)=(0.27628,0.42118,0.89123); rgb(33pt)=(0.27667,0.43134,0.90254); rgb(34pt)=(0.27691,0.44145,0.91328); rgb(35pt)=(0.27701,0.45152,0.92347); rgb(36pt)=(0.27698,0.46153,0.93309); rgb(37pt)=(0.2768,0.47151,0.94214); rgb(38pt)=(0.27648,0.48144,0.95064); rgb(39pt)=(0.27603,0.49132,0.95857); rgb(40pt)=(0.27543,0.50115,0.96594); rgb(41pt)=(0.27469,0.51094,0.97275); rgb(42pt)=(0.27381,0.52069,0.97899); rgb(43pt)=(0.27273,0.5304,0.98461); rgb(44pt)=(0.27106,0.54015,0.9893); rgb(45pt)=(0.26878,0.54995,0.99303); rgb(46pt)=(0.26592,0.55979,0.99583); rgb(47pt)=(0.26252,0.56967,0.99773); rgb(48pt)=(0.25862,0.57958,0.99876); rgb(49pt)=(0.25425,0.5895,0.99896); rgb(50pt)=(0.24946,0.59943,0.99835); rgb(51pt)=(0.24427,0.60937,0.99697); rgb(52pt)=(0.23874,0.61931,0.99485); rgb(53pt)=(0.23288,0.62923,0.99202); rgb(54pt)=(0.22676,0.63913,0.98851); rgb(55pt)=(0.22039,0.64901,0.98436); rgb(56pt)=(0.21382,0.65886,0.97959); rgb(57pt)=(0.20708,0.66866,0.97423); rgb(58pt)=(0.20021,0.67842,0.96833); rgb(59pt)=(0.19326,0.68812,0.9619); rgb(60pt)=(0.18625,0.69775,0.95498); rgb(61pt)=(0.17923,0.70732,0.94761); rgb(62pt)=(0.17223,0.7168,0.93981); rgb(63pt)=(0.16529,0.7262,0.93161); rgb(64pt)=(0.15844,0.73551,0.92305); rgb(65pt)=(0.15173,0.74472,0.91416); rgb(66pt)=(0.14519,0.75381,0.90496); rgb(67pt)=(0.13886,0.76279,0.8955); rgb(68pt)=(0.13278,0.77165,0.8858); rgb(69pt)=(0.12698,0.78037,0.8759); rgb(70pt)=(0.12151,0.78896,0.86581); rgb(71pt)=(0.11639,0.7974,0.85559); rgb(72pt)=(0.11167,0.80569,0.84525); rgb(73pt)=(0.10738,0.81381,0.83484); rgb(74pt)=(0.10357,0.82177,0.82437); rgb(75pt)=(0.10026,0.82955,0.81389); rgb(76pt)=(0.0975,0.83714,0.80342); rgb(77pt)=(0.09532,0.84455,0.79299); rgb(78pt)=(0.09377,0.85175,0.78264); rgb(79pt)=(0.09287,0.85875,0.7724); rgb(80pt)=(0.09267,0.86554,0.7623); rgb(81pt)=(0.0932,0.87211,0.75237); rgb(82pt)=(0.09451,0.87844,0.74265); rgb(83pt)=(0.09662,0.88454,0.73316); rgb(84pt)=(0.09958,0.8904,0.72393); rgb(85pt)=(0.10342,0.896,0.715); rgb(86pt)=(0.10815,0.90142,0.70599); rgb(87pt)=(0.11374,0.90673,0.69651); rgb(88pt)=(0.12014,0.91193,0.6866); rgb(89pt)=(0.12733,0.91701,0.67627); rgb(90pt)=(0.13526,0.92197,0.66556); rgb(91pt)=(0.14391,0.9268,0.65448); rgb(92pt)=(0.15323,0.93151,0.64308); rgb(93pt)=(0.16319,0.93609,0.63137); rgb(94pt)=(0.17377,0.94053,0.61938); rgb(95pt)=(0.18491,0.94484,0.60713); rgb(96pt)=(0.19659,0.94901,0.59466); rgb(97pt)=(0.20877,0.95304,0.58199); rgb(98pt)=(0.22142,0.95692,0.56914); rgb(99pt)=(0.23449,0.96065,0.55614); rgb(100pt)=(0.24797,0.96423,0.54303); rgb(101pt)=(0.2618,0.96765,0.52981); rgb(102pt)=(0.27597,0.97092,0.51653); rgb(103pt)=(0.29042,0.97403,0.50321); rgb(104pt)=(0.30513,0.97697,0.48987); rgb(105pt)=(0.32006,0.97974,0.47654); rgb(106pt)=(0.33517,0.98234,0.46325); rgb(107pt)=(0.35043,0.98477,0.45002); rgb(108pt)=(0.36581,0.98702,0.43688); rgb(109pt)=(0.38127,0.98909,0.42386); rgb(110pt)=(0.39678,0.99098,0.41098); rgb(111pt)=(0.41229,0.99268,0.39826); rgb(112pt)=(0.42778,0.99419,0.38575); rgb(113pt)=(0.44321,0.99551,0.37345); rgb(114pt)=(0.45854,0.99663,0.3614); rgb(115pt)=(0.47375,0.99755,0.34963); rgb(116pt)=(0.48879,0.99828,0.33816); rgb(117pt)=(0.50362,0.99879,0.32701); rgb(118pt)=(0.51822,0.9991,0.31622); rgb(119pt)=(0.53255,0.99919,0.30581); rgb(120pt)=(0.54658,0.99907,0.29581); rgb(121pt)=(0.56026,0.99873,0.28623); rgb(122pt)=(0.57357,0.99817,0.27712); rgb(123pt)=(0.58646,0.99739,0.26849); rgb(124pt)=(0.59891,0.99638,0.26038); rgb(125pt)=(0.61088,0.99514,0.2528); rgb(126pt)=(0.62233,0.99366,0.24579); rgb(127pt)=(0.63323,0.99195,0.23937); rgb(128pt)=(0.64362,0.98999,0.23356); rgb(129pt)=(0.65394,0.98775,0.22835); rgb(130pt)=(0.66428,0.98524,0.2237); rgb(131pt)=(0.67462,0.98246,0.2196); rgb(132pt)=(0.68494,0.97941,0.21602); rgb(133pt)=(0.69525,0.9761,0.21294); rgb(134pt)=(0.70553,0.97255,0.21032); rgb(135pt)=(0.71577,0.96875,0.20815); rgb(136pt)=(0.72596,0.9647,0.2064); rgb(137pt)=(0.7361,0.96043,0.20504); rgb(138pt)=(0.74617,0.95593,0.20406); rgb(139pt)=(0.75617,0.95121,0.20343); rgb(140pt)=(0.76608,0.94627,0.20311); rgb(141pt)=(0.77591,0.94113,0.2031); rgb(142pt)=(0.78563,0.93579,0.20336); rgb(143pt)=(0.79524,0.93025,0.20386); rgb(144pt)=(0.80473,0.92452,0.20459); rgb(145pt)=(0.8141,0.91861,0.20552); rgb(146pt)=(0.82333,0.91253,0.20663); rgb(147pt)=(0.83241,0.90627,0.20788); rgb(148pt)=(0.84133,0.89986,0.20926); rgb(149pt)=(0.8501,0.89328,0.21074); rgb(150pt)=(0.85868,0.88655,0.2123); rgb(151pt)=(0.86709,0.87968,0.21391); rgb(152pt)=(0.8753,0.87267,0.21555); rgb(153pt)=(0.88331,0.86553,0.21719); rgb(154pt)=(0.89112,0.85826,0.2188); rgb(155pt)=(0.8987,0.85087,0.22038); rgb(156pt)=(0.90605,0.84337,0.22188); rgb(157pt)=(0.91317,0.83576,0.22328); rgb(158pt)=(0.92004,0.82806,0.22456); rgb(159pt)=(0.92666,0.82025,0.2257); rgb(160pt)=(0.93301,0.81236,0.22667); rgb(161pt)=(0.93909,0.80439,0.22744); rgb(162pt)=(0.94489,0.79634,0.228); rgb(163pt)=(0.95039,0.78823,0.22831); rgb(164pt)=(0.9556,0.78005,0.22836); rgb(165pt)=(0.96049,0.77181,0.22811); rgb(166pt)=(0.96507,0.76352,0.22754); rgb(167pt)=(0.96931,0.75519,0.22663); rgb(168pt)=(0.97323,0.74682,0.22536); rgb(169pt)=(0.97679,0.73842,0.22369); rgb(170pt)=(0.98,0.73,0.22161); rgb(171pt)=(0.98289,0.7214,0.21918); rgb(172pt)=(0.98549,0.7125,0.2165); rgb(173pt)=(0.98781,0.7033,0.21358); rgb(174pt)=(0.98986,0.69382,0.21043); rgb(175pt)=(0.99163,0.68408,0.20706); rgb(176pt)=(0.99314,0.67408,0.20348); rgb(177pt)=(0.99438,0.66386,0.19971); rgb(178pt)=(0.99535,0.65341,0.19577); rgb(179pt)=(0.99607,0.64277,0.19165); rgb(180pt)=(0.99654,0.63193,0.18738); rgb(181pt)=(0.99675,0.62093,0.18297); rgb(182pt)=(0.99672,0.60977,0.17842); rgb(183pt)=(0.99644,0.59846,0.17376); rgb(184pt)=(0.99593,0.58703,0.16899); rgb(185pt)=(0.99517,0.57549,0.16412); rgb(186pt)=(0.99419,0.56386,0.15918); rgb(187pt)=(0.99297,0.55214,0.15417); rgb(188pt)=(0.99153,0.54036,0.1491); rgb(189pt)=(0.98987,0.52854,0.14398); rgb(190pt)=(0.98799,0.51667,0.13883); rgb(191pt)=(0.9859,0.50479,0.13367); rgb(192pt)=(0.9836,0.49291,0.12849); rgb(193pt)=(0.98108,0.48104,0.12332); rgb(194pt)=(0.97837,0.4692,0.11817); rgb(195pt)=(0.97545,0.4574,0.11305); rgb(196pt)=(0.97234,0.44565,0.10797); rgb(197pt)=(0.96904,0.43399,0.10294); rgb(198pt)=(0.96555,0.42241,0.09798); rgb(199pt)=(0.96187,0.41093,0.0931); rgb(200pt)=(0.95801,0.39958,0.08831); rgb(201pt)=(0.95398,0.38836,0.08362); rgb(202pt)=(0.94977,0.37729,0.07905); rgb(203pt)=(0.94538,0.36638,0.07461); rgb(204pt)=(0.94084,0.35566,0.07031); rgb(205pt)=(0.93612,0.34513,0.06616); rgb(206pt)=(0.93125,0.33482,0.06218); rgb(207pt)=(0.92623,0.32473,0.05837); rgb(208pt)=(0.92105,0.31489,0.05475); rgb(209pt)=(0.91572,0.3053,0.05134); rgb(210pt)=(0.91024,0.29599,0.04814); rgb(211pt)=(0.90463,0.28696,0.04516); rgb(212pt)=(0.89888,0.27824,0.04243); rgb(213pt)=(0.89298,0.26981,0.03993); rgb(214pt)=(0.88691,0.26152,0.03753); rgb(215pt)=(0.88066,0.25334,0.03521); rgb(216pt)=(0.87422,0.24526,0.03297); rgb(217pt)=(0.8676,0.2373,0.03082); rgb(218pt)=(0.86079,0.22945,0.02875); rgb(219pt)=(0.8538,0.2217,0.02677); rgb(220pt)=(0.84662,0.21407,0.02487); rgb(221pt)=(0.83926,0.20654,0.02305); rgb(222pt)=(0.83172,0.19912,0.02131); rgb(223pt)=(0.82399,0.19182,0.01966); rgb(224pt)=(0.81608,0.18462,0.01809); rgb(225pt)=(0.80799,0.17753,0.0166); rgb(226pt)=(0.79971,0.17055,0.0152); rgb(227pt)=(0.79125,0.16368,0.01387); rgb(228pt)=(0.7826,0.15693,0.01264); rgb(229pt)=(0.77377,0.15028,0.01148); rgb(230pt)=(0.76476,0.14374,0.01041); rgb(231pt)=(0.75556,0.13731,0.00942); rgb(232pt)=(0.74617,0.13098,0.00851); rgb(233pt)=(0.73661,0.12477,0.00769); rgb(234pt)=(0.72686,0.11867,0.00695); rgb(235pt)=(0.71692,0.11268,0.00629); rgb(236pt)=(0.7068,0.1068,0.00571); rgb(237pt)=(0.6965,0.10102,0.00522); rgb(238pt)=(0.68602,0.09536,0.00481); rgb(239pt)=(0.67535,0.0898,0.00449); rgb(240pt)=(0.66449,0.08436,0.00424); rgb(241pt)=(0.65345,0.07902,0.00408); rgb(242pt)=(0.64223,0.0738,0.00401); rgb(243pt)=(0.63082,0.06868,0.00401); rgb(244pt)=(0.61923,0.06367,0.0041); rgb(245pt)=(0.60746,0.05878,0.00427); rgb(246pt)=(0.5955,0.05399,0.00453); rgb(247pt)=(0.58336,0.04931,0.00486); rgb(248pt)=(0.57103,0.04474,0.00529); rgb(249pt)=(0.55852,0.04028,0.00579); rgb(250pt)=(0.54583,0.03593,0.00638); rgb(251pt)=(0.53295,0.03169,0.00705); rgb(252pt)=(0.51989,0.02756,0.0078); rgb(253pt)=(0.50664,0.02354,0.00863); rgb(254pt)=(0.49321,0.01963,0.00955); rgb(255pt)=(0.4796,0.01583,0.01055)}, mesh/rows=11]
table[row sep=crcr, point meta=\thisrow{c}] {%
x	y	c\\
0	7.15552451014913	0.0436830670101369\\
0	7.15552451014913	0.0436830670101369\\
0.05	7.15432875545331	0.0432128142760893\\
0.05	7.15432875545331	0.0432128142760893\\
0.1	7.14989292301115	0.0427785276339362\\
0.1	7.14989292301115	0.0427785276339362\\
0.15	7.1423168213149	0.0423775444819262\\
0.15	7.1423168213149	0.0423775444819262\\
0.2	7.13169151812742	0.0420075322171686\\
0.2	7.13169151812742	0.0420075322171686\\
0.25	7.11810021999505	0.0416664438770201\\
0.25	7.11810021999505	0.0416664438770201\\
0.3	7.10161902626081	0.0413524813228966\\
0.3	7.10161902626081	0.0413524813228966\\
0.35	7.08231757883152	0.0410640645465885\\
0.35	7.08231757883152	0.0410640645465885\\
0.4	7.06025962481303	0.0407998059847833\\
0.4	7.06025962481303	0.0407998059847833\\
0.45	7.0355035059079	0.0405584889611939\\
0.45	7.0355035059079	0.0405584889611939\\
0.5	7.00810258594797	0.0403390495559979\\
0.5	7.00810258594797	0.0403390495559979\\
};
\end{axis}

\begin{axis}[%
width=2in,
height=2in,
at={(3.1in,0.431in)},
scale only axis,
point meta min=0.0403390495559979,
point meta max=0.0436830670101369,
xmin=-0.05,
xmax=0.55,
xtick = {0,0.1,0.2,0.3,0.4,0.5},
xticklabels={0,0.1,0.2,0.3,0.4,0.5},
xlabel style={font=\color{white!15!black}},
xlabel={$R_v$ [$\Omega$]},
ymin=-2,
ymax=14,
ylabel style={font=\color{white!15!black}},
ylabel={Phase Margin [Deg]},
axis background/.style={fill=white},
xmajorgrids,
ymajorgrids,
grid style={dashed, opacity=0.5},
colormap={mymap}{[1pt] rgb(0pt)=(0.18995,0.07176,0.23217); rgb(1pt)=(0.19483,0.08339,0.26149); rgb(2pt)=(0.19956,0.09498,0.29024); rgb(3pt)=(0.20415,0.10652,0.31844); rgb(4pt)=(0.2086,0.11802,0.34607); rgb(5pt)=(0.21291,0.12947,0.37314); rgb(6pt)=(0.21708,0.14087,0.39964); rgb(7pt)=(0.22111,0.15223,0.42558); rgb(8pt)=(0.225,0.16354,0.45096); rgb(9pt)=(0.22875,0.17481,0.47578); rgb(10pt)=(0.23236,0.18603,0.50004); rgb(11pt)=(0.23582,0.1972,0.52373); rgb(12pt)=(0.23915,0.20833,0.54686); rgb(13pt)=(0.24234,0.21941,0.56942); rgb(14pt)=(0.24539,0.23044,0.59142); rgb(15pt)=(0.2483,0.24143,0.61286); rgb(16pt)=(0.25107,0.25237,0.63374); rgb(17pt)=(0.25369,0.26327,0.65406); rgb(18pt)=(0.25618,0.27412,0.67381); rgb(19pt)=(0.25853,0.28492,0.693); rgb(20pt)=(0.26074,0.29568,0.71162); rgb(21pt)=(0.2628,0.30639,0.72968); rgb(22pt)=(0.26473,0.31706,0.74718); rgb(23pt)=(0.26652,0.32768,0.76412); rgb(24pt)=(0.26816,0.33825,0.7805); rgb(25pt)=(0.26967,0.34878,0.79631); rgb(26pt)=(0.27103,0.35926,0.81156); rgb(27pt)=(0.27226,0.3697,0.82624); rgb(28pt)=(0.27334,0.38008,0.84037); rgb(29pt)=(0.27429,0.39043,0.85393); rgb(30pt)=(0.27509,0.40072,0.86692); rgb(31pt)=(0.27576,0.41097,0.87936); rgb(32pt)=(0.27628,0.42118,0.89123); rgb(33pt)=(0.27667,0.43134,0.90254); rgb(34pt)=(0.27691,0.44145,0.91328); rgb(35pt)=(0.27701,0.45152,0.92347); rgb(36pt)=(0.27698,0.46153,0.93309); rgb(37pt)=(0.2768,0.47151,0.94214); rgb(38pt)=(0.27648,0.48144,0.95064); rgb(39pt)=(0.27603,0.49132,0.95857); rgb(40pt)=(0.27543,0.50115,0.96594); rgb(41pt)=(0.27469,0.51094,0.97275); rgb(42pt)=(0.27381,0.52069,0.97899); rgb(43pt)=(0.27273,0.5304,0.98461); rgb(44pt)=(0.27106,0.54015,0.9893); rgb(45pt)=(0.26878,0.54995,0.99303); rgb(46pt)=(0.26592,0.55979,0.99583); rgb(47pt)=(0.26252,0.56967,0.99773); rgb(48pt)=(0.25862,0.57958,0.99876); rgb(49pt)=(0.25425,0.5895,0.99896); rgb(50pt)=(0.24946,0.59943,0.99835); rgb(51pt)=(0.24427,0.60937,0.99697); rgb(52pt)=(0.23874,0.61931,0.99485); rgb(53pt)=(0.23288,0.62923,0.99202); rgb(54pt)=(0.22676,0.63913,0.98851); rgb(55pt)=(0.22039,0.64901,0.98436); rgb(56pt)=(0.21382,0.65886,0.97959); rgb(57pt)=(0.20708,0.66866,0.97423); rgb(58pt)=(0.20021,0.67842,0.96833); rgb(59pt)=(0.19326,0.68812,0.9619); rgb(60pt)=(0.18625,0.69775,0.95498); rgb(61pt)=(0.17923,0.70732,0.94761); rgb(62pt)=(0.17223,0.7168,0.93981); rgb(63pt)=(0.16529,0.7262,0.93161); rgb(64pt)=(0.15844,0.73551,0.92305); rgb(65pt)=(0.15173,0.74472,0.91416); rgb(66pt)=(0.14519,0.75381,0.90496); rgb(67pt)=(0.13886,0.76279,0.8955); rgb(68pt)=(0.13278,0.77165,0.8858); rgb(69pt)=(0.12698,0.78037,0.8759); rgb(70pt)=(0.12151,0.78896,0.86581); rgb(71pt)=(0.11639,0.7974,0.85559); rgb(72pt)=(0.11167,0.80569,0.84525); rgb(73pt)=(0.10738,0.81381,0.83484); rgb(74pt)=(0.10357,0.82177,0.82437); rgb(75pt)=(0.10026,0.82955,0.81389); rgb(76pt)=(0.0975,0.83714,0.80342); rgb(77pt)=(0.09532,0.84455,0.79299); rgb(78pt)=(0.09377,0.85175,0.78264); rgb(79pt)=(0.09287,0.85875,0.7724); rgb(80pt)=(0.09267,0.86554,0.7623); rgb(81pt)=(0.0932,0.87211,0.75237); rgb(82pt)=(0.09451,0.87844,0.74265); rgb(83pt)=(0.09662,0.88454,0.73316); rgb(84pt)=(0.09958,0.8904,0.72393); rgb(85pt)=(0.10342,0.896,0.715); rgb(86pt)=(0.10815,0.90142,0.70599); rgb(87pt)=(0.11374,0.90673,0.69651); rgb(88pt)=(0.12014,0.91193,0.6866); rgb(89pt)=(0.12733,0.91701,0.67627); rgb(90pt)=(0.13526,0.92197,0.66556); rgb(91pt)=(0.14391,0.9268,0.65448); rgb(92pt)=(0.15323,0.93151,0.64308); rgb(93pt)=(0.16319,0.93609,0.63137); rgb(94pt)=(0.17377,0.94053,0.61938); rgb(95pt)=(0.18491,0.94484,0.60713); rgb(96pt)=(0.19659,0.94901,0.59466); rgb(97pt)=(0.20877,0.95304,0.58199); rgb(98pt)=(0.22142,0.95692,0.56914); rgb(99pt)=(0.23449,0.96065,0.55614); rgb(100pt)=(0.24797,0.96423,0.54303); rgb(101pt)=(0.2618,0.96765,0.52981); rgb(102pt)=(0.27597,0.97092,0.51653); rgb(103pt)=(0.29042,0.97403,0.50321); rgb(104pt)=(0.30513,0.97697,0.48987); rgb(105pt)=(0.32006,0.97974,0.47654); rgb(106pt)=(0.33517,0.98234,0.46325); rgb(107pt)=(0.35043,0.98477,0.45002); rgb(108pt)=(0.36581,0.98702,0.43688); rgb(109pt)=(0.38127,0.98909,0.42386); rgb(110pt)=(0.39678,0.99098,0.41098); rgb(111pt)=(0.41229,0.99268,0.39826); rgb(112pt)=(0.42778,0.99419,0.38575); rgb(113pt)=(0.44321,0.99551,0.37345); rgb(114pt)=(0.45854,0.99663,0.3614); rgb(115pt)=(0.47375,0.99755,0.34963); rgb(116pt)=(0.48879,0.99828,0.33816); rgb(117pt)=(0.50362,0.99879,0.32701); rgb(118pt)=(0.51822,0.9991,0.31622); rgb(119pt)=(0.53255,0.99919,0.30581); rgb(120pt)=(0.54658,0.99907,0.29581); rgb(121pt)=(0.56026,0.99873,0.28623); rgb(122pt)=(0.57357,0.99817,0.27712); rgb(123pt)=(0.58646,0.99739,0.26849); rgb(124pt)=(0.59891,0.99638,0.26038); rgb(125pt)=(0.61088,0.99514,0.2528); rgb(126pt)=(0.62233,0.99366,0.24579); rgb(127pt)=(0.63323,0.99195,0.23937); rgb(128pt)=(0.64362,0.98999,0.23356); rgb(129pt)=(0.65394,0.98775,0.22835); rgb(130pt)=(0.66428,0.98524,0.2237); rgb(131pt)=(0.67462,0.98246,0.2196); rgb(132pt)=(0.68494,0.97941,0.21602); rgb(133pt)=(0.69525,0.9761,0.21294); rgb(134pt)=(0.70553,0.97255,0.21032); rgb(135pt)=(0.71577,0.96875,0.20815); rgb(136pt)=(0.72596,0.9647,0.2064); rgb(137pt)=(0.7361,0.96043,0.20504); rgb(138pt)=(0.74617,0.95593,0.20406); rgb(139pt)=(0.75617,0.95121,0.20343); rgb(140pt)=(0.76608,0.94627,0.20311); rgb(141pt)=(0.77591,0.94113,0.2031); rgb(142pt)=(0.78563,0.93579,0.20336); rgb(143pt)=(0.79524,0.93025,0.20386); rgb(144pt)=(0.80473,0.92452,0.20459); rgb(145pt)=(0.8141,0.91861,0.20552); rgb(146pt)=(0.82333,0.91253,0.20663); rgb(147pt)=(0.83241,0.90627,0.20788); rgb(148pt)=(0.84133,0.89986,0.20926); rgb(149pt)=(0.8501,0.89328,0.21074); rgb(150pt)=(0.85868,0.88655,0.2123); rgb(151pt)=(0.86709,0.87968,0.21391); rgb(152pt)=(0.8753,0.87267,0.21555); rgb(153pt)=(0.88331,0.86553,0.21719); rgb(154pt)=(0.89112,0.85826,0.2188); rgb(155pt)=(0.8987,0.85087,0.22038); rgb(156pt)=(0.90605,0.84337,0.22188); rgb(157pt)=(0.91317,0.83576,0.22328); rgb(158pt)=(0.92004,0.82806,0.22456); rgb(159pt)=(0.92666,0.82025,0.2257); rgb(160pt)=(0.93301,0.81236,0.22667); rgb(161pt)=(0.93909,0.80439,0.22744); rgb(162pt)=(0.94489,0.79634,0.228); rgb(163pt)=(0.95039,0.78823,0.22831); rgb(164pt)=(0.9556,0.78005,0.22836); rgb(165pt)=(0.96049,0.77181,0.22811); rgb(166pt)=(0.96507,0.76352,0.22754); rgb(167pt)=(0.96931,0.75519,0.22663); rgb(168pt)=(0.97323,0.74682,0.22536); rgb(169pt)=(0.97679,0.73842,0.22369); rgb(170pt)=(0.98,0.73,0.22161); rgb(171pt)=(0.98289,0.7214,0.21918); rgb(172pt)=(0.98549,0.7125,0.2165); rgb(173pt)=(0.98781,0.7033,0.21358); rgb(174pt)=(0.98986,0.69382,0.21043); rgb(175pt)=(0.99163,0.68408,0.20706); rgb(176pt)=(0.99314,0.67408,0.20348); rgb(177pt)=(0.99438,0.66386,0.19971); rgb(178pt)=(0.99535,0.65341,0.19577); rgb(179pt)=(0.99607,0.64277,0.19165); rgb(180pt)=(0.99654,0.63193,0.18738); rgb(181pt)=(0.99675,0.62093,0.18297); rgb(182pt)=(0.99672,0.60977,0.17842); rgb(183pt)=(0.99644,0.59846,0.17376); rgb(184pt)=(0.99593,0.58703,0.16899); rgb(185pt)=(0.99517,0.57549,0.16412); rgb(186pt)=(0.99419,0.56386,0.15918); rgb(187pt)=(0.99297,0.55214,0.15417); rgb(188pt)=(0.99153,0.54036,0.1491); rgb(189pt)=(0.98987,0.52854,0.14398); rgb(190pt)=(0.98799,0.51667,0.13883); rgb(191pt)=(0.9859,0.50479,0.13367); rgb(192pt)=(0.9836,0.49291,0.12849); rgb(193pt)=(0.98108,0.48104,0.12332); rgb(194pt)=(0.97837,0.4692,0.11817); rgb(195pt)=(0.97545,0.4574,0.11305); rgb(196pt)=(0.97234,0.44565,0.10797); rgb(197pt)=(0.96904,0.43399,0.10294); rgb(198pt)=(0.96555,0.42241,0.09798); rgb(199pt)=(0.96187,0.41093,0.0931); rgb(200pt)=(0.95801,0.39958,0.08831); rgb(201pt)=(0.95398,0.38836,0.08362); rgb(202pt)=(0.94977,0.37729,0.07905); rgb(203pt)=(0.94538,0.36638,0.07461); rgb(204pt)=(0.94084,0.35566,0.07031); rgb(205pt)=(0.93612,0.34513,0.06616); rgb(206pt)=(0.93125,0.33482,0.06218); rgb(207pt)=(0.92623,0.32473,0.05837); rgb(208pt)=(0.92105,0.31489,0.05475); rgb(209pt)=(0.91572,0.3053,0.05134); rgb(210pt)=(0.91024,0.29599,0.04814); rgb(211pt)=(0.90463,0.28696,0.04516); rgb(212pt)=(0.89888,0.27824,0.04243); rgb(213pt)=(0.89298,0.26981,0.03993); rgb(214pt)=(0.88691,0.26152,0.03753); rgb(215pt)=(0.88066,0.25334,0.03521); rgb(216pt)=(0.87422,0.24526,0.03297); rgb(217pt)=(0.8676,0.2373,0.03082); rgb(218pt)=(0.86079,0.22945,0.02875); rgb(219pt)=(0.8538,0.2217,0.02677); rgb(220pt)=(0.84662,0.21407,0.02487); rgb(221pt)=(0.83926,0.20654,0.02305); rgb(222pt)=(0.83172,0.19912,0.02131); rgb(223pt)=(0.82399,0.19182,0.01966); rgb(224pt)=(0.81608,0.18462,0.01809); rgb(225pt)=(0.80799,0.17753,0.0166); rgb(226pt)=(0.79971,0.17055,0.0152); rgb(227pt)=(0.79125,0.16368,0.01387); rgb(228pt)=(0.7826,0.15693,0.01264); rgb(229pt)=(0.77377,0.15028,0.01148); rgb(230pt)=(0.76476,0.14374,0.01041); rgb(231pt)=(0.75556,0.13731,0.00942); rgb(232pt)=(0.74617,0.13098,0.00851); rgb(233pt)=(0.73661,0.12477,0.00769); rgb(234pt)=(0.72686,0.11867,0.00695); rgb(235pt)=(0.71692,0.11268,0.00629); rgb(236pt)=(0.7068,0.1068,0.00571); rgb(237pt)=(0.6965,0.10102,0.00522); rgb(238pt)=(0.68602,0.09536,0.00481); rgb(239pt)=(0.67535,0.0898,0.00449); rgb(240pt)=(0.66449,0.08436,0.00424); rgb(241pt)=(0.65345,0.07902,0.00408); rgb(242pt)=(0.64223,0.0738,0.00401); rgb(243pt)=(0.63082,0.06868,0.00401); rgb(244pt)=(0.61923,0.06367,0.0041); rgb(245pt)=(0.60746,0.05878,0.00427); rgb(246pt)=(0.5955,0.05399,0.00453); rgb(247pt)=(0.58336,0.04931,0.00486); rgb(248pt)=(0.57103,0.04474,0.00529); rgb(249pt)=(0.55852,0.04028,0.00579); rgb(250pt)=(0.54583,0.03593,0.00638); rgb(251pt)=(0.53295,0.03169,0.00705); rgb(252pt)=(0.51989,0.02756,0.0078); rgb(253pt)=(0.50664,0.02354,0.00863); rgb(254pt)=(0.49321,0.01963,0.00955); rgb(255pt)=(0.4796,0.01583,0.01055)},
colorbar,
colorbar style={ylabel style={rotate=-90,font=\color{white!15!black},at={(0.5,1)}, anchor=south,},scaled ticks=false,ytick={0.0410,0.0420,0.0430},yticklabels={0.041,0.042,0.043}, ylabel={$\bm{AQI}$}}
]
\draw[fill=black, thick,opacity=0.1] (axis cs: -0.1,-10) -- (axis cs: 1,-10) -- (axis cs: 1,0) -- (axis cs: -0.1,0) -- cycle;
\node[color=black, fill=none] at (axis cs: 0.4,-1) {$Unstable$};
\node[color=black, fill=none] at (axis cs: 0.4,1) {$Stable$};
\addplot[%
mesh,
shader=flat,
    mark=none,
    line join=round,
    line cap=round,
    line width=2pt,
    point meta=explicit, colormap={mymap}{[1pt] rgb(0pt)=(0.18995,0.07176,0.23217); rgb(1pt)=(0.19483,0.08339,0.26149); rgb(2pt)=(0.19956,0.09498,0.29024); rgb(3pt)=(0.20415,0.10652,0.31844); rgb(4pt)=(0.2086,0.11802,0.34607); rgb(5pt)=(0.21291,0.12947,0.37314); rgb(6pt)=(0.21708,0.14087,0.39964); rgb(7pt)=(0.22111,0.15223,0.42558); rgb(8pt)=(0.225,0.16354,0.45096); rgb(9pt)=(0.22875,0.17481,0.47578); rgb(10pt)=(0.23236,0.18603,0.50004); rgb(11pt)=(0.23582,0.1972,0.52373); rgb(12pt)=(0.23915,0.20833,0.54686); rgb(13pt)=(0.24234,0.21941,0.56942); rgb(14pt)=(0.24539,0.23044,0.59142); rgb(15pt)=(0.2483,0.24143,0.61286); rgb(16pt)=(0.25107,0.25237,0.63374); rgb(17pt)=(0.25369,0.26327,0.65406); rgb(18pt)=(0.25618,0.27412,0.67381); rgb(19pt)=(0.25853,0.28492,0.693); rgb(20pt)=(0.26074,0.29568,0.71162); rgb(21pt)=(0.2628,0.30639,0.72968); rgb(22pt)=(0.26473,0.31706,0.74718); rgb(23pt)=(0.26652,0.32768,0.76412); rgb(24pt)=(0.26816,0.33825,0.7805); rgb(25pt)=(0.26967,0.34878,0.79631); rgb(26pt)=(0.27103,0.35926,0.81156); rgb(27pt)=(0.27226,0.3697,0.82624); rgb(28pt)=(0.27334,0.38008,0.84037); rgb(29pt)=(0.27429,0.39043,0.85393); rgb(30pt)=(0.27509,0.40072,0.86692); rgb(31pt)=(0.27576,0.41097,0.87936); rgb(32pt)=(0.27628,0.42118,0.89123); rgb(33pt)=(0.27667,0.43134,0.90254); rgb(34pt)=(0.27691,0.44145,0.91328); rgb(35pt)=(0.27701,0.45152,0.92347); rgb(36pt)=(0.27698,0.46153,0.93309); rgb(37pt)=(0.2768,0.47151,0.94214); rgb(38pt)=(0.27648,0.48144,0.95064); rgb(39pt)=(0.27603,0.49132,0.95857); rgb(40pt)=(0.27543,0.50115,0.96594); rgb(41pt)=(0.27469,0.51094,0.97275); rgb(42pt)=(0.27381,0.52069,0.97899); rgb(43pt)=(0.27273,0.5304,0.98461); rgb(44pt)=(0.27106,0.54015,0.9893); rgb(45pt)=(0.26878,0.54995,0.99303); rgb(46pt)=(0.26592,0.55979,0.99583); rgb(47pt)=(0.26252,0.56967,0.99773); rgb(48pt)=(0.25862,0.57958,0.99876); rgb(49pt)=(0.25425,0.5895,0.99896); rgb(50pt)=(0.24946,0.59943,0.99835); rgb(51pt)=(0.24427,0.60937,0.99697); rgb(52pt)=(0.23874,0.61931,0.99485); rgb(53pt)=(0.23288,0.62923,0.99202); rgb(54pt)=(0.22676,0.63913,0.98851); rgb(55pt)=(0.22039,0.64901,0.98436); rgb(56pt)=(0.21382,0.65886,0.97959); rgb(57pt)=(0.20708,0.66866,0.97423); rgb(58pt)=(0.20021,0.67842,0.96833); rgb(59pt)=(0.19326,0.68812,0.9619); rgb(60pt)=(0.18625,0.69775,0.95498); rgb(61pt)=(0.17923,0.70732,0.94761); rgb(62pt)=(0.17223,0.7168,0.93981); rgb(63pt)=(0.16529,0.7262,0.93161); rgb(64pt)=(0.15844,0.73551,0.92305); rgb(65pt)=(0.15173,0.74472,0.91416); rgb(66pt)=(0.14519,0.75381,0.90496); rgb(67pt)=(0.13886,0.76279,0.8955); rgb(68pt)=(0.13278,0.77165,0.8858); rgb(69pt)=(0.12698,0.78037,0.8759); rgb(70pt)=(0.12151,0.78896,0.86581); rgb(71pt)=(0.11639,0.7974,0.85559); rgb(72pt)=(0.11167,0.80569,0.84525); rgb(73pt)=(0.10738,0.81381,0.83484); rgb(74pt)=(0.10357,0.82177,0.82437); rgb(75pt)=(0.10026,0.82955,0.81389); rgb(76pt)=(0.0975,0.83714,0.80342); rgb(77pt)=(0.09532,0.84455,0.79299); rgb(78pt)=(0.09377,0.85175,0.78264); rgb(79pt)=(0.09287,0.85875,0.7724); rgb(80pt)=(0.09267,0.86554,0.7623); rgb(81pt)=(0.0932,0.87211,0.75237); rgb(82pt)=(0.09451,0.87844,0.74265); rgb(83pt)=(0.09662,0.88454,0.73316); rgb(84pt)=(0.09958,0.8904,0.72393); rgb(85pt)=(0.10342,0.896,0.715); rgb(86pt)=(0.10815,0.90142,0.70599); rgb(87pt)=(0.11374,0.90673,0.69651); rgb(88pt)=(0.12014,0.91193,0.6866); rgb(89pt)=(0.12733,0.91701,0.67627); rgb(90pt)=(0.13526,0.92197,0.66556); rgb(91pt)=(0.14391,0.9268,0.65448); rgb(92pt)=(0.15323,0.93151,0.64308); rgb(93pt)=(0.16319,0.93609,0.63137); rgb(94pt)=(0.17377,0.94053,0.61938); rgb(95pt)=(0.18491,0.94484,0.60713); rgb(96pt)=(0.19659,0.94901,0.59466); rgb(97pt)=(0.20877,0.95304,0.58199); rgb(98pt)=(0.22142,0.95692,0.56914); rgb(99pt)=(0.23449,0.96065,0.55614); rgb(100pt)=(0.24797,0.96423,0.54303); rgb(101pt)=(0.2618,0.96765,0.52981); rgb(102pt)=(0.27597,0.97092,0.51653); rgb(103pt)=(0.29042,0.97403,0.50321); rgb(104pt)=(0.30513,0.97697,0.48987); rgb(105pt)=(0.32006,0.97974,0.47654); rgb(106pt)=(0.33517,0.98234,0.46325); rgb(107pt)=(0.35043,0.98477,0.45002); rgb(108pt)=(0.36581,0.98702,0.43688); rgb(109pt)=(0.38127,0.98909,0.42386); rgb(110pt)=(0.39678,0.99098,0.41098); rgb(111pt)=(0.41229,0.99268,0.39826); rgb(112pt)=(0.42778,0.99419,0.38575); rgb(113pt)=(0.44321,0.99551,0.37345); rgb(114pt)=(0.45854,0.99663,0.3614); rgb(115pt)=(0.47375,0.99755,0.34963); rgb(116pt)=(0.48879,0.99828,0.33816); rgb(117pt)=(0.50362,0.99879,0.32701); rgb(118pt)=(0.51822,0.9991,0.31622); rgb(119pt)=(0.53255,0.99919,0.30581); rgb(120pt)=(0.54658,0.99907,0.29581); rgb(121pt)=(0.56026,0.99873,0.28623); rgb(122pt)=(0.57357,0.99817,0.27712); rgb(123pt)=(0.58646,0.99739,0.26849); rgb(124pt)=(0.59891,0.99638,0.26038); rgb(125pt)=(0.61088,0.99514,0.2528); rgb(126pt)=(0.62233,0.99366,0.24579); rgb(127pt)=(0.63323,0.99195,0.23937); rgb(128pt)=(0.64362,0.98999,0.23356); rgb(129pt)=(0.65394,0.98775,0.22835); rgb(130pt)=(0.66428,0.98524,0.2237); rgb(131pt)=(0.67462,0.98246,0.2196); rgb(132pt)=(0.68494,0.97941,0.21602); rgb(133pt)=(0.69525,0.9761,0.21294); rgb(134pt)=(0.70553,0.97255,0.21032); rgb(135pt)=(0.71577,0.96875,0.20815); rgb(136pt)=(0.72596,0.9647,0.2064); rgb(137pt)=(0.7361,0.96043,0.20504); rgb(138pt)=(0.74617,0.95593,0.20406); rgb(139pt)=(0.75617,0.95121,0.20343); rgb(140pt)=(0.76608,0.94627,0.20311); rgb(141pt)=(0.77591,0.94113,0.2031); rgb(142pt)=(0.78563,0.93579,0.20336); rgb(143pt)=(0.79524,0.93025,0.20386); rgb(144pt)=(0.80473,0.92452,0.20459); rgb(145pt)=(0.8141,0.91861,0.20552); rgb(146pt)=(0.82333,0.91253,0.20663); rgb(147pt)=(0.83241,0.90627,0.20788); rgb(148pt)=(0.84133,0.89986,0.20926); rgb(149pt)=(0.8501,0.89328,0.21074); rgb(150pt)=(0.85868,0.88655,0.2123); rgb(151pt)=(0.86709,0.87968,0.21391); rgb(152pt)=(0.8753,0.87267,0.21555); rgb(153pt)=(0.88331,0.86553,0.21719); rgb(154pt)=(0.89112,0.85826,0.2188); rgb(155pt)=(0.8987,0.85087,0.22038); rgb(156pt)=(0.90605,0.84337,0.22188); rgb(157pt)=(0.91317,0.83576,0.22328); rgb(158pt)=(0.92004,0.82806,0.22456); rgb(159pt)=(0.92666,0.82025,0.2257); rgb(160pt)=(0.93301,0.81236,0.22667); rgb(161pt)=(0.93909,0.80439,0.22744); rgb(162pt)=(0.94489,0.79634,0.228); rgb(163pt)=(0.95039,0.78823,0.22831); rgb(164pt)=(0.9556,0.78005,0.22836); rgb(165pt)=(0.96049,0.77181,0.22811); rgb(166pt)=(0.96507,0.76352,0.22754); rgb(167pt)=(0.96931,0.75519,0.22663); rgb(168pt)=(0.97323,0.74682,0.22536); rgb(169pt)=(0.97679,0.73842,0.22369); rgb(170pt)=(0.98,0.73,0.22161); rgb(171pt)=(0.98289,0.7214,0.21918); rgb(172pt)=(0.98549,0.7125,0.2165); rgb(173pt)=(0.98781,0.7033,0.21358); rgb(174pt)=(0.98986,0.69382,0.21043); rgb(175pt)=(0.99163,0.68408,0.20706); rgb(176pt)=(0.99314,0.67408,0.20348); rgb(177pt)=(0.99438,0.66386,0.19971); rgb(178pt)=(0.99535,0.65341,0.19577); rgb(179pt)=(0.99607,0.64277,0.19165); rgb(180pt)=(0.99654,0.63193,0.18738); rgb(181pt)=(0.99675,0.62093,0.18297); rgb(182pt)=(0.99672,0.60977,0.17842); rgb(183pt)=(0.99644,0.59846,0.17376); rgb(184pt)=(0.99593,0.58703,0.16899); rgb(185pt)=(0.99517,0.57549,0.16412); rgb(186pt)=(0.99419,0.56386,0.15918); rgb(187pt)=(0.99297,0.55214,0.15417); rgb(188pt)=(0.99153,0.54036,0.1491); rgb(189pt)=(0.98987,0.52854,0.14398); rgb(190pt)=(0.98799,0.51667,0.13883); rgb(191pt)=(0.9859,0.50479,0.13367); rgb(192pt)=(0.9836,0.49291,0.12849); rgb(193pt)=(0.98108,0.48104,0.12332); rgb(194pt)=(0.97837,0.4692,0.11817); rgb(195pt)=(0.97545,0.4574,0.11305); rgb(196pt)=(0.97234,0.44565,0.10797); rgb(197pt)=(0.96904,0.43399,0.10294); rgb(198pt)=(0.96555,0.42241,0.09798); rgb(199pt)=(0.96187,0.41093,0.0931); rgb(200pt)=(0.95801,0.39958,0.08831); rgb(201pt)=(0.95398,0.38836,0.08362); rgb(202pt)=(0.94977,0.37729,0.07905); rgb(203pt)=(0.94538,0.36638,0.07461); rgb(204pt)=(0.94084,0.35566,0.07031); rgb(205pt)=(0.93612,0.34513,0.06616); rgb(206pt)=(0.93125,0.33482,0.06218); rgb(207pt)=(0.92623,0.32473,0.05837); rgb(208pt)=(0.92105,0.31489,0.05475); rgb(209pt)=(0.91572,0.3053,0.05134); rgb(210pt)=(0.91024,0.29599,0.04814); rgb(211pt)=(0.90463,0.28696,0.04516); rgb(212pt)=(0.89888,0.27824,0.04243); rgb(213pt)=(0.89298,0.26981,0.03993); rgb(214pt)=(0.88691,0.26152,0.03753); rgb(215pt)=(0.88066,0.25334,0.03521); rgb(216pt)=(0.87422,0.24526,0.03297); rgb(217pt)=(0.8676,0.2373,0.03082); rgb(218pt)=(0.86079,0.22945,0.02875); rgb(219pt)=(0.8538,0.2217,0.02677); rgb(220pt)=(0.84662,0.21407,0.02487); rgb(221pt)=(0.83926,0.20654,0.02305); rgb(222pt)=(0.83172,0.19912,0.02131); rgb(223pt)=(0.82399,0.19182,0.01966); rgb(224pt)=(0.81608,0.18462,0.01809); rgb(225pt)=(0.80799,0.17753,0.0166); rgb(226pt)=(0.79971,0.17055,0.0152); rgb(227pt)=(0.79125,0.16368,0.01387); rgb(228pt)=(0.7826,0.15693,0.01264); rgb(229pt)=(0.77377,0.15028,0.01148); rgb(230pt)=(0.76476,0.14374,0.01041); rgb(231pt)=(0.75556,0.13731,0.00942); rgb(232pt)=(0.74617,0.13098,0.00851); rgb(233pt)=(0.73661,0.12477,0.00769); rgb(234pt)=(0.72686,0.11867,0.00695); rgb(235pt)=(0.71692,0.11268,0.00629); rgb(236pt)=(0.7068,0.1068,0.00571); rgb(237pt)=(0.6965,0.10102,0.00522); rgb(238pt)=(0.68602,0.09536,0.00481); rgb(239pt)=(0.67535,0.0898,0.00449); rgb(240pt)=(0.66449,0.08436,0.00424); rgb(241pt)=(0.65345,0.07902,0.00408); rgb(242pt)=(0.64223,0.0738,0.00401); rgb(243pt)=(0.63082,0.06868,0.00401); rgb(244pt)=(0.61923,0.06367,0.0041); rgb(245pt)=(0.60746,0.05878,0.00427); rgb(246pt)=(0.5955,0.05399,0.00453); rgb(247pt)=(0.58336,0.04931,0.00486); rgb(248pt)=(0.57103,0.04474,0.00529); rgb(249pt)=(0.55852,0.04028,0.00579); rgb(250pt)=(0.54583,0.03593,0.00638); rgb(251pt)=(0.53295,0.03169,0.00705); rgb(252pt)=(0.51989,0.02756,0.0078); rgb(253pt)=(0.50664,0.02354,0.00863); rgb(254pt)=(0.49321,0.01963,0.00955); rgb(255pt)=(0.4796,0.01583,0.01055)}, mesh/rows=11]
table[row sep=crcr, point meta=\thisrow{c}] {%
x	y	c\\
0	-1.26862410477304	0.0436830670101369\\
0	-1.26862410477304	0.0436830670101369\\
0.05	0.143107967841246	0.0432128142760893\\
0.05	0.143107967841246	0.0432128142760893\\
0.1	1.53979116564881	0.0427785276339362\\
0.1	1.53979116564881	0.0427785276339362\\
0.15	2.92244720130006	0.0423775444819262\\
0.15	2.92244720130006	0.0423775444819262\\
0.2	4.2920238277712	0.0420075322171686\\
0.2	4.2920238277712	0.0420075322171686\\
0.25	5.64940388497089	0.0416664438770201\\
0.25	5.64940388497089	0.0416664438770201\\
0.3	6.99541304383635	0.0413524813228966\\
0.3	6.99541304383635	0.0413524813228966\\
0.35	8.33082647418064	0.0410640645465885\\
0.35	8.33082647418064	0.0410640645465885\\
0.4	9.65637461724862	0.0407998059847833\\
0.4	9.65637461724862	0.0407998059847833\\
0.45	10.9727482087178	0.0405584889611939\\
0.45	10.9727482087178	0.0405584889611939\\
0.5	12.2806026702523	0.0403390495559979\\
0.5	12.2806026702523	0.0403390495559979\\
};
\addplot[
    only marks,
    mark=*,
    mark size=5pt,
    color=red, 
    mark options={
        fill=red,
        fill opacity=0.5, 
        draw opacity=1     
    },
]
coordinates {(0.0,-1.267)};
\node[color=black, fill=none] at (axis cs: 0.15,-0.7) {\textcolor{red}{Case 1b}};
\draw[line width=0.5mm,draw=black!80,->] (axis cs:0.02,-1)--(axis cs:0.08,-0.8);

\addplot[
    only marks,
    mark=*,
    mark size=5pt,
    color=blue, 
    mark options={
        fill=blue,
        fill opacity=0.5, 
        draw opacity=1     
    },
]
coordinates {(0.5,12.28)};
\node[color=black, fill=none] at (axis cs: 0.34,11.5) {\textcolor{blue}{Case 1a}};
\draw[line width=0.5mm,draw=black!80,->] (axis cs:0.47,12)--(axis cs:0.42,11.5);
\end{axis}
\end{tikzpicture}%

%% file: updated_images/Case_study2_dd.tex
%
%
\definecolor{mycolor1}{rgb}{1.00000,0.00000,1.00000}%
\begin{tikzpicture}

\begin{axis}[%
width=2.5in,
height=1in,
at={(0.758in,2.0in)},
scale only axis,
xmode=log,
xmin=1,
xmax=10000,
xminorticks=true,
xlabel style={font=\color{white!15!black}},
title style={font=\Large\bfseries},
title={$d-$ axis},
ylabel style={font=\large\color{white!15!black}},
ylabel={Magnitude [dB]},
axis background/.style={fill=white},
xmajorgrids,
xminorgrids,
ymajorgrids,
legend pos=north west,
legend style={legend cell align=left, align=left,font=\normalsize, draw=white!15!black}
]
\addplot [color=blue, line width=2.0pt]
  table[row sep=crcr]{%
1	-14.1265388868542\\
3.0888435964775	-4.37041169462748\\
4.49843266896945	-1.14719256079181\\
6.5512855685955	2.03100972220813\\
7.90604321090773	3.59021663996637\\
9.54095476349988	5.11866267466905\\
11.5139539932645	6.6052675977873\\
13.8949549437315	8.03712272193023\\
16.76832936811	9.40225477682612\\
20.2358964772516	10.6982773112825\\
24.4205309454863	11.9569185461506\\
29.4705170255181	13.3051455336572\\
35.5648030622315	15.0579946867956\\
42.9193426012876	17.626363750133\\
51.7947467923122	41.3709157775903\\
62.5055192527392	33.2969004358244\\
75.4312006335461	36.8035514434949\\
91.0298177991526	38.5216498579302\\
109.854114198755	30.6443657453008\\
132.571136559011	25.4869464170281\\
159.985871960607	21.8379725556592\\
193.069772888325	18.9701416959763\\
232.995181051538	16.6051929972213\\
281.176869797421	14.6639568932891\\
339.322177189533	13.1746341055894\\
409.491506238045	7.30196780909744\\
494.171336132382	8.14026108065946\\
596.362331659466	9.1236546212367\\
719.685673001147	10.0689666247234\\
868.511373751352	11.2210112575336\\
1048.11313415469	13.0345767043647\\
1264.85521685529	14.9354260153598\\
1526.41796717524	17.5007718570247\\
1842.06996932673	19.8388874485618\\
2222.99648252619	21.9499664814446\\
2682.69579527974	23.8988231698188\\
3237.45754281763	25.7403212973987\\
3906.93993705462	27.5117702983364\\
4714.86636345743	29.2373237733535\\
6866.48845004303	32.6075407583873\\
10000	35.9212027595052\\
};
\addlegendentry{${Z}_{s}$}
\addplot [color=mycolor1, line width=2.0pt]
  table[row sep=crcr]{%
1	-31.9883327381597\\
11.5139539932645	-10.7706611479113\\
10000	48.0047971937903\\
};
\addlegendentry{${Z}_{g}$}
\end{axis}

\begin{axis}[%
width=2.5in,
height=1in,
at={(0.758in,0.7in)},
scale only axis,
xmode=log,
xmin=1,
xmax=10000,
xminorticks=true,
xlabel style={font=\large\color{white!15!black}},
xlabel={Frequency [Hz]},
ymin=-200,
ymax=200,
ylabel style={font=\large\color{white!15!black}},
ylabel={Phase [deg]},
axis background/.style={fill=white},
xmajorgrids,
xminorgrids,
ymajorgrids,
legend pos=north east,
legend style={legend cell align=left, align=left,font=\scriptsize, draw=white!15!black}
]
\addplot [color=blue, line width=2.0pt]
  table[row sep=crcr]{%
1	87.5972489300958\\
1.20679264063936	87.9622075403612\\
1.45634847750121	88.2556358940384\\
1.7575106248548	88.4882102088597\\
2.12095088792024	88.6687626289259\\
2.55954792269949	88.8047864557123\\
3.0888435964775	88.9030575849387\\
3.72759372031504	88.9705087869905\\
4.49843266896938	89.0155908683911\\
5.4286754393239	89.0505233130967\\
6.55128556859571	89.0951275707533\\
7.9060432109076	89.1834343011348\\
9.54095476350004	89.3751078996605\\
11.5139539932641	89.7752005722355\\
13.8949549437312	90.5683486356116\\
16.7683293681103	92.0782740860244\\
20.235896477251	94.8714711218288\\
24.4205309454863	99.928578240303\\
29.4705170255186	108.860011600548\\
35.5648030622303	124.101049205998\\
42.9193426012876	151.313118925969\\
51.7947467923131	93.0672163489101\\
62.5055192527382	97.527780179954\\
75.4312006335461	122.731510980412\\
91.0298177991541	-151.143214117518\\
109.854114198753	-107.289752682296\\
132.571136559011	-91.6827076316362\\
159.98587196061	-80.8347822975059\\
193.069772888321	-70.7708267167647\\
232.995181051538	-60.157833226043\\
281.176869797431	-48.3673304403352\\
339.322177189527	-35.3088264754622\\
409.491506238045	-5.86916541732012\\
494.171336132398	10.2277851432162\\
596.362331659456	23.6141887477338\\
719.685673001159	36.2404031560446\\
1048.11313415467	63.9523659750321\\
1264.85521685531	70.7500532881484\\
1526.41796717519	78.3534912698476\\
1842.0699693267	82.3085264916286\\
2222.99648252623	84.5181171062686\\
2682.69579527965	85.8783860430693\\
3237.45754281763	86.7922073505886\\
3906.93993705468	87.4484278224937\\
4714.86636345727	87.9422212390469\\
5689.86602901828	88.3256851380515\\
6866.48845004314	88.6297752153947\\
8286.42772854666	88.8743071434291\\
10000	89.0727922558462\\
};

\addplot [color=mycolor1, line width=2.0pt]
  table[row sep=crcr]{%
1	87.721475271378\\
1.20679264063932	88.1116050725981\\
1.45634847750126	88.4350176475496\\
1.7575106248548	88.7030876292847\\
2.12095088792018	88.925265424599\\
2.55954792269958	89.109396232265\\
3.0888435964775	89.2619889998965\\
3.72759372031492	89.3884419205782\\
4.49843266896938	89.4932307869265\\
5.4286754393239	89.5800659227522\\
6.5512855685955	89.6520227093844\\
7.9060432109076	89.7116500220891\\
9.54095476350004	89.7610602426426\\
11.5139539932645	89.8020039372539\\
13.8949549437312	89.8359317889065\\
16.7683293681103	89.8640459461813\\
20.2358964772516	89.8873425896094\\
24.4205309454863	89.9066472133739\\
29.4705170255186	89.9226438665871\\
35.5648030622315	89.935899386907\\
42.9193426012876	89.9468834832808\\
51.7947467923114	89.9559853783503\\
62.5055192527403	89.9635275996219\\
91.0298177991511	89.974956266618\\
132.571136559011	89.9828037488199\\
193.069772888328	89.9881922138014\\
339.322177189539	89.9932815277787\\
719.685673001159	89.9968323301197\\
2222.99648252623	89.9989744803244\\
10000	89.9997720273368\\
};

\end{axis}
\end{tikzpicture}%

%% file: updated_images/Case_study2_qq.tex
%
%
\definecolor{mycolor1}{rgb}{1.00000,0.00000,1.00000}%
\begin{tikzpicture}

\begin{axis}[%
width=2.5in,
height=1in,
at={(0.758in,2.0in)},
scale only axis,
xmode=log,
xmin=1,
xmax=10000,
xminorticks=true,
xlabel style={font=\color{white!15!black}},
title style={font=\Large\bfseries},
title={$q-$ axis},
ylabel style={font=\large\color{white!15!black}},
ylabel={Magnitude [dB]},
axis background/.style={fill=white},
xmajorgrids,
xminorgrids,
ymajorgrids,
legend pos=north west,
legend style={legend cell align=left, align=left,font=\normalsize, draw=white!15!black}
]
\addplot [color=blue, line width=2.0pt, forget plot]
  table[row sep=crcr]{%
1	-17.0612287073617\\
3.0888435964775	-7.31118777160544\\
4.49843266896945	-4.09495012911324\\
6.5512855685955	-0.929135973653111\\
7.90604321090773	0.622105353796215\\
9.54095476349988	2.14365551704447\\
11.5139539932645	3.63063154142775\\
16.76832936811	6.53818972639807\\
20.2358964772516	8.07562822509966\\
24.4205309454863	9.9233586310585\\
29.4705170255181	12.5359076010385\\
35.5648030622315	16.6450102984931\\
42.9193426012876	24.1190317513396\\
51.7947467923122	17.2631111814373\\
62.5055192527392	14.8015584123324\\
75.4312006335461	13.7105538809778\\
109.854114198755	12.3272454109488\\
132.571136559011	11.503066842815\\
159.985871960607	10.5114580140097\\
232.995181051538	8.25593551205257\\
281.176869797421	7.34389801569971\\
339.322177189533	6.97121810786711\\
409.491506238045	12.1826215340011\\
494.171336132382	11.6262805478776\\
596.362331659466	11.2887680051748\\
719.685673001147	10.9921293602396\\
868.511373751352	11.1012093646377\\
1048.11313415469	12.4932021573577\\
1264.85521685529	15.3778131797807\\
1526.41796717524	17.7679489919152\\
1842.06996932673	19.9887449136557\\
2222.99648252619	22.0329110905724\\
2682.69579527974	23.9448212140707\\
3237.45754281763	25.7659599006882\\
3906.93993705462	27.526132601491\\
4714.86636345743	29.245402800113\\
6866.48845004303	32.6101198289617\\
10000	35.9220319496448\\
};
\addplot [color=mycolor1, line width=2.0pt, forget plot]
  table[row sep=crcr]{%
1	-31.9883327381597\\
11.5139539932645	-10.7706611479113\\
10000	48.0047971937903\\
};
\draw[line width=0.5mm,draw=black!70,loosely dashed] (axis cs:143,-50)--(axis cs:143,70);
\node[color=black, fill=none,font=\large] at (axis cs: 400,-30) {\SI{143}{\hertz}};
\end{axis}

\begin{axis}[%
width=2.5in,
height=1in,
at={(0.758in,0.7in)},
scale only axis,
xmode=log,
xmin=1,
xmax=10000,
xminorticks=true,
xlabel style={font=\large\color{white!15!black}},
xlabel={Frequency [Hz]},
ymin=-200,
ymax=200,
ylabel style={font=\large\color{white!15!black}},
ylabel={Phase [deg]},
axis background/.style={fill=white},
xmajorgrids,
xminorgrids,
ymajorgrids,
legend pos=north east,
legend style={legend cell align=left, align=left,font=\scriptsize, draw=white!15!black}
]
\addplot [color=blue, line width=2.0pt, forget plot]
  table[row sep=crcr]{%
1	-92.3681811056009\\
1.20679264063936	-91.9958011076251\\
1.45634847750121	-91.6932102806008\\
1.7575106248548	-91.4492161429148\\
2.12095088792024	-91.2542453965389\\
2.55954792269949	-91.0997030526253\\
3.0888435964775	-90.9771203822912\\
3.72759372031504	-90.8768858348166\\
4.49843266896938	-90.786205979981\\
5.4286754393239	-90.6856928451438\\
6.55128556859571	-90.5435558511509\\
7.9060432109076	-90.3057096944957\\
9.54095476350004	-89.879149511917\\
11.5139539932641	-89.1049370819609\\
13.8949549437312	-87.7176296944091\\
16.7683293681103	-85.2966088492512\\
20.235896477251	-81.2562946197556\\
24.4205309454863	-75.0665439952595\\
29.4705170255186	-67.1641718330291\\
35.5648030622303	-60.6760896438127\\
42.9193426012876	-63.926733234998\\
51.7947467923131	-159.342379554745\\
62.5055192527382	-138.58821045197\\
75.4312006335461	-125.895572714913\\
91.0298177991541	-115.627627259338\\
132.571136559011	-96.7683705508288\\
159.98587196061	-86.4636614356876\\
193.069772888321	-74.6177088733882\\
232.995181051538	-60.4717614836829\\
281.176869797431	-43.545205410002\\
339.322177189527	-24.5534735553162\\
409.491506238045	-21.5824814005714\\
494.171336132398	-8.19873641381932\\
596.362331659456	4.57903391046355\\
719.685673001159	18.5767399269451\\
868.51137375138	36.7979952479091\\
1048.11313415467	56.670307798414\\
1264.85521685531	73.9455548404431\\
1526.41796717519	79.7539553534928\\
1842.0699693267	82.9771104296285\\
2222.99648252623	84.8678183053287\\
2682.69579527965	86.0757151674168\\
3237.45754281763	86.9102113805704\\
3906.93993705468	87.5220514106023\\
4714.86636345727	87.9895546141572\\
5689.86602901828	88.3567560108457\\
6866.48845004314	88.6504640600423\\
8286.42772854666	88.8882178508887\\
10000	89.0822078311732\\
};
\addplot [color=mycolor1, line width=2.0pt, forget plot]
  table[row sep=crcr]{%
1	87.721475271378\\
1.20679264063932	88.1116050725981\\
1.45634847750126	88.4350176475496\\
1.7575106248548	88.7030876292847\\
2.12095088792018	88.925265424599\\
2.55954792269958	89.109396232265\\
3.0888435964775	89.2619889998965\\
3.72759372031492	89.3884419205782\\
4.49843266896938	89.4932307869265\\
5.4286754393239	89.5800659227522\\
6.5512855685955	89.6520227093844\\
7.9060432109076	89.7116500220891\\
9.54095476350004	89.7610602426426\\
11.5139539932645	89.8020039372539\\
13.8949549437312	89.8359317889065\\
16.7683293681103	89.8640459461813\\
20.2358964772516	89.8873425896094\\
24.4205309454863	89.9066472133739\\
29.4705170255186	89.9226438665871\\
35.5648030622315	89.935899386907\\
42.9193426012876	89.9468834832808\\
51.7947467923114	89.9559853783503\\
62.5055192527403	89.9635275996219\\
91.0298177991511	89.974956266618\\
132.571136559011	89.9828037488199\\
193.069772888328	89.9881922138014\\
339.322177189539	89.9932815277787\\
719.685673001159	89.9968323301197\\
2222.99648252623	89.9989744803244\\
10000	89.9997720273368\\
};
\draw[line width=0.5mm,draw=black!70,loosely dashed] (axis cs:143,-200)--(axis cs:143,200);
\node[color=black, fill=none, font=\normalsize] at (axis cs: 40,0) {$\Delta\phi\approx$\SI{183}{\degree}};
\end{axis}
\end{tikzpicture}%

%% file: updated_images/PLLbw_Change.tex
%
%
\begin{tikzpicture}

\begin{axis}[%
width=2in,
height=2in,
at={(0.494in,0.431in)},
scale only axis,
point meta min=0.0336515997698858,
point meta max=0.0550274408791649,
xmin=0.9,
xmax=1.6,
xtick = {1,1.1,1.2,1.3,1.4,1.5},
xticklabels={1,1.1,1.2,1.3,1.4,1.5},
xlabel style={font=\color{white!15!black}},
xlabel={PLL Bandwidth Multiplication},
ymin=142,
ymax=154,
ylabel style={align=center,font=\color{white!15!black}},
ylabel={Potential \\ Resonance Frequency [Hz]},
axis background/.style={fill=white},
xmajorgrids,
ymajorgrids,
grid style={dashed, opacity=0.5}
]

\addplot[%
mesh,
    shader=flat,
    mark=none,
    line join=round,
    line cap=round,
    line width=2pt,
    point meta=explicit, colormap={mymap}{[1pt] rgb(0pt)=(0.18995,0.07176,0.23217); rgb(1pt)=(0.19483,0.08339,0.26149); rgb(2pt)=(0.19956,0.09498,0.29024); rgb(3pt)=(0.20415,0.10652,0.31844); rgb(4pt)=(0.2086,0.11802,0.34607); rgb(5pt)=(0.21291,0.12947,0.37314); rgb(6pt)=(0.21708,0.14087,0.39964); rgb(7pt)=(0.22111,0.15223,0.42558); rgb(8pt)=(0.225,0.16354,0.45096); rgb(9pt)=(0.22875,0.17481,0.47578); rgb(10pt)=(0.23236,0.18603,0.50004); rgb(11pt)=(0.23582,0.1972,0.52373); rgb(12pt)=(0.23915,0.20833,0.54686); rgb(13pt)=(0.24234,0.21941,0.56942); rgb(14pt)=(0.24539,0.23044,0.59142); rgb(15pt)=(0.2483,0.24143,0.61286); rgb(16pt)=(0.25107,0.25237,0.63374); rgb(17pt)=(0.25369,0.26327,0.65406); rgb(18pt)=(0.25618,0.27412,0.67381); rgb(19pt)=(0.25853,0.28492,0.693); rgb(20pt)=(0.26074,0.29568,0.71162); rgb(21pt)=(0.2628,0.30639,0.72968); rgb(22pt)=(0.26473,0.31706,0.74718); rgb(23pt)=(0.26652,0.32768,0.76412); rgb(24pt)=(0.26816,0.33825,0.7805); rgb(25pt)=(0.26967,0.34878,0.79631); rgb(26pt)=(0.27103,0.35926,0.81156); rgb(27pt)=(0.27226,0.3697,0.82624); rgb(28pt)=(0.27334,0.38008,0.84037); rgb(29pt)=(0.27429,0.39043,0.85393); rgb(30pt)=(0.27509,0.40072,0.86692); rgb(31pt)=(0.27576,0.41097,0.87936); rgb(32pt)=(0.27628,0.42118,0.89123); rgb(33pt)=(0.27667,0.43134,0.90254); rgb(34pt)=(0.27691,0.44145,0.91328); rgb(35pt)=(0.27701,0.45152,0.92347); rgb(36pt)=(0.27698,0.46153,0.93309); rgb(37pt)=(0.2768,0.47151,0.94214); rgb(38pt)=(0.27648,0.48144,0.95064); rgb(39pt)=(0.27603,0.49132,0.95857); rgb(40pt)=(0.27543,0.50115,0.96594); rgb(41pt)=(0.27469,0.51094,0.97275); rgb(42pt)=(0.27381,0.52069,0.97899); rgb(43pt)=(0.27273,0.5304,0.98461); rgb(44pt)=(0.27106,0.54015,0.9893); rgb(45pt)=(0.26878,0.54995,0.99303); rgb(46pt)=(0.26592,0.55979,0.99583); rgb(47pt)=(0.26252,0.56967,0.99773); rgb(48pt)=(0.25862,0.57958,0.99876); rgb(49pt)=(0.25425,0.5895,0.99896); rgb(50pt)=(0.24946,0.59943,0.99835); rgb(51pt)=(0.24427,0.60937,0.99697); rgb(52pt)=(0.23874,0.61931,0.99485); rgb(53pt)=(0.23288,0.62923,0.99202); rgb(54pt)=(0.22676,0.63913,0.98851); rgb(55pt)=(0.22039,0.64901,0.98436); rgb(56pt)=(0.21382,0.65886,0.97959); rgb(57pt)=(0.20708,0.66866,0.97423); rgb(58pt)=(0.20021,0.67842,0.96833); rgb(59pt)=(0.19326,0.68812,0.9619); rgb(60pt)=(0.18625,0.69775,0.95498); rgb(61pt)=(0.17923,0.70732,0.94761); rgb(62pt)=(0.17223,0.7168,0.93981); rgb(63pt)=(0.16529,0.7262,0.93161); rgb(64pt)=(0.15844,0.73551,0.92305); rgb(65pt)=(0.15173,0.74472,0.91416); rgb(66pt)=(0.14519,0.75381,0.90496); rgb(67pt)=(0.13886,0.76279,0.8955); rgb(68pt)=(0.13278,0.77165,0.8858); rgb(69pt)=(0.12698,0.78037,0.8759); rgb(70pt)=(0.12151,0.78896,0.86581); rgb(71pt)=(0.11639,0.7974,0.85559); rgb(72pt)=(0.11167,0.80569,0.84525); rgb(73pt)=(0.10738,0.81381,0.83484); rgb(74pt)=(0.10357,0.82177,0.82437); rgb(75pt)=(0.10026,0.82955,0.81389); rgb(76pt)=(0.0975,0.83714,0.80342); rgb(77pt)=(0.09532,0.84455,0.79299); rgb(78pt)=(0.09377,0.85175,0.78264); rgb(79pt)=(0.09287,0.85875,0.7724); rgb(80pt)=(0.09267,0.86554,0.7623); rgb(81pt)=(0.0932,0.87211,0.75237); rgb(82pt)=(0.09451,0.87844,0.74265); rgb(83pt)=(0.09662,0.88454,0.73316); rgb(84pt)=(0.09958,0.8904,0.72393); rgb(85pt)=(0.10342,0.896,0.715); rgb(86pt)=(0.10815,0.90142,0.70599); rgb(87pt)=(0.11374,0.90673,0.69651); rgb(88pt)=(0.12014,0.91193,0.6866); rgb(89pt)=(0.12733,0.91701,0.67627); rgb(90pt)=(0.13526,0.92197,0.66556); rgb(91pt)=(0.14391,0.9268,0.65448); rgb(92pt)=(0.15323,0.93151,0.64308); rgb(93pt)=(0.16319,0.93609,0.63137); rgb(94pt)=(0.17377,0.94053,0.61938); rgb(95pt)=(0.18491,0.94484,0.60713); rgb(96pt)=(0.19659,0.94901,0.59466); rgb(97pt)=(0.20877,0.95304,0.58199); rgb(98pt)=(0.22142,0.95692,0.56914); rgb(99pt)=(0.23449,0.96065,0.55614); rgb(100pt)=(0.24797,0.96423,0.54303); rgb(101pt)=(0.2618,0.96765,0.52981); rgb(102pt)=(0.27597,0.97092,0.51653); rgb(103pt)=(0.29042,0.97403,0.50321); rgb(104pt)=(0.30513,0.97697,0.48987); rgb(105pt)=(0.32006,0.97974,0.47654); rgb(106pt)=(0.33517,0.98234,0.46325); rgb(107pt)=(0.35043,0.98477,0.45002); rgb(108pt)=(0.36581,0.98702,0.43688); rgb(109pt)=(0.38127,0.98909,0.42386); rgb(110pt)=(0.39678,0.99098,0.41098); rgb(111pt)=(0.41229,0.99268,0.39826); rgb(112pt)=(0.42778,0.99419,0.38575); rgb(113pt)=(0.44321,0.99551,0.37345); rgb(114pt)=(0.45854,0.99663,0.3614); rgb(115pt)=(0.47375,0.99755,0.34963); rgb(116pt)=(0.48879,0.99828,0.33816); rgb(117pt)=(0.50362,0.99879,0.32701); rgb(118pt)=(0.51822,0.9991,0.31622); rgb(119pt)=(0.53255,0.99919,0.30581); rgb(120pt)=(0.54658,0.99907,0.29581); rgb(121pt)=(0.56026,0.99873,0.28623); rgb(122pt)=(0.57357,0.99817,0.27712); rgb(123pt)=(0.58646,0.99739,0.26849); rgb(124pt)=(0.59891,0.99638,0.26038); rgb(125pt)=(0.61088,0.99514,0.2528); rgb(126pt)=(0.62233,0.99366,0.24579); rgb(127pt)=(0.63323,0.99195,0.23937); rgb(128pt)=(0.64362,0.98999,0.23356); rgb(129pt)=(0.65394,0.98775,0.22835); rgb(130pt)=(0.66428,0.98524,0.2237); rgb(131pt)=(0.67462,0.98246,0.2196); rgb(132pt)=(0.68494,0.97941,0.21602); rgb(133pt)=(0.69525,0.9761,0.21294); rgb(134pt)=(0.70553,0.97255,0.21032); rgb(135pt)=(0.71577,0.96875,0.20815); rgb(136pt)=(0.72596,0.9647,0.2064); rgb(137pt)=(0.7361,0.96043,0.20504); rgb(138pt)=(0.74617,0.95593,0.20406); rgb(139pt)=(0.75617,0.95121,0.20343); rgb(140pt)=(0.76608,0.94627,0.20311); rgb(141pt)=(0.77591,0.94113,0.2031); rgb(142pt)=(0.78563,0.93579,0.20336); rgb(143pt)=(0.79524,0.93025,0.20386); rgb(144pt)=(0.80473,0.92452,0.20459); rgb(145pt)=(0.8141,0.91861,0.20552); rgb(146pt)=(0.82333,0.91253,0.20663); rgb(147pt)=(0.83241,0.90627,0.20788); rgb(148pt)=(0.84133,0.89986,0.20926); rgb(149pt)=(0.8501,0.89328,0.21074); rgb(150pt)=(0.85868,0.88655,0.2123); rgb(151pt)=(0.86709,0.87968,0.21391); rgb(152pt)=(0.8753,0.87267,0.21555); rgb(153pt)=(0.88331,0.86553,0.21719); rgb(154pt)=(0.89112,0.85826,0.2188); rgb(155pt)=(0.8987,0.85087,0.22038); rgb(156pt)=(0.90605,0.84337,0.22188); rgb(157pt)=(0.91317,0.83576,0.22328); rgb(158pt)=(0.92004,0.82806,0.22456); rgb(159pt)=(0.92666,0.82025,0.2257); rgb(160pt)=(0.93301,0.81236,0.22667); rgb(161pt)=(0.93909,0.80439,0.22744); rgb(162pt)=(0.94489,0.79634,0.228); rgb(163pt)=(0.95039,0.78823,0.22831); rgb(164pt)=(0.9556,0.78005,0.22836); rgb(165pt)=(0.96049,0.77181,0.22811); rgb(166pt)=(0.96507,0.76352,0.22754); rgb(167pt)=(0.96931,0.75519,0.22663); rgb(168pt)=(0.97323,0.74682,0.22536); rgb(169pt)=(0.97679,0.73842,0.22369); rgb(170pt)=(0.98,0.73,0.22161); rgb(171pt)=(0.98289,0.7214,0.21918); rgb(172pt)=(0.98549,0.7125,0.2165); rgb(173pt)=(0.98781,0.7033,0.21358); rgb(174pt)=(0.98986,0.69382,0.21043); rgb(175pt)=(0.99163,0.68408,0.20706); rgb(176pt)=(0.99314,0.67408,0.20348); rgb(177pt)=(0.99438,0.66386,0.19971); rgb(178pt)=(0.99535,0.65341,0.19577); rgb(179pt)=(0.99607,0.64277,0.19165); rgb(180pt)=(0.99654,0.63193,0.18738); rgb(181pt)=(0.99675,0.62093,0.18297); rgb(182pt)=(0.99672,0.60977,0.17842); rgb(183pt)=(0.99644,0.59846,0.17376); rgb(184pt)=(0.99593,0.58703,0.16899); rgb(185pt)=(0.99517,0.57549,0.16412); rgb(186pt)=(0.99419,0.56386,0.15918); rgb(187pt)=(0.99297,0.55214,0.15417); rgb(188pt)=(0.99153,0.54036,0.1491); rgb(189pt)=(0.98987,0.52854,0.14398); rgb(190pt)=(0.98799,0.51667,0.13883); rgb(191pt)=(0.9859,0.50479,0.13367); rgb(192pt)=(0.9836,0.49291,0.12849); rgb(193pt)=(0.98108,0.48104,0.12332); rgb(194pt)=(0.97837,0.4692,0.11817); rgb(195pt)=(0.97545,0.4574,0.11305); rgb(196pt)=(0.97234,0.44565,0.10797); rgb(197pt)=(0.96904,0.43399,0.10294); rgb(198pt)=(0.96555,0.42241,0.09798); rgb(199pt)=(0.96187,0.41093,0.0931); rgb(200pt)=(0.95801,0.39958,0.08831); rgb(201pt)=(0.95398,0.38836,0.08362); rgb(202pt)=(0.94977,0.37729,0.07905); rgb(203pt)=(0.94538,0.36638,0.07461); rgb(204pt)=(0.94084,0.35566,0.07031); rgb(205pt)=(0.93612,0.34513,0.06616); rgb(206pt)=(0.93125,0.33482,0.06218); rgb(207pt)=(0.92623,0.32473,0.05837); rgb(208pt)=(0.92105,0.31489,0.05475); rgb(209pt)=(0.91572,0.3053,0.05134); rgb(210pt)=(0.91024,0.29599,0.04814); rgb(211pt)=(0.90463,0.28696,0.04516); rgb(212pt)=(0.89888,0.27824,0.04243); rgb(213pt)=(0.89298,0.26981,0.03993); rgb(214pt)=(0.88691,0.26152,0.03753); rgb(215pt)=(0.88066,0.25334,0.03521); rgb(216pt)=(0.87422,0.24526,0.03297); rgb(217pt)=(0.8676,0.2373,0.03082); rgb(218pt)=(0.86079,0.22945,0.02875); rgb(219pt)=(0.8538,0.2217,0.02677); rgb(220pt)=(0.84662,0.21407,0.02487); rgb(221pt)=(0.83926,0.20654,0.02305); rgb(222pt)=(0.83172,0.19912,0.02131); rgb(223pt)=(0.82399,0.19182,0.01966); rgb(224pt)=(0.81608,0.18462,0.01809); rgb(225pt)=(0.80799,0.17753,0.0166); rgb(226pt)=(0.79971,0.17055,0.0152); rgb(227pt)=(0.79125,0.16368,0.01387); rgb(228pt)=(0.7826,0.15693,0.01264); rgb(229pt)=(0.77377,0.15028,0.01148); rgb(230pt)=(0.76476,0.14374,0.01041); rgb(231pt)=(0.75556,0.13731,0.00942); rgb(232pt)=(0.74617,0.13098,0.00851); rgb(233pt)=(0.73661,0.12477,0.00769); rgb(234pt)=(0.72686,0.11867,0.00695); rgb(235pt)=(0.71692,0.11268,0.00629); rgb(236pt)=(0.7068,0.1068,0.00571); rgb(237pt)=(0.6965,0.10102,0.00522); rgb(238pt)=(0.68602,0.09536,0.00481); rgb(239pt)=(0.67535,0.0898,0.00449); rgb(240pt)=(0.66449,0.08436,0.00424); rgb(241pt)=(0.65345,0.07902,0.00408); rgb(242pt)=(0.64223,0.0738,0.00401); rgb(243pt)=(0.63082,0.06868,0.00401); rgb(244pt)=(0.61923,0.06367,0.0041); rgb(245pt)=(0.60746,0.05878,0.00427); rgb(246pt)=(0.5955,0.05399,0.00453); rgb(247pt)=(0.58336,0.04931,0.00486); rgb(248pt)=(0.57103,0.04474,0.00529); rgb(249pt)=(0.55852,0.04028,0.00579); rgb(250pt)=(0.54583,0.03593,0.00638); rgb(251pt)=(0.53295,0.03169,0.00705); rgb(252pt)=(0.51989,0.02756,0.0078); rgb(253pt)=(0.50664,0.02354,0.00863); rgb(254pt)=(0.49321,0.01963,0.00955); rgb(255pt)=(0.4796,0.01583,0.01055)}, mesh/rows=11]
table[row sep=crcr, point meta=\thisrow{c}] {%
x	y	c\\
1	153.819367838882	0.0336515997698858\\
1	153.819367838882	0.0336515997698858\\
1.05	152.765093896341	0.0355451425539663\\
1.05	152.765093896341	0.0355451425539663\\
1.1	151.703140869073	0.0374961668515161\\
1.1	151.703140869073	0.0374961668515161\\
1.15	150.633475292218	0.0395037508608466\\
1.15	150.633475292218	0.0395037508608466\\
1.2	149.556151272062	0.0415668464270833\\
1.2	149.556151272062	0.0415668464270833\\
1.25	148.471324909778	0.0436842463462985\\
1.25	148.471324909778	0.0436842463462985\\
1.3	147.379270774133	0.0458545476256419\\
1.3	147.379270774133	0.0458545476256419\\
1.35	146.280400587252	0.0480761105050804\\
1.35	146.280400587252	0.0480761105050804\\
1.4	145.175284235634	0.0503470131940121\\
1.4	145.175284235634	0.0503470131940121\\
1.45	144.0646731317	0.0526650025028645\\
1.45	144.0646731317	0.0526650025028645\\
1.5	142.949525817057	0.0550274408791649\\
1.5	142.949525817057	0.0550274408791649\\
};
\end{axis}

\begin{axis}[%
width=2in,
height=2in,
at={(3.1in,0.431in)},
scale only axis,
point meta min=0.0336515997698858,
point meta max=0.0550274408791649,
xmin=0.9,
xmax=1.6,
xtick = {1,1.1,1.2,1.3,1.4,1.5},
xticklabels={1,1.1,1.2,1.3,1.4,1.5},
xlabel style={font=\color{white!15!black}},
xlabel={PLL Bandwidth Multiplication},
ymin=-3,
ymax=6,
ylabel style={font=\color{white!15!black}},
ylabel={Phase Margin [Deg]},
axis background/.style={fill=white},
xmajorgrids,
ymajorgrids,
grid style={dashed, opacity=0.5},
colormap={mymap}{[1pt] rgb(0pt)=(0.18995,0.07176,0.23217); rgb(1pt)=(0.19483,0.08339,0.26149); rgb(2pt)=(0.19956,0.09498,0.29024); rgb(3pt)=(0.20415,0.10652,0.31844); rgb(4pt)=(0.2086,0.11802,0.34607); rgb(5pt)=(0.21291,0.12947,0.37314); rgb(6pt)=(0.21708,0.14087,0.39964); rgb(7pt)=(0.22111,0.15223,0.42558); rgb(8pt)=(0.225,0.16354,0.45096); rgb(9pt)=(0.22875,0.17481,0.47578); rgb(10pt)=(0.23236,0.18603,0.50004); rgb(11pt)=(0.23582,0.1972,0.52373); rgb(12pt)=(0.23915,0.20833,0.54686); rgb(13pt)=(0.24234,0.21941,0.56942); rgb(14pt)=(0.24539,0.23044,0.59142); rgb(15pt)=(0.2483,0.24143,0.61286); rgb(16pt)=(0.25107,0.25237,0.63374); rgb(17pt)=(0.25369,0.26327,0.65406); rgb(18pt)=(0.25618,0.27412,0.67381); rgb(19pt)=(0.25853,0.28492,0.693); rgb(20pt)=(0.26074,0.29568,0.71162); rgb(21pt)=(0.2628,0.30639,0.72968); rgb(22pt)=(0.26473,0.31706,0.74718); rgb(23pt)=(0.26652,0.32768,0.76412); rgb(24pt)=(0.26816,0.33825,0.7805); rgb(25pt)=(0.26967,0.34878,0.79631); rgb(26pt)=(0.27103,0.35926,0.81156); rgb(27pt)=(0.27226,0.3697,0.82624); rgb(28pt)=(0.27334,0.38008,0.84037); rgb(29pt)=(0.27429,0.39043,0.85393); rgb(30pt)=(0.27509,0.40072,0.86692); rgb(31pt)=(0.27576,0.41097,0.87936); rgb(32pt)=(0.27628,0.42118,0.89123); rgb(33pt)=(0.27667,0.43134,0.90254); rgb(34pt)=(0.27691,0.44145,0.91328); rgb(35pt)=(0.27701,0.45152,0.92347); rgb(36pt)=(0.27698,0.46153,0.93309); rgb(37pt)=(0.2768,0.47151,0.94214); rgb(38pt)=(0.27648,0.48144,0.95064); rgb(39pt)=(0.27603,0.49132,0.95857); rgb(40pt)=(0.27543,0.50115,0.96594); rgb(41pt)=(0.27469,0.51094,0.97275); rgb(42pt)=(0.27381,0.52069,0.97899); rgb(43pt)=(0.27273,0.5304,0.98461); rgb(44pt)=(0.27106,0.54015,0.9893); rgb(45pt)=(0.26878,0.54995,0.99303); rgb(46pt)=(0.26592,0.55979,0.99583); rgb(47pt)=(0.26252,0.56967,0.99773); rgb(48pt)=(0.25862,0.57958,0.99876); rgb(49pt)=(0.25425,0.5895,0.99896); rgb(50pt)=(0.24946,0.59943,0.99835); rgb(51pt)=(0.24427,0.60937,0.99697); rgb(52pt)=(0.23874,0.61931,0.99485); rgb(53pt)=(0.23288,0.62923,0.99202); rgb(54pt)=(0.22676,0.63913,0.98851); rgb(55pt)=(0.22039,0.64901,0.98436); rgb(56pt)=(0.21382,0.65886,0.97959); rgb(57pt)=(0.20708,0.66866,0.97423); rgb(58pt)=(0.20021,0.67842,0.96833); rgb(59pt)=(0.19326,0.68812,0.9619); rgb(60pt)=(0.18625,0.69775,0.95498); rgb(61pt)=(0.17923,0.70732,0.94761); rgb(62pt)=(0.17223,0.7168,0.93981); rgb(63pt)=(0.16529,0.7262,0.93161); rgb(64pt)=(0.15844,0.73551,0.92305); rgb(65pt)=(0.15173,0.74472,0.91416); rgb(66pt)=(0.14519,0.75381,0.90496); rgb(67pt)=(0.13886,0.76279,0.8955); rgb(68pt)=(0.13278,0.77165,0.8858); rgb(69pt)=(0.12698,0.78037,0.8759); rgb(70pt)=(0.12151,0.78896,0.86581); rgb(71pt)=(0.11639,0.7974,0.85559); rgb(72pt)=(0.11167,0.80569,0.84525); rgb(73pt)=(0.10738,0.81381,0.83484); rgb(74pt)=(0.10357,0.82177,0.82437); rgb(75pt)=(0.10026,0.82955,0.81389); rgb(76pt)=(0.0975,0.83714,0.80342); rgb(77pt)=(0.09532,0.84455,0.79299); rgb(78pt)=(0.09377,0.85175,0.78264); rgb(79pt)=(0.09287,0.85875,0.7724); rgb(80pt)=(0.09267,0.86554,0.7623); rgb(81pt)=(0.0932,0.87211,0.75237); rgb(82pt)=(0.09451,0.87844,0.74265); rgb(83pt)=(0.09662,0.88454,0.73316); rgb(84pt)=(0.09958,0.8904,0.72393); rgb(85pt)=(0.10342,0.896,0.715); rgb(86pt)=(0.10815,0.90142,0.70599); rgb(87pt)=(0.11374,0.90673,0.69651); rgb(88pt)=(0.12014,0.91193,0.6866); rgb(89pt)=(0.12733,0.91701,0.67627); rgb(90pt)=(0.13526,0.92197,0.66556); rgb(91pt)=(0.14391,0.9268,0.65448); rgb(92pt)=(0.15323,0.93151,0.64308); rgb(93pt)=(0.16319,0.93609,0.63137); rgb(94pt)=(0.17377,0.94053,0.61938); rgb(95pt)=(0.18491,0.94484,0.60713); rgb(96pt)=(0.19659,0.94901,0.59466); rgb(97pt)=(0.20877,0.95304,0.58199); rgb(98pt)=(0.22142,0.95692,0.56914); rgb(99pt)=(0.23449,0.96065,0.55614); rgb(100pt)=(0.24797,0.96423,0.54303); rgb(101pt)=(0.2618,0.96765,0.52981); rgb(102pt)=(0.27597,0.97092,0.51653); rgb(103pt)=(0.29042,0.97403,0.50321); rgb(104pt)=(0.30513,0.97697,0.48987); rgb(105pt)=(0.32006,0.97974,0.47654); rgb(106pt)=(0.33517,0.98234,0.46325); rgb(107pt)=(0.35043,0.98477,0.45002); rgb(108pt)=(0.36581,0.98702,0.43688); rgb(109pt)=(0.38127,0.98909,0.42386); rgb(110pt)=(0.39678,0.99098,0.41098); rgb(111pt)=(0.41229,0.99268,0.39826); rgb(112pt)=(0.42778,0.99419,0.38575); rgb(113pt)=(0.44321,0.99551,0.37345); rgb(114pt)=(0.45854,0.99663,0.3614); rgb(115pt)=(0.47375,0.99755,0.34963); rgb(116pt)=(0.48879,0.99828,0.33816); rgb(117pt)=(0.50362,0.99879,0.32701); rgb(118pt)=(0.51822,0.9991,0.31622); rgb(119pt)=(0.53255,0.99919,0.30581); rgb(120pt)=(0.54658,0.99907,0.29581); rgb(121pt)=(0.56026,0.99873,0.28623); rgb(122pt)=(0.57357,0.99817,0.27712); rgb(123pt)=(0.58646,0.99739,0.26849); rgb(124pt)=(0.59891,0.99638,0.26038); rgb(125pt)=(0.61088,0.99514,0.2528); rgb(126pt)=(0.62233,0.99366,0.24579); rgb(127pt)=(0.63323,0.99195,0.23937); rgb(128pt)=(0.64362,0.98999,0.23356); rgb(129pt)=(0.65394,0.98775,0.22835); rgb(130pt)=(0.66428,0.98524,0.2237); rgb(131pt)=(0.67462,0.98246,0.2196); rgb(132pt)=(0.68494,0.97941,0.21602); rgb(133pt)=(0.69525,0.9761,0.21294); rgb(134pt)=(0.70553,0.97255,0.21032); rgb(135pt)=(0.71577,0.96875,0.20815); rgb(136pt)=(0.72596,0.9647,0.2064); rgb(137pt)=(0.7361,0.96043,0.20504); rgb(138pt)=(0.74617,0.95593,0.20406); rgb(139pt)=(0.75617,0.95121,0.20343); rgb(140pt)=(0.76608,0.94627,0.20311); rgb(141pt)=(0.77591,0.94113,0.2031); rgb(142pt)=(0.78563,0.93579,0.20336); rgb(143pt)=(0.79524,0.93025,0.20386); rgb(144pt)=(0.80473,0.92452,0.20459); rgb(145pt)=(0.8141,0.91861,0.20552); rgb(146pt)=(0.82333,0.91253,0.20663); rgb(147pt)=(0.83241,0.90627,0.20788); rgb(148pt)=(0.84133,0.89986,0.20926); rgb(149pt)=(0.8501,0.89328,0.21074); rgb(150pt)=(0.85868,0.88655,0.2123); rgb(151pt)=(0.86709,0.87968,0.21391); rgb(152pt)=(0.8753,0.87267,0.21555); rgb(153pt)=(0.88331,0.86553,0.21719); rgb(154pt)=(0.89112,0.85826,0.2188); rgb(155pt)=(0.8987,0.85087,0.22038); rgb(156pt)=(0.90605,0.84337,0.22188); rgb(157pt)=(0.91317,0.83576,0.22328); rgb(158pt)=(0.92004,0.82806,0.22456); rgb(159pt)=(0.92666,0.82025,0.2257); rgb(160pt)=(0.93301,0.81236,0.22667); rgb(161pt)=(0.93909,0.80439,0.22744); rgb(162pt)=(0.94489,0.79634,0.228); rgb(163pt)=(0.95039,0.78823,0.22831); rgb(164pt)=(0.9556,0.78005,0.22836); rgb(165pt)=(0.96049,0.77181,0.22811); rgb(166pt)=(0.96507,0.76352,0.22754); rgb(167pt)=(0.96931,0.75519,0.22663); rgb(168pt)=(0.97323,0.74682,0.22536); rgb(169pt)=(0.97679,0.73842,0.22369); rgb(170pt)=(0.98,0.73,0.22161); rgb(171pt)=(0.98289,0.7214,0.21918); rgb(172pt)=(0.98549,0.7125,0.2165); rgb(173pt)=(0.98781,0.7033,0.21358); rgb(174pt)=(0.98986,0.69382,0.21043); rgb(175pt)=(0.99163,0.68408,0.20706); rgb(176pt)=(0.99314,0.67408,0.20348); rgb(177pt)=(0.99438,0.66386,0.19971); rgb(178pt)=(0.99535,0.65341,0.19577); rgb(179pt)=(0.99607,0.64277,0.19165); rgb(180pt)=(0.99654,0.63193,0.18738); rgb(181pt)=(0.99675,0.62093,0.18297); rgb(182pt)=(0.99672,0.60977,0.17842); rgb(183pt)=(0.99644,0.59846,0.17376); rgb(184pt)=(0.99593,0.58703,0.16899); rgb(185pt)=(0.99517,0.57549,0.16412); rgb(186pt)=(0.99419,0.56386,0.15918); rgb(187pt)=(0.99297,0.55214,0.15417); rgb(188pt)=(0.99153,0.54036,0.1491); rgb(189pt)=(0.98987,0.52854,0.14398); rgb(190pt)=(0.98799,0.51667,0.13883); rgb(191pt)=(0.9859,0.50479,0.13367); rgb(192pt)=(0.9836,0.49291,0.12849); rgb(193pt)=(0.98108,0.48104,0.12332); rgb(194pt)=(0.97837,0.4692,0.11817); rgb(195pt)=(0.97545,0.4574,0.11305); rgb(196pt)=(0.97234,0.44565,0.10797); rgb(197pt)=(0.96904,0.43399,0.10294); rgb(198pt)=(0.96555,0.42241,0.09798); rgb(199pt)=(0.96187,0.41093,0.0931); rgb(200pt)=(0.95801,0.39958,0.08831); rgb(201pt)=(0.95398,0.38836,0.08362); rgb(202pt)=(0.94977,0.37729,0.07905); rgb(203pt)=(0.94538,0.36638,0.07461); rgb(204pt)=(0.94084,0.35566,0.07031); rgb(205pt)=(0.93612,0.34513,0.06616); rgb(206pt)=(0.93125,0.33482,0.06218); rgb(207pt)=(0.92623,0.32473,0.05837); rgb(208pt)=(0.92105,0.31489,0.05475); rgb(209pt)=(0.91572,0.3053,0.05134); rgb(210pt)=(0.91024,0.29599,0.04814); rgb(211pt)=(0.90463,0.28696,0.04516); rgb(212pt)=(0.89888,0.27824,0.04243); rgb(213pt)=(0.89298,0.26981,0.03993); rgb(214pt)=(0.88691,0.26152,0.03753); rgb(215pt)=(0.88066,0.25334,0.03521); rgb(216pt)=(0.87422,0.24526,0.03297); rgb(217pt)=(0.8676,0.2373,0.03082); rgb(218pt)=(0.86079,0.22945,0.02875); rgb(219pt)=(0.8538,0.2217,0.02677); rgb(220pt)=(0.84662,0.21407,0.02487); rgb(221pt)=(0.83926,0.20654,0.02305); rgb(222pt)=(0.83172,0.19912,0.02131); rgb(223pt)=(0.82399,0.19182,0.01966); rgb(224pt)=(0.81608,0.18462,0.01809); rgb(225pt)=(0.80799,0.17753,0.0166); rgb(226pt)=(0.79971,0.17055,0.0152); rgb(227pt)=(0.79125,0.16368,0.01387); rgb(228pt)=(0.7826,0.15693,0.01264); rgb(229pt)=(0.77377,0.15028,0.01148); rgb(230pt)=(0.76476,0.14374,0.01041); rgb(231pt)=(0.75556,0.13731,0.00942); rgb(232pt)=(0.74617,0.13098,0.00851); rgb(233pt)=(0.73661,0.12477,0.00769); rgb(234pt)=(0.72686,0.11867,0.00695); rgb(235pt)=(0.71692,0.11268,0.00629); rgb(236pt)=(0.7068,0.1068,0.00571); rgb(237pt)=(0.6965,0.10102,0.00522); rgb(238pt)=(0.68602,0.09536,0.00481); rgb(239pt)=(0.67535,0.0898,0.00449); rgb(240pt)=(0.66449,0.08436,0.00424); rgb(241pt)=(0.65345,0.07902,0.00408); rgb(242pt)=(0.64223,0.0738,0.00401); rgb(243pt)=(0.63082,0.06868,0.00401); rgb(244pt)=(0.61923,0.06367,0.0041); rgb(245pt)=(0.60746,0.05878,0.00427); rgb(246pt)=(0.5955,0.05399,0.00453); rgb(247pt)=(0.58336,0.04931,0.00486); rgb(248pt)=(0.57103,0.04474,0.00529); rgb(249pt)=(0.55852,0.04028,0.00579); rgb(250pt)=(0.54583,0.03593,0.00638); rgb(251pt)=(0.53295,0.03169,0.00705); rgb(252pt)=(0.51989,0.02756,0.0078); rgb(253pt)=(0.50664,0.02354,0.00863); rgb(254pt)=(0.49321,0.01963,0.00955); rgb(255pt)=(0.4796,0.01583,0.01055)},
colorbar,
colorbar style={ylabel style={rotate=-90,font=\color{white!15!black},at={(0.5,1)}, anchor=south,},scaled ticks=false,ytick={0.04,0.045,0.05},yticklabels={0.04,0.045,0.05}, ylabel={$\bm{AQI}$}}
]
\draw[fill=black, thick,opacity=0.1] (axis cs: 0,-90) -- (axis cs: 16,-90) -- (axis cs: 16,0) -- (axis cs: 0,0) -- cycle;
\node[color=black, fill=none] at (axis cs: 1,-0.5) {$Unstable$};
\node[color=black, fill=none] at (axis cs: 1,0.5) {$Stable$};
\addplot[%
mesh,
    shader=flat,
    mark=none,
    line join=round,
    line cap=round,
    line width=2pt,
    point meta=explicit,colormap={mymap}{[1pt] rgb(0pt)=(0.18995,0.07176,0.23217); rgb(1pt)=(0.19483,0.08339,0.26149); rgb(2pt)=(0.19956,0.09498,0.29024); rgb(3pt)=(0.20415,0.10652,0.31844); rgb(4pt)=(0.2086,0.11802,0.34607); rgb(5pt)=(0.21291,0.12947,0.37314); rgb(6pt)=(0.21708,0.14087,0.39964); rgb(7pt)=(0.22111,0.15223,0.42558); rgb(8pt)=(0.225,0.16354,0.45096); rgb(9pt)=(0.22875,0.17481,0.47578); rgb(10pt)=(0.23236,0.18603,0.50004); rgb(11pt)=(0.23582,0.1972,0.52373); rgb(12pt)=(0.23915,0.20833,0.54686); rgb(13pt)=(0.24234,0.21941,0.56942); rgb(14pt)=(0.24539,0.23044,0.59142); rgb(15pt)=(0.2483,0.24143,0.61286); rgb(16pt)=(0.25107,0.25237,0.63374); rgb(17pt)=(0.25369,0.26327,0.65406); rgb(18pt)=(0.25618,0.27412,0.67381); rgb(19pt)=(0.25853,0.28492,0.693); rgb(20pt)=(0.26074,0.29568,0.71162); rgb(21pt)=(0.2628,0.30639,0.72968); rgb(22pt)=(0.26473,0.31706,0.74718); rgb(23pt)=(0.26652,0.32768,0.76412); rgb(24pt)=(0.26816,0.33825,0.7805); rgb(25pt)=(0.26967,0.34878,0.79631); rgb(26pt)=(0.27103,0.35926,0.81156); rgb(27pt)=(0.27226,0.3697,0.82624); rgb(28pt)=(0.27334,0.38008,0.84037); rgb(29pt)=(0.27429,0.39043,0.85393); rgb(30pt)=(0.27509,0.40072,0.86692); rgb(31pt)=(0.27576,0.41097,0.87936); rgb(32pt)=(0.27628,0.42118,0.89123); rgb(33pt)=(0.27667,0.43134,0.90254); rgb(34pt)=(0.27691,0.44145,0.91328); rgb(35pt)=(0.27701,0.45152,0.92347); rgb(36pt)=(0.27698,0.46153,0.93309); rgb(37pt)=(0.2768,0.47151,0.94214); rgb(38pt)=(0.27648,0.48144,0.95064); rgb(39pt)=(0.27603,0.49132,0.95857); rgb(40pt)=(0.27543,0.50115,0.96594); rgb(41pt)=(0.27469,0.51094,0.97275); rgb(42pt)=(0.27381,0.52069,0.97899); rgb(43pt)=(0.27273,0.5304,0.98461); rgb(44pt)=(0.27106,0.54015,0.9893); rgb(45pt)=(0.26878,0.54995,0.99303); rgb(46pt)=(0.26592,0.55979,0.99583); rgb(47pt)=(0.26252,0.56967,0.99773); rgb(48pt)=(0.25862,0.57958,0.99876); rgb(49pt)=(0.25425,0.5895,0.99896); rgb(50pt)=(0.24946,0.59943,0.99835); rgb(51pt)=(0.24427,0.60937,0.99697); rgb(52pt)=(0.23874,0.61931,0.99485); rgb(53pt)=(0.23288,0.62923,0.99202); rgb(54pt)=(0.22676,0.63913,0.98851); rgb(55pt)=(0.22039,0.64901,0.98436); rgb(56pt)=(0.21382,0.65886,0.97959); rgb(57pt)=(0.20708,0.66866,0.97423); rgb(58pt)=(0.20021,0.67842,0.96833); rgb(59pt)=(0.19326,0.68812,0.9619); rgb(60pt)=(0.18625,0.69775,0.95498); rgb(61pt)=(0.17923,0.70732,0.94761); rgb(62pt)=(0.17223,0.7168,0.93981); rgb(63pt)=(0.16529,0.7262,0.93161); rgb(64pt)=(0.15844,0.73551,0.92305); rgb(65pt)=(0.15173,0.74472,0.91416); rgb(66pt)=(0.14519,0.75381,0.90496); rgb(67pt)=(0.13886,0.76279,0.8955); rgb(68pt)=(0.13278,0.77165,0.8858); rgb(69pt)=(0.12698,0.78037,0.8759); rgb(70pt)=(0.12151,0.78896,0.86581); rgb(71pt)=(0.11639,0.7974,0.85559); rgb(72pt)=(0.11167,0.80569,0.84525); rgb(73pt)=(0.10738,0.81381,0.83484); rgb(74pt)=(0.10357,0.82177,0.82437); rgb(75pt)=(0.10026,0.82955,0.81389); rgb(76pt)=(0.0975,0.83714,0.80342); rgb(77pt)=(0.09532,0.84455,0.79299); rgb(78pt)=(0.09377,0.85175,0.78264); rgb(79pt)=(0.09287,0.85875,0.7724); rgb(80pt)=(0.09267,0.86554,0.7623); rgb(81pt)=(0.0932,0.87211,0.75237); rgb(82pt)=(0.09451,0.87844,0.74265); rgb(83pt)=(0.09662,0.88454,0.73316); rgb(84pt)=(0.09958,0.8904,0.72393); rgb(85pt)=(0.10342,0.896,0.715); rgb(86pt)=(0.10815,0.90142,0.70599); rgb(87pt)=(0.11374,0.90673,0.69651); rgb(88pt)=(0.12014,0.91193,0.6866); rgb(89pt)=(0.12733,0.91701,0.67627); rgb(90pt)=(0.13526,0.92197,0.66556); rgb(91pt)=(0.14391,0.9268,0.65448); rgb(92pt)=(0.15323,0.93151,0.64308); rgb(93pt)=(0.16319,0.93609,0.63137); rgb(94pt)=(0.17377,0.94053,0.61938); rgb(95pt)=(0.18491,0.94484,0.60713); rgb(96pt)=(0.19659,0.94901,0.59466); rgb(97pt)=(0.20877,0.95304,0.58199); rgb(98pt)=(0.22142,0.95692,0.56914); rgb(99pt)=(0.23449,0.96065,0.55614); rgb(100pt)=(0.24797,0.96423,0.54303); rgb(101pt)=(0.2618,0.96765,0.52981); rgb(102pt)=(0.27597,0.97092,0.51653); rgb(103pt)=(0.29042,0.97403,0.50321); rgb(104pt)=(0.30513,0.97697,0.48987); rgb(105pt)=(0.32006,0.97974,0.47654); rgb(106pt)=(0.33517,0.98234,0.46325); rgb(107pt)=(0.35043,0.98477,0.45002); rgb(108pt)=(0.36581,0.98702,0.43688); rgb(109pt)=(0.38127,0.98909,0.42386); rgb(110pt)=(0.39678,0.99098,0.41098); rgb(111pt)=(0.41229,0.99268,0.39826); rgb(112pt)=(0.42778,0.99419,0.38575); rgb(113pt)=(0.44321,0.99551,0.37345); rgb(114pt)=(0.45854,0.99663,0.3614); rgb(115pt)=(0.47375,0.99755,0.34963); rgb(116pt)=(0.48879,0.99828,0.33816); rgb(117pt)=(0.50362,0.99879,0.32701); rgb(118pt)=(0.51822,0.9991,0.31622); rgb(119pt)=(0.53255,0.99919,0.30581); rgb(120pt)=(0.54658,0.99907,0.29581); rgb(121pt)=(0.56026,0.99873,0.28623); rgb(122pt)=(0.57357,0.99817,0.27712); rgb(123pt)=(0.58646,0.99739,0.26849); rgb(124pt)=(0.59891,0.99638,0.26038); rgb(125pt)=(0.61088,0.99514,0.2528); rgb(126pt)=(0.62233,0.99366,0.24579); rgb(127pt)=(0.63323,0.99195,0.23937); rgb(128pt)=(0.64362,0.98999,0.23356); rgb(129pt)=(0.65394,0.98775,0.22835); rgb(130pt)=(0.66428,0.98524,0.2237); rgb(131pt)=(0.67462,0.98246,0.2196); rgb(132pt)=(0.68494,0.97941,0.21602); rgb(133pt)=(0.69525,0.9761,0.21294); rgb(134pt)=(0.70553,0.97255,0.21032); rgb(135pt)=(0.71577,0.96875,0.20815); rgb(136pt)=(0.72596,0.9647,0.2064); rgb(137pt)=(0.7361,0.96043,0.20504); rgb(138pt)=(0.74617,0.95593,0.20406); rgb(139pt)=(0.75617,0.95121,0.20343); rgb(140pt)=(0.76608,0.94627,0.20311); rgb(141pt)=(0.77591,0.94113,0.2031); rgb(142pt)=(0.78563,0.93579,0.20336); rgb(143pt)=(0.79524,0.93025,0.20386); rgb(144pt)=(0.80473,0.92452,0.20459); rgb(145pt)=(0.8141,0.91861,0.20552); rgb(146pt)=(0.82333,0.91253,0.20663); rgb(147pt)=(0.83241,0.90627,0.20788); rgb(148pt)=(0.84133,0.89986,0.20926); rgb(149pt)=(0.8501,0.89328,0.21074); rgb(150pt)=(0.85868,0.88655,0.2123); rgb(151pt)=(0.86709,0.87968,0.21391); rgb(152pt)=(0.8753,0.87267,0.21555); rgb(153pt)=(0.88331,0.86553,0.21719); rgb(154pt)=(0.89112,0.85826,0.2188); rgb(155pt)=(0.8987,0.85087,0.22038); rgb(156pt)=(0.90605,0.84337,0.22188); rgb(157pt)=(0.91317,0.83576,0.22328); rgb(158pt)=(0.92004,0.82806,0.22456); rgb(159pt)=(0.92666,0.82025,0.2257); rgb(160pt)=(0.93301,0.81236,0.22667); rgb(161pt)=(0.93909,0.80439,0.22744); rgb(162pt)=(0.94489,0.79634,0.228); rgb(163pt)=(0.95039,0.78823,0.22831); rgb(164pt)=(0.9556,0.78005,0.22836); rgb(165pt)=(0.96049,0.77181,0.22811); rgb(166pt)=(0.96507,0.76352,0.22754); rgb(167pt)=(0.96931,0.75519,0.22663); rgb(168pt)=(0.97323,0.74682,0.22536); rgb(169pt)=(0.97679,0.73842,0.22369); rgb(170pt)=(0.98,0.73,0.22161); rgb(171pt)=(0.98289,0.7214,0.21918); rgb(172pt)=(0.98549,0.7125,0.2165); rgb(173pt)=(0.98781,0.7033,0.21358); rgb(174pt)=(0.98986,0.69382,0.21043); rgb(175pt)=(0.99163,0.68408,0.20706); rgb(176pt)=(0.99314,0.67408,0.20348); rgb(177pt)=(0.99438,0.66386,0.19971); rgb(178pt)=(0.99535,0.65341,0.19577); rgb(179pt)=(0.99607,0.64277,0.19165); rgb(180pt)=(0.99654,0.63193,0.18738); rgb(181pt)=(0.99675,0.62093,0.18297); rgb(182pt)=(0.99672,0.60977,0.17842); rgb(183pt)=(0.99644,0.59846,0.17376); rgb(184pt)=(0.99593,0.58703,0.16899); rgb(185pt)=(0.99517,0.57549,0.16412); rgb(186pt)=(0.99419,0.56386,0.15918); rgb(187pt)=(0.99297,0.55214,0.15417); rgb(188pt)=(0.99153,0.54036,0.1491); rgb(189pt)=(0.98987,0.52854,0.14398); rgb(190pt)=(0.98799,0.51667,0.13883); rgb(191pt)=(0.9859,0.50479,0.13367); rgb(192pt)=(0.9836,0.49291,0.12849); rgb(193pt)=(0.98108,0.48104,0.12332); rgb(194pt)=(0.97837,0.4692,0.11817); rgb(195pt)=(0.97545,0.4574,0.11305); rgb(196pt)=(0.97234,0.44565,0.10797); rgb(197pt)=(0.96904,0.43399,0.10294); rgb(198pt)=(0.96555,0.42241,0.09798); rgb(199pt)=(0.96187,0.41093,0.0931); rgb(200pt)=(0.95801,0.39958,0.08831); rgb(201pt)=(0.95398,0.38836,0.08362); rgb(202pt)=(0.94977,0.37729,0.07905); rgb(203pt)=(0.94538,0.36638,0.07461); rgb(204pt)=(0.94084,0.35566,0.07031); rgb(205pt)=(0.93612,0.34513,0.06616); rgb(206pt)=(0.93125,0.33482,0.06218); rgb(207pt)=(0.92623,0.32473,0.05837); rgb(208pt)=(0.92105,0.31489,0.05475); rgb(209pt)=(0.91572,0.3053,0.05134); rgb(210pt)=(0.91024,0.29599,0.04814); rgb(211pt)=(0.90463,0.28696,0.04516); rgb(212pt)=(0.89888,0.27824,0.04243); rgb(213pt)=(0.89298,0.26981,0.03993); rgb(214pt)=(0.88691,0.26152,0.03753); rgb(215pt)=(0.88066,0.25334,0.03521); rgb(216pt)=(0.87422,0.24526,0.03297); rgb(217pt)=(0.8676,0.2373,0.03082); rgb(218pt)=(0.86079,0.22945,0.02875); rgb(219pt)=(0.8538,0.2217,0.02677); rgb(220pt)=(0.84662,0.21407,0.02487); rgb(221pt)=(0.83926,0.20654,0.02305); rgb(222pt)=(0.83172,0.19912,0.02131); rgb(223pt)=(0.82399,0.19182,0.01966); rgb(224pt)=(0.81608,0.18462,0.01809); rgb(225pt)=(0.80799,0.17753,0.0166); rgb(226pt)=(0.79971,0.17055,0.0152); rgb(227pt)=(0.79125,0.16368,0.01387); rgb(228pt)=(0.7826,0.15693,0.01264); rgb(229pt)=(0.77377,0.15028,0.01148); rgb(230pt)=(0.76476,0.14374,0.01041); rgb(231pt)=(0.75556,0.13731,0.00942); rgb(232pt)=(0.74617,0.13098,0.00851); rgb(233pt)=(0.73661,0.12477,0.00769); rgb(234pt)=(0.72686,0.11867,0.00695); rgb(235pt)=(0.71692,0.11268,0.00629); rgb(236pt)=(0.7068,0.1068,0.00571); rgb(237pt)=(0.6965,0.10102,0.00522); rgb(238pt)=(0.68602,0.09536,0.00481); rgb(239pt)=(0.67535,0.0898,0.00449); rgb(240pt)=(0.66449,0.08436,0.00424); rgb(241pt)=(0.65345,0.07902,0.00408); rgb(242pt)=(0.64223,0.0738,0.00401); rgb(243pt)=(0.63082,0.06868,0.00401); rgb(244pt)=(0.61923,0.06367,0.0041); rgb(245pt)=(0.60746,0.05878,0.00427); rgb(246pt)=(0.5955,0.05399,0.00453); rgb(247pt)=(0.58336,0.04931,0.00486); rgb(248pt)=(0.57103,0.04474,0.00529); rgb(249pt)=(0.55852,0.04028,0.00579); rgb(250pt)=(0.54583,0.03593,0.00638); rgb(251pt)=(0.53295,0.03169,0.00705); rgb(252pt)=(0.51989,0.02756,0.0078); rgb(253pt)=(0.50664,0.02354,0.00863); rgb(254pt)=(0.49321,0.01963,0.00955); rgb(255pt)=(0.4796,0.01583,0.01055)}, mesh/rows=11]
table[row sep=crcr, point meta=\thisrow{c}] {%
x	y	c\\
1	5.71789287554148	0.0336515997698858\\
1	5.71789287554148	0.0336515997698858\\
1.05	5.0047239686042	0.0355451425539663\\
1.05	5.0047239686042	0.0355451425539663\\
1.1	4.27575082945035	0.0374961668515161\\
1.1	4.27575082945035	0.0374961668515161\\
1.15	3.53022681067537	0.0395037508608466\\
1.15	3.53022681067537	0.0395037508608466\\
1.2	2.76736448246083	0.0415668464270833\\
1.2	2.76736448246083	0.0415668464270833\\
1.25	1.98633925647644	0.0436842463462985\\
1.25	1.98633925647644	0.0436842463462985\\
1.3	1.18629460027304	0.0458545476256419\\
1.3	1.18629460027304	0.0458545476256419\\
1.35	0.366349253612952	0.0480761105050804\\
1.35	0.366349253612952	0.0480761105050804\\
1.4	-0.474393072888233	0.0503470131940121\\
1.4	-0.474393072888233	0.0503470131940121\\
1.45	-1.33683096881555	0.0526650025028645\\
1.45	-1.33683096881555	0.0526650025028645\\
1.5	-2.22184895853354	0.0550274408791649\\
1.5	-2.22184895853354	0.0550274408791649\\
};
\addplot[
    only marks,
    mark=*,
    mark size=5pt,
    color=red, 
    mark options={
        fill=red,
        fill opacity=0.5, 
        draw opacity=1     
    },
]
coordinates {(1.5,-2.22)};
\node[color=black, fill=none] at (axis cs: 1.3,-2) {\textcolor{red}{Case 2b}};
\draw[line width=0.5mm,draw=black!80,->] (axis cs:1.47,-2.3)--(axis cs:1.39,-2);

\addplot[
    only marks,
    mark=*,
    mark size=5pt,
    color=blue, 
    mark options={
        fill=blue,
        fill opacity=0.5, 
        draw opacity=1     
    },
]
coordinates {(1,5.7)};
\node[color=black, fill=none] at (axis cs: 1.2,5.5) {\textcolor{blue}{Case 2a}};
\draw[line width=0.5mm,draw=black!80,->] (axis cs:1.03,5.6)--(axis cs:1.12,5.5);
\end{axis}
\end{tikzpicture}%

%% file: updated_images/Id_Change.tex
%
%
\begin{tikzpicture}

\begin{axis}[%
width=2in,
height=2in,
at={(0.494in,0.431in)},
scale only axis,
point meta min=0.00157771821517024,
point meta max=0.0550274408791649,
xmin=0,
xmax=45,
xtick = {0,10,20,30,40,50},
xticklabels={0,10,20,30,40,50},
xlabel style={font=\color{white!15!black}},
xlabel={${I}_{{d}}$ [A]},
ymin=142,
ymax=158,
ylabel style={align=center,font=\color{white!15!black}},
ylabel={Potential \\ Resonance Frequency [Hz]},
axis background/.style={fill=white},
xmajorgrids,
ymajorgrids,
grid style={dashed, opacity=0.5}
]

\addplot[%
mesh,
    shader=flat,
    mark=none,
    line join=round,
    line cap=round,
    line width=2pt,
    point meta=explicit, colormap={mymap}{[1pt] rgb(0pt)=(0.18995,0.07176,0.23217); rgb(1pt)=(0.19483,0.08339,0.26149); rgb(2pt)=(0.19956,0.09498,0.29024); rgb(3pt)=(0.20415,0.10652,0.31844); rgb(4pt)=(0.2086,0.11802,0.34607); rgb(5pt)=(0.21291,0.12947,0.37314); rgb(6pt)=(0.21708,0.14087,0.39964); rgb(7pt)=(0.22111,0.15223,0.42558); rgb(8pt)=(0.225,0.16354,0.45096); rgb(9pt)=(0.22875,0.17481,0.47578); rgb(10pt)=(0.23236,0.18603,0.50004); rgb(11pt)=(0.23582,0.1972,0.52373); rgb(12pt)=(0.23915,0.20833,0.54686); rgb(13pt)=(0.24234,0.21941,0.56942); rgb(14pt)=(0.24539,0.23044,0.59142); rgb(15pt)=(0.2483,0.24143,0.61286); rgb(16pt)=(0.25107,0.25237,0.63374); rgb(17pt)=(0.25369,0.26327,0.65406); rgb(18pt)=(0.25618,0.27412,0.67381); rgb(19pt)=(0.25853,0.28492,0.693); rgb(20pt)=(0.26074,0.29568,0.71162); rgb(21pt)=(0.2628,0.30639,0.72968); rgb(22pt)=(0.26473,0.31706,0.74718); rgb(23pt)=(0.26652,0.32768,0.76412); rgb(24pt)=(0.26816,0.33825,0.7805); rgb(25pt)=(0.26967,0.34878,0.79631); rgb(26pt)=(0.27103,0.35926,0.81156); rgb(27pt)=(0.27226,0.3697,0.82624); rgb(28pt)=(0.27334,0.38008,0.84037); rgb(29pt)=(0.27429,0.39043,0.85393); rgb(30pt)=(0.27509,0.40072,0.86692); rgb(31pt)=(0.27576,0.41097,0.87936); rgb(32pt)=(0.27628,0.42118,0.89123); rgb(33pt)=(0.27667,0.43134,0.90254); rgb(34pt)=(0.27691,0.44145,0.91328); rgb(35pt)=(0.27701,0.45152,0.92347); rgb(36pt)=(0.27698,0.46153,0.93309); rgb(37pt)=(0.2768,0.47151,0.94214); rgb(38pt)=(0.27648,0.48144,0.95064); rgb(39pt)=(0.27603,0.49132,0.95857); rgb(40pt)=(0.27543,0.50115,0.96594); rgb(41pt)=(0.27469,0.51094,0.97275); rgb(42pt)=(0.27381,0.52069,0.97899); rgb(43pt)=(0.27273,0.5304,0.98461); rgb(44pt)=(0.27106,0.54015,0.9893); rgb(45pt)=(0.26878,0.54995,0.99303); rgb(46pt)=(0.26592,0.55979,0.99583); rgb(47pt)=(0.26252,0.56967,0.99773); rgb(48pt)=(0.25862,0.57958,0.99876); rgb(49pt)=(0.25425,0.5895,0.99896); rgb(50pt)=(0.24946,0.59943,0.99835); rgb(51pt)=(0.24427,0.60937,0.99697); rgb(52pt)=(0.23874,0.61931,0.99485); rgb(53pt)=(0.23288,0.62923,0.99202); rgb(54pt)=(0.22676,0.63913,0.98851); rgb(55pt)=(0.22039,0.64901,0.98436); rgb(56pt)=(0.21382,0.65886,0.97959); rgb(57pt)=(0.20708,0.66866,0.97423); rgb(58pt)=(0.20021,0.67842,0.96833); rgb(59pt)=(0.19326,0.68812,0.9619); rgb(60pt)=(0.18625,0.69775,0.95498); rgb(61pt)=(0.17923,0.70732,0.94761); rgb(62pt)=(0.17223,0.7168,0.93981); rgb(63pt)=(0.16529,0.7262,0.93161); rgb(64pt)=(0.15844,0.73551,0.92305); rgb(65pt)=(0.15173,0.74472,0.91416); rgb(66pt)=(0.14519,0.75381,0.90496); rgb(67pt)=(0.13886,0.76279,0.8955); rgb(68pt)=(0.13278,0.77165,0.8858); rgb(69pt)=(0.12698,0.78037,0.8759); rgb(70pt)=(0.12151,0.78896,0.86581); rgb(71pt)=(0.11639,0.7974,0.85559); rgb(72pt)=(0.11167,0.80569,0.84525); rgb(73pt)=(0.10738,0.81381,0.83484); rgb(74pt)=(0.10357,0.82177,0.82437); rgb(75pt)=(0.10026,0.82955,0.81389); rgb(76pt)=(0.0975,0.83714,0.80342); rgb(77pt)=(0.09532,0.84455,0.79299); rgb(78pt)=(0.09377,0.85175,0.78264); rgb(79pt)=(0.09287,0.85875,0.7724); rgb(80pt)=(0.09267,0.86554,0.7623); rgb(81pt)=(0.0932,0.87211,0.75237); rgb(82pt)=(0.09451,0.87844,0.74265); rgb(83pt)=(0.09662,0.88454,0.73316); rgb(84pt)=(0.09958,0.8904,0.72393); rgb(85pt)=(0.10342,0.896,0.715); rgb(86pt)=(0.10815,0.90142,0.70599); rgb(87pt)=(0.11374,0.90673,0.69651); rgb(88pt)=(0.12014,0.91193,0.6866); rgb(89pt)=(0.12733,0.91701,0.67627); rgb(90pt)=(0.13526,0.92197,0.66556); rgb(91pt)=(0.14391,0.9268,0.65448); rgb(92pt)=(0.15323,0.93151,0.64308); rgb(93pt)=(0.16319,0.93609,0.63137); rgb(94pt)=(0.17377,0.94053,0.61938); rgb(95pt)=(0.18491,0.94484,0.60713); rgb(96pt)=(0.19659,0.94901,0.59466); rgb(97pt)=(0.20877,0.95304,0.58199); rgb(98pt)=(0.22142,0.95692,0.56914); rgb(99pt)=(0.23449,0.96065,0.55614); rgb(100pt)=(0.24797,0.96423,0.54303); rgb(101pt)=(0.2618,0.96765,0.52981); rgb(102pt)=(0.27597,0.97092,0.51653); rgb(103pt)=(0.29042,0.97403,0.50321); rgb(104pt)=(0.30513,0.97697,0.48987); rgb(105pt)=(0.32006,0.97974,0.47654); rgb(106pt)=(0.33517,0.98234,0.46325); rgb(107pt)=(0.35043,0.98477,0.45002); rgb(108pt)=(0.36581,0.98702,0.43688); rgb(109pt)=(0.38127,0.98909,0.42386); rgb(110pt)=(0.39678,0.99098,0.41098); rgb(111pt)=(0.41229,0.99268,0.39826); rgb(112pt)=(0.42778,0.99419,0.38575); rgb(113pt)=(0.44321,0.99551,0.37345); rgb(114pt)=(0.45854,0.99663,0.3614); rgb(115pt)=(0.47375,0.99755,0.34963); rgb(116pt)=(0.48879,0.99828,0.33816); rgb(117pt)=(0.50362,0.99879,0.32701); rgb(118pt)=(0.51822,0.9991,0.31622); rgb(119pt)=(0.53255,0.99919,0.30581); rgb(120pt)=(0.54658,0.99907,0.29581); rgb(121pt)=(0.56026,0.99873,0.28623); rgb(122pt)=(0.57357,0.99817,0.27712); rgb(123pt)=(0.58646,0.99739,0.26849); rgb(124pt)=(0.59891,0.99638,0.26038); rgb(125pt)=(0.61088,0.99514,0.2528); rgb(126pt)=(0.62233,0.99366,0.24579); rgb(127pt)=(0.63323,0.99195,0.23937); rgb(128pt)=(0.64362,0.98999,0.23356); rgb(129pt)=(0.65394,0.98775,0.22835); rgb(130pt)=(0.66428,0.98524,0.2237); rgb(131pt)=(0.67462,0.98246,0.2196); rgb(132pt)=(0.68494,0.97941,0.21602); rgb(133pt)=(0.69525,0.9761,0.21294); rgb(134pt)=(0.70553,0.97255,0.21032); rgb(135pt)=(0.71577,0.96875,0.20815); rgb(136pt)=(0.72596,0.9647,0.2064); rgb(137pt)=(0.7361,0.96043,0.20504); rgb(138pt)=(0.74617,0.95593,0.20406); rgb(139pt)=(0.75617,0.95121,0.20343); rgb(140pt)=(0.76608,0.94627,0.20311); rgb(141pt)=(0.77591,0.94113,0.2031); rgb(142pt)=(0.78563,0.93579,0.20336); rgb(143pt)=(0.79524,0.93025,0.20386); rgb(144pt)=(0.80473,0.92452,0.20459); rgb(145pt)=(0.8141,0.91861,0.20552); rgb(146pt)=(0.82333,0.91253,0.20663); rgb(147pt)=(0.83241,0.90627,0.20788); rgb(148pt)=(0.84133,0.89986,0.20926); rgb(149pt)=(0.8501,0.89328,0.21074); rgb(150pt)=(0.85868,0.88655,0.2123); rgb(151pt)=(0.86709,0.87968,0.21391); rgb(152pt)=(0.8753,0.87267,0.21555); rgb(153pt)=(0.88331,0.86553,0.21719); rgb(154pt)=(0.89112,0.85826,0.2188); rgb(155pt)=(0.8987,0.85087,0.22038); rgb(156pt)=(0.90605,0.84337,0.22188); rgb(157pt)=(0.91317,0.83576,0.22328); rgb(158pt)=(0.92004,0.82806,0.22456); rgb(159pt)=(0.92666,0.82025,0.2257); rgb(160pt)=(0.93301,0.81236,0.22667); rgb(161pt)=(0.93909,0.80439,0.22744); rgb(162pt)=(0.94489,0.79634,0.228); rgb(163pt)=(0.95039,0.78823,0.22831); rgb(164pt)=(0.9556,0.78005,0.22836); rgb(165pt)=(0.96049,0.77181,0.22811); rgb(166pt)=(0.96507,0.76352,0.22754); rgb(167pt)=(0.96931,0.75519,0.22663); rgb(168pt)=(0.97323,0.74682,0.22536); rgb(169pt)=(0.97679,0.73842,0.22369); rgb(170pt)=(0.98,0.73,0.22161); rgb(171pt)=(0.98289,0.7214,0.21918); rgb(172pt)=(0.98549,0.7125,0.2165); rgb(173pt)=(0.98781,0.7033,0.21358); rgb(174pt)=(0.98986,0.69382,0.21043); rgb(175pt)=(0.99163,0.68408,0.20706); rgb(176pt)=(0.99314,0.67408,0.20348); rgb(177pt)=(0.99438,0.66386,0.19971); rgb(178pt)=(0.99535,0.65341,0.19577); rgb(179pt)=(0.99607,0.64277,0.19165); rgb(180pt)=(0.99654,0.63193,0.18738); rgb(181pt)=(0.99675,0.62093,0.18297); rgb(182pt)=(0.99672,0.60977,0.17842); rgb(183pt)=(0.99644,0.59846,0.17376); rgb(184pt)=(0.99593,0.58703,0.16899); rgb(185pt)=(0.99517,0.57549,0.16412); rgb(186pt)=(0.99419,0.56386,0.15918); rgb(187pt)=(0.99297,0.55214,0.15417); rgb(188pt)=(0.99153,0.54036,0.1491); rgb(189pt)=(0.98987,0.52854,0.14398); rgb(190pt)=(0.98799,0.51667,0.13883); rgb(191pt)=(0.9859,0.50479,0.13367); rgb(192pt)=(0.9836,0.49291,0.12849); rgb(193pt)=(0.98108,0.48104,0.12332); rgb(194pt)=(0.97837,0.4692,0.11817); rgb(195pt)=(0.97545,0.4574,0.11305); rgb(196pt)=(0.97234,0.44565,0.10797); rgb(197pt)=(0.96904,0.43399,0.10294); rgb(198pt)=(0.96555,0.42241,0.09798); rgb(199pt)=(0.96187,0.41093,0.0931); rgb(200pt)=(0.95801,0.39958,0.08831); rgb(201pt)=(0.95398,0.38836,0.08362); rgb(202pt)=(0.94977,0.37729,0.07905); rgb(203pt)=(0.94538,0.36638,0.07461); rgb(204pt)=(0.94084,0.35566,0.07031); rgb(205pt)=(0.93612,0.34513,0.06616); rgb(206pt)=(0.93125,0.33482,0.06218); rgb(207pt)=(0.92623,0.32473,0.05837); rgb(208pt)=(0.92105,0.31489,0.05475); rgb(209pt)=(0.91572,0.3053,0.05134); rgb(210pt)=(0.91024,0.29599,0.04814); rgb(211pt)=(0.90463,0.28696,0.04516); rgb(212pt)=(0.89888,0.27824,0.04243); rgb(213pt)=(0.89298,0.26981,0.03993); rgb(214pt)=(0.88691,0.26152,0.03753); rgb(215pt)=(0.88066,0.25334,0.03521); rgb(216pt)=(0.87422,0.24526,0.03297); rgb(217pt)=(0.8676,0.2373,0.03082); rgb(218pt)=(0.86079,0.22945,0.02875); rgb(219pt)=(0.8538,0.2217,0.02677); rgb(220pt)=(0.84662,0.21407,0.02487); rgb(221pt)=(0.83926,0.20654,0.02305); rgb(222pt)=(0.83172,0.19912,0.02131); rgb(223pt)=(0.82399,0.19182,0.01966); rgb(224pt)=(0.81608,0.18462,0.01809); rgb(225pt)=(0.80799,0.17753,0.0166); rgb(226pt)=(0.79971,0.17055,0.0152); rgb(227pt)=(0.79125,0.16368,0.01387); rgb(228pt)=(0.7826,0.15693,0.01264); rgb(229pt)=(0.77377,0.15028,0.01148); rgb(230pt)=(0.76476,0.14374,0.01041); rgb(231pt)=(0.75556,0.13731,0.00942); rgb(232pt)=(0.74617,0.13098,0.00851); rgb(233pt)=(0.73661,0.12477,0.00769); rgb(234pt)=(0.72686,0.11867,0.00695); rgb(235pt)=(0.71692,0.11268,0.00629); rgb(236pt)=(0.7068,0.1068,0.00571); rgb(237pt)=(0.6965,0.10102,0.00522); rgb(238pt)=(0.68602,0.09536,0.00481); rgb(239pt)=(0.67535,0.0898,0.00449); rgb(240pt)=(0.66449,0.08436,0.00424); rgb(241pt)=(0.65345,0.07902,0.00408); rgb(242pt)=(0.64223,0.0738,0.00401); rgb(243pt)=(0.63082,0.06868,0.00401); rgb(244pt)=(0.61923,0.06367,0.0041); rgb(245pt)=(0.60746,0.05878,0.00427); rgb(246pt)=(0.5955,0.05399,0.00453); rgb(247pt)=(0.58336,0.04931,0.00486); rgb(248pt)=(0.57103,0.04474,0.00529); rgb(249pt)=(0.55852,0.04028,0.00579); rgb(250pt)=(0.54583,0.03593,0.00638); rgb(251pt)=(0.53295,0.03169,0.00705); rgb(252pt)=(0.51989,0.02756,0.0078); rgb(253pt)=(0.50664,0.02354,0.00863); rgb(254pt)=(0.49321,0.01963,0.00955); rgb(255pt)=(0.4796,0.01583,0.01055)}, mesh/rows=8]
table[row sep=crcr, point meta=\thisrow{c}] {%
x	y	c\\
5	157.403032753358	0.00157771821517024\\
5	157.403032753358	0.00157771821517024\\
10	156.299921850468	0.00438385121126728\\
10	156.299921850468	0.00438385121126728\\
15	154.839388405939	0.00896349075801616\\
15	154.839388405939	0.00896349075801616\\
20	153.038862873537	0.0152275830015485\\
20	153.038862873537	0.0152275830015485\\
25	150.921939983421	0.023076510558497\\
25	150.921939983421	0.023076510558497\\
30	148.515876135357	0.03240255189655\\
30	148.515876135357	0.03240255189655\\
35	145.849150368999	0.0430921113873273\\
35	145.849150368999	0.0430921113873273\\
40	142.949525817057	0.0550274408791649\\
40	142.949525817057	0.0550274408791649\\
};
\end{axis}

\begin{axis}[%
width=2in,
height=2in,
at={(3.1in,0.431in)},
scale only axis,
point meta min=0.00157771821517024,
point meta max=0.0550274408791649,
xmin=0,
xmax=45,
xtick = {0,10,20,30,40,50},
xticklabels={0,10,20,30,40,50},
xlabel style={font=\color{white!15!black}},
xlabel={${I}_{d}$[A]},
ymin=-3,
ymax=5,
ylabel style={font=\color{white!15!black}},
ylabel={Phase Margin [Deg]},
axis background/.style={fill=white},
xmajorgrids,
ymajorgrids,
grid style={dashed, opacity=0.5},
colormap={mymap}{[1pt] rgb(0pt)=(0.18995,0.07176,0.23217); rgb(1pt)=(0.19483,0.08339,0.26149); rgb(2pt)=(0.19956,0.09498,0.29024); rgb(3pt)=(0.20415,0.10652,0.31844); rgb(4pt)=(0.2086,0.11802,0.34607); rgb(5pt)=(0.21291,0.12947,0.37314); rgb(6pt)=(0.21708,0.14087,0.39964); rgb(7pt)=(0.22111,0.15223,0.42558); rgb(8pt)=(0.225,0.16354,0.45096); rgb(9pt)=(0.22875,0.17481,0.47578); rgb(10pt)=(0.23236,0.18603,0.50004); rgb(11pt)=(0.23582,0.1972,0.52373); rgb(12pt)=(0.23915,0.20833,0.54686); rgb(13pt)=(0.24234,0.21941,0.56942); rgb(14pt)=(0.24539,0.23044,0.59142); rgb(15pt)=(0.2483,0.24143,0.61286); rgb(16pt)=(0.25107,0.25237,0.63374); rgb(17pt)=(0.25369,0.26327,0.65406); rgb(18pt)=(0.25618,0.27412,0.67381); rgb(19pt)=(0.25853,0.28492,0.693); rgb(20pt)=(0.26074,0.29568,0.71162); rgb(21pt)=(0.2628,0.30639,0.72968); rgb(22pt)=(0.26473,0.31706,0.74718); rgb(23pt)=(0.26652,0.32768,0.76412); rgb(24pt)=(0.26816,0.33825,0.7805); rgb(25pt)=(0.26967,0.34878,0.79631); rgb(26pt)=(0.27103,0.35926,0.81156); rgb(27pt)=(0.27226,0.3697,0.82624); rgb(28pt)=(0.27334,0.38008,0.84037); rgb(29pt)=(0.27429,0.39043,0.85393); rgb(30pt)=(0.27509,0.40072,0.86692); rgb(31pt)=(0.27576,0.41097,0.87936); rgb(32pt)=(0.27628,0.42118,0.89123); rgb(33pt)=(0.27667,0.43134,0.90254); rgb(34pt)=(0.27691,0.44145,0.91328); rgb(35pt)=(0.27701,0.45152,0.92347); rgb(36pt)=(0.27698,0.46153,0.93309); rgb(37pt)=(0.2768,0.47151,0.94214); rgb(38pt)=(0.27648,0.48144,0.95064); rgb(39pt)=(0.27603,0.49132,0.95857); rgb(40pt)=(0.27543,0.50115,0.96594); rgb(41pt)=(0.27469,0.51094,0.97275); rgb(42pt)=(0.27381,0.52069,0.97899); rgb(43pt)=(0.27273,0.5304,0.98461); rgb(44pt)=(0.27106,0.54015,0.9893); rgb(45pt)=(0.26878,0.54995,0.99303); rgb(46pt)=(0.26592,0.55979,0.99583); rgb(47pt)=(0.26252,0.56967,0.99773); rgb(48pt)=(0.25862,0.57958,0.99876); rgb(49pt)=(0.25425,0.5895,0.99896); rgb(50pt)=(0.24946,0.59943,0.99835); rgb(51pt)=(0.24427,0.60937,0.99697); rgb(52pt)=(0.23874,0.61931,0.99485); rgb(53pt)=(0.23288,0.62923,0.99202); rgb(54pt)=(0.22676,0.63913,0.98851); rgb(55pt)=(0.22039,0.64901,0.98436); rgb(56pt)=(0.21382,0.65886,0.97959); rgb(57pt)=(0.20708,0.66866,0.97423); rgb(58pt)=(0.20021,0.67842,0.96833); rgb(59pt)=(0.19326,0.68812,0.9619); rgb(60pt)=(0.18625,0.69775,0.95498); rgb(61pt)=(0.17923,0.70732,0.94761); rgb(62pt)=(0.17223,0.7168,0.93981); rgb(63pt)=(0.16529,0.7262,0.93161); rgb(64pt)=(0.15844,0.73551,0.92305); rgb(65pt)=(0.15173,0.74472,0.91416); rgb(66pt)=(0.14519,0.75381,0.90496); rgb(67pt)=(0.13886,0.76279,0.8955); rgb(68pt)=(0.13278,0.77165,0.8858); rgb(69pt)=(0.12698,0.78037,0.8759); rgb(70pt)=(0.12151,0.78896,0.86581); rgb(71pt)=(0.11639,0.7974,0.85559); rgb(72pt)=(0.11167,0.80569,0.84525); rgb(73pt)=(0.10738,0.81381,0.83484); rgb(74pt)=(0.10357,0.82177,0.82437); rgb(75pt)=(0.10026,0.82955,0.81389); rgb(76pt)=(0.0975,0.83714,0.80342); rgb(77pt)=(0.09532,0.84455,0.79299); rgb(78pt)=(0.09377,0.85175,0.78264); rgb(79pt)=(0.09287,0.85875,0.7724); rgb(80pt)=(0.09267,0.86554,0.7623); rgb(81pt)=(0.0932,0.87211,0.75237); rgb(82pt)=(0.09451,0.87844,0.74265); rgb(83pt)=(0.09662,0.88454,0.73316); rgb(84pt)=(0.09958,0.8904,0.72393); rgb(85pt)=(0.10342,0.896,0.715); rgb(86pt)=(0.10815,0.90142,0.70599); rgb(87pt)=(0.11374,0.90673,0.69651); rgb(88pt)=(0.12014,0.91193,0.6866); rgb(89pt)=(0.12733,0.91701,0.67627); rgb(90pt)=(0.13526,0.92197,0.66556); rgb(91pt)=(0.14391,0.9268,0.65448); rgb(92pt)=(0.15323,0.93151,0.64308); rgb(93pt)=(0.16319,0.93609,0.63137); rgb(94pt)=(0.17377,0.94053,0.61938); rgb(95pt)=(0.18491,0.94484,0.60713); rgb(96pt)=(0.19659,0.94901,0.59466); rgb(97pt)=(0.20877,0.95304,0.58199); rgb(98pt)=(0.22142,0.95692,0.56914); rgb(99pt)=(0.23449,0.96065,0.55614); rgb(100pt)=(0.24797,0.96423,0.54303); rgb(101pt)=(0.2618,0.96765,0.52981); rgb(102pt)=(0.27597,0.97092,0.51653); rgb(103pt)=(0.29042,0.97403,0.50321); rgb(104pt)=(0.30513,0.97697,0.48987); rgb(105pt)=(0.32006,0.97974,0.47654); rgb(106pt)=(0.33517,0.98234,0.46325); rgb(107pt)=(0.35043,0.98477,0.45002); rgb(108pt)=(0.36581,0.98702,0.43688); rgb(109pt)=(0.38127,0.98909,0.42386); rgb(110pt)=(0.39678,0.99098,0.41098); rgb(111pt)=(0.41229,0.99268,0.39826); rgb(112pt)=(0.42778,0.99419,0.38575); rgb(113pt)=(0.44321,0.99551,0.37345); rgb(114pt)=(0.45854,0.99663,0.3614); rgb(115pt)=(0.47375,0.99755,0.34963); rgb(116pt)=(0.48879,0.99828,0.33816); rgb(117pt)=(0.50362,0.99879,0.32701); rgb(118pt)=(0.51822,0.9991,0.31622); rgb(119pt)=(0.53255,0.99919,0.30581); rgb(120pt)=(0.54658,0.99907,0.29581); rgb(121pt)=(0.56026,0.99873,0.28623); rgb(122pt)=(0.57357,0.99817,0.27712); rgb(123pt)=(0.58646,0.99739,0.26849); rgb(124pt)=(0.59891,0.99638,0.26038); rgb(125pt)=(0.61088,0.99514,0.2528); rgb(126pt)=(0.62233,0.99366,0.24579); rgb(127pt)=(0.63323,0.99195,0.23937); rgb(128pt)=(0.64362,0.98999,0.23356); rgb(129pt)=(0.65394,0.98775,0.22835); rgb(130pt)=(0.66428,0.98524,0.2237); rgb(131pt)=(0.67462,0.98246,0.2196); rgb(132pt)=(0.68494,0.97941,0.21602); rgb(133pt)=(0.69525,0.9761,0.21294); rgb(134pt)=(0.70553,0.97255,0.21032); rgb(135pt)=(0.71577,0.96875,0.20815); rgb(136pt)=(0.72596,0.9647,0.2064); rgb(137pt)=(0.7361,0.96043,0.20504); rgb(138pt)=(0.74617,0.95593,0.20406); rgb(139pt)=(0.75617,0.95121,0.20343); rgb(140pt)=(0.76608,0.94627,0.20311); rgb(141pt)=(0.77591,0.94113,0.2031); rgb(142pt)=(0.78563,0.93579,0.20336); rgb(143pt)=(0.79524,0.93025,0.20386); rgb(144pt)=(0.80473,0.92452,0.20459); rgb(145pt)=(0.8141,0.91861,0.20552); rgb(146pt)=(0.82333,0.91253,0.20663); rgb(147pt)=(0.83241,0.90627,0.20788); rgb(148pt)=(0.84133,0.89986,0.20926); rgb(149pt)=(0.8501,0.89328,0.21074); rgb(150pt)=(0.85868,0.88655,0.2123); rgb(151pt)=(0.86709,0.87968,0.21391); rgb(152pt)=(0.8753,0.87267,0.21555); rgb(153pt)=(0.88331,0.86553,0.21719); rgb(154pt)=(0.89112,0.85826,0.2188); rgb(155pt)=(0.8987,0.85087,0.22038); rgb(156pt)=(0.90605,0.84337,0.22188); rgb(157pt)=(0.91317,0.83576,0.22328); rgb(158pt)=(0.92004,0.82806,0.22456); rgb(159pt)=(0.92666,0.82025,0.2257); rgb(160pt)=(0.93301,0.81236,0.22667); rgb(161pt)=(0.93909,0.80439,0.22744); rgb(162pt)=(0.94489,0.79634,0.228); rgb(163pt)=(0.95039,0.78823,0.22831); rgb(164pt)=(0.9556,0.78005,0.22836); rgb(165pt)=(0.96049,0.77181,0.22811); rgb(166pt)=(0.96507,0.76352,0.22754); rgb(167pt)=(0.96931,0.75519,0.22663); rgb(168pt)=(0.97323,0.74682,0.22536); rgb(169pt)=(0.97679,0.73842,0.22369); rgb(170pt)=(0.98,0.73,0.22161); rgb(171pt)=(0.98289,0.7214,0.21918); rgb(172pt)=(0.98549,0.7125,0.2165); rgb(173pt)=(0.98781,0.7033,0.21358); rgb(174pt)=(0.98986,0.69382,0.21043); rgb(175pt)=(0.99163,0.68408,0.20706); rgb(176pt)=(0.99314,0.67408,0.20348); rgb(177pt)=(0.99438,0.66386,0.19971); rgb(178pt)=(0.99535,0.65341,0.19577); rgb(179pt)=(0.99607,0.64277,0.19165); rgb(180pt)=(0.99654,0.63193,0.18738); rgb(181pt)=(0.99675,0.62093,0.18297); rgb(182pt)=(0.99672,0.60977,0.17842); rgb(183pt)=(0.99644,0.59846,0.17376); rgb(184pt)=(0.99593,0.58703,0.16899); rgb(185pt)=(0.99517,0.57549,0.16412); rgb(186pt)=(0.99419,0.56386,0.15918); rgb(187pt)=(0.99297,0.55214,0.15417); rgb(188pt)=(0.99153,0.54036,0.1491); rgb(189pt)=(0.98987,0.52854,0.14398); rgb(190pt)=(0.98799,0.51667,0.13883); rgb(191pt)=(0.9859,0.50479,0.13367); rgb(192pt)=(0.9836,0.49291,0.12849); rgb(193pt)=(0.98108,0.48104,0.12332); rgb(194pt)=(0.97837,0.4692,0.11817); rgb(195pt)=(0.97545,0.4574,0.11305); rgb(196pt)=(0.97234,0.44565,0.10797); rgb(197pt)=(0.96904,0.43399,0.10294); rgb(198pt)=(0.96555,0.42241,0.09798); rgb(199pt)=(0.96187,0.41093,0.0931); rgb(200pt)=(0.95801,0.39958,0.08831); rgb(201pt)=(0.95398,0.38836,0.08362); rgb(202pt)=(0.94977,0.37729,0.07905); rgb(203pt)=(0.94538,0.36638,0.07461); rgb(204pt)=(0.94084,0.35566,0.07031); rgb(205pt)=(0.93612,0.34513,0.06616); rgb(206pt)=(0.93125,0.33482,0.06218); rgb(207pt)=(0.92623,0.32473,0.05837); rgb(208pt)=(0.92105,0.31489,0.05475); rgb(209pt)=(0.91572,0.3053,0.05134); rgb(210pt)=(0.91024,0.29599,0.04814); rgb(211pt)=(0.90463,0.28696,0.04516); rgb(212pt)=(0.89888,0.27824,0.04243); rgb(213pt)=(0.89298,0.26981,0.03993); rgb(214pt)=(0.88691,0.26152,0.03753); rgb(215pt)=(0.88066,0.25334,0.03521); rgb(216pt)=(0.87422,0.24526,0.03297); rgb(217pt)=(0.8676,0.2373,0.03082); rgb(218pt)=(0.86079,0.22945,0.02875); rgb(219pt)=(0.8538,0.2217,0.02677); rgb(220pt)=(0.84662,0.21407,0.02487); rgb(221pt)=(0.83926,0.20654,0.02305); rgb(222pt)=(0.83172,0.19912,0.02131); rgb(223pt)=(0.82399,0.19182,0.01966); rgb(224pt)=(0.81608,0.18462,0.01809); rgb(225pt)=(0.80799,0.17753,0.0166); rgb(226pt)=(0.79971,0.17055,0.0152); rgb(227pt)=(0.79125,0.16368,0.01387); rgb(228pt)=(0.7826,0.15693,0.01264); rgb(229pt)=(0.77377,0.15028,0.01148); rgb(230pt)=(0.76476,0.14374,0.01041); rgb(231pt)=(0.75556,0.13731,0.00942); rgb(232pt)=(0.74617,0.13098,0.00851); rgb(233pt)=(0.73661,0.12477,0.00769); rgb(234pt)=(0.72686,0.11867,0.00695); rgb(235pt)=(0.71692,0.11268,0.00629); rgb(236pt)=(0.7068,0.1068,0.00571); rgb(237pt)=(0.6965,0.10102,0.00522); rgb(238pt)=(0.68602,0.09536,0.00481); rgb(239pt)=(0.67535,0.0898,0.00449); rgb(240pt)=(0.66449,0.08436,0.00424); rgb(241pt)=(0.65345,0.07902,0.00408); rgb(242pt)=(0.64223,0.0738,0.00401); rgb(243pt)=(0.63082,0.06868,0.00401); rgb(244pt)=(0.61923,0.06367,0.0041); rgb(245pt)=(0.60746,0.05878,0.00427); rgb(246pt)=(0.5955,0.05399,0.00453); rgb(247pt)=(0.58336,0.04931,0.00486); rgb(248pt)=(0.57103,0.04474,0.00529); rgb(249pt)=(0.55852,0.04028,0.00579); rgb(250pt)=(0.54583,0.03593,0.00638); rgb(251pt)=(0.53295,0.03169,0.00705); rgb(252pt)=(0.51989,0.02756,0.0078); rgb(253pt)=(0.50664,0.02354,0.00863); rgb(254pt)=(0.49321,0.01963,0.00955); rgb(255pt)=(0.4796,0.01583,0.01055)},
colorbar,
colorbar style={ylabel style={rotate=-90,font=\color{white!15!black},at={(0.5,1)}, anchor=south,},scaled ticks=false,ytick={0.01,0.03,0.05},yticklabels={0.01,0.03,0.05}, ylabel={$\bm{AQI}$}}
]
\draw[fill=black, thick,opacity=0.1] (axis cs: 0,-90) -- (axis cs: 50,-90) -- (axis cs: 50,0) -- (axis cs: 0,0) -- cycle;
\node[color=black, fill=none] at (axis cs: 7,-0.5) {$Unstable$};
\node[color=black, fill=none] at (axis cs: 7,0.5) {$Stable$};
\addplot[%
mesh,
    shader=flat,
    mark=none,
    line join=round,
    line cap=round,
    line width=2pt,
    point meta=explicit, colormap={mymap}{[1pt] rgb(0pt)=(0.18995,0.07176,0.23217); rgb(1pt)=(0.19483,0.08339,0.26149); rgb(2pt)=(0.19956,0.09498,0.29024); rgb(3pt)=(0.20415,0.10652,0.31844); rgb(4pt)=(0.2086,0.11802,0.34607); rgb(5pt)=(0.21291,0.12947,0.37314); rgb(6pt)=(0.21708,0.14087,0.39964); rgb(7pt)=(0.22111,0.15223,0.42558); rgb(8pt)=(0.225,0.16354,0.45096); rgb(9pt)=(0.22875,0.17481,0.47578); rgb(10pt)=(0.23236,0.18603,0.50004); rgb(11pt)=(0.23582,0.1972,0.52373); rgb(12pt)=(0.23915,0.20833,0.54686); rgb(13pt)=(0.24234,0.21941,0.56942); rgb(14pt)=(0.24539,0.23044,0.59142); rgb(15pt)=(0.2483,0.24143,0.61286); rgb(16pt)=(0.25107,0.25237,0.63374); rgb(17pt)=(0.25369,0.26327,0.65406); rgb(18pt)=(0.25618,0.27412,0.67381); rgb(19pt)=(0.25853,0.28492,0.693); rgb(20pt)=(0.26074,0.29568,0.71162); rgb(21pt)=(0.2628,0.30639,0.72968); rgb(22pt)=(0.26473,0.31706,0.74718); rgb(23pt)=(0.26652,0.32768,0.76412); rgb(24pt)=(0.26816,0.33825,0.7805); rgb(25pt)=(0.26967,0.34878,0.79631); rgb(26pt)=(0.27103,0.35926,0.81156); rgb(27pt)=(0.27226,0.3697,0.82624); rgb(28pt)=(0.27334,0.38008,0.84037); rgb(29pt)=(0.27429,0.39043,0.85393); rgb(30pt)=(0.27509,0.40072,0.86692); rgb(31pt)=(0.27576,0.41097,0.87936); rgb(32pt)=(0.27628,0.42118,0.89123); rgb(33pt)=(0.27667,0.43134,0.90254); rgb(34pt)=(0.27691,0.44145,0.91328); rgb(35pt)=(0.27701,0.45152,0.92347); rgb(36pt)=(0.27698,0.46153,0.93309); rgb(37pt)=(0.2768,0.47151,0.94214); rgb(38pt)=(0.27648,0.48144,0.95064); rgb(39pt)=(0.27603,0.49132,0.95857); rgb(40pt)=(0.27543,0.50115,0.96594); rgb(41pt)=(0.27469,0.51094,0.97275); rgb(42pt)=(0.27381,0.52069,0.97899); rgb(43pt)=(0.27273,0.5304,0.98461); rgb(44pt)=(0.27106,0.54015,0.9893); rgb(45pt)=(0.26878,0.54995,0.99303); rgb(46pt)=(0.26592,0.55979,0.99583); rgb(47pt)=(0.26252,0.56967,0.99773); rgb(48pt)=(0.25862,0.57958,0.99876); rgb(49pt)=(0.25425,0.5895,0.99896); rgb(50pt)=(0.24946,0.59943,0.99835); rgb(51pt)=(0.24427,0.60937,0.99697); rgb(52pt)=(0.23874,0.61931,0.99485); rgb(53pt)=(0.23288,0.62923,0.99202); rgb(54pt)=(0.22676,0.63913,0.98851); rgb(55pt)=(0.22039,0.64901,0.98436); rgb(56pt)=(0.21382,0.65886,0.97959); rgb(57pt)=(0.20708,0.66866,0.97423); rgb(58pt)=(0.20021,0.67842,0.96833); rgb(59pt)=(0.19326,0.68812,0.9619); rgb(60pt)=(0.18625,0.69775,0.95498); rgb(61pt)=(0.17923,0.70732,0.94761); rgb(62pt)=(0.17223,0.7168,0.93981); rgb(63pt)=(0.16529,0.7262,0.93161); rgb(64pt)=(0.15844,0.73551,0.92305); rgb(65pt)=(0.15173,0.74472,0.91416); rgb(66pt)=(0.14519,0.75381,0.90496); rgb(67pt)=(0.13886,0.76279,0.8955); rgb(68pt)=(0.13278,0.77165,0.8858); rgb(69pt)=(0.12698,0.78037,0.8759); rgb(70pt)=(0.12151,0.78896,0.86581); rgb(71pt)=(0.11639,0.7974,0.85559); rgb(72pt)=(0.11167,0.80569,0.84525); rgb(73pt)=(0.10738,0.81381,0.83484); rgb(74pt)=(0.10357,0.82177,0.82437); rgb(75pt)=(0.10026,0.82955,0.81389); rgb(76pt)=(0.0975,0.83714,0.80342); rgb(77pt)=(0.09532,0.84455,0.79299); rgb(78pt)=(0.09377,0.85175,0.78264); rgb(79pt)=(0.09287,0.85875,0.7724); rgb(80pt)=(0.09267,0.86554,0.7623); rgb(81pt)=(0.0932,0.87211,0.75237); rgb(82pt)=(0.09451,0.87844,0.74265); rgb(83pt)=(0.09662,0.88454,0.73316); rgb(84pt)=(0.09958,0.8904,0.72393); rgb(85pt)=(0.10342,0.896,0.715); rgb(86pt)=(0.10815,0.90142,0.70599); rgb(87pt)=(0.11374,0.90673,0.69651); rgb(88pt)=(0.12014,0.91193,0.6866); rgb(89pt)=(0.12733,0.91701,0.67627); rgb(90pt)=(0.13526,0.92197,0.66556); rgb(91pt)=(0.14391,0.9268,0.65448); rgb(92pt)=(0.15323,0.93151,0.64308); rgb(93pt)=(0.16319,0.93609,0.63137); rgb(94pt)=(0.17377,0.94053,0.61938); rgb(95pt)=(0.18491,0.94484,0.60713); rgb(96pt)=(0.19659,0.94901,0.59466); rgb(97pt)=(0.20877,0.95304,0.58199); rgb(98pt)=(0.22142,0.95692,0.56914); rgb(99pt)=(0.23449,0.96065,0.55614); rgb(100pt)=(0.24797,0.96423,0.54303); rgb(101pt)=(0.2618,0.96765,0.52981); rgb(102pt)=(0.27597,0.97092,0.51653); rgb(103pt)=(0.29042,0.97403,0.50321); rgb(104pt)=(0.30513,0.97697,0.48987); rgb(105pt)=(0.32006,0.97974,0.47654); rgb(106pt)=(0.33517,0.98234,0.46325); rgb(107pt)=(0.35043,0.98477,0.45002); rgb(108pt)=(0.36581,0.98702,0.43688); rgb(109pt)=(0.38127,0.98909,0.42386); rgb(110pt)=(0.39678,0.99098,0.41098); rgb(111pt)=(0.41229,0.99268,0.39826); rgb(112pt)=(0.42778,0.99419,0.38575); rgb(113pt)=(0.44321,0.99551,0.37345); rgb(114pt)=(0.45854,0.99663,0.3614); rgb(115pt)=(0.47375,0.99755,0.34963); rgb(116pt)=(0.48879,0.99828,0.33816); rgb(117pt)=(0.50362,0.99879,0.32701); rgb(118pt)=(0.51822,0.9991,0.31622); rgb(119pt)=(0.53255,0.99919,0.30581); rgb(120pt)=(0.54658,0.99907,0.29581); rgb(121pt)=(0.56026,0.99873,0.28623); rgb(122pt)=(0.57357,0.99817,0.27712); rgb(123pt)=(0.58646,0.99739,0.26849); rgb(124pt)=(0.59891,0.99638,0.26038); rgb(125pt)=(0.61088,0.99514,0.2528); rgb(126pt)=(0.62233,0.99366,0.24579); rgb(127pt)=(0.63323,0.99195,0.23937); rgb(128pt)=(0.64362,0.98999,0.23356); rgb(129pt)=(0.65394,0.98775,0.22835); rgb(130pt)=(0.66428,0.98524,0.2237); rgb(131pt)=(0.67462,0.98246,0.2196); rgb(132pt)=(0.68494,0.97941,0.21602); rgb(133pt)=(0.69525,0.9761,0.21294); rgb(134pt)=(0.70553,0.97255,0.21032); rgb(135pt)=(0.71577,0.96875,0.20815); rgb(136pt)=(0.72596,0.9647,0.2064); rgb(137pt)=(0.7361,0.96043,0.20504); rgb(138pt)=(0.74617,0.95593,0.20406); rgb(139pt)=(0.75617,0.95121,0.20343); rgb(140pt)=(0.76608,0.94627,0.20311); rgb(141pt)=(0.77591,0.94113,0.2031); rgb(142pt)=(0.78563,0.93579,0.20336); rgb(143pt)=(0.79524,0.93025,0.20386); rgb(144pt)=(0.80473,0.92452,0.20459); rgb(145pt)=(0.8141,0.91861,0.20552); rgb(146pt)=(0.82333,0.91253,0.20663); rgb(147pt)=(0.83241,0.90627,0.20788); rgb(148pt)=(0.84133,0.89986,0.20926); rgb(149pt)=(0.8501,0.89328,0.21074); rgb(150pt)=(0.85868,0.88655,0.2123); rgb(151pt)=(0.86709,0.87968,0.21391); rgb(152pt)=(0.8753,0.87267,0.21555); rgb(153pt)=(0.88331,0.86553,0.21719); rgb(154pt)=(0.89112,0.85826,0.2188); rgb(155pt)=(0.8987,0.85087,0.22038); rgb(156pt)=(0.90605,0.84337,0.22188); rgb(157pt)=(0.91317,0.83576,0.22328); rgb(158pt)=(0.92004,0.82806,0.22456); rgb(159pt)=(0.92666,0.82025,0.2257); rgb(160pt)=(0.93301,0.81236,0.22667); rgb(161pt)=(0.93909,0.80439,0.22744); rgb(162pt)=(0.94489,0.79634,0.228); rgb(163pt)=(0.95039,0.78823,0.22831); rgb(164pt)=(0.9556,0.78005,0.22836); rgb(165pt)=(0.96049,0.77181,0.22811); rgb(166pt)=(0.96507,0.76352,0.22754); rgb(167pt)=(0.96931,0.75519,0.22663); rgb(168pt)=(0.97323,0.74682,0.22536); rgb(169pt)=(0.97679,0.73842,0.22369); rgb(170pt)=(0.98,0.73,0.22161); rgb(171pt)=(0.98289,0.7214,0.21918); rgb(172pt)=(0.98549,0.7125,0.2165); rgb(173pt)=(0.98781,0.7033,0.21358); rgb(174pt)=(0.98986,0.69382,0.21043); rgb(175pt)=(0.99163,0.68408,0.20706); rgb(176pt)=(0.99314,0.67408,0.20348); rgb(177pt)=(0.99438,0.66386,0.19971); rgb(178pt)=(0.99535,0.65341,0.19577); rgb(179pt)=(0.99607,0.64277,0.19165); rgb(180pt)=(0.99654,0.63193,0.18738); rgb(181pt)=(0.99675,0.62093,0.18297); rgb(182pt)=(0.99672,0.60977,0.17842); rgb(183pt)=(0.99644,0.59846,0.17376); rgb(184pt)=(0.99593,0.58703,0.16899); rgb(185pt)=(0.99517,0.57549,0.16412); rgb(186pt)=(0.99419,0.56386,0.15918); rgb(187pt)=(0.99297,0.55214,0.15417); rgb(188pt)=(0.99153,0.54036,0.1491); rgb(189pt)=(0.98987,0.52854,0.14398); rgb(190pt)=(0.98799,0.51667,0.13883); rgb(191pt)=(0.9859,0.50479,0.13367); rgb(192pt)=(0.9836,0.49291,0.12849); rgb(193pt)=(0.98108,0.48104,0.12332); rgb(194pt)=(0.97837,0.4692,0.11817); rgb(195pt)=(0.97545,0.4574,0.11305); rgb(196pt)=(0.97234,0.44565,0.10797); rgb(197pt)=(0.96904,0.43399,0.10294); rgb(198pt)=(0.96555,0.42241,0.09798); rgb(199pt)=(0.96187,0.41093,0.0931); rgb(200pt)=(0.95801,0.39958,0.08831); rgb(201pt)=(0.95398,0.38836,0.08362); rgb(202pt)=(0.94977,0.37729,0.07905); rgb(203pt)=(0.94538,0.36638,0.07461); rgb(204pt)=(0.94084,0.35566,0.07031); rgb(205pt)=(0.93612,0.34513,0.06616); rgb(206pt)=(0.93125,0.33482,0.06218); rgb(207pt)=(0.92623,0.32473,0.05837); rgb(208pt)=(0.92105,0.31489,0.05475); rgb(209pt)=(0.91572,0.3053,0.05134); rgb(210pt)=(0.91024,0.29599,0.04814); rgb(211pt)=(0.90463,0.28696,0.04516); rgb(212pt)=(0.89888,0.27824,0.04243); rgb(213pt)=(0.89298,0.26981,0.03993); rgb(214pt)=(0.88691,0.26152,0.03753); rgb(215pt)=(0.88066,0.25334,0.03521); rgb(216pt)=(0.87422,0.24526,0.03297); rgb(217pt)=(0.8676,0.2373,0.03082); rgb(218pt)=(0.86079,0.22945,0.02875); rgb(219pt)=(0.8538,0.2217,0.02677); rgb(220pt)=(0.84662,0.21407,0.02487); rgb(221pt)=(0.83926,0.20654,0.02305); rgb(222pt)=(0.83172,0.19912,0.02131); rgb(223pt)=(0.82399,0.19182,0.01966); rgb(224pt)=(0.81608,0.18462,0.01809); rgb(225pt)=(0.80799,0.17753,0.0166); rgb(226pt)=(0.79971,0.17055,0.0152); rgb(227pt)=(0.79125,0.16368,0.01387); rgb(228pt)=(0.7826,0.15693,0.01264); rgb(229pt)=(0.77377,0.15028,0.01148); rgb(230pt)=(0.76476,0.14374,0.01041); rgb(231pt)=(0.75556,0.13731,0.00942); rgb(232pt)=(0.74617,0.13098,0.00851); rgb(233pt)=(0.73661,0.12477,0.00769); rgb(234pt)=(0.72686,0.11867,0.00695); rgb(235pt)=(0.71692,0.11268,0.00629); rgb(236pt)=(0.7068,0.1068,0.00571); rgb(237pt)=(0.6965,0.10102,0.00522); rgb(238pt)=(0.68602,0.09536,0.00481); rgb(239pt)=(0.67535,0.0898,0.00449); rgb(240pt)=(0.66449,0.08436,0.00424); rgb(241pt)=(0.65345,0.07902,0.00408); rgb(242pt)=(0.64223,0.0738,0.00401); rgb(243pt)=(0.63082,0.06868,0.00401); rgb(244pt)=(0.61923,0.06367,0.0041); rgb(245pt)=(0.60746,0.05878,0.00427); rgb(246pt)=(0.5955,0.05399,0.00453); rgb(247pt)=(0.58336,0.04931,0.00486); rgb(248pt)=(0.57103,0.04474,0.00529); rgb(249pt)=(0.55852,0.04028,0.00579); rgb(250pt)=(0.54583,0.03593,0.00638); rgb(251pt)=(0.53295,0.03169,0.00705); rgb(252pt)=(0.51989,0.02756,0.0078); rgb(253pt)=(0.50664,0.02354,0.00863); rgb(254pt)=(0.49321,0.01963,0.00955); rgb(255pt)=(0.4796,0.01583,0.01055)}, mesh/rows=8]
table[row sep=crcr, point meta=\thisrow{c}] {%
x	y	c\\
5	4.35664300942338	0.00157771821517024\\
5	4.35664300942338	0.00157771821517024\\
10	3.73194318445354	0.00438385121126728\\
10	3.73194318445354	0.00438385121126728\\
15	2.97486286374294	0.00896349075801616\\
15	2.97486286374294	0.00896349075801616\\
20	2.10234794558662	0.0152275830015485\\
20	2.10234794558662	0.0152275830015485\\
25	1.13159163321035	0.023076510558497\\
25	1.13159163321035	0.023076510558497\\
30	0.0785535692707526	0.03240255189655\\
30	0.0785535692707526	0.03240255189655\\
35	-1.04298311731759	0.0430921113873273\\
35	-1.04298311731759	0.0430921113873273\\
40	-2.22184895853354	0.0550274408791649\\
40	-2.22184895853354	0.0550274408791649\\
};
\addplot[
    only marks,
    mark=*,
    mark size=5pt,
    color=red, 
    mark options={
        fill=red,
        fill opacity=0.5, 
        draw opacity=1     
    },
]
coordinates {(40,-2.22)};
\node[color=black, fill=none] at (axis cs: 29,-2) {\textcolor{red}{Case 2b}};
\draw[line width=0.5mm,draw=black!80,->] (axis cs:38,-2.3)--(axis cs:34.5,-2.2);
\end{axis}
\end{tikzpicture}%

%% file: updated_images/Smart_Grid_Lab_Setup.pdf_tex
\begingroup%
  \makeatletter%
  \providecommand\color[2][]{%
    \errmessage{(Inkscape) Color is used for the text in Inkscape, but the package 'color.sty' is not loaded}%
    \renewcommand\color[2][]{}%
  }%
  \providecommand\transparent[1]{%
    \errmessage{(Inkscape) Transparency is used (non-zero) for the text in Inkscape, but the package 'transparent.sty' is not loaded}%
    \renewcommand\transparent[1]{}%
  }%
  \providecommand\rotatebox[2]{#2}%
  \newcommand*\fsize{\dimexpr\f@size pt\relax}%
  \newcommand*\lineheight[1]{\fontsize{\fsize}{#1\fsize}\selectfont}%
  \ifx\svgwidth\undefined%
    \setlength{\unitlength}{594.69599686bp}%
    \ifx\svgscale\undefined%
      \relax%
    \else%
      \setlength{\unitlength}{\unitlength * \real{\svgscale}}%
    \fi%
  \else%
    \setlength{\unitlength}{\svgwidth}%
  \fi%
  \global\let\svgwidth\undefined%
  \global\let\svgscale\undefined%
  \makeatother%
  \begin{picture}(1,0.52839017)%
    \lineheight{1}%
    \setlength\tabcolsep{0pt}%
    \put(0,0){\includegraphics[width=\unitlength,page=1]{updated_images/Smart_Grid_Lab_Setup.pdf}}%
    \put(0.055,0.3433141){\makebox(0,0)[lt]{\lineheight{1.25}\smash{\begin{tabular}[t]{l}\textbf{\textit{1 \& 6}}\end{tabular}}}}%
    \put(0,0){\includegraphics[width=\unitlength,page=2]{updated_images/Smart_Grid_Lab_Setup.pdf}}%
    \put(0.66744545,0.23368271){\makebox(0,0)[lt]{\lineheight{1.25}\smash{\begin{tabular}[t]{l}\textbf{\textit{2}}\end{tabular}}}}%
    \put(0,0){\includegraphics[width=\unitlength,page=3]{updated_images/Smart_Grid_Lab_Setup.pdf}}%
    \put(0.89571337,0.27){\makebox(0,0)[lt]{\lineheight{1.25}\smash{\begin{tabular}[t]{l}\textbf{\textit{3}}\end{tabular}}}}%
    \put(0,0){\includegraphics[width=\unitlength,page=3]{updated_images/Smart_Grid_Lab_Setup.pdf}}%
    \put(0.04727567,0.16495011){\makebox(0,0)[lt]{\lineheight{1.25}\smash{\begin{tabular}[t]{l}\textbf{\textit{5}}\end{tabular}}}}%
    \put(0,0){\includegraphics[width=\unitlength,page=3]{updated_images/Smart_Grid_Lab_Setup.pdf}}%
    \put(0.45967126,0.40425317){\makebox(0,0)[lt]{\lineheight{1.25}\smash{\begin{tabular}[t]{l}\textbf{\textit{4}}\end{tabular}}}}%
  \end{picture}%
\endgroup%

%% file: updated_images/Bode_freq_sweep_Z_dd.tex
%
%
\definecolor{mycolor1}{rgb}{0.00000,0.44700,0.74100}%
\begin{tikzpicture}

\begin{axis}[%
width=2.5in,
height=1in,
at={(0.758in,2.0in)},
scale only axis,
xmode=log,
xmin=1,
xmax=10000,
xminorticks=true,
xlabel style={font=\color{white!15!black}},
ymin=-40,
ymax=40,
ylabel style={font=\large\color{white!15!black}},
ylabel={Magnitude [dB]},
axis background/.style={fill=white},
title style={font=\Large\bfseries},
title={$Z_{s}^{dd}$},
xmajorgrids,
xminorgrids,
ymajorgrids,
legend pos=south west,
legend style={legend cell align=left, align=left,font=\normalsize, draw=white!15!black}
]
\addplot [color=mycolor1, line width=2.0pt]
  table[row sep=crcr]{%
1	37.7758854759384\\
1.20679264063934	34.8851434744472\\
1.7575106248548	28.9362127071834\\
2.55954792269953	22.9612183786556\\
3.0888435964775	20.0413351228973\\
3.72759372031492	17.2119178585147\\
4.49843266896945	14.5101738671955\\
5.4286754393239	11.9742066983069\\
6.5512855685955	9.63820848968709\\
7.90604321090773	7.52545489696477\\
9.54095476349988	5.65026092918867\\
11.5139539932645	4.02365005365168\\
13.8949549437315	2.64858155214466\\
16.76832936811	1.51654198748961\\
20.2358964772516	0.608359378958731\\
24.4205309454863	-0.103756806561158\\
29.4705170255181	-0.653306394359007\\
35.5648030622315	-1.0761446536909\\
42.9193426012876	-1.40771782306647\\
51.7947467923122	-1.68174180695954\\
62.5055192527392	-1.93062188164163\\
75.4312006335461	-2.1875823181255\\
91.0298177991526	-2.49058380332669\\
109.854114198755	-2.88858407852053\\
132.571136559011	-3.4512547787908\\
159.985871960607	-4.28180821337984\\
193.069772888325	-5.51103150558205\\
232.995181051538	-7.05852893920746\\
281.176869797421	-7.13800584583103\\
339.322177189533	-3.18134014750726\\
409.491506238045	2.00134727723577\\
494.171336132382	6.55643820902294\\
596.362331659466	10.1287342636353\\
719.685673001147	12.6366442770523\\
868.511373751352	14.3110296412513\\
1048.11313415469	15.5576341300611\\
1264.85521685529	16.6846563665568\\
1526.41796717524	17.8432648776221\\
1842.06996932673	19.0854781840514\\
2222.99648252619	20.4172041854396\\
2682.69579527974	21.8268085576402\\
3237.45754281763	23.2981909326979\\
3906.93993705462	24.8161407569795\\
4714.86636345743	26.3681117480485\\
6866.48845004303	29.5379777569939\\
10000	32.7575276005247\\
};
\addlegendentry{${Z}_{{analy}}$}

\addplot [color=red, line width=1.0pt, only marks, mark size=1pt, mark=*, mark options={solid, red}]
  table[row sep=crcr]{%
5	12.4206600629849\\
6	10.0436677801104\\
7	8.00752775401422\\
8	6.78861689725057\\
10	4.5947624721906\\
11	3.72638958465562\\
13	2.56750804464769\\
15	1.63372265381035\\
16.9999999999999	0.892758736789013\\
20	0.218457025052164\\
23	-0.309155910136155\\
27	-0.827510327552581\\
30.9999999999999	-1.15250682355192\\
35	-1.4705831615505\\
41.0000000000002	-1.72311317600582\\
47.0000000000001	-1.9719564724783\\
54.0000000000001	-2.12880162079466\\
63.0000000000001	-2.41862545006008\\
72.0000000000001	-2.66663341426175\\
83.0000000000003	-2.97996412662444\\
96.0000000000001	-3.37879393704844\\
111	-3.73107925583646\\
127	-3.87946252296027\\
147	-4.30569189191614\\
169	-5.01945427977315\\
195	-5.94087740002415\\
224	-6.79818543803237\\
257.999999999999	-6.94305474658298\\
297.999999999999	-5.3058990717239\\
343	-2.03243760548202\\
395	1.13763249955328\\
455	4.12392166210887\\
524.000000000001	6.78421975335008\\
603.000000000001	9.1117326193257\\
693.999999999999	11.2023281816097\\
798.999999999998	12.7861176506886\\
921.000000000001	13.9054017049783\\
1060	14.7332889180802\\
1221	15.3452361665539\\
1405.00000000001	15.9911717613267\\
1618	16.8208407759972\\
1863	18.0754494548999\\
2145	19.981111718425\\
2470	20.9047001432675\\
2844.00000000001	20.9724700698598\\
};
\addlegendentry{${Z}_{{meas}}$}

\end{axis}

\begin{axis}[%
width=2.5in,
height=1in,
at={(0.758in,0.7in)},
scale only axis,
xmode=log,
xmin=1,
xmax=5000,
xminorticks=true,
xlabel style={font=\large\color{white!15!black}},
xlabel={Frequency [Hz]},
ymin=-200,
ymax=200,
ylabel style={font=\large\color{white!15!black}},
ylabel={Phase [deg]},
axis background/.style={fill=white},
xmajorgrids,
xminorgrids,
ymajorgrids,
]
\addplot [color=mycolor1, line width=2.0pt, forget plot]
  table[row sep=crcr]{%
1	-139.307189525071\\
1.20679264063936	-141.161371349455\\
1.45634847750121	-142.03654672252\\
1.7575106248548	-141.849550251656\\
2.12095088792024	-140.54071004926\\
2.55954792269949	-138.067125085448\\
3.0888435964775	-134.402731852808\\
3.72759372031504	-129.54591343871\\
4.49843266896938	-123.533651682167\\
5.4286754393239	-116.460335276362\\
6.55128556859571	-108.492702308838\\
7.9060432109076	-99.8450656825833\\
9.54095476350004	-90.7238982365052\\
13.8949549437312	-71.9544854190487\\
16.7683293681103	-62.7957672144972\\
20.235896477251	-54.0840971038375\\
24.4205309454863	-45.9850688095268\\
29.4705170255186	-38.6014133091042\\
35.5648030622303	-31.9692146306114\\
42.9193426012876	-26.064227820802\\
51.7947467923131	-20.8136747849112\\
62.5055192527382	-16.106947448849\\
75.4312006335461	-11.7990539157299\\
91.0298177991541	-7.70002497234231\\
109.854114198753	-3.53733583445873\\
132.571136559011	1.14323687791872\\
159.98587196061	7.24345319548567\\
193.069772888321	16.8861511531471\\
232.995181051538	35.5222889328372\\
281.176869797431	70.484594509935\\
339.322177189527	102.990065581708\\
409.491506238045	115.490135261132\\
494.171336132398	115.352670252529\\
596.362331659456	108.442793235177\\
719.685673001159	99.1614659524606\\
868.51137375138	91.148562405937\\
1048.11313415467	85.8982353670086\\
1264.85521685531	83.1510909295041\\
1526.41796717519	82.1223725814889\\
1842.0699693267	82.1162789215103\\
2222.99648252623	82.6510007418278\\
2682.69579527965	83.4266879201342\\
3237.45754281763	84.2697939177751\\
3906.93993705468	85.087164607528\\
4714.86636345727	85.834021727449\\
5689.86602901828	86.4932901386507\\
6866.48845004314	87.0629017152444\\
8286.42772854666	87.54830444818\\
10000	87.9582062943939\\
};
\addplot [color=red, line width=1.0pt, only marks, mark size=1pt, mark=*, mark options={solid, red}, forget plot]
  table[row sep=crcr]{%
5	-119.854137697754\\
6	-112.737333733382\\
6.99999999999989	-106.954161772114\\
8	-100.046984207168\\
10	-89.4734476459435\\
11.0000000000001	-84.5524556915784\\
12.9999999999999	-75.9993244715293\\
15	-68.6981829981601\\
16.9999999999998	-62.4961824678401\\
20	-54.796045888112\\
23	-49.0584928384364\\
27	-43.0216767023556\\
31.0000000000004	-37.5113024470087\\
35.0000000000006	-33.3493317842171\\
40.9999999999995	-28.6573974380893\\
46.9999999999993	-24.330262484349\\
54.0000000000001	-20.6655098182048\\
62.999999999999	-17.0481523197614\\
72.0000000000001	-13.9839897561686\\
82.9999999999989	-11.0627504825445\\
96.0000000000001	-7.98109670597738\\
111	-1.9009906560526\\
126.999999999999	2.3675957601363\\
147	6.55819739208052\\
168.999999999997	11.8516400908756\\
194.999999999999	20.5325300049103\\
223.999999999997	34.9184873919905\\
258.000000000002	57.6699019030272\\
298.000000000004	79.0861327572626\\
342.999999999995	96.5442219273534\\
395.000000000004	105.566741680402\\
455.000000000004	107.301912830782\\
524.000000000006	106.021054230075\\
603.000000000006	103.199750598938\\
694.00000000001	97.5066808573525\\
799.000000000005	92.9132158805248\\
921.000000000001	87.6296732549026\\
1060	82.5161726433151\\
1221.00000000001	80.1479876978863\\
1405.00000000002	81.0936700687029\\
1617.99999999999	84.1699107296261\\
1863	85.5164693852623\\
2145	86.7507815923369\\
2470.00000000004	82.851211950852\\
2844.00000000003	101.693573718893\\
};
\end{axis}
\end{tikzpicture}%

%% file: updated_images/Bode_freq_sweep_Z_dq.tex
%
%
\definecolor{mycolor1}{rgb}{0.00000,0.44700,0.74100}%
\begin{tikzpicture}

\begin{axis}[%
width=2.5in,
height=1in,
at={(0.758in,2.0in)},
scale only axis,
xmode=log,
xmin=1,
xmax=5000,
xminorticks=true,
xlabel style={font=\color{white!15!black}},
ymin=-40,
ymax=40,
ylabel style={font=\large\color{white!15!black}},
ylabel={Magnitude [dB]},
axis background/.style={fill=white},
title style={font=\Large\bfseries},
title={$Z_{s}^{dq}$},
xmajorgrids,
xminorgrids,
ymajorgrids,
legend pos=north west,
legend style={legend cell align=left, align=left,font=\normalsize, draw=white!15!black}
]
\addplot [color=mycolor1, line width=2.0pt, forget plot]
  table[row sep=crcr]{%
1	-36.6208136225504\\
1.20679264063934	-36.4204232020886\\
1.45634847750124	-36.3827424671594\\
1.7575106248548	-36.5354688723947\\
2.12095088792018	-36.9286304639429\\
2.55954792269953	-37.5827170777822\\
3.0888435964775	-38.1305786768189\\
3.72759372031492	-36.8874701289251\\
4.49843266896945	-32.8891737614227\\
5.4286754393239	-28.2976157996266\\
6.5512855685955	-24.6236200077383\\
7.90604321090773	-22.4047418291447\\
9.54095476349988	-21.496999867799\\
11.5139539932645	-21.3504465346562\\
13.8949549437315	-21.5200837587656\\
16.76832936811	-21.7771474335237\\
20.2358964772516	-22.0172320632018\\
24.4205309454863	-22.1908879900612\\
29.4705170255181	-22.2723921705286\\
35.5648030622315	-22.2458456457113\\
42.9193426012876	-22.0981171762898\\
51.7947467923122	-21.815548301518\\
62.5055192527392	-21.3836892074885\\
75.4312006335461	-20.7894662382349\\
91.0298177991526	-20.0245693846188\\
109.854114198755	-19.0882939635807\\
132.571136559011	-17.9882962656085\\
159.985871960607	-16.7389673760522\\
193.069772888325	-15.3587486985723\\
232.995181051538	-13.8689249449983\\
281.176869797421	-12.2975136975526\\
339.322177189533	-10.6941918621828\\
409.491506238045	-9.16549615832475\\
494.171336132382	-7.9296270936347\\
596.362331659466	-7.31889724921692\\
719.685673001147	-7.57469820402745\\
868.511373751352	-8.5874748720211\\
1048.11313415469	-10.022681307578\\
1526.41796717524	-13.1831083879345\\
1842.06996932673	-14.6921950557817\\
2222.99648252619	-16.0966032020272\\
2682.69579527974	-17.373859328915\\
3237.45754281763	-18.5064963109471\\
3906.93993705462	-19.4821222997605\\
4714.86636345743	-20.2959005612256\\
5689.86602901828	-20.9524111251176\\
6866.48845004303	-21.4652520595563\\
8286.42772854679	-21.85438259924\\
10000	-22.1424457673478\\
};
\addplot [color=red, line width=1.0pt, only marks, mark size=1pt, mark=*, mark options={solid, red}, forget plot]
  table[row sep=crcr]{%
5	-14.0774207918934\\
6	-17.7816501767443\\
7	-16.1462093527213\\
8	-18.3142645477685\\
10	-20.2624516428788\\
11	-20.8808683041448\\
13	-22.6104590855323\\
15	-21.0177605025345\\
16.9999999999999	-21.011134341325\\
20	-21.1174757879979\\
23	-20.7339761009637\\
27	-20.5616516829557\\
30.9999999999999	-20.6558868786251\\
35	-20.3042108804536\\
41.0000000000002	-20.5899967870463\\
47.0000000000001	-20.4321675779424\\
54.0000000000001	-20.5390742811417\\
63.0000000000001	-20.1477724599122\\
72.0000000000001	-19.7387338688184\\
83.0000000000003	-19.148162770474\\
96.0000000000001	-18.3414509990921\\
111	-18.2527609928521\\
127	-18.3815637284539\\
147	-17.8228154893824\\
169	-16.5998033314321\\
195	-15.9116493498603\\
224	-15.0501687531999\\
257.999999999999	-13.9737776081124\\
297.999999999999	-13.8033885284505\\
343	-11.6619101562581\\
395	-10.7363129045228\\
455	-9.53883401097708\\
524.000000000001	-9.41509119971636\\
603.000000000001	-9.04828933727352\\
693.999999999999	-9.29817643289021\\
798.999999999998	-8.4757136731702\\
921.000000000001	-8.52388853707296\\
1060	-9.82057803811417\\
1221	-11.0635214975342\\
1405.00000000001	-12.7719796146294\\
1618	-17.0744123607183\\
1863	-8.38925427667832\\
2145	-7.17480633559845\\
2470	4.99361146853521\\
2844.00000000001	19.3801116108358\\
};
\end{axis}

\begin{axis}[%
width=2.5in,
height=1in,
at={(0.758in,0.7in)},
scale only axis,
xmode=log,
xmin=1,
xmax=5000,
xminorticks=true,
xlabel style={font=\large\color{white!15!black}},
xlabel={Frequency [Hz]},
ymin=-200,
ymax=200,
ylabel style={font=\large\color{white!15!black}},
ylabel={Phase [deg]},
axis background/.style={fill=white},
xmajorgrids,
xminorgrids,
ymajorgrids,
]
\addplot [color=mycolor1, line width=2.0pt, forget plot]
  table[row sep=crcr]{%
1	25.7662082941026\\
1.20679264063936	20.1188087094725\\
1.45634847750121	14.1141506979433\\
1.7575106248548	7.0895497760516\\
2.12095088792024	-2.28689803105496\\
2.55954792269949	-16.7885908356654\\
3.0888435964775	-41.7380631974989\\
3.72759372031504	-79.5796761379613\\
4.49843266896938	-115.591851548544\\
5.4286754393239	-142.261274292076\\
6.55128556859571	-164.539230414345\\
7.9060432109076	176.373871126456\\
9.54095476350004	162.812408630105\\
11.5139539932641	155.664976684832\\
13.8949549437312	153.719211863604\\
16.7683293681103	155.338471585512\\
20.235896477251	159.219174810063\\
24.4205309454863	164.457893342727\\
29.4705170255186	170.464656453836\\
35.5648030622303	176.871075816543\\
42.9193426012876	-176.54958220392\\
51.7947467923131	-169.950959493004\\
62.5055192527382	-163.468639704478\\
75.4312006335461	-157.257053915696\\
91.0298177991541	-151.504545340577\\
109.854114198753	-146.428092612371\\
132.571136559011	-142.25721163876\\
159.98587196061	-139.222025164158\\
193.069772888321	-137.559000572042\\
232.995181051538	-137.541191111803\\
281.176869797431	-139.53208223488\\
339.322177189527	-144.047194042988\\
409.491506238045	-151.75072825091\\
494.171336132398	-163.153576832311\\
596.362331659456	-177.695529413587\\
719.685673001159	167.060556969426\\
868.51137375138	154.082044293239\\
1048.11313415467	144.704453946902\\
1264.85521685531	138.707239493452\\
1526.41796717519	135.401545486196\\
1842.0699693267	134.173238266945\\
2222.99648252623	134.568430201269\\
2682.69579527965	136.242566481056\\
3237.45754281763	138.900883355027\\
3906.93993705468	142.261907897167\\
4714.86636345727	146.048325757968\\
5689.86602901828	149.999541585664\\
6866.48845004314	153.893577960597\\
8286.42772854666	157.56463884886\\
10000	160.908354872689\\
};
\addplot [color=red, line width=1.0pt, only marks, mark size=1pt, mark=*, mark options={solid, red}, forget plot]
  table[row sep=crcr]{%
5	-126.094643517759\\
6	-139.830342760601\\
6.99999999999989	-158.673450357298\\
8	-172.3690132208\\
10	171.903638432893\\
11.0000000000001	166.982014793755\\
13.0000000000003	169.436743995653\\
15	158.023008608346\\
16.9999999999998	161.132339998477\\
20	161.893546731326\\
23	163.725965171056\\
27	163.378659968003\\
31.0000000000004	165.056586612721\\
34.9999999999994	172.077773776891\\
41.0000000000008	176.438728585155\\
46.9999999999993	173.294328974495\\
54.0000000000001	177.647512037369\\
62.999999999999	-170.033659307611\\
72.0000000000001	-171.921301556216\\
83.0000000000017	-165.007450143852\\
96.0000000000001	-160.216442420624\\
111	-159.96188018917\\
127.000000000003	-155.024545479037\\
146.999999999995	-153.332549556684\\
168.999999999997	-151.49674406356\\
195.000000000005	-149.693261567421\\
223.999999999997	-148.088702657438\\
258.000000000002	-146.9843768574\\
298.000000000004	-146.611894122141\\
343.000000000006	-151.35986931568\\
395.000000000004	-155.416458783852\\
455.000000000004	-162.770327986119\\
524.000000000006	-171.606628324329\\
602.999999999986	-178.189098678167\\
693.999999999988	175.37583539791\\
798.999999999978	158.598261049956\\
920.999999999971	142.150828389638\\
1060.00000000003	147.88784042416\\
1221.00000000001	134.267350984556\\
1405.00000000002	137.108694876934\\
1617.99999999999	165.614453374823\\
1863	177.856918366186\\
2144.99999999993	-131.781970754092\\
2470.00000000004	96.7084958414204\\
2844.00000000003	4.11097387320621\\
};
\end{axis}
\end{tikzpicture}%

%% file: updated_images/Bode_freq_sweep_Z_qd.tex
%
%
\definecolor{mycolor1}{rgb}{0.00000,0.44700,0.74100}%
\begin{tikzpicture}

\begin{axis}[%
width=2.5in,
height=1in,
at={(0.758in,2.0in)},
scale only axis,
xmode=log,
xmin=1,
xmax=5000,
xminorticks=true,
xlabel style={font=\color{white!15!black}},
ymin=-40,
ymax=40,
ylabel style={font=\large\color{white!15!black}},
ylabel={Magnitude [dB]},
axis background/.style={fill=white},
title style={font=\Large\bfseries},
title={$Z_{s}^{qd}$},
xmajorgrids,
xminorgrids,
ymajorgrids,
legend pos=north west,
legend style={legend cell align=left, align=left,font=\normalsize, draw=white!15!black}
]
\addplot [color=mycolor1, line width=2.0pt, forget plot]
  table[row sep=crcr]{%
1	-27.9105030029856\\
1.20679264063933	-26.3628107337559\\
1.45634847750124	-24.8510794064788\\
1.7575106248548	-23.3891550975049\\
2.12095088792019	-21.9953793270547\\
2.55954792269953	-20.6943248701163\\
3.08884359647747	-19.5214510227517\\
3.72759372031495	-18.5372044949203\\
4.49843266896945	-17.8667351023406\\
5.42867543932386	-17.7892260313393\\
6.5512855685955	-18.8350082638115\\
7.90604321090767	-21.6106128533073\\
9.54095476349996	-26.2117355818413\\
11.5139539932645	-30.5997614035542\\
13.8949549437314	-29.8731668147745\\
16.76832936811	-27.5822879218591\\
20.2358964772516	-25.9681768187002\\
24.4205309454865	-24.8940426311338\\
29.4705170255181	-24.1276892274404\\
35.5648030622312	-23.5188307711512\\
42.919342601288	-22.969133009886\\
51.7947467923122	-22.4087207498934\\
62.5055192527398	-21.7849085664832\\
75.4312006335461	-21.0582212968585\\
91.0298177991519	-20.2020252351735\\
109.854114198756	-19.2028968158076\\
132.571136559011	-18.0596819417568\\
159.985871960606	-16.780688688336\\
193.069772888325	-15.3801815424912\\
232.995181051538	-13.8766123598646\\
281.176869797424	-12.296141412594\\
339.322177189533	-10.6873439287983\\
409.491506238042	-9.15627608507548\\
494.171336132382	-7.92102260351462\\
596.362331659466	-7.31335694251242\\
719.685673001153	-7.57289007495528\\
868.511373751352	-8.58821154341126\\
1048.11313415468	-10.0244474023653\\
1526.41796717524	-13.184708149649\\
1842.06996932672	-14.6934399116423\\
2222.99648252619	-16.0975253588086\\
2682.69579527972	-17.3745235706875\\
3237.45754281765	-18.5069667729977\\
3906.93993705462	-19.4824520150872\\
4714.86636345739	-20.2961300725304\\
5689.86602901828	-20.9525701750714\\
6866.48845004303	-21.4653619538947\\
8286.42772854686	-21.8544583787929\\
10000	-22.1424979521081\\
};
\addplot [color=red, line width=1.0pt, only marks, mark size=1pt, mark=*, mark options={solid, red}, forget plot]
  table[row sep=crcr]{%
5	-17.7004061162603\\
6	-18.425586627589\\
7	-20.4756568536025\\
8	-22.918053095651\\
10	-33.6319065967684\\
10.9999999999999	-30.8877614444138\\
12.9999999999999	-26.1317832112824\\
15	-25.8147303238568\\
17.0000000000001	-24.5469053696877\\
20	-23.7190529321138\\
23	-24.0679039443627\\
27	-23.5661681572511\\
30.9999999999999	-23.4096208182477\\
35	-23.1979992534359\\
41.0000000000002	-22.2956901986741\\
47.0000000000001	-22.2861831134346\\
54.0000000000001	-21.5700605293405\\
63.0000000000001	-21.9789497246918\\
72.0000000000001	-21.6700832148689\\
83.0000000000003	-20.7910693398692\\
96.0000000000001	-20.1460068274816\\
111	-19.3434286886751\\
127.000000000001	-18.2572111269985\\
147	-17.8447099189915\\
169	-17.1961734807641\\
194.999999999999	-15.8945580017148\\
224	-15.0220156194129\\
258.000000000002	-13.9395768632991\\
297.999999999999	-13.8273842539287\\
343	-11.918489865087\\
394.999999999997	-10.90403503282\\
455.000000000004	-9.98611588964243\\
523.999999999997	-9.08497242561082\\
602.999999999996	-8.86535846199119\\
693.999999999999	-7.90551007023525\\
799.000000000005	-9.98826391200509\\
921.000000000001	-10.8267067339818\\
1060	-17.4558909875388\\
1220.99999999999	-12.332326900015\\
1404.99999999999	-18.9236432437465\\
1617.99999999999	-14.4244456155105\\
1863	-23.7139994318419\\
2145	-1.23352876612363\\
2470	-11.604361412584\\
2843.99999999998	-8.30620832017059\\
};
\end{axis}

\begin{axis}[%
width=2.5in,
height=1in,
at={(0.758in,0.7in)},
scale only axis,
xmode=log,
xmin=1,
xmax=5000,
xminorticks=true,
xlabel style={font=\large\color{white!15!black}},
xlabel={Frequency [Hz]},
ymin=-200,
ymax=200,
ylabel style={font=\large\color{white!15!black}},
ylabel={Phase [deg]},
axis background/.style={fill=white},
xmajorgrids,
xminorgrids,
ymajorgrids,
]
\addplot [color=mycolor1, line width=2.0pt, forget plot]
  table[row sep=crcr]{%
1	-97.7848270115569\\
1.20679264063936	-99.3504119566268\\
1.45634847750121	-101.212995162226\\
1.7575106248548	-103.423088393506\\
2.12095088792024	-106.045170537262\\
2.55954792269949	-109.175530234214\\
3.0888435964775	-112.982382600899\\
3.72759372031504	-117.777038843866\\
4.49843266896938	-124.077068546604\\
5.4286754393239	-132.396026202184\\
6.55128556859571	-141.981970083404\\
7.9060432109076	-148.298270476333\\
9.54095476350004	-141.570639414371\\
11.5139539932641	-106.755785110341\\
13.8949549437312	-63.4859337314069\\
16.7683293681103	-41.3707819949734\\
20.235896477251	-28.6743415736723\\
24.4205309454863	-19.1558095164064\\
29.4705170255186	-10.9510183801441\\
35.5648030622303	-3.41408548857498\\
42.9193426012876	3.72110852797056\\
51.7947467923131	10.563269202897\\
62.5055192527382	17.1221037481186\\
75.4312006335461	23.3243805925783\\
91.0298177991541	29.0267333450481\\
109.854114198753	34.0365265619603\\
132.571136559011	38.1368760864555\\
159.98587196061	41.1036991625845\\
193.069772888321	42.7028928355267\\
232.995181051538	42.6617427346897\\
281.176869797431	40.6160547846541\\
339.322177189527	36.0493515383315\\
409.491506238045	28.2974735037419\\
494.171336132398	16.8528443403268\\
596.362331659456	2.28346817058843\\
719.685673001159	-12.968091419245\\
868.51137375138	-25.9404386242491\\
1048.11313415467	-35.3084382015361\\
1264.85521685531	-41.2980395835542\\
1526.41796717519	-44.5990039482191\\
1842.0699693267	-45.8247966985666\\
2222.99648252623	-45.428486452697\\
2682.69579527965	-43.7540254714258\\
3237.45754281763	-41.0957988852979\\
3906.93993705468	-37.735059113814\\
4714.86636345727	-33.9489985303103\\
5689.86602901828	-29.9981484159977\\
6866.48845004314	-26.1044549482146\\
8286.42772854666	-22.433700970706\\
10000	-19.0902522488884\\
};
\addplot [color=red, line width=1.0pt, only marks, mark size=1pt, mark=*, mark options={solid, red}, forget plot]
  table[row sep=crcr]{%
5	-115.168890417883\\
6	-107.133648660113\\
6.99999999999989	-169.415281153665\\
8	-147.30126128099\\
10	-124.116300834962\\
11.0000000000001	-94.6301162437985\\
13.0000000000003	-57.8507165132869\\
15	-48.9488226814345\\
16.9999999999998	-39.8853557236622\\
20	-32.7049919791993\\
23	-21.3910790068273\\
27	-12.3885200147691\\
31.0000000000004	-9.49727815310362\\
34.9999999999994	-5.14371638834885\\
41.0000000000008	0.131412620700644\\
46.9999999999993	3.50213716415411\\
54.0000000000001	4.82215763498618\\
62.999999999999	15.2221711036412\\
72.0000000000001	16.5035102082807\\
83.0000000000017	21.172601325273\\
96.0000000000001	25.2953773194147\\
111	29.1111445469136\\
127.000000000003	32.0170301794781\\
146.999999999995	33.1810109181292\\
168.999999999997	35.7452711776284\\
195.000000000005	36.5679304370867\\
223.999999999997	36.2109220283333\\
258.000000000002	36.7733807913554\\
298.000000000004	31.9649933275122\\
343.000000000006	30.4175845704599\\
395.000000000004	27.101441152284\\
455.000000000004	20.3636878987245\\
524.000000000006	12.3481661945196\\
602.999999999986	2.8479594515772\\
693.999999999988	-8.42718143534447\\
798.999999999978	-19.3332222529642\\
920.999999999971	-33.3937535905633\\
1060.00000000003	-64.3824913253885\\
1221.00000000001	-38.2513853122274\\
1405.00000000002	-35.6207951196893\\
1617.99999999999	-18.1820762585307\\
1863	20.2314217644471\\
2144.99999999993	-87.5087147754223\\
2470.00000000004	33.7712340182757\\
2844.00000000003	94.1224992251528\\
};
\end{axis}
\end{tikzpicture}%

%% file: updated_images/Bode_freq_sweep_Z_qq.tex
%
%
\definecolor{mycolor1}{rgb}{0.00000,0.44700,0.74100}%
\begin{tikzpicture}

\begin{axis}[%
width=2.5in,
height=1in,
at={(0.758in,2.0in)},
scale only axis,
xmode=log,
xmin=1,
xmax=5000,
xminorticks=true,
xlabel style={font=\color{white!15!black}},
ymin=-40,
ymax=40,
ylabel style={font=\large\color{white!15!black}},
ylabel={Magnitude [dB]},
axis background/.style={fill=white},
title style={font=\Large\bfseries},
title={$Z_{s}^{qq}$},
xmajorgrids,
xminorgrids,
ymajorgrids,
legend pos=north west,
legend style={legend cell align=left, align=left,font=\normalsize, draw=white!15!black}
]
\addplot [color=mycolor1, line width=2.0pt, forget plot]
  table[row sep=crcr]{%
1	10.3777968696914\\
1.20679264063934	10.3432093160236\\
1.45634847750124	10.2946068089265\\
1.7575106248548	10.2270663569187\\
2.12095088792018	10.1341612726684\\
2.55954792269953	10.0066352327529\\
3.0888435964775	9.82768101265199\\
3.72759372031492	9.55787588604912\\
4.49843266896945	9.09323819147391\\
5.4286754393239	8.17244058122606\\
6.5512855685955	6.27579130157763\\
7.90604321090773	2.79726489476855\\
9.54095476349988	-2.36908652719953\\
11.5139539932645	-7.20377559791947\\
13.8949549437315	-6.82995782732826\\
16.76832936811	-4.82497719327481\\
20.2358964772516	-3.45717612668176\\
24.4205309454863	-2.61673181309461\\
29.4705170255181	-2.09861326713654\\
35.5648030622315	-1.78098384726291\\
42.9193426012876	-1.59630270031964\\
51.7947467923122	-1.50832629799496\\
62.5055192527392	-1.50043844317968\\
75.4312006335461	-1.5705940881745\\
91.0298177991526	-1.73036392616339\\
109.854114198755	-2.00736867857842\\
132.571136559011	-2.45148057914018\\
159.985871960607	-3.14459452059491\\
193.069772888325	-4.20049901678466\\
232.995181051538	-5.62533303075531\\
281.176869797421	-6.27708414731033\\
339.322177189533	-3.25727077063168\\
409.491506238045	1.78877776854203\\
494.171336132382	6.41281409355776\\
596.362331659466	10.0499540278513\\
719.685673001147	12.6000043641151\\
868.511373751352	14.2985527767189\\
1048.11313415469	15.557371810582\\
1264.85521685529	16.6896190738206\\
1526.41796717524	17.8497306048207\\
1842.06996932673	19.0916604656469\\
2222.99648252619	20.4224011117026\\
2682.69579527974	21.8308780257527\\
3237.45754281763	23.3012411583438\\
3906.93993705462	24.8183629778324\\
4714.86636345743	26.3697003725689\\
6866.48845004303	29.5387623014416\\
10000	32.7579056347102\\
};
\addplot [color=red, line width=1.0pt, only marks, mark size=1pt, mark=*, mark options={solid, red}, forget plot]
  table[row sep=crcr]{%
5	8.27595241816946\\
6	6.63657152394038\\
7	4.3909235320237\\
8	1.69813477296249\\
10	-4.31607022912949\\
11	-6.77094181804771\\
13	-7.89193266944766\\
15	-6.40675194818817\\
16.9999999999999	-5.10347667203184\\
20	-3.92882942683289\\
23	-3.18021515894364\\
27	-2.6326397484449\\
30.9999999999999	-2.3209044476938\\
35	-2.17896567246534\\
41.0000000000002	-1.93259447656063\\
47.0000000000001	-1.91717537052589\\
54.0000000000001	-1.85150038158895\\
63.0000000000001	-1.85038836700305\\
72.0000000000001	-1.9118926334482\\
83.0000000000003	-1.97795371365492\\
96.0000000000001	-2.15900487407675\\
111	-2.4034105165093\\
127	-2.73320155672571\\
147	-3.13241484539622\\
169	-3.7240766473568\\
195	-4.3154271277853\\
224	-4.84222889838477\\
257.999999999999	-4.85863852642957\\
297.999999999999	-3.74925653862399\\
343	-1.37299863160291\\
395	1.46439405466029\\
455	4.22844448647234\\
524.000000000001	6.86715495397257\\
603.000000000001	8.94621641265428\\
693.999999999999	11.1333323372644\\
798.999999999998	12.5394449664914\\
921.000000000001	13.7566456012381\\
1060	14.6904171336279\\
1221	15.3368345024415\\
1405.00000000001	15.9773256622361\\
1618	16.7682197281278\\
1863	18.2652252455245\\
2145	19.3453749080353\\
2470	19.4853114301362\\
2844.00000000001	22.8211178099503\\
};
\end{axis}

\begin{axis}[%
width=2.5in,
height=1in,
at={(0.758in,0.7in)},
scale only axis,
xmode=log,
xmin=1,
xmax=5000,
xminorticks=true,
xlabel style={font=\large\color{white!15!black}},
xlabel={Frequency [Hz]},
ymin=-200,
ymax=200,
ylabel style={font=\large\color{white!15!black}},
ylabel={Phase [deg]},
axis background/.style={fill=white},
xmajorgrids,
xminorgrids,
ymajorgrids,
]
\addplot [color=mycolor1, line width=2.0pt, forget plot]
  table[row sep=crcr]{%
1	-179.069769626083\\
1.20679264063936	-178.876543316259\\
1.45634847750121	-178.648187374282\\
1.7575106248548	-178.388559673316\\
2.12095088792024	-178.1198702561\\
2.55954792269949	-177.908633802665\\
3.0888435964775	-177.917214941812\\
3.72759372031504	-178.488569357009\\
4.49843266896938	179.778458634754\\
5.4286754393239	176.234647458614\\
6.55128556859571	171.450300743245\\
7.9060432109076	169.757113311535\\
9.54095476350004	-179.268571472044\\
11.5139539932641	-140.747926282343\\
13.8949549437312	-94.4302658473297\\
16.7683293681103	-69.9948810186976\\
20.235896477251	-55.7363582769343\\
24.4205309454863	-45.415556041197\\
29.4705170255186	-37.1474489318691\\
35.5648030622303	-30.2426143227384\\
42.9193426012876	-24.3612670912784\\
51.7947467923131	-19.277339627357\\
62.5055192527382	-14.8034933914597\\
75.4312006335461	-10.7597805122443\\
109.854114198753	-3.11478184714625\\
132.571136559011	1.1382293345562\\
159.98587196061	6.55916558960288\\
193.069772888321	14.8322208679478\\
232.995181051538	30.198976266596\\
281.176869797431	60.2947887069309\\
339.322177189527	94.9228335106783\\
409.491506238045	111.011707725027\\
494.171336132398	112.594357976898\\
596.362331659456	106.508728188586\\
719.685673001159	97.6760647249241\\
868.51137375138	89.9395152331605\\
1048.11313415467	84.8835799490463\\
1264.85521685531	82.2891816538477\\
1526.41796717519	81.3884649941168\\
1842.0699693267	81.4927840367897\\
2222.99648252623	82.1235038193179\\
2682.69579527965	82.9824029127059\\
3237.45754281763	83.8970982198027\\
3906.93993705468	84.7755495726612\\
4714.86636345727	85.5741365528265\\
5689.86602901828	86.2769547406356\\
6866.48845004314	86.8830633781768\\
8286.42772854666	87.3989507353884\\
10000	87.8342542958019\\
};
\addplot [color=red, line width=1.0pt, only marks, mark size=1pt, mark=*, mark options={solid, red}, forget plot]
  table[row sep=crcr]{%
5	173.816613616268\\
6	170.677199955041\\
6.99999999999989	166.829660853828\\
8	167.567134673197\\
10	-173.743109844974\\
11.0000000000001	-153.792763013203\\
13.0000000000003	-107.827762084256\\
15	-81.4436092362275\\
16.9999999999998	-67.7206002795107\\
20	-55.1116505980781\\
23	-47.2460711105519\\
27	-39.6302026464507\\
31.0000000000004	-34.0668815137589\\
34.9999999999994	-29.6181459574523\\
41.0000000000008	-25.0849844573601\\
46.9999999999993	-20.9300060366732\\
54.0000000000001	-17.0745325487018\\
62.999999999999	-13.7081377110408\\
72.0000000000001	-10.7981455400615\\
83.0000000000017	-7.71711298674882\\
96.0000000000001	-4.81721092723848\\
111	-1.47745078922782\\
127.000000000003	2.61554001942196\\
146.999999999995	7.0778842956224\\
168.999999999997	12.8803757440978\\
195.000000000005	21.4600604948165\\
223.999999999997	33.5137298058444\\
258.000000000002	51.3280503329772\\
298.000000000004	68.780559695392\\
343.000000000006	86.8145925594086\\
395.000000000004	95.7308916197008\\
455.000000000004	101.278708300337\\
524.000000000006	101.538083576304\\
602.999999999986	99.7483416047578\\
693.999999999988	95.9794665345684\\
798.999999999978	90.3465372213549\\
920.999999999971	85.2081262842174\\
1060.00000000003	81.3447179267723\\
1221.00000000001	80.1533877987321\\
1405.00000000002	80.7325669419232\\
1617.99999999999	84.2180096043072\\
1863	86.0061884607203\\
2144.99999999993	85.2117406768238\\
2470.00000000004	85.2027828048911\\
2844.00000000003	83.9722299563733\\
};
\end{axis}
\end{tikzpicture}%

%% file: updated_images/test1_IL1dq.tex
%
%
\definecolor{mycolor1}{rgb}{0.85098,0.32549,0.09804}%
\begin{tikzpicture}

\begin{axis}[%
width=6in,
height=2in,
at={(0.925in,3.1in)},
scale only axis,
grid=both,
unbounded coords=jump,
xmin=0,
xmax=1.5,
xlabel style={font=\Large\color{white!15!black}},
xlabel={Time [s]},
ymin=-5,
ymax=15,
ylabel style={font=\Large\color{white!15!black}},
ylabel={$I_{dq}\text{ [A]}$},
yticklabel style={font=\Large},
axis background/.style={fill=white},
xmajorgrids,
ymajorgrids,
legend style={at={(0.99,0.6)}, anchor=east,legend cell align=left, align=left,font=\normalsize, draw=white!15!black}
]
\draw[line width=0.5mm,draw=black!80] (axis cs:0.975,0)--(axis cs:0.975,5);
\draw[line width=0.5mm,draw=black!80] (axis cs:1.065,0)--(axis cs:1.065,5);
\draw[line width=0.5mm,draw=black!80,->] (axis cs:0.85,4)--(axis cs:0.968,4);
\draw[line width=0.5mm,draw=black!80,->] (axis cs:1.15,4)--(axis cs:1.055,4);
\node[color=black, fill=white,font=\large] at (axis cs: 1.02,6) {$\sim$ \SI{0.09}{\second} $~\approx$ \SI{11.11}{\hertz}};
\addplot [color=red, line width=2.0pt]
  table[row sep=crcr]{%
0	5.09116882454314\\
0.00250000000005457	4.58205194208883\\
0.00499999999999545	4.83661038331599\\
0.00750000000005002	4.83661038331599\\
0.00999999999999091	5.09116882454314\\
0.0125000000000455	5.09116882454314\\
0.0149999999999864	4.83661038331599\\
0.0175000000000409	5.09116882454314\\
0.0199999999999818	5.09116882454314\\
0.0225000000000364	4.83661038331599\\
0.0275000000000318	4.83661038331599\\
0.0300000000000864	5.09116882454314\\
0.0350000000000819	5.09116882454314\\
0.0375000000000227	5.3457272657703\\
0.0425000000000182	4.83661038331599\\
0.0475000000000136	4.83661038331599\\
0.0500000000000682	5.09116882454314\\
0.0600000000000591	5.09116882454314\\
0.0625	4.83661038331599\\
0.0674999999999955	4.83661038331599\\
0.07000000000005	5.09116882454314\\
0.0774999999999864	5.09116882454314\\
0.0800000000000409	5.3457272657703\\
0.0824999999999818	4.83661038331599\\
0.0875000000000909	4.83661038331599\\
0.0900000000000318	5.09116882454314\\
0.0925000000000864	5.09116882454314\\
0.0950000000000273	4.83661038331599\\
0.0975000000000819	5.09116882454314\\
0.100000000000023	5.09116882454314\\
0.102500000000077	4.83661038331599\\
0.110000000000014	4.83661038331599\\
0.112500000000068	5.09116882454314\\
0.115000000000009	4.83661038331599\\
0.117500000000064	5.09116882454314\\
0.120000000000005	4.83661038331599\\
0.127500000000055	4.83661038331599\\
0.129999999999995	5.09116882454314\\
0.134999999999991	5.09116882454314\\
0.137500000000045	5.3457272657703\\
0.142500000000041	4.83661038331599\\
0.144999999999982	5.09116882454314\\
0.147500000000036	4.83661038331599\\
0.150000000000091	5.09116882454314\\
0.152500000000032	5.09116882454314\\
0.155000000000086	4.83661038331599\\
0.157500000000027	5.09116882454314\\
0.160000000000082	5.09116882454314\\
0.162500000000023	4.83661038331599\\
0.167500000000018	4.83661038331599\\
0.170000000000073	5.09116882454314\\
0.172500000000014	5.09116882454314\\
0.175000000000068	4.83661038331599\\
0.177500000000009	5.3457272657703\\
0.182500000000005	4.83661038331599\\
0.1875	4.83661038331599\\
0.190000000000055	5.09116882454314\\
0.200000000000045	5.09116882454314\\
0.202499999999986	4.83661038331599\\
0.207499999999982	4.83661038331599\\
0.210000000000036	5.09116882454314\\
0.212500000000091	5.09116882454314\\
0.215000000000032	4.83661038331599\\
0.217500000000086	5.09116882454314\\
0.220000000000027	5.09116882454314\\
0.222500000000082	4.58205194208883\\
0.225000000000023	5.09116882454314\\
0.227500000000077	4.83661038331599\\
0.230000000000018	5.09116882454314\\
0.235000000000014	5.09116882454314\\
0.237500000000068	5.3457272657703\\
0.240000000000009	5.09116882454314\\
0.245000000000005	5.09116882454314\\
0.247500000000059	4.83661038331599\\
0.25	5.09116882454314\\
0.252500000000055	5.09116882454314\\
0.254999999999995	4.83661038331599\\
0.25750000000005	5.09116882454314\\
0.259999999999991	5.09116882454314\\
0.262500000000045	4.83661038331599\\
0.267500000000041	4.83661038331599\\
0.269999999999982	5.09116882454314\\
0.272500000000036	5.09116882454314\\
0.275000000000091	4.83661038331599\\
0.277500000000032	5.09116882454314\\
0.280000000000086	5.09116882454314\\
0.282500000000027	4.83661038331599\\
0.285000000000082	5.09116882454314\\
0.287500000000023	4.83661038331599\\
0.290000000000077	5.09116882454314\\
0.300000000000068	5.09116882454314\\
0.302500000000009	4.83661038331599\\
0.307500000000005	4.83661038331599\\
0.310000000000059	5.09116882454314\\
0.3125	5.09116882454314\\
0.315000000000055	4.83661038331599\\
0.317499999999995	5.09116882454314\\
0.32000000000005	5.09116882454314\\
0.322499999999991	4.83661038331599\\
0.327499999999986	4.83661038331599\\
0.332499999999982	5.3457272657703\\
0.335000000000036	5.09116882454314\\
0.340000000000032	5.09116882454314\\
0.342500000000086	4.83661038331599\\
0.345000000000027	4.83661038331599\\
0.347500000000082	5.09116882454314\\
0.350000000000023	5.09116882454314\\
0.352500000000077	5.3457272657703\\
0.355000000000018	4.83661038331599\\
0.357500000000073	5.09116882454314\\
0.360000000000014	5.09116882454314\\
0.362500000000068	4.83661038331599\\
0.367500000000064	4.83661038331599\\
0.370000000000005	5.09116882454314\\
0.372500000000059	5.09116882454314\\
0.375	4.83661038331599\\
0.377500000000055	5.09116882454314\\
0.379999999999995	4.83661038331599\\
0.384999999999991	4.83661038331599\\
0.387500000000045	5.09116882454314\\
0.400000000000091	5.09116882454314\\
0.402500000000032	4.83661038331599\\
0.407500000000027	4.83661038331599\\
0.410000000000082	5.09116882454314\\
0.420000000000073	5.09116882454314\\
0.422500000000014	4.83661038331599\\
0.427500000000009	4.83661038331599\\
0.430000000000064	5.09116882454314\\
0.440000000000055	5.09116882454314\\
0.442499999999995	4.83661038331599\\
0.44500000000005	4.83661038331599\\
0.450000000000045	5.3457272657703\\
0.452499999999986	5.09116882454314\\
0.460000000000036	5.09116882454314\\
0.462500000000091	4.83661038331599\\
0.470000000000027	4.83661038331599\\
0.472500000000082	5.09116882454314\\
0.475000000000023	4.83661038331599\\
0.477500000000077	5.09116882454314\\
0.480000000000018	5.09116882454314\\
0.482500000000073	4.58205194208883\\
0.485000000000014	4.83661038331599\\
0.487500000000068	4.83661038331599\\
0.490000000000009	5.09116882454314\\
0.495000000000005	5.09116882454314\\
0.497500000000059	5.3457272657703\\
0.5	5.09116882454314\\
0.502500000000055	9.16410388417766\\
0.504999999999995	10.6914545315406\\
0.509999999999991	11.7096882964492\\
0.512500000000045	11.7096882964492\\
0.514999999999986	12.09152595829\\
0.517500000000041	12.6006428407443\\
0.519999999999982	12.6006428407443\\
0.522500000000036	12.09152595829\\
0.525000000000091	12.6006428407443\\
0.530000000000086	13.1097597231986\\
0.532500000000027	12.8552012819714\\
0.537500000000023	13.3643181644257\\
0.540000000000077	12.8552012819714\\
0.542500000000018	12.6006428407443\\
0.547500000000014	12.6006428407443\\
0.552500000000009	12.09152595829\\
0.555000000000064	11.9642467376764\\
0.557500000000005	11.9642467376764\\
0.5625	10.9460129727678\\
0.575000000000045	10.9460129727678\\
0.577499999999986	11.2005714139949\\
0.580000000000041	11.2005714139949\\
0.582499999999982	10.9460129727678\\
0.590000000000032	11.7096882964492\\
0.592500000000086	11.7096882964492\\
0.595000000000027	12.09152595829\\
0.597500000000082	12.6006428407443\\
0.600000000000023	12.6006428407443\\
0.602500000000077	12.3460843995171\\
0.605000000000018	12.6006428407443\\
0.607500000000073	13.1097597231986\\
0.610000000000014	13.1097597231986\\
0.612500000000068	12.8552012819714\\
0.617500000000064	13.3643181644257\\
0.625	12.6006428407443\\
0.627500000000055	12.8552012819714\\
0.63250000000005	12.3460843995171\\
0.637500000000045	12.3460843995171\\
0.639999999999986	11.9642467376764\\
0.644999999999982	11.4551298552221\\
0.650000000000091	11.4551298552221\\
0.652500000000032	10.9460129727678\\
0.655000000000086	10.9460129727678\\
0.657500000000027	11.2005714139949\\
0.660000000000082	11.2005714139949\\
0.662500000000023	10.6914545315406\\
0.670000000000073	11.4551298552221\\
0.672500000000014	11.2005714139949\\
0.675000000000068	11.4551298552221\\
0.677500000000009	11.9642467376764\\
0.680000000000064	11.9642467376764\\
0.682500000000005	11.4551298552221\\
0.685000000000059	11.9642467376764\\
0.6875	12.3460843995171\\
0.690000000000055	12.6006428407443\\
0.692499999999995	12.3460843995171\\
0.69500000000005	12.6006428407443\\
0.697499999999991	13.1097597231986\\
0.700000000000045	13.1097597231986\\
0.702499999999986	12.8552012819714\\
0.705000000000041	13.1097597231986\\
0.710000000000036	13.1097597231986\\
0.712500000000091	12.8552012819714\\
0.717500000000086	12.8552012819714\\
0.720000000000027	12.6006428407443\\
0.722500000000082	11.9642467376764\\
0.725000000000023	12.09152595829\\
0.727500000000077	11.9642467376764\\
0.730000000000018	11.9642467376764\\
0.735000000000014	11.4551298552221\\
0.737500000000068	11.4551298552221\\
0.740000000000009	11.2005714139949\\
0.742500000000064	10.6914545315406\\
0.747500000000059	11.2005714139949\\
0.754999999999995	11.2005714139949\\
0.75750000000005	11.4551298552221\\
0.759999999999991	11.4551298552221\\
0.762500000000045	11.2005714139949\\
0.764999999999986	11.7096882964492\\
0.767500000000041	11.9642467376764\\
0.769999999999982	12.09152595829\\
0.775000000000091	12.09152595829\\
0.777500000000032	12.6006428407443\\
0.780000000000086	12.6006428407443\\
0.782500000000027	12.09152595829\\
0.785000000000082	12.6006428407443\\
0.790000000000077	13.1097597231986\\
0.792500000000018	12.8552012819714\\
0.800000000000068	12.8552012819714\\
0.805000000000064	12.3460843995171\\
0.807500000000005	12.6006428407443\\
0.810000000000059	12.6006428407443\\
0.8125	12.09152595829\\
0.815000000000055	12.09152595829\\
0.817499999999995	11.9642467376764\\
0.82000000000005	11.7096882964492\\
0.822499999999991	11.2005714139949\\
0.825000000000045	11.2005714139949\\
0.827499999999986	11.4551298552221\\
0.830000000000041	11.4551298552221\\
0.832499999999982	10.9460129727678\\
0.835000000000036	11.2005714139949\\
0.840000000000032	11.2005714139949\\
0.842500000000086	10.6914545315406\\
0.845000000000027	11.2005714139949\\
0.850000000000023	11.7096882964492\\
0.855000000000018	11.7096882964492\\
0.857500000000073	12.09152595829\\
0.860000000000014	12.09152595829\\
0.862500000000068	11.9642467376764\\
0.865000000000009	12.09152595829\\
0.867500000000064	12.6006428407443\\
0.870000000000005	12.8552012819714\\
0.872500000000059	12.6006428407443\\
0.875	12.8552012819714\\
0.879999999999995	12.8552012819714\\
0.88250000000005	12.6006428407443\\
0.887500000000045	12.6006428407443\\
0.889999999999986	12.8552012819714\\
0.894999999999982	12.3460843995171\\
0.897500000000036	12.3460843995171\\
0.900000000000091	12.09152595829\\
0.902500000000032	11.7096882964492\\
0.907500000000027	11.7096882964492\\
0.910000000000082	11.9642467376764\\
0.912500000000023	11.4551298552221\\
0.917500000000018	11.4551298552221\\
0.922500000000014	10.9460129727678\\
0.925000000000068	10.9460129727678\\
0.927500000000009	11.4551298552221\\
0.935000000000059	11.4551298552221\\
0.9375	11.7096882964492\\
0.940000000000055	11.7096882964492\\
0.942499999999995	11.2005714139949\\
0.94500000000005	11.7096882964492\\
0.947499999999991	12.09152595829\\
0.950000000000045	12.3460843995171\\
0.955000000000041	12.3460843995171\\
0.957499999999982	12.8552012819714\\
0.962500000000091	12.3460843995171\\
0.970000000000027	13.1097597231986\\
0.972500000000082	12.8552012819714\\
0.977500000000077	12.8552012819714\\
0.980000000000018	12.3460843995171\\
0.982500000000073	12.09152595829\\
0.985000000000014	12.3460843995171\\
0.987500000000068	12.09152595829\\
0.990000000000009	12.09152595829\\
0.992500000000064	11.9642467376764\\
0.995000000000005	11.7096882964492\\
0.997500000000059	11.9642467376764\\
1.00250000000005	10.9460129727678\\
1.005	10.9460129727678\\
1.00750000000005	11.4551298552221\\
1.00999999999999	11.4551298552221\\
1.01250000000005	11.2005714139949\\
1.01750000000004	11.7096882964492\\
1.01999999999998	11.7096882964492\\
1.02250000000004	11.2005714139949\\
1.02500000000009	11.2005714139949\\
1.02750000000003	11.7096882964492\\
1.03000000000009	11.9642467376764\\
1.03500000000008	11.9642467376764\\
1.03750000000002	12.3460843995171\\
1.04000000000008	12.3460843995171\\
1.04250000000002	11.7096882964492\\
1.04500000000007	12.09152595829\\
1.04750000000001	12.3460843995171\\
1.05000000000007	12.8552012819714\\
1.05250000000001	12.6006428407443\\
1.0575	13.1097597231986\\
1.06000000000006	12.8552012819714\\
1.0625	12.3460843995171\\
1.06500000000005	12.6006428407443\\
1.07000000000005	12.6006428407443\\
1.07249999999999	12.3460843995171\\
1.07749999999999	12.3460843995171\\
1.08000000000004	11.9642467376764\\
1.08249999999998	11.7096882964492\\
1.09000000000003	11.7096882964492\\
1.09250000000009	11.2005714139949\\
1.09500000000003	11.2005714139949\\
1.09750000000008	11.4551298552221\\
1.10000000000002	11.4551298552221\\
1.10250000000008	10.9460129727678\\
1.10750000000007	11.4551298552221\\
1.11250000000007	11.4551298552221\\
1.11500000000001	11.7096882964492\\
1.11750000000006	12.09152595829\\
1.12	12.09152595829\\
1.12250000000006	11.7096882964492\\
1.125	11.9642467376764\\
1.12750000000005	12.09152595829\\
1.13	12.3460843995171\\
1.13250000000005	12.3460843995171\\
1.13750000000005	12.8552012819714\\
1.13999999999999	12.6006428407443\\
1.14250000000004	12.6006428407443\\
1.14499999999998	12.3460843995171\\
1.14750000000004	12.6006428407443\\
1.15000000000009	12.6006428407443\\
1.15250000000003	12.3460843995171\\
1.15500000000009	12.6006428407443\\
1.15750000000003	12.6006428407443\\
1.16250000000002	12.09152595829\\
1.16500000000008	11.9642467376764\\
1.16750000000002	12.09152595829\\
1.17000000000007	12.09152595829\\
1.17250000000001	11.7096882964492\\
1.17500000000007	11.7096882964492\\
1.17750000000001	11.9642467376764\\
1.18000000000006	11.4551298552221\\
1.1825	11.2005714139949\\
1.18500000000006	11.2005714139949\\
1.1875	11.4551298552221\\
1.19000000000005	11.4551298552221\\
1.1925	11.2005714139949\\
1.19500000000005	11.2005714139949\\
1.19749999999999	11.7096882964492\\
1.20000000000005	11.7096882964492\\
1.20249999999999	11.2005714139949\\
1.20500000000004	11.4551298552221\\
1.20749999999998	11.9642467376764\\
1.21250000000009	11.9642467376764\\
1.21500000000003	12.09152595829\\
1.22000000000003	12.6006428407443\\
1.22250000000008	12.09152595829\\
1.22500000000002	12.6006428407443\\
1.22750000000008	12.8552012819714\\
1.23000000000002	12.8552012819714\\
1.23250000000007	12.6006428407443\\
1.23500000000001	12.6006428407443\\
1.23750000000007	12.8552012819714\\
1.24250000000006	12.3460843995171\\
1.245	12.3460843995171\\
1.24750000000006	12.09152595829\\
1.25	12.3460843995171\\
1.25250000000005	12.09152595829\\
1.255	11.9642467376764\\
1.25750000000005	12.09152595829\\
1.25999999999999	11.9642467376764\\
1.26250000000005	11.4551298552221\\
1.26499999999999	11.4551298552221\\
1.26750000000004	11.7096882964492\\
1.26999999999998	11.7096882964492\\
1.27250000000004	11.2005714139949\\
1.27500000000009	11.2005714139949\\
1.27750000000003	11.7096882964492\\
1.28000000000009	11.7096882964492\\
1.28250000000003	11.2005714139949\\
1.28750000000002	11.7096882964492\\
1.29250000000002	11.7096882964492\\
1.29500000000007	11.4551298552221\\
1.29750000000001	11.9642467376764\\
1.30000000000007	12.09152595829\\
1.30250000000001	11.7096882964492\\
1.30500000000006	11.9642467376764\\
1.3075	12.3460843995171\\
1.31500000000005	12.3460843995171\\
1.3175	12.8552012819714\\
1.32249999999999	12.3460843995171\\
1.32500000000005	12.3460843995171\\
1.32749999999999	12.8552012819714\\
1.33249999999998	12.3460843995171\\
1.33750000000009	12.3460843995171\\
1.34000000000003	12.09152595829\\
1.34250000000009	11.7096882964492\\
1.34500000000003	11.9642467376764\\
1.35000000000002	11.9642467376764\\
1.35250000000008	11.4551298552221\\
1.35500000000002	11.7096882964492\\
1.35750000000007	11.7096882964492\\
1.36250000000007	11.2005714139949\\
1.36500000000001	11.2005714139949\\
1.37	11.7096882964492\\
1.37250000000006	11.4551298552221\\
1.375	11.4551298552221\\
1.38	11.9642467376764\\
1.38250000000005	11.4551298552221\\
1.38499999999999	11.7096882964492\\
1.38750000000005	12.09152595829\\
1.38999999999999	12.3460843995171\\
1.39250000000004	12.09152595829\\
1.39499999999998	12.09152595829\\
1.39750000000004	12.6006428407443\\
1.40250000000003	12.09152595829\\
1.41000000000008	12.8552012819714\\
1.41250000000002	12.6006428407443\\
1.41750000000002	12.6006428407443\\
1.42250000000001	12.09152595829\\
1.42500000000007	12.09152595829\\
1.42750000000001	12.3460843995171\\
1.43000000000006	12.3460843995171\\
1.4325	12.09152595829\\
1.43500000000006	11.9642467376764\\
1.4375	12.09152595829\\
1.44000000000005	11.7096882964492\\
1.44500000000005	11.2005714139949\\
1.44749999999999	11.7096882964492\\
1.45000000000005	11.7096882964492\\
1.45249999999999	11.2005714139949\\
1.45500000000004	11.4551298552221\\
1.46000000000004	11.4551298552221\\
1.46250000000009	10.9460129727678\\
1.46500000000003	11.4551298552221\\
1.47000000000003	11.9642467376764\\
1.47500000000002	11.9642467376764\\
1.47750000000008	12.09152595829\\
1.48000000000002	12.3460843995171\\
1.48250000000007	11.9642467376764\\
1.48500000000001	12.09152595829\\
1.48750000000007	12.6006428407443\\
1.49000000000001	12.8552012819714\\
1.49250000000006	12.6006428407443\\
1.495	12.6006428407443\\
1.49750000000006	12.8552012819714\\
1.50250000000005	12.3460843995171\\
};
\addlegendentry{$I_{L1,d}$}
\addplot [color=mycolor1, line width=2.0pt]
  table[row sep=crcr]{%
0	-0.127279220613579\\
0.00499999999999545	0.127279220613579\\
0.00750000000005002	-0.127279220613579\\
0.00999999999999091	-0\\
0.0175000000000409	-0\\
0.0199999999999818	-0.127279220613579\\
0.0225000000000364	-0\\
0.0250000000000909	-0\\
0.0275000000000318	-0.127279220613579\\
0.0300000000000864	-0\\
0.0375000000000227	-0\\
0.0400000000000773	-0.127279220613579\\
0.0425000000000182	-0\\
0.0450000000000728	-0\\
0.0475000000000136	-0.127279220613579\\
0.0500000000000682	-0\\
0.0575000000000045	-0\\
0.0600000000000591	-0.127279220613579\\
0.0650000000000546	0.127279220613579\\
0.0674999999999955	-0.127279220613579\\
0.0724999999999909	0.127279220613579\\
0.0750000000000455	-0.127279220613579\\
0.0774999999999864	-0\\
0.0824999999999818	-0\\
0.0850000000000364	0.127279220613579\\
0.0875000000000909	-0.127279220613579\\
0.0925000000000864	0.127279220613579\\
0.0950000000000273	-0\\
0.0975000000000819	-0\\
0.100000000000023	-0.127279220613579\\
0.105000000000018	0.127279220613579\\
0.107500000000073	-0.127279220613579\\
0.110000000000014	-0\\
0.117500000000064	-0\\
0.120000000000005	-0.127279220613579\\
0.122500000000059	-0\\
0.125	-0\\
0.127500000000055	-0.127279220613579\\
0.129999999999995	-0\\
0.137500000000045	-0\\
0.139999999999986	-0.127279220613579\\
0.142500000000041	-0\\
0.144999999999982	-0\\
0.147500000000036	-0.127279220613579\\
0.150000000000091	0.127279220613579\\
0.152500000000032	-0\\
0.157500000000027	-0\\
0.160000000000082	-0.127279220613579\\
0.165000000000077	0.127279220613579\\
0.167500000000018	-0.127279220613579\\
0.172500000000014	0.127279220613579\\
0.175000000000068	-0\\
0.182500000000005	-0\\
0.185000000000059	0.127279220613579\\
0.1875	-0.127279220613579\\
0.192499999999995	0.127279220613579\\
0.19500000000005	-0.127279220613579\\
0.197499999999991	-0\\
0.200000000000045	-0.127279220613579\\
0.205000000000041	0.127279220613579\\
0.207499999999982	-0.127279220613579\\
0.210000000000036	-0\\
0.217500000000086	-0\\
0.220000000000027	-0.127279220613579\\
0.222500000000082	-0\\
0.225000000000023	-0\\
0.227500000000077	-0.127279220613579\\
0.230000000000018	-0\\
0.237500000000068	-0\\
0.240000000000009	-0.127279220613579\\
0.242500000000064	-0\\
0.245000000000005	-0\\
0.247500000000059	-0.127279220613579\\
0.25	0.127279220613579\\
0.252500000000055	-0\\
0.254999999999995	-0\\
0.25750000000005	0.127279220613579\\
0.259999999999991	-0.127279220613579\\
0.264999999999986	0.127279220613579\\
0.267500000000041	-0\\
0.269999999999982	-0\\
0.272500000000036	0.127279220613579\\
0.275000000000091	-0\\
0.277500000000032	-0\\
0.280000000000086	-0.127279220613579\\
0.285000000000082	0.127279220613579\\
0.287500000000023	-0.127279220613579\\
0.290000000000077	-0\\
0.292500000000018	-0\\
0.295000000000073	-0.127279220613579\\
0.297500000000014	-0\\
0.300000000000068	-0.127279220613579\\
0.305000000000064	0.127279220613579\\
0.307500000000005	-0.127279220613579\\
0.310000000000059	-0\\
0.317499999999995	-0\\
0.32000000000005	-0.127279220613579\\
0.325000000000045	0.127279220613579\\
0.327499999999986	-0.127279220613579\\
0.330000000000041	-0\\
0.337500000000091	-0\\
0.340000000000032	-0.127279220613579\\
0.342500000000086	0.127279220613579\\
0.345000000000027	0.127279220613579\\
0.347500000000082	-0.127279220613579\\
0.350000000000023	0.127279220613579\\
0.352500000000077	-0\\
0.357500000000073	-0\\
0.360000000000014	-0.127279220613579\\
0.362500000000068	0.127279220613579\\
0.365000000000009	0.127279220613579\\
0.367500000000064	-0.127279220613579\\
0.370000000000005	-0\\
0.377500000000055	-0\\
0.379999999999995	-0.127279220613579\\
0.384999999999991	0.127279220613579\\
0.387500000000045	-0.127279220613579\\
0.389999999999986	-0\\
0.392500000000041	-0\\
0.394999999999982	-0.127279220613579\\
0.397500000000036	-0\\
0.400000000000091	-0.127279220613579\\
0.405000000000086	0.127279220613579\\
0.407500000000027	-0.127279220613579\\
0.410000000000082	-0\\
0.412500000000023	-0\\
0.415000000000077	-0.127279220613579\\
0.417500000000018	-0\\
0.420000000000073	-0.127279220613579\\
0.425000000000068	0.127279220613579\\
0.427500000000009	-0.127279220613579\\
0.430000000000064	0.127279220613579\\
0.432500000000005	-0\\
0.4375	-0\\
0.440000000000055	-0.127279220613579\\
0.442499999999995	0.127279220613579\\
0.44500000000005	0.127279220613579\\
0.447499999999991	-0.127279220613579\\
0.450000000000045	0.127279220613579\\
0.452499999999986	-0\\
0.455000000000041	-0\\
0.457499999999982	0.127279220613579\\
0.460000000000036	-0.127279220613579\\
0.465000000000032	0.127279220613579\\
0.467500000000086	-0.127279220613579\\
0.470000000000027	-0\\
0.472500000000082	-0\\
0.475000000000023	-0.127279220613579\\
0.477500000000077	-0\\
0.480000000000018	-0.127279220613579\\
0.485000000000014	0.127279220613579\\
0.487500000000068	-0.127279220613579\\
0.490000000000009	-0\\
0.492500000000064	-0\\
0.495000000000005	-0.127279220613579\\
0.497500000000059	-0\\
0.5	-0.127279220613579\\
0.502500000000055	-1.40007142674936\\
0.504999999999995	-1.90918830920368\\
0.50750000000005	-1.90918830920368\\
0.509999999999991	-1.65462986797652\\
0.512500000000045	-1.14551298552221\\
0.517500000000041	-0.636396103067893\\
0.519999999999982	-0.127279220613579\\
0.522500000000036	0.127279220613579\\
0.525000000000091	0.89095454429505\\
0.527500000000032	0.636396103067893\\
0.530000000000086	0.89095454429505\\
0.532500000000027	1.40007142674936\\
0.535000000000082	1.65462986797652\\
0.540000000000077	1.14551298552221\\
0.542500000000018	1.65462986797652\\
0.545000000000073	1.90918830920368\\
0.547500000000014	1.40007142674936\\
0.552500000000009	0.89095454429505\\
0.555000000000064	0.89095454429505\\
0.557500000000005	0.127279220613579\\
0.560000000000059	-0.127279220613579\\
0.5625	-0.127279220613579\\
0.565000000000055	-0.381837661840736\\
0.567499999999995	-1.14551298552221\\
0.57000000000005	-1.14551298552221\\
0.572499999999991	-1.40007142674936\\
0.575000000000045	-1.90918830920368\\
0.577499999999986	-1.90918830920368\\
0.580000000000041	-2.16374675043084\\
0.582499999999982	-2.16374675043084\\
0.585000000000036	-1.90918830920368\\
0.587500000000091	-1.90918830920368\\
0.592500000000086	-1.40007142674936\\
0.595000000000027	-1.40007142674936\\
0.597500000000082	-1.14551298552221\\
0.600000000000023	-0.636396103067893\\
0.602500000000077	-0.381837661840736\\
0.605000000000018	0.127279220613579\\
0.607500000000073	0.127279220613579\\
0.610000000000014	0.381837661840736\\
0.612500000000068	1.14551298552221\\
0.615000000000009	1.40007142674936\\
0.620000000000005	1.40007142674936\\
0.625	2.41830519165799\\
0.627500000000055	1.65462986797652\\
0.629999999999995	1.65462986797652\\
0.63250000000005	1.90918830920368\\
0.634999999999991	1.90918830920368\\
0.637500000000045	1.14551298552221\\
0.639999999999986	0.636396103067893\\
0.642500000000041	0.89095454429505\\
0.644999999999982	0.89095454429505\\
0.647500000000036	0.127279220613579\\
0.650000000000091	-0\\
0.652500000000032	-0.381837661840736\\
0.655000000000086	-0.636396103067893\\
0.657500000000027	-1.14551298552221\\
0.660000000000082	-1.40007142674936\\
0.662500000000023	-1.40007142674936\\
0.665000000000077	-1.65462986797652\\
0.677500000000009	-1.65462986797652\\
0.690000000000055	-0.381837661840736\\
0.692499999999995	-0\\
0.69500000000005	0.127279220613579\\
0.697499999999991	0.381837661840736\\
0.700000000000045	0.381837661840736\\
0.705000000000041	1.90918830920368\\
0.707499999999982	1.40007142674936\\
0.710000000000036	1.40007142674936\\
0.712500000000091	1.90918830920368\\
0.715000000000032	1.90918830920368\\
0.717500000000086	1.65462986797652\\
0.720000000000027	1.14551298552221\\
0.725000000000023	1.65462986797652\\
0.727500000000077	0.89095454429505\\
0.730000000000018	0.636396103067893\\
0.732500000000073	0.636396103067893\\
0.735000000000014	0.381837661840736\\
0.740000000000009	-0.381837661840736\\
0.742500000000064	-0.381837661840736\\
0.745000000000005	-0.636396103067893\\
0.747500000000059	-1.40007142674936\\
0.75	-1.14551298552221\\
0.754999999999995	-1.65462986797652\\
0.759999999999991	-1.65462986797652\\
0.762500000000045	-1.40007142674936\\
0.767500000000041	-1.40007142674936\\
0.769999999999982	-0.89095454429505\\
0.772500000000036	-0.89095454429505\\
0.775000000000091	-0.636396103067893\\
0.777500000000032	-0.636396103067893\\
0.780000000000086	-0.381837661840736\\
0.785000000000082	0.636396103067893\\
0.787500000000023	0.636396103067893\\
0.792500000000018	1.14551298552221\\
0.795000000000073	1.65462986797652\\
0.797500000000014	1.14551298552221\\
0.800000000000068	1.14551298552221\\
0.805000000000064	2.16374675043084\\
0.807500000000005	1.14551298552221\\
0.810000000000059	1.14551298552221\\
0.8125	1.40007142674936\\
0.815000000000055	1.14551298552221\\
0.82000000000005	0.127279220613579\\
0.822499999999991	0.381837661840736\\
0.825000000000045	0.381837661840736\\
0.827499999999986	-0.381837661840736\\
0.832499999999982	-0.381837661840736\\
0.835000000000036	-1.14551298552221\\
0.837500000000091	-1.14551298552221\\
0.840000000000032	-1.40007142674936\\
0.845000000000027	-1.40007142674936\\
0.847500000000082	-1.65462986797652\\
0.850000000000023	-1.40007142674936\\
0.852500000000077	-1.40007142674936\\
0.855000000000018	-1.14551298552221\\
0.857500000000073	-1.14551298552221\\
0.860000000000014	-0.89095454429505\\
0.862500000000068	-0.381837661840736\\
0.865000000000009	-0.127279220613579\\
0.867500000000064	-0.127279220613579\\
0.872500000000059	0.127279220613579\\
0.875	0.636396103067893\\
0.879999999999995	0.636396103067893\\
0.88250000000005	1.40007142674936\\
0.884999999999991	1.65462986797652\\
0.889999999999986	1.14551298552221\\
0.892500000000041	1.65462986797652\\
0.894999999999982	1.65462986797652\\
0.897500000000036	0.89095454429505\\
0.902500000000032	0.89095454429505\\
0.905000000000086	1.40007142674936\\
0.907500000000027	0.381837661840736\\
0.910000000000082	0.127279220613579\\
0.912500000000023	0.127279220613579\\
0.920000000000073	-0.636396103067893\\
0.925000000000068	-0.636396103067893\\
0.927500000000009	-1.40007142674936\\
0.930000000000064	-1.40007142674936\\
0.932500000000005	-1.14551298552221\\
0.935000000000059	-1.40007142674936\\
0.9375	-1.40007142674936\\
0.940000000000055	-1.14551298552221\\
0.942499999999995	-1.14551298552221\\
0.94500000000005	-0.636396103067893\\
0.947499999999991	-0.89095454429505\\
0.955000000000041	-0.127279220613579\\
0.957499999999982	-0.127279220613579\\
0.960000000000036	-0\\
0.962500000000091	0.636396103067893\\
0.965000000000032	1.14551298552221\\
0.967500000000086	0.89095454429505\\
0.970000000000027	0.89095454429505\\
0.975000000000023	1.40007142674936\\
0.977500000000077	0.89095454429505\\
0.980000000000018	0.89095454429505\\
0.982500000000073	1.40007142674936\\
0.985000000000014	1.65462986797652\\
0.987500000000068	0.89095454429505\\
0.992500000000064	0.89095454429505\\
0.995000000000005	0.636396103067893\\
0.997500000000059	0.127279220613579\\
1	-0\\
1.005	-0\\
1.00750000000005	-0.636396103067893\\
1.00999999999999	-0.89095454429505\\
1.01250000000005	-0.636396103067893\\
1.01499999999999	-1.14551298552221\\
1.01750000000004	-1.40007142674936\\
1.01999999999998	-1.14551298552221\\
1.02250000000004	-1.14551298552221\\
1.02500000000009	-0.89095454429505\\
1.02750000000003	-1.40007142674936\\
1.03000000000009	-1.14551298552221\\
1.03250000000003	-0.636396103067893\\
1.03500000000008	-0.89095454429505\\
1.03750000000002	-0.89095454429505\\
1.04000000000008	-0.381837661840736\\
1.04500000000007	0.127279220613579\\
1.04750000000001	-0\\
1.05000000000007	0.381837661840736\\
1.05500000000006	0.89095454429505\\
1.0575	0.89095454429505\\
1.06000000000006	0.636396103067893\\
1.0625	1.40007142674936\\
1.06500000000005	1.65462986797652\\
1.0675	1.14551298552221\\
1.07500000000005	1.14551298552221\\
1.08000000000004	0.127279220613579\\
1.08249999999998	0.89095454429505\\
1.08500000000004	0.636396103067893\\
1.08750000000009	-0\\
1.09000000000003	-0\\
1.09250000000009	-0.127279220613579\\
1.09500000000003	-0.381837661840736\\
1.09750000000008	-0.89095454429505\\
1.10500000000002	-0.89095454429505\\
1.10750000000007	-1.14551298552221\\
1.11000000000001	-1.14551298552221\\
1.11250000000007	-0.89095454429505\\
1.11500000000001	-1.14551298552221\\
1.11750000000006	-1.14551298552221\\
1.12	-0.89095454429505\\
1.12250000000006	-0.89095454429505\\
1.125	-0.381837661840736\\
1.13	-0.381837661840736\\
1.13250000000005	-0\\
1.13499999999999	-0\\
1.13750000000005	0.127279220613579\\
1.13999999999999	0.127279220613579\\
1.14250000000004	0.89095454429505\\
1.14499999999998	1.40007142674936\\
1.14750000000004	0.636396103067893\\
1.15500000000009	1.40007142674936\\
1.15750000000003	0.89095454429505\\
1.16000000000008	0.636396103067893\\
1.16250000000002	1.14551298552221\\
1.16500000000008	1.14551298552221\\
1.16750000000002	0.636396103067893\\
1.17000000000007	0.636396103067893\\
1.17250000000001	0.381837661840736\\
1.17500000000007	0.381837661840736\\
1.18000000000006	-0.381837661840736\\
1.1825	-0.127279220613579\\
1.18500000000006	-0.381837661840736\\
1.1875	-0.89095454429505\\
1.1925	-0.89095454429505\\
1.19500000000005	-1.14551298552221\\
1.20249999999999	-1.14551298552221\\
1.20500000000004	-0.89095454429505\\
1.21000000000004	-0.89095454429505\\
1.21250000000009	-0.381837661840736\\
1.21500000000003	-0.381837661840736\\
1.21750000000009	-0.636396103067893\\
1.22000000000003	-0.127279220613579\\
1.22250000000008	-0\\
1.22500000000002	0.381837661840736\\
1.22750000000008	0.127279220613579\\
1.23000000000002	0.381837661840736\\
1.23250000000007	0.89095454429505\\
1.23750000000007	0.89095454429505\\
1.24000000000001	0.636396103067893\\
1.245	1.65462986797652\\
1.24750000000006	0.636396103067893\\
1.25	0.89095454429505\\
1.255	0.89095454429505\\
1.25750000000005	0.381837661840736\\
1.25999999999999	-0\\
1.26250000000005	0.381837661840736\\
1.26499999999999	0.381837661840736\\
1.26750000000004	-0.127279220613579\\
1.26999999999998	-0.127279220613579\\
1.28000000000009	-1.14551298552221\\
1.28250000000003	-0.89095454429505\\
1.28500000000008	-0.89095454429505\\
1.28750000000002	-1.14551298552221\\
1.29000000000008	-0.89095454429505\\
1.30000000000007	-0.89095454429505\\
1.30250000000001	-0.381837661840736\\
1.30500000000006	-0.127279220613579\\
1.3075	-0.127279220613579\\
1.3125	0.127279220613579\\
1.31500000000005	0.381837661840736\\
1.3175	0.127279220613579\\
1.32000000000005	0.381837661840736\\
1.32500000000005	1.40007142674936\\
1.32749999999999	0.636396103067893\\
1.33249999999998	1.14551298552221\\
1.33500000000004	1.14551298552221\\
1.33750000000009	0.636396103067893\\
1.34000000000003	0.636396103067893\\
1.34500000000003	1.14551298552221\\
1.34750000000008	0.127279220613579\\
1.35250000000008	0.127279220613579\\
1.35750000000007	-0.127279220613579\\
1.36000000000001	-0.636396103067893\\
1.36250000000007	-0.381837661840736\\
1.36500000000001	-0.381837661840736\\
1.36750000000006	-0.89095454429505\\
1.375	-0.89095454429505\\
1.37750000000005	-1.14551298552221\\
1.38	-1.14551298552221\\
1.38250000000005	-0.636396103067893\\
1.38750000000005	-0.636396103067893\\
1.38999999999999	-0.381837661840736\\
1.39250000000004	-0.381837661840736\\
1.39499999999998	-0.127279220613579\\
1.40000000000009	-0.127279220613579\\
1.40500000000009	0.89095454429505\\
1.40750000000003	0.381837661840736\\
1.41000000000008	0.381837661840736\\
1.41250000000002	0.89095454429505\\
1.41500000000008	0.89095454429505\\
1.41750000000002	0.636396103067893\\
1.42000000000007	0.636396103067893\\
1.42250000000001	0.89095454429505\\
1.42500000000007	1.40007142674936\\
1.42750000000001	0.636396103067893\\
1.43000000000006	0.636396103067893\\
1.4325	0.89095454429505\\
1.43500000000006	0.636396103067893\\
1.4375	0.127279220613579\\
1.44000000000005	-0\\
1.4425	0.127279220613579\\
1.44500000000005	0.127279220613579\\
1.44749999999999	-0.381837661840736\\
1.45249999999999	-0.381837661840736\\
1.45500000000004	-0.89095454429505\\
1.46500000000003	-0.89095454429505\\
1.46750000000009	-1.14551298552221\\
1.47250000000008	-0.636396103067893\\
1.47500000000002	-0.89095454429505\\
1.47750000000008	-0.89095454429505\\
1.48000000000002	-0.636396103067893\\
1.48250000000007	-0\\
1.48750000000007	-0\\
1.49000000000001	0.127279220613579\\
1.495	0.636396103067893\\
1.5	0.127279220613579\\
1.50250000000005	1.14551298552221\\
};
\addlegendentry{$I_{L1,q}$}
\draw[line width=0.5mm,draw=black!70,loosely dashed] (axis cs:0.5,-5)--(axis cs:0.5,15);
\node[color=black, fill=none, font=\large] at (axis cs: 0.4,9) {\textcolor{blue}{Case 3a}};
\node[color=black, fill=none, font=\large] at (axis cs: 0.6,9) {\textcolor{red}{Case 3b}};
\end{axis}
\end{tikzpicture}%

%% file: updated_images/test1_Zdd.tex
%
%
\definecolor{mycolor1}{rgb}{1.00000,0.00000,1.00000}%
\begin{tikzpicture}

\begin{axis}[%
width=2.5in,
height=1in,
at={(0.758in,2.0in)},
scale only axis,
xmode=log,
xmin=1,
xmax=10000,
xminorticks=true,
xlabel style={font=\color{white!15!black}},
title style={font=\Large\bfseries},
title={$d-$ axis},
ymin=-30,
ymax=40,
ylabel style={font=\large\color{white!15!black}},
ylabel={Magnitude [dB]},
axis background/.style={fill=white},
xmajorgrids,
xminorgrids,
ymajorgrids,
legend pos=north west,
legend style={legend cell align=left, align=left,font=\normalsize, draw=white!15!black}
]
\addplot [color=blue, line width=2.0pt]
  table[row sep=crcr]{%
1	-0.777600531123646\\
1.20450354025879	-2.17635046101596\\
1.32194114846604	-2.84456022430529\\
1.45082877849593	-3.48765049541343\\
1.59228279334109	-4.10197265583993\\
1.74752840000768	-4.68365881059648\\
1.91791026167249	-5.22877204418098\\
2.10490414451203	-5.73355457789466\\
2.31012970008318	-6.19482304394661\\
2.5353644939701	-6.61058279955324\\
2.78255940220711	-6.9809585360424\\
3.05385550883341	-7.30954591317744\\
3.35160265093885	-7.60521517404359\\
4.03701725859658	-8.17075343474045\\
4.43062145758385	-8.49641676227149\\
4.86260158006533	-8.89200124128912\\
5.3366992312063	-9.37677146126696\\
5.85702081805667	-9.94978479293027\\
6.42807311728435	-10.5920519955751\\
7.05480231071869	-11.2778920767639\\
9.32603346883218	-13.3846941127186\\
10.2353102189903	-13.9440877001493\\
11.2332403297803	-14.0705790482217\\
12.3284673944207	-13.3540742692533\\
13.530477745798	-11.9487568118662\\
19.6304065004028	-5.51844261047385\\
21.544346900319	-3.90613140632814\\
23.6448941264543	-2.24070804257307\\
25.9502421139972	-0.485739229350195\\
28.4803586843579	1.40479767881634\\
31.2571584968824	3.49529118504997\\
34.3046928631493	5.88532756111101\\
37.6493580679249	8.74618190227709\\
41.3201240011537	12.4098712930411\\
45.3487850812855	17.5642725594655\\
49.7702356433209	23.347967306611\\
54.6227721768434	19.2761831414702\\
59.9484250318942	14.7016522251672\\
65.7933224657571	11.7459857103477\\
72.2080901838551	9.66505447746421\\
86.974900261778	7.08212132719162\\
95.4548456661833	6.023304838071\\
104.761575278967	5.11291649051397\\
114.975699539774	4.30324528425322\\
126.185688306603	3.55746097896665\\
151.991108295293	2.13257865324875\\
166.810053720006	1.39055711228521\\
183.073828029537	0.578598414047057\\
200.923300256505	-0.355730510020365\\
220.513073990306	-1.48807652511352\\
242.012826479436	-2.9278999474131\\
291.505306282517	-6.65491841582597\\
319.926713779738	-6.41135533995507\\
351.119173421514	-4.66493337868437\\
385.352859371055	-1.38901493576444\\
422.924287438953	2.07958593924575\\
509.413801481636	8.62109240219696\\
559.081018251222	11.9235650230595\\
613.590727341319	15.5858739787848\\
673.415065775086	20.1904798384556\\
739.072203352584	27.6071640585822\\
811.130830789682	36.8117270802423\\
890.215085445036	24.9444924033806\\
977.009957299225	19.9807727454063\\
1072.26722201033	16.842452011775\\
1176.811952435	14.4990952591849\\
1291.54966501489	12.5913169320229\\
1417.4741629268	10.9553225677135\\
1555.67614393047	9.5042972707349\\
1707.35264747069	8.18736285772489\\
1873.81742286039	6.97241786172587\\
2056.51230834866	5.83803040172219\\
2257.01971963394	4.76926882565179\\
2477.07635599169	3.75541779000666\\
2718.58824273293	2.78866193096474\\
2983.64724028333	1.86329237679445\\
3274.54916287773	0.975206971773588\\
3593.81366380464	0.121580103218257\\
3944.20605943768	-0.699368266520956\\
4328.76128108303	-1.48854628520679\\
4750.81016210277	-2.24613295166922\\
5214.00828799967	-2.97170731742204\\
5722.36765935022	-3.66437590571625\\
6280.29144183427	-4.3229058364836\\
6892.61210434974	-4.94586718653711\\
7564.63327554635	-5.53178398485151\\
8302.1756813197	-6.07928897479022\\
9111.62756115487	-6.58727335726235\\
10000	-7.05501996816884\\
};
\addlegendentry{${Z}_{s}$}
\addplot [color=mycolor1, line width=2.0pt]
  table[row sep=crcr]{%
1	-30.0487992977232\\
1.09749876549306	-29.9697408274943\\
1.20450354025879	-29.8763867067046\\
1.32194114846604	-29.7665421787921\\
1.45082877849593	-29.6378198629483\\
1.59228279334109	-29.4876741990202\\
1.74752840000768	-29.31345534283\\
1.91791026167249	-29.1124840157991\\
2.10490414451203	-28.8821465470381\\
2.31012970008318	-28.6200062678016\\
2.5353644939701	-28.3239239105169\\
2.78255940220711	-27.9921764266608\\
3.05385550883341	-27.6235615727801\\
3.35160265093885	-27.2174755677073\\
3.67837977182865	-26.7739535422922\\
4.03701725859658	-26.2936671593176\\
4.43062145758385	-25.7778797393478\\
4.86260158006533	-25.2283651071853\\
5.3366992312063	-24.6473008200917\\
5.85702081805667	-24.0371485863083\\
6.42807311728435	-23.4005343846088\\
7.05480231071869	-22.7401385579806\\
8.4975343590864	-21.3584578254532\\
10.2353102189903	-19.9116155365664\\
12.3284673944207	-18.4161250646543\\
16.2975083462064	-16.1095656167057\\
21.544346900319	-13.7541285990073\\
34.3046928631493	-9.7702748983663\\
72.2080901838551	-3.33442904453316\\
509.413801481636	13.6270062540885\\
10000	39.4854229883082\\
};
\addlegendentry{${Z}_{g}$}
\end{axis}

\begin{axis}[%
width=2.5in,
height=1in,
at={(0.758in,0.7in)},
scale only axis,
xmode=log,
xmin=1,
xmax=10000,
xminorticks=true,
xlabel style={font=\large\color{white!15!black}},
xlabel={Frequency [Hz]},
ymin=-200,
ymax=200,
ylabel style={font=\large\color{white!15!black}},
ylabel={Phase [deg]},
axis background/.style={fill=white},
xmajorgrids,
xminorgrids,
ymajorgrids,
legend pos=north east,
legend style={legend cell align=left, align=left,font=\scriptsize, draw=white!15!black}
]
\addplot [color=blue, line width=2.0pt]
  table[row sep=crcr]{%
1	-75.2624452496319\\
1.09749876549306	-71.8222609370975\\
1.20450354025879	-68.2798679729518\\
1.32194114846604	-64.6220355050732\\
1.45082877849596	-60.8389582018793\\
1.59228279334111	-56.9253008454017\\
1.74752840000771	-52.881464822799\\
2.10490414451206	-44.4428037869469\\
3.05385550883351	-27.0372419423286\\
3.35160265093874	-22.9004376027152\\
3.67837977182853	-18.9845938868042\\
4.03701725859645	-15.3198761830729\\
4.86260158006525	-8.44701023551471\\
5.33669923120621	-4.8138147542237\\
5.85702081805658	-0.650538857239553\\
6.42807311728424	4.30972934804529\\
7.05480231071857	10.255607579118\\
7.74263682681121	17.3440848196917\\
8.4975343590864	25.8090260107257\\
9.32603346883218	36.1086248016009\\
10.2353102189903	49.0270154518366\\
11.2332403297803	65.0698820921956\\
12.3284673944207	82.0334170854851\\
13.5304777457982	95.9314569919348\\
14.8496826225448	106.146943935822\\
16.2975083462067	113.876285295976\\
17.8864952905746	120.032197876628\\
19.6304065004031	125.112262983735\\
21.5443469003193	129.379654496088\\
23.6448941264547	132.972904141324\\
25.9502421139981	135.954997333523\\
28.4803586843589	138.327619833309\\
31.2571584968834	140.017302012469\\
34.3046928631481	140.816307323241\\
37.6493580679236	140.199153972929\\
41.3201240011523	136.658147989029\\
45.3487850812848	124.357557047181\\
49.7702356433201	74.6412592526131\\
54.6227721768425	16.673677637975\\
59.9484250318932	1.33640393429036\\
65.793322465756	-3.66560576647473\\
72.208090183854	-5.66850771013148\\
79.2482898353912	-8.52730140105351\\
86.974900261778	-8.61614595824636\\
95.4548456661833	-8.42714193124971\\
104.761575278967	-8.08195320764165\\
114.975699539774	-7.63727293186531\\
126.185688306603	-7.11335288376296\\
138.488637139389	-6.50439033048806\\
151.991108295295	-5.77926596622373\\
166.810053720008	-4.87390135738244\\
183.07382802954	-3.67104702330181\\
200.923300256509	-1.95303171799003\\
220.51307399031	0.717538254118153\\
242.012826479444	5.33685604953982\\
265.608778294676	14.4308352345751\\
291.505306282527	33.3522690509454\\
319.926713779749	58.8631075277774\\
351.119173421503	86.8750110976754\\
385.352859371042	105.29547117556\\
422.92428743894	114.919667487919\\
464.158883361268	119.627662233198\\
509.413801481628	121.72560901897\\
559.081018251213	122.472130523406\\
613.590727341309	122.760451867434\\
673.415065775075	123.803790758633\\
739.072203352571	130.588295926575\\
811.130830789682	-114.277740847375\\
890.215085445036	-79.0916705687191\\
977.009957299225	-76.0474900270152\\
1072.26722201033	-75.7000665905898\\
1176.811952435	-75.9524122876751\\
1291.54966501489	-76.3022908620543\\
1417.47416292682	-76.574959712492\\
1555.67614393049	-76.7014134272081\\
1707.35264747072	-76.6554512349612\\
1873.81742286042	-76.4300384380388\\
2056.51230834869	-76.0267371602671\\
2257.01971963397	-75.4505070081982\\
2477.07635599177	-74.7070769852557\\
2718.58824273302	-73.8016830148663\\
2983.64724028343	-72.7385790107973\\
3274.54916287762	-71.521003100978\\
3593.81366380452	-70.1514179131725\\
3944.20605943755	-68.631918040047\\
4328.76128108296	-66.964738495946\\
4750.8101621027	-65.1528186003916\\
5214.00828799959	-63.2003833288757\\
5722.36765935013	-61.1135033150781\\
6280.29144183417	-58.9005892037281\\
6892.61210434962	-56.5727698966126\\
7564.63327554623	-54.144101678358\\
9111.62756115487	-49.0547845240932\\
10000	-46.4355641380668\\
};

\addplot [color=mycolor1, line width=2.0pt]
  table[row sep=crcr]{%
1	17.4405944905119\\
1.09749876549306	19.0236509170512\\
1.20450354025879	20.7269406109583\\
1.32194114846604	22.5531147578928\\
1.45082877849596	24.5030859551488\\
1.59228279334111	26.57561062271\\
1.74752840000771	28.7668846321705\\
1.91791026167246	31.0701942593807\\
2.104904144512	33.4756730511183\\
2.31012970008314	35.9702178580836\\
2.78255940220711	41.1588766418261\\
4.03701725859658	51.7450018268022\\
4.43062145758392	54.3053358630758\\
4.86260158006541	56.7909130207518\\
5.33669923120639	59.1858417165851\\
5.85702081805658	61.4773694426211\\
6.42807311728424	63.6559805049931\\
7.05480231071857	65.7152922429851\\
7.74263682681121	67.6517977566608\\
8.4975343590864	69.4645084940665\\
9.32603346883218	71.1545465840312\\
10.2353102189903	72.7247278899632\\
11.2332403297803	74.1791656430354\\
12.3284673944207	75.5229136291528\\
13.5304777457982	76.7616586962948\\
14.8496826225448	77.9014653967253\\
16.2975083462067	78.9485708674951\\
17.8864952905746	79.9092252442814\\
19.6304065004024	80.7895715419321\\
21.5443469003186	81.595558574158\\
23.6448941264539	82.3328807513925\\
25.9502421139972	83.0069392083154\\
28.4803586843579	83.6228194756272\\
31.2571584968824	84.1852816994935\\
34.3046928631493	84.698760154107\\
37.6493580679249	85.167369451558\\
41.3201240011537	85.5949154153687\\
45.3487850812863	85.9849090501778\\
49.7702356433217	86.3405824181285\\
54.6227721768443	86.6649055337841\\
59.9484250318932	86.9606036259778\\
65.793322465756	87.2301742984396\\
72.208090183854	87.4759042616293\\
79.2482898353912	87.6998854147776\\
86.974900261778	87.9040301370122\\
95.4548456661833	88.090085705607\\
104.761575278967	88.2596478026005\\
114.975699539774	88.4141731020762\\
126.185688306603	88.5549909522262\\
138.488637139389	88.6833141812286\\
151.991108295295	88.800249065725\\
166.810053720008	88.9068045066338\\
183.073828029534	89.0039004602134\\
200.923300256502	89.0923756734661\\
220.513073990302	89.1729947727437\\
242.012826479436	89.2464547532073\\
265.608778294667	89.3133909149365\\
291.505306282517	89.3743822892211\\
319.926713779738	89.4299565960627\\
351.119173421514	89.4805947713063\\
385.352859371055	89.5267350991935\\
422.924287438953	89.5687769835397\\
464.158883361283	89.6070843882421\\
509.413801481645	89.6419889754341\\
559.081018251231	89.6737929673452\\
613.590727341309	89.7027717558056\\
673.415065775075	89.7291762813555\\
739.072203352571	89.7532352020792\\
811.130830789682	89.7751568705822\\
890.215085445036	89.7951311359558\\
977.009957299225	89.8133309861281\\
1072.26722201033	89.8299140446656\\
1176.811952435	89.8450239348703\\
1291.54966501489	89.8587915228968\\
1555.67614393049	89.8827661677828\\
1873.81742286036	89.9026703707073\\
2257.0197196339	89.9191952077093\\
2718.58824273293	89.9329144280302\\
3274.54916287773	89.9443043716737\\
3944.20605943768	89.9537605063697\\
5214.00828799976	89.9650215155356\\
6892.61210434962	89.9735400578431\\
9111.62756115487	89.9799840235004\\
10000	89.9817621875603\\
};

\end{axis}
\end{tikzpicture}%

%% file: updated_images/test1_Zqq.tex
%
%
\definecolor{mycolor1}{rgb}{1.00000,0.00000,1.00000}%
\begin{tikzpicture}

\begin{axis}[%
width=2.5in,
height=1in,
at={(0.758in,2.0in)},
scale only axis,
xmode=log,
xmin=1,
xmax=10000,
xminorticks=true,
xlabel style={font=\color{white!15!black}},
title style={font=\Large\bfseries},
title={$q-$ axis},
ymin=-30,
ymax=40,
ylabel style={font=\large\color{white!15!black}},
ylabel={Magnitude [dB]},
axis background/.style={fill=white},
xmajorgrids,
xminorgrids,
ymajorgrids,
legend pos=north west,
legend style={legend cell align=left, align=left,font=\normalsize, draw=white!15!black}
]
\addplot [color=blue, line width=2.0pt, forget plot]
  table[row sep=crcr]{%
1	-2.55792365094786\\
1.20450354025878	-3.85296517262805\\
1.45082877849594	-5.10852338735214\\
1.59228279334109	-5.70973756568359\\
1.74752840000768	-6.28675131385181\\
1.91791026167249	-6.83462452213466\\
2.10490414451201	-7.34850708871421\\
2.31012970008316	-7.82389120013428\\
2.53536449397012	-8.25704489282736\\
2.78255940220711	-8.64572561120982\\
3.05385550883341	-8.9903102465667\\
3.35160265093885	-9.29548522910291\\
3.67837977182862	-9.57251198823571\\
4.03701725859655	-9.84160068962066\\
4.43062145758389	-10.1328806991255\\
4.86260158006537	-10.4833581757293\\
5.3366992312063	-10.928339762207\\
5.85702081805667	-11.4911063944776\\
6.42807311728435	-12.1795046832152\\
7.05480231071863	-12.9939421842536\\
7.74263682681128	-13.9417880062298\\
8.49753435908647	-15.0501915914904\\
9.32603346883218	-16.3682010222952\\
10.2353102189903	-17.9144250812888\\
11.2332403297803	-19.3759494870927\\
12.3284673944206	-19.7005498856548\\
13.5304777457981	-18.6368749800078\\
14.8496826225447	-17.1885315390964\\
16.2975083462064	-15.8442059008427\\
17.8864952905744	-14.6700832205463\\
19.6304065004028	-13.6411827765955\\
21.5443469003188	-12.7267171554622\\
23.6448941264541	-11.9032050910839\\
25.9502421139974	-11.1541557858416\\
28.4803586843579	-10.4681146898053\\
31.2571584968824	-9.83706493012293\\
34.3046928631493	-9.25534236249714\\
37.6493580679246	-8.71893502105026\\
41.3201240011533	-8.22503784424617\\
45.3487850812859	-7.77177265203554\\
49.7702356433209	-7.35801635357244\\
54.6227721768434	-6.98330280883794\\
59.9484250318942	-6.64777816661795\\
65.7933224657566	-6.35219860756014\\
72.2080901838546	-6.09796514615732\\
79.2482898353919	-7.17661969641103\\
86.9749002617787	-6.70249623131856\\
95.4548456661833	-6.37380615063438\\
104.761575278967	-6.15552073361298\\
114.975699539774	-6.03093539767464\\
126.185688306602	-5.99216999897849\\
138.488637139387	-6.03492022082187\\
151.991108295294	-6.15367443017721\\
166.810053720006	-6.33451128578756\\
183.073828029537	-6.54208152528263\\
200.923300256505	-6.69884890042229\\
220.513073990304	-6.67017643867545\\
242.012826479438	-6.31388288618128\\
265.608778294669	-5.61471162301874\\
291.505306282517	-4.5012170532933\\
319.926713779738	-2.39355422969215\\
351.119173421514	0.275779283411808\\
385.352859371052	3.07142830444561\\
422.92428743895	5.96109222853305\\
464.158883361279	9.00447525209989\\
509.413801481636	12.3509091575813\\
559.081018251222	16.3148973871223\\
613.590727341319	21.6852882325905\\
673.41506577508	31.0363259662663\\
739.072203352578	28.7438160912466\\
811.130830789689	21.9423640887013\\
890.215085445036	18.2166414331716\\
977.009957299225	15.6561203327196\\
1072.26722201033	13.6690839171597\\
1176.81195243499	12.0114062192318\\
1291.54966501488	10.5630721890969\\
1417.47416292681	9.25835206892767\\
1555.67614393048	8.05838915904582\\
1707.35264747069	6.93875169089522\\
1873.81742286039	5.88323042466214\\
2056.51230834866	4.88056165178878\\
2257.01971963392	3.92262910454718\\
2477.07635599171	3.0034500333779\\
2718.58824273295	2.11858881921153\\
2983.64724028333	1.2648071283805\\
3274.54916287773	0.439845568092959\\
3593.81366380464	-0.357722120915231\\
3944.20605943765	-1.12859627671388\\
4328.76128108306	-1.87284029248151\\
4750.81016210281	-2.58996433851886\\
5214.00828799967	-3.2790126463333\\
5722.36765935022	-3.93866042727643\\
6280.29144183427	-4.56732408141616\\
6892.61210434968	-5.16328563729442\\
7564.63327554629	-5.72482912433861\\
8302.17568131977	-6.25038298992416\\
9111.62756115487	-6.73865928511878\\
10000	-7.18877797479187\\
};
\addplot [color=mycolor1, line width=2.0pt, forget plot]
  table[row sep=crcr]{%
1	-30.0487992977232\\
1.09749876549306	-29.9697408274943\\
1.20450354025879	-29.8763867067046\\
1.32194114846604	-29.7665421787921\\
1.45082877849593	-29.6378198629483\\
1.59228279334109	-29.4876741990202\\
1.74752840000768	-29.31345534283\\
1.91791026167249	-29.1124840157991\\
2.10490414451203	-28.8821465470381\\
2.31012970008318	-28.6200062678016\\
2.5353644939701	-28.3239239105169\\
2.78255940220711	-27.9921764266608\\
3.05385550883341	-27.6235615727801\\
3.35160265093885	-27.2174755677073\\
3.67837977182865	-26.7739535422922\\
4.03701725859658	-26.2936671593176\\
4.43062145758385	-25.7778797393478\\
4.86260158006533	-25.2283651071853\\
5.3366992312063	-24.6473008200917\\
5.85702081805667	-24.0371485863083\\
6.42807311728435	-23.4005343846088\\
7.05480231071869	-22.7401385579806\\
8.4975343590864	-21.3584578254532\\
10.2353102189903	-19.9116155365664\\
12.3284673944207	-18.4161250646543\\
16.2975083462064	-16.1095656167057\\
21.544346900319	-13.7541285990073\\
34.3046928631493	-9.7702748983663\\
72.2080901838551	-3.33442904453316\\
509.413801481636	13.6270062540885\\
10000	39.4854229883082\\
};
\draw[line width=0.5mm,draw=black!70,dashed] (axis cs:11.1,-50)--(axis cs:11.1,70);
\node[color=black, fill=none,font=\large] at (axis cs: 3.5,35) {\SI{11.1}{\hertz}};
\end{axis}

\begin{axis}[%
width=2.5in,
height=1in,
at={(0.758in,0.7in)},
scale only axis,
xmode=log,
xmin=1,
xmax=10000,
xminorticks=true,
xlabel style={font=\large\color{white!15!black}},
xlabel={Frequency [Hz]},
ymin=-200,
ymax=200,
ylabel style={font=\large\color{white!15!black}},
ylabel={Phase [deg]},
axis background/.style={fill=white},
xmajorgrids,
xminorgrids,
ymajorgrids,
legend pos=north east,
legend style={legend cell align=left, align=left,font=\scriptsize, draw=white!15!black}
]
\addplot [color=blue, line width=2.0pt, forget plot]
  table[row sep=crcr]{%
1	137.812087817805\\
1.09749876549306	138.563580322546\\
1.20450354025879	139.663043228835\\
1.32194114846604	141.1102361296\\
1.45082877849596	142.902633623376\\
1.59228279334111	145.034587639091\\
1.74752840000771	147.496186425503\\
1.91791026167252	150.271769482577\\
2.10490414451206	153.338058997925\\
2.31012970008321	156.661890809105\\
2.53536449397018	160.197607298173\\
3.05385550883351	167.644312684415\\
3.35160265093874	171.382945010507\\
3.67837977182853	174.996083876867\\
4.03701725859645	178.388105764465\\
4.43062145758378	-178.493680543884\\
4.86260158006525	-175.614862247187\\
5.33669923120621	-172.827260116038\\
5.85702081805658	-169.8946848425\\
6.42807311728424	-166.566252615164\\
7.05480231071857	-162.635822955506\\
7.74263682681121	-157.925894289744\\
8.4975343590864	-152.168227530036\\
9.32603346883218	-144.72478318482\\
10.2353102189903	-133.990560047648\\
11.2332403297803	-116.987869736209\\
12.3284673944207	-94.2903081861013\\
13.5304777457982	-74.8170888292261\\
14.8496826225448	-61.6931476137694\\
16.2975083462067	-52.5722770547912\\
17.8864952905746	-45.6467102834564\\
19.6304065004031	-40.0274367424616\\
21.5443469003193	-35.2762863084428\\
23.6448941264547	-31.1576915238778\\
25.9502421139981	-27.5314730341717\\
28.4803586843589	-24.3056529802884\\
31.2571584968834	-21.4144633315218\\
34.3046928631481	-18.8074304658591\\
37.6493580679236	-16.4435897548405\\
41.3201240011523	-14.2881849442639\\
45.3487850812848	-12.3106102606803\\
49.7702356433201	-10.4829793419378\\
54.6227721768425	-8.77899382645515\\
59.9484250318932	-7.17291715119688\\
65.793322465756	-5.63851365870443\\
72.208090183854	-4.14782253172837\\
79.2482898353912	10.9233682416829\\
86.974900261778	9.58259777686089\\
95.4548456661833	9.32844329980946\\
104.761575278967	9.81708024095079\\
114.975699539774	10.9189106831724\\
126.185688306603	12.6254332322372\\
138.488637139389	15.0200109510276\\
151.991108295295	18.276862682097\\
166.810053720008	22.6723652532869\\
183.07382802954	28.5855959129748\\
200.923300256509	36.4255482496152\\
220.51307399031	46.3623807938574\\
242.012826479444	57.9092730497933\\
265.608778294676	70.3726381635815\\
291.505306282527	84.410568357918\\
319.926713779749	101.852065726151\\
351.119173421503	110.641387623244\\
385.352859371042	116.553120104546\\
422.92428743894	120.469395133138\\
464.158883361268	123.123041771201\\
509.413801481628	125.187016995202\\
559.081018251213	127.607598780737\\
613.590727341309	133.027807371854\\
673.415065775075	161.399013352771\\
739.072203352571	-94.891803424151\\
811.130830789682	-78.7948321402654\\
890.215085445036	-75.6058129290493\\
977.009957299225	-74.9276516477787\\
1072.26722201033	-75.0360151569994\\
1176.811952435	-75.3876316573179\\
1291.54966501489	-75.7543543216398\\
1417.47416292682	-76.0276503388773\\
1555.67614393049	-76.1549299526334\\
1707.35264747072	-76.1127805321404\\
1873.81742286042	-75.89349123474\\
2056.51230834869	-75.4975024401916\\
2257.01971963397	-74.9289949352926\\
2477.07635599177	-74.1933449749941\\
2718.58824273302	-73.2957545062644\\
2983.64724028343	-72.2406361365651\\
3274.54916287762	-71.0314824205944\\
3593.81366380452	-69.671042989333\\
3944.20605943755	-68.1616944638787\\
4328.76128108296	-66.5059272390289\\
4750.8101621027	-64.7068958638463\\
5214.00828799959	-62.768990044543\\
5722.36765935013	-60.6983848627439\\
6280.29144183417	-58.5035252711961\\
6892.61210434962	-56.195495258617\\
7564.63327554623	-53.7882205261355\\
9111.62756115487	-48.745550137616\\
10000	-46.1509160471809\\
};
\addplot [color=mycolor1, line width=2.0pt, forget plot]
  table[row sep=crcr]{%
1	17.4405944905119\\
1.09749876549306	19.0236509170512\\
1.20450354025879	20.7269406109583\\
1.32194114846604	22.5531147578928\\
1.45082877849596	24.5030859551488\\
1.59228279334111	26.57561062271\\
1.74752840000771	28.7668846321705\\
1.91791026167246	31.0701942593807\\
2.104904144512	33.4756730511183\\
2.31012970008314	35.9702178580836\\
2.78255940220711	41.1588766418261\\
4.03701725859658	51.7450018268022\\
4.43062145758392	54.3053358630758\\
4.86260158006541	56.7909130207518\\
5.33669923120639	59.1858417165851\\
5.85702081805658	61.4773694426211\\
6.42807311728424	63.6559805049931\\
7.05480231071857	65.7152922429851\\
7.74263682681121	67.6517977566608\\
8.4975343590864	69.4645084940665\\
9.32603346883218	71.1545465840312\\
10.2353102189903	72.7247278899632\\
11.2332403297803	74.1791656430354\\
12.3284673944207	75.5229136291528\\
13.5304777457982	76.7616586962948\\
14.8496826225448	77.9014653967253\\
16.2975083462067	78.9485708674951\\
17.8864952905746	79.9092252442814\\
19.6304065004024	80.7895715419321\\
21.5443469003186	81.595558574158\\
23.6448941264539	82.3328807513925\\
25.9502421139972	83.0069392083154\\
28.4803586843579	83.6228194756272\\
31.2571584968824	84.1852816994935\\
34.3046928631493	84.698760154107\\
37.6493580679249	85.167369451558\\
41.3201240011537	85.5949154153687\\
45.3487850812863	85.9849090501778\\
49.7702356433217	86.3405824181285\\
54.6227721768443	86.6649055337841\\
59.9484250318932	86.9606036259778\\
65.793322465756	87.2301742984396\\
72.208090183854	87.4759042616293\\
79.2482898353912	87.6998854147776\\
86.974900261778	87.9040301370122\\
95.4548456661833	88.090085705607\\
104.761575278967	88.2596478026005\\
114.975699539774	88.4141731020762\\
126.185688306603	88.5549909522262\\
138.488637139389	88.6833141812286\\
151.991108295295	88.800249065725\\
166.810053720008	88.9068045066338\\
183.073828029534	89.0039004602134\\
200.923300256502	89.0923756734661\\
220.513073990302	89.1729947727437\\
242.012826479436	89.2464547532073\\
265.608778294667	89.3133909149365\\
291.505306282517	89.3743822892211\\
319.926713779738	89.4299565960627\\
351.119173421514	89.4805947713063\\
385.352859371055	89.5267350991935\\
422.924287438953	89.5687769835397\\
464.158883361283	89.6070843882421\\
509.413801481645	89.6419889754341\\
559.081018251231	89.6737929673452\\
613.590727341309	89.7027717558056\\
673.415065775075	89.7291762813555\\
739.072203352571	89.7532352020792\\
811.130830789682	89.7751568705822\\
890.215085445036	89.7951311359558\\
977.009957299225	89.8133309861281\\
1072.26722201033	89.8299140446656\\
1176.811952435	89.8450239348703\\
1291.54966501489	89.8587915228968\\
1555.67614393049	89.8827661677828\\
1873.81742286036	89.9026703707073\\
2257.0197196339	89.9191952077093\\
2718.58824273293	89.9329144280302\\
3274.54916287773	89.9443043716737\\
3944.20605943768	89.9537605063697\\
5214.00828799976	89.9650215155356\\
6892.61210434962	89.9735400578431\\
9111.62756115487	89.9799840235004\\
10000	89.9817621875603\\
};
\draw[decorate,decoration={brace,mirror, amplitude=5pt}](axis cs:11.1,-110) -- (axis cs:11.1,73);
\draw[line width=0.5mm,draw=black!80,->] (axis cs:15,-17)--(axis cs:30,-85);
\draw[line width=0.5mm,draw=black!70,loosely dashed] (axis cs:11.1,-200)--(axis cs:11.1,200);
\node[color=black, fill=none, font=\normalsize] at (axis cs: 100,-100) {$\Delta\phi\approx$\SI{193}{\degree}};
\end{axis}
\end{tikzpicture}%

%% file: updated_images/test1_Id_Change.tex
%
%
\begin{tikzpicture}

\begin{axis}[%
width=2in,
height=2in,
at={(0.494in,0.431in)},
scale only axis,
scale only axis,
point meta min=0.316885056645923,
point meta max=0.521957484490431,
xmin=4,
xmax=13,
xtick = {5,7,9,11},
xticklabels={5,7,9,11},
xlabel style={font=\color{white!15!black}},
xlabel={${I}_{{ref}}^d$ [A]},
ymin=11,
ymax=12.3 ,
ylabel style={align=center,font=\color{white!15!black}},
ylabel={Potential \\ Resonance Frequency [Hz]},
axis background/.style={fill=white},
xmajorgrids,
ymajorgrids,
grid style={dashed, opacity=0.5}
]

\addplot[%
mesh,
    shader=flat,
    mark=none,
    line join=round,
    line cap=round,
    line width=2pt,
    point meta=explicit, colormap={mymap}{[1pt] rgb(0pt)=(0.18995,0.07176,0.23217); rgb(1pt)=(0.19483,0.08339,0.26149); rgb(2pt)=(0.19956,0.09498,0.29024); rgb(3pt)=(0.20415,0.10652,0.31844); rgb(4pt)=(0.2086,0.11802,0.34607); rgb(5pt)=(0.21291,0.12947,0.37314); rgb(6pt)=(0.21708,0.14087,0.39964); rgb(7pt)=(0.22111,0.15223,0.42558); rgb(8pt)=(0.225,0.16354,0.45096); rgb(9pt)=(0.22875,0.17481,0.47578); rgb(10pt)=(0.23236,0.18603,0.50004); rgb(11pt)=(0.23582,0.1972,0.52373); rgb(12pt)=(0.23915,0.20833,0.54686); rgb(13pt)=(0.24234,0.21941,0.56942); rgb(14pt)=(0.24539,0.23044,0.59142); rgb(15pt)=(0.2483,0.24143,0.61286); rgb(16pt)=(0.25107,0.25237,0.63374); rgb(17pt)=(0.25369,0.26327,0.65406); rgb(18pt)=(0.25618,0.27412,0.67381); rgb(19pt)=(0.25853,0.28492,0.693); rgb(20pt)=(0.26074,0.29568,0.71162); rgb(21pt)=(0.2628,0.30639,0.72968); rgb(22pt)=(0.26473,0.31706,0.74718); rgb(23pt)=(0.26652,0.32768,0.76412); rgb(24pt)=(0.26816,0.33825,0.7805); rgb(25pt)=(0.26967,0.34878,0.79631); rgb(26pt)=(0.27103,0.35926,0.81156); rgb(27pt)=(0.27226,0.3697,0.82624); rgb(28pt)=(0.27334,0.38008,0.84037); rgb(29pt)=(0.27429,0.39043,0.85393); rgb(30pt)=(0.27509,0.40072,0.86692); rgb(31pt)=(0.27576,0.41097,0.87936); rgb(32pt)=(0.27628,0.42118,0.89123); rgb(33pt)=(0.27667,0.43134,0.90254); rgb(34pt)=(0.27691,0.44145,0.91328); rgb(35pt)=(0.27701,0.45152,0.92347); rgb(36pt)=(0.27698,0.46153,0.93309); rgb(37pt)=(0.2768,0.47151,0.94214); rgb(38pt)=(0.27648,0.48144,0.95064); rgb(39pt)=(0.27603,0.49132,0.95857); rgb(40pt)=(0.27543,0.50115,0.96594); rgb(41pt)=(0.27469,0.51094,0.97275); rgb(42pt)=(0.27381,0.52069,0.97899); rgb(43pt)=(0.27273,0.5304,0.98461); rgb(44pt)=(0.27106,0.54015,0.9893); rgb(45pt)=(0.26878,0.54995,0.99303); rgb(46pt)=(0.26592,0.55979,0.99583); rgb(47pt)=(0.26252,0.56967,0.99773); rgb(48pt)=(0.25862,0.57958,0.99876); rgb(49pt)=(0.25425,0.5895,0.99896); rgb(50pt)=(0.24946,0.59943,0.99835); rgb(51pt)=(0.24427,0.60937,0.99697); rgb(52pt)=(0.23874,0.61931,0.99485); rgb(53pt)=(0.23288,0.62923,0.99202); rgb(54pt)=(0.22676,0.63913,0.98851); rgb(55pt)=(0.22039,0.64901,0.98436); rgb(56pt)=(0.21382,0.65886,0.97959); rgb(57pt)=(0.20708,0.66866,0.97423); rgb(58pt)=(0.20021,0.67842,0.96833); rgb(59pt)=(0.19326,0.68812,0.9619); rgb(60pt)=(0.18625,0.69775,0.95498); rgb(61pt)=(0.17923,0.70732,0.94761); rgb(62pt)=(0.17223,0.7168,0.93981); rgb(63pt)=(0.16529,0.7262,0.93161); rgb(64pt)=(0.15844,0.73551,0.92305); rgb(65pt)=(0.15173,0.74472,0.91416); rgb(66pt)=(0.14519,0.75381,0.90496); rgb(67pt)=(0.13886,0.76279,0.8955); rgb(68pt)=(0.13278,0.77165,0.8858); rgb(69pt)=(0.12698,0.78037,0.8759); rgb(70pt)=(0.12151,0.78896,0.86581); rgb(71pt)=(0.11639,0.7974,0.85559); rgb(72pt)=(0.11167,0.80569,0.84525); rgb(73pt)=(0.10738,0.81381,0.83484); rgb(74pt)=(0.10357,0.82177,0.82437); rgb(75pt)=(0.10026,0.82955,0.81389); rgb(76pt)=(0.0975,0.83714,0.80342); rgb(77pt)=(0.09532,0.84455,0.79299); rgb(78pt)=(0.09377,0.85175,0.78264); rgb(79pt)=(0.09287,0.85875,0.7724); rgb(80pt)=(0.09267,0.86554,0.7623); rgb(81pt)=(0.0932,0.87211,0.75237); rgb(82pt)=(0.09451,0.87844,0.74265); rgb(83pt)=(0.09662,0.88454,0.73316); rgb(84pt)=(0.09958,0.8904,0.72393); rgb(85pt)=(0.10342,0.896,0.715); rgb(86pt)=(0.10815,0.90142,0.70599); rgb(87pt)=(0.11374,0.90673,0.69651); rgb(88pt)=(0.12014,0.91193,0.6866); rgb(89pt)=(0.12733,0.91701,0.67627); rgb(90pt)=(0.13526,0.92197,0.66556); rgb(91pt)=(0.14391,0.9268,0.65448); rgb(92pt)=(0.15323,0.93151,0.64308); rgb(93pt)=(0.16319,0.93609,0.63137); rgb(94pt)=(0.17377,0.94053,0.61938); rgb(95pt)=(0.18491,0.94484,0.60713); rgb(96pt)=(0.19659,0.94901,0.59466); rgb(97pt)=(0.20877,0.95304,0.58199); rgb(98pt)=(0.22142,0.95692,0.56914); rgb(99pt)=(0.23449,0.96065,0.55614); rgb(100pt)=(0.24797,0.96423,0.54303); rgb(101pt)=(0.2618,0.96765,0.52981); rgb(102pt)=(0.27597,0.97092,0.51653); rgb(103pt)=(0.29042,0.97403,0.50321); rgb(104pt)=(0.30513,0.97697,0.48987); rgb(105pt)=(0.32006,0.97974,0.47654); rgb(106pt)=(0.33517,0.98234,0.46325); rgb(107pt)=(0.35043,0.98477,0.45002); rgb(108pt)=(0.36581,0.98702,0.43688); rgb(109pt)=(0.38127,0.98909,0.42386); rgb(110pt)=(0.39678,0.99098,0.41098); rgb(111pt)=(0.41229,0.99268,0.39826); rgb(112pt)=(0.42778,0.99419,0.38575); rgb(113pt)=(0.44321,0.99551,0.37345); rgb(114pt)=(0.45854,0.99663,0.3614); rgb(115pt)=(0.47375,0.99755,0.34963); rgb(116pt)=(0.48879,0.99828,0.33816); rgb(117pt)=(0.50362,0.99879,0.32701); rgb(118pt)=(0.51822,0.9991,0.31622); rgb(119pt)=(0.53255,0.99919,0.30581); rgb(120pt)=(0.54658,0.99907,0.29581); rgb(121pt)=(0.56026,0.99873,0.28623); rgb(122pt)=(0.57357,0.99817,0.27712); rgb(123pt)=(0.58646,0.99739,0.26849); rgb(124pt)=(0.59891,0.99638,0.26038); rgb(125pt)=(0.61088,0.99514,0.2528); rgb(126pt)=(0.62233,0.99366,0.24579); rgb(127pt)=(0.63323,0.99195,0.23937); rgb(128pt)=(0.64362,0.98999,0.23356); rgb(129pt)=(0.65394,0.98775,0.22835); rgb(130pt)=(0.66428,0.98524,0.2237); rgb(131pt)=(0.67462,0.98246,0.2196); rgb(132pt)=(0.68494,0.97941,0.21602); rgb(133pt)=(0.69525,0.9761,0.21294); rgb(134pt)=(0.70553,0.97255,0.21032); rgb(135pt)=(0.71577,0.96875,0.20815); rgb(136pt)=(0.72596,0.9647,0.2064); rgb(137pt)=(0.7361,0.96043,0.20504); rgb(138pt)=(0.74617,0.95593,0.20406); rgb(139pt)=(0.75617,0.95121,0.20343); rgb(140pt)=(0.76608,0.94627,0.20311); rgb(141pt)=(0.77591,0.94113,0.2031); rgb(142pt)=(0.78563,0.93579,0.20336); rgb(143pt)=(0.79524,0.93025,0.20386); rgb(144pt)=(0.80473,0.92452,0.20459); rgb(145pt)=(0.8141,0.91861,0.20552); rgb(146pt)=(0.82333,0.91253,0.20663); rgb(147pt)=(0.83241,0.90627,0.20788); rgb(148pt)=(0.84133,0.89986,0.20926); rgb(149pt)=(0.8501,0.89328,0.21074); rgb(150pt)=(0.85868,0.88655,0.2123); rgb(151pt)=(0.86709,0.87968,0.21391); rgb(152pt)=(0.8753,0.87267,0.21555); rgb(153pt)=(0.88331,0.86553,0.21719); rgb(154pt)=(0.89112,0.85826,0.2188); rgb(155pt)=(0.8987,0.85087,0.22038); rgb(156pt)=(0.90605,0.84337,0.22188); rgb(157pt)=(0.91317,0.83576,0.22328); rgb(158pt)=(0.92004,0.82806,0.22456); rgb(159pt)=(0.92666,0.82025,0.2257); rgb(160pt)=(0.93301,0.81236,0.22667); rgb(161pt)=(0.93909,0.80439,0.22744); rgb(162pt)=(0.94489,0.79634,0.228); rgb(163pt)=(0.95039,0.78823,0.22831); rgb(164pt)=(0.9556,0.78005,0.22836); rgb(165pt)=(0.96049,0.77181,0.22811); rgb(166pt)=(0.96507,0.76352,0.22754); rgb(167pt)=(0.96931,0.75519,0.22663); rgb(168pt)=(0.97323,0.74682,0.22536); rgb(169pt)=(0.97679,0.73842,0.22369); rgb(170pt)=(0.98,0.73,0.22161); rgb(171pt)=(0.98289,0.7214,0.21918); rgb(172pt)=(0.98549,0.7125,0.2165); rgb(173pt)=(0.98781,0.7033,0.21358); rgb(174pt)=(0.98986,0.69382,0.21043); rgb(175pt)=(0.99163,0.68408,0.20706); rgb(176pt)=(0.99314,0.67408,0.20348); rgb(177pt)=(0.99438,0.66386,0.19971); rgb(178pt)=(0.99535,0.65341,0.19577); rgb(179pt)=(0.99607,0.64277,0.19165); rgb(180pt)=(0.99654,0.63193,0.18738); rgb(181pt)=(0.99675,0.62093,0.18297); rgb(182pt)=(0.99672,0.60977,0.17842); rgb(183pt)=(0.99644,0.59846,0.17376); rgb(184pt)=(0.99593,0.58703,0.16899); rgb(185pt)=(0.99517,0.57549,0.16412); rgb(186pt)=(0.99419,0.56386,0.15918); rgb(187pt)=(0.99297,0.55214,0.15417); rgb(188pt)=(0.99153,0.54036,0.1491); rgb(189pt)=(0.98987,0.52854,0.14398); rgb(190pt)=(0.98799,0.51667,0.13883); rgb(191pt)=(0.9859,0.50479,0.13367); rgb(192pt)=(0.9836,0.49291,0.12849); rgb(193pt)=(0.98108,0.48104,0.12332); rgb(194pt)=(0.97837,0.4692,0.11817); rgb(195pt)=(0.97545,0.4574,0.11305); rgb(196pt)=(0.97234,0.44565,0.10797); rgb(197pt)=(0.96904,0.43399,0.10294); rgb(198pt)=(0.96555,0.42241,0.09798); rgb(199pt)=(0.96187,0.41093,0.0931); rgb(200pt)=(0.95801,0.39958,0.08831); rgb(201pt)=(0.95398,0.38836,0.08362); rgb(202pt)=(0.94977,0.37729,0.07905); rgb(203pt)=(0.94538,0.36638,0.07461); rgb(204pt)=(0.94084,0.35566,0.07031); rgb(205pt)=(0.93612,0.34513,0.06616); rgb(206pt)=(0.93125,0.33482,0.06218); rgb(207pt)=(0.92623,0.32473,0.05837); rgb(208pt)=(0.92105,0.31489,0.05475); rgb(209pt)=(0.91572,0.3053,0.05134); rgb(210pt)=(0.91024,0.29599,0.04814); rgb(211pt)=(0.90463,0.28696,0.04516); rgb(212pt)=(0.89888,0.27824,0.04243); rgb(213pt)=(0.89298,0.26981,0.03993); rgb(214pt)=(0.88691,0.26152,0.03753); rgb(215pt)=(0.88066,0.25334,0.03521); rgb(216pt)=(0.87422,0.24526,0.03297); rgb(217pt)=(0.8676,0.2373,0.03082); rgb(218pt)=(0.86079,0.22945,0.02875); rgb(219pt)=(0.8538,0.2217,0.02677); rgb(220pt)=(0.84662,0.21407,0.02487); rgb(221pt)=(0.83926,0.20654,0.02305); rgb(222pt)=(0.83172,0.19912,0.02131); rgb(223pt)=(0.82399,0.19182,0.01966); rgb(224pt)=(0.81608,0.18462,0.01809); rgb(225pt)=(0.80799,0.17753,0.0166); rgb(226pt)=(0.79971,0.17055,0.0152); rgb(227pt)=(0.79125,0.16368,0.01387); rgb(228pt)=(0.7826,0.15693,0.01264); rgb(229pt)=(0.77377,0.15028,0.01148); rgb(230pt)=(0.76476,0.14374,0.01041); rgb(231pt)=(0.75556,0.13731,0.00942); rgb(232pt)=(0.74617,0.13098,0.00851); rgb(233pt)=(0.73661,0.12477,0.00769); rgb(234pt)=(0.72686,0.11867,0.00695); rgb(235pt)=(0.71692,0.11268,0.00629); rgb(236pt)=(0.7068,0.1068,0.00571); rgb(237pt)=(0.6965,0.10102,0.00522); rgb(238pt)=(0.68602,0.09536,0.00481); rgb(239pt)=(0.67535,0.0898,0.00449); rgb(240pt)=(0.66449,0.08436,0.00424); rgb(241pt)=(0.65345,0.07902,0.00408); rgb(242pt)=(0.64223,0.0738,0.00401); rgb(243pt)=(0.63082,0.06868,0.00401); rgb(244pt)=(0.61923,0.06367,0.0041); rgb(245pt)=(0.60746,0.05878,0.00427); rgb(246pt)=(0.5955,0.05399,0.00453); rgb(247pt)=(0.58336,0.04931,0.00486); rgb(248pt)=(0.57103,0.04474,0.00529); rgb(249pt)=(0.55852,0.04028,0.00579); rgb(250pt)=(0.54583,0.03593,0.00638); rgb(251pt)=(0.53295,0.03169,0.00705); rgb(252pt)=(0.51989,0.02756,0.0078); rgb(253pt)=(0.50664,0.02354,0.00863); rgb(254pt)=(0.49321,0.01963,0.00955); rgb(255pt)=(0.4796,0.01583,0.01055)}, mesh/rows=8]
table[row sep=crcr, point meta=\thisrow{c}] {%
x	y	c\\
5	12.1425009158361	0.316885056645923\\
5	12.1425009158361	0.316885056645923\\
6	11.7919329510718	0.381091730319001\\
6	11.7919329510718	0.381091730319001\\
7	11.6035095765315	0.419809012586647\\
7	11.6035095765315	0.419809012586647\\
8	11.4688849428243	0.448442192057085\\
8	11.4688849428243	0.448442192057085\\
9	11.3618315111059	0.471353685405731\\
9	11.3618315111059	0.471353685405731\\
10	11.2714484533799	0.490565789183854\\
10	11.2714484533799	0.490565789183854\\
11	11.1921418844036	0.50720309524711\\
11	11.1921418844036	0.50720309524711\\
12	11.1206732264371	0.521957484490431\\
12	11.1206732264371	0.521957484490431\\
};
\end{axis}

\begin{axis}[%
width=2in,
height=2in,
at={(3.1in,0.431in)},
scale only axis,
point meta min=0.316885056645923,
point meta max=0.521957484490431,
xmin=4,
xmax=13,
xtick = {5,7,9,11},
xticklabels={5,7,9,11},
xlabel style={font=\color{white!15!black}},
xlabel={${I}_{{ref}}^d$ [A]},
ymin=-15,
ymax=8,
ylabel style={font=\color{white!15!black}},
ylabel={Phase Margin [Deg]},
axis background/.style={fill=white},
xmajorgrids,
ymajorgrids,
grid style={dashed, opacity=0.5},
colormap={mymap}{[1pt] rgb(0pt)=(0.18995,0.07176,0.23217); rgb(1pt)=(0.19483,0.08339,0.26149); rgb(2pt)=(0.19956,0.09498,0.29024); rgb(3pt)=(0.20415,0.10652,0.31844); rgb(4pt)=(0.2086,0.11802,0.34607); rgb(5pt)=(0.21291,0.12947,0.37314); rgb(6pt)=(0.21708,0.14087,0.39964); rgb(7pt)=(0.22111,0.15223,0.42558); rgb(8pt)=(0.225,0.16354,0.45096); rgb(9pt)=(0.22875,0.17481,0.47578); rgb(10pt)=(0.23236,0.18603,0.50004); rgb(11pt)=(0.23582,0.1972,0.52373); rgb(12pt)=(0.23915,0.20833,0.54686); rgb(13pt)=(0.24234,0.21941,0.56942); rgb(14pt)=(0.24539,0.23044,0.59142); rgb(15pt)=(0.2483,0.24143,0.61286); rgb(16pt)=(0.25107,0.25237,0.63374); rgb(17pt)=(0.25369,0.26327,0.65406); rgb(18pt)=(0.25618,0.27412,0.67381); rgb(19pt)=(0.25853,0.28492,0.693); rgb(20pt)=(0.26074,0.29568,0.71162); rgb(21pt)=(0.2628,0.30639,0.72968); rgb(22pt)=(0.26473,0.31706,0.74718); rgb(23pt)=(0.26652,0.32768,0.76412); rgb(24pt)=(0.26816,0.33825,0.7805); rgb(25pt)=(0.26967,0.34878,0.79631); rgb(26pt)=(0.27103,0.35926,0.81156); rgb(27pt)=(0.27226,0.3697,0.82624); rgb(28pt)=(0.27334,0.38008,0.84037); rgb(29pt)=(0.27429,0.39043,0.85393); rgb(30pt)=(0.27509,0.40072,0.86692); rgb(31pt)=(0.27576,0.41097,0.87936); rgb(32pt)=(0.27628,0.42118,0.89123); rgb(33pt)=(0.27667,0.43134,0.90254); rgb(34pt)=(0.27691,0.44145,0.91328); rgb(35pt)=(0.27701,0.45152,0.92347); rgb(36pt)=(0.27698,0.46153,0.93309); rgb(37pt)=(0.2768,0.47151,0.94214); rgb(38pt)=(0.27648,0.48144,0.95064); rgb(39pt)=(0.27603,0.49132,0.95857); rgb(40pt)=(0.27543,0.50115,0.96594); rgb(41pt)=(0.27469,0.51094,0.97275); rgb(42pt)=(0.27381,0.52069,0.97899); rgb(43pt)=(0.27273,0.5304,0.98461); rgb(44pt)=(0.27106,0.54015,0.9893); rgb(45pt)=(0.26878,0.54995,0.99303); rgb(46pt)=(0.26592,0.55979,0.99583); rgb(47pt)=(0.26252,0.56967,0.99773); rgb(48pt)=(0.25862,0.57958,0.99876); rgb(49pt)=(0.25425,0.5895,0.99896); rgb(50pt)=(0.24946,0.59943,0.99835); rgb(51pt)=(0.24427,0.60937,0.99697); rgb(52pt)=(0.23874,0.61931,0.99485); rgb(53pt)=(0.23288,0.62923,0.99202); rgb(54pt)=(0.22676,0.63913,0.98851); rgb(55pt)=(0.22039,0.64901,0.98436); rgb(56pt)=(0.21382,0.65886,0.97959); rgb(57pt)=(0.20708,0.66866,0.97423); rgb(58pt)=(0.20021,0.67842,0.96833); rgb(59pt)=(0.19326,0.68812,0.9619); rgb(60pt)=(0.18625,0.69775,0.95498); rgb(61pt)=(0.17923,0.70732,0.94761); rgb(62pt)=(0.17223,0.7168,0.93981); rgb(63pt)=(0.16529,0.7262,0.93161); rgb(64pt)=(0.15844,0.73551,0.92305); rgb(65pt)=(0.15173,0.74472,0.91416); rgb(66pt)=(0.14519,0.75381,0.90496); rgb(67pt)=(0.13886,0.76279,0.8955); rgb(68pt)=(0.13278,0.77165,0.8858); rgb(69pt)=(0.12698,0.78037,0.8759); rgb(70pt)=(0.12151,0.78896,0.86581); rgb(71pt)=(0.11639,0.7974,0.85559); rgb(72pt)=(0.11167,0.80569,0.84525); rgb(73pt)=(0.10738,0.81381,0.83484); rgb(74pt)=(0.10357,0.82177,0.82437); rgb(75pt)=(0.10026,0.82955,0.81389); rgb(76pt)=(0.0975,0.83714,0.80342); rgb(77pt)=(0.09532,0.84455,0.79299); rgb(78pt)=(0.09377,0.85175,0.78264); rgb(79pt)=(0.09287,0.85875,0.7724); rgb(80pt)=(0.09267,0.86554,0.7623); rgb(81pt)=(0.0932,0.87211,0.75237); rgb(82pt)=(0.09451,0.87844,0.74265); rgb(83pt)=(0.09662,0.88454,0.73316); rgb(84pt)=(0.09958,0.8904,0.72393); rgb(85pt)=(0.10342,0.896,0.715); rgb(86pt)=(0.10815,0.90142,0.70599); rgb(87pt)=(0.11374,0.90673,0.69651); rgb(88pt)=(0.12014,0.91193,0.6866); rgb(89pt)=(0.12733,0.91701,0.67627); rgb(90pt)=(0.13526,0.92197,0.66556); rgb(91pt)=(0.14391,0.9268,0.65448); rgb(92pt)=(0.15323,0.93151,0.64308); rgb(93pt)=(0.16319,0.93609,0.63137); rgb(94pt)=(0.17377,0.94053,0.61938); rgb(95pt)=(0.18491,0.94484,0.60713); rgb(96pt)=(0.19659,0.94901,0.59466); rgb(97pt)=(0.20877,0.95304,0.58199); rgb(98pt)=(0.22142,0.95692,0.56914); rgb(99pt)=(0.23449,0.96065,0.55614); rgb(100pt)=(0.24797,0.96423,0.54303); rgb(101pt)=(0.2618,0.96765,0.52981); rgb(102pt)=(0.27597,0.97092,0.51653); rgb(103pt)=(0.29042,0.97403,0.50321); rgb(104pt)=(0.30513,0.97697,0.48987); rgb(105pt)=(0.32006,0.97974,0.47654); rgb(106pt)=(0.33517,0.98234,0.46325); rgb(107pt)=(0.35043,0.98477,0.45002); rgb(108pt)=(0.36581,0.98702,0.43688); rgb(109pt)=(0.38127,0.98909,0.42386); rgb(110pt)=(0.39678,0.99098,0.41098); rgb(111pt)=(0.41229,0.99268,0.39826); rgb(112pt)=(0.42778,0.99419,0.38575); rgb(113pt)=(0.44321,0.99551,0.37345); rgb(114pt)=(0.45854,0.99663,0.3614); rgb(115pt)=(0.47375,0.99755,0.34963); rgb(116pt)=(0.48879,0.99828,0.33816); rgb(117pt)=(0.50362,0.99879,0.32701); rgb(118pt)=(0.51822,0.9991,0.31622); rgb(119pt)=(0.53255,0.99919,0.30581); rgb(120pt)=(0.54658,0.99907,0.29581); rgb(121pt)=(0.56026,0.99873,0.28623); rgb(122pt)=(0.57357,0.99817,0.27712); rgb(123pt)=(0.58646,0.99739,0.26849); rgb(124pt)=(0.59891,0.99638,0.26038); rgb(125pt)=(0.61088,0.99514,0.2528); rgb(126pt)=(0.62233,0.99366,0.24579); rgb(127pt)=(0.63323,0.99195,0.23937); rgb(128pt)=(0.64362,0.98999,0.23356); rgb(129pt)=(0.65394,0.98775,0.22835); rgb(130pt)=(0.66428,0.98524,0.2237); rgb(131pt)=(0.67462,0.98246,0.2196); rgb(132pt)=(0.68494,0.97941,0.21602); rgb(133pt)=(0.69525,0.9761,0.21294); rgb(134pt)=(0.70553,0.97255,0.21032); rgb(135pt)=(0.71577,0.96875,0.20815); rgb(136pt)=(0.72596,0.9647,0.2064); rgb(137pt)=(0.7361,0.96043,0.20504); rgb(138pt)=(0.74617,0.95593,0.20406); rgb(139pt)=(0.75617,0.95121,0.20343); rgb(140pt)=(0.76608,0.94627,0.20311); rgb(141pt)=(0.77591,0.94113,0.2031); rgb(142pt)=(0.78563,0.93579,0.20336); rgb(143pt)=(0.79524,0.93025,0.20386); rgb(144pt)=(0.80473,0.92452,0.20459); rgb(145pt)=(0.8141,0.91861,0.20552); rgb(146pt)=(0.82333,0.91253,0.20663); rgb(147pt)=(0.83241,0.90627,0.20788); rgb(148pt)=(0.84133,0.89986,0.20926); rgb(149pt)=(0.8501,0.89328,0.21074); rgb(150pt)=(0.85868,0.88655,0.2123); rgb(151pt)=(0.86709,0.87968,0.21391); rgb(152pt)=(0.8753,0.87267,0.21555); rgb(153pt)=(0.88331,0.86553,0.21719); rgb(154pt)=(0.89112,0.85826,0.2188); rgb(155pt)=(0.8987,0.85087,0.22038); rgb(156pt)=(0.90605,0.84337,0.22188); rgb(157pt)=(0.91317,0.83576,0.22328); rgb(158pt)=(0.92004,0.82806,0.22456); rgb(159pt)=(0.92666,0.82025,0.2257); rgb(160pt)=(0.93301,0.81236,0.22667); rgb(161pt)=(0.93909,0.80439,0.22744); rgb(162pt)=(0.94489,0.79634,0.228); rgb(163pt)=(0.95039,0.78823,0.22831); rgb(164pt)=(0.9556,0.78005,0.22836); rgb(165pt)=(0.96049,0.77181,0.22811); rgb(166pt)=(0.96507,0.76352,0.22754); rgb(167pt)=(0.96931,0.75519,0.22663); rgb(168pt)=(0.97323,0.74682,0.22536); rgb(169pt)=(0.97679,0.73842,0.22369); rgb(170pt)=(0.98,0.73,0.22161); rgb(171pt)=(0.98289,0.7214,0.21918); rgb(172pt)=(0.98549,0.7125,0.2165); rgb(173pt)=(0.98781,0.7033,0.21358); rgb(174pt)=(0.98986,0.69382,0.21043); rgb(175pt)=(0.99163,0.68408,0.20706); rgb(176pt)=(0.99314,0.67408,0.20348); rgb(177pt)=(0.99438,0.66386,0.19971); rgb(178pt)=(0.99535,0.65341,0.19577); rgb(179pt)=(0.99607,0.64277,0.19165); rgb(180pt)=(0.99654,0.63193,0.18738); rgb(181pt)=(0.99675,0.62093,0.18297); rgb(182pt)=(0.99672,0.60977,0.17842); rgb(183pt)=(0.99644,0.59846,0.17376); rgb(184pt)=(0.99593,0.58703,0.16899); rgb(185pt)=(0.99517,0.57549,0.16412); rgb(186pt)=(0.99419,0.56386,0.15918); rgb(187pt)=(0.99297,0.55214,0.15417); rgb(188pt)=(0.99153,0.54036,0.1491); rgb(189pt)=(0.98987,0.52854,0.14398); rgb(190pt)=(0.98799,0.51667,0.13883); rgb(191pt)=(0.9859,0.50479,0.13367); rgb(192pt)=(0.9836,0.49291,0.12849); rgb(193pt)=(0.98108,0.48104,0.12332); rgb(194pt)=(0.97837,0.4692,0.11817); rgb(195pt)=(0.97545,0.4574,0.11305); rgb(196pt)=(0.97234,0.44565,0.10797); rgb(197pt)=(0.96904,0.43399,0.10294); rgb(198pt)=(0.96555,0.42241,0.09798); rgb(199pt)=(0.96187,0.41093,0.0931); rgb(200pt)=(0.95801,0.39958,0.08831); rgb(201pt)=(0.95398,0.38836,0.08362); rgb(202pt)=(0.94977,0.37729,0.07905); rgb(203pt)=(0.94538,0.36638,0.07461); rgb(204pt)=(0.94084,0.35566,0.07031); rgb(205pt)=(0.93612,0.34513,0.06616); rgb(206pt)=(0.93125,0.33482,0.06218); rgb(207pt)=(0.92623,0.32473,0.05837); rgb(208pt)=(0.92105,0.31489,0.05475); rgb(209pt)=(0.91572,0.3053,0.05134); rgb(210pt)=(0.91024,0.29599,0.04814); rgb(211pt)=(0.90463,0.28696,0.04516); rgb(212pt)=(0.89888,0.27824,0.04243); rgb(213pt)=(0.89298,0.26981,0.03993); rgb(214pt)=(0.88691,0.26152,0.03753); rgb(215pt)=(0.88066,0.25334,0.03521); rgb(216pt)=(0.87422,0.24526,0.03297); rgb(217pt)=(0.8676,0.2373,0.03082); rgb(218pt)=(0.86079,0.22945,0.02875); rgb(219pt)=(0.8538,0.2217,0.02677); rgb(220pt)=(0.84662,0.21407,0.02487); rgb(221pt)=(0.83926,0.20654,0.02305); rgb(222pt)=(0.83172,0.19912,0.02131); rgb(223pt)=(0.82399,0.19182,0.01966); rgb(224pt)=(0.81608,0.18462,0.01809); rgb(225pt)=(0.80799,0.17753,0.0166); rgb(226pt)=(0.79971,0.17055,0.0152); rgb(227pt)=(0.79125,0.16368,0.01387); rgb(228pt)=(0.7826,0.15693,0.01264); rgb(229pt)=(0.77377,0.15028,0.01148); rgb(230pt)=(0.76476,0.14374,0.01041); rgb(231pt)=(0.75556,0.13731,0.00942); rgb(232pt)=(0.74617,0.13098,0.00851); rgb(233pt)=(0.73661,0.12477,0.00769); rgb(234pt)=(0.72686,0.11867,0.00695); rgb(235pt)=(0.71692,0.11268,0.00629); rgb(236pt)=(0.7068,0.1068,0.00571); rgb(237pt)=(0.6965,0.10102,0.00522); rgb(238pt)=(0.68602,0.09536,0.00481); rgb(239pt)=(0.67535,0.0898,0.00449); rgb(240pt)=(0.66449,0.08436,0.00424); rgb(241pt)=(0.65345,0.07902,0.00408); rgb(242pt)=(0.64223,0.0738,0.00401); rgb(243pt)=(0.63082,0.06868,0.00401); rgb(244pt)=(0.61923,0.06367,0.0041); rgb(245pt)=(0.60746,0.05878,0.00427); rgb(246pt)=(0.5955,0.05399,0.00453); rgb(247pt)=(0.58336,0.04931,0.00486); rgb(248pt)=(0.57103,0.04474,0.00529); rgb(249pt)=(0.55852,0.04028,0.00579); rgb(250pt)=(0.54583,0.03593,0.00638); rgb(251pt)=(0.53295,0.03169,0.00705); rgb(252pt)=(0.51989,0.02756,0.0078); rgb(253pt)=(0.50664,0.02354,0.00863); rgb(254pt)=(0.49321,0.01963,0.00955); rgb(255pt)=(0.4796,0.01583,0.01055)},
colorbar,
colorbar style={ylabel style={rotate=-90,font=\color{white!15!black},at={(0.5,1)}, anchor=south,},scaled ticks=false,ytick={0.35,0.4,0.45,0.5},yticklabels={0.35,0.4,0.45,0.5}, ylabel={$\bm{AQI}$}}
]
\draw[fill=black, thick,opacity=0.1] (axis cs: 0,-90) -- (axis cs: 16,-90) -- (axis cs: 16,0) -- (axis cs: 0,0) -- cycle;
\node[color=black, fill=none] at (axis cs: 11.3,-1) {$Unstable$};
\node[color=black, fill=none] at (axis cs: 11.3,1) {$Stable$};
\addplot[%
mesh,
    shader=flat,
    mark=none,
    line join=round,
    line cap=round,
    line width=2pt,
    point meta=explicit, colormap={mymap}{[1pt] rgb(0pt)=(0.18995,0.07176,0.23217); rgb(1pt)=(0.19483,0.08339,0.26149); rgb(2pt)=(0.19956,0.09498,0.29024); rgb(3pt)=(0.20415,0.10652,0.31844); rgb(4pt)=(0.2086,0.11802,0.34607); rgb(5pt)=(0.21291,0.12947,0.37314); rgb(6pt)=(0.21708,0.14087,0.39964); rgb(7pt)=(0.22111,0.15223,0.42558); rgb(8pt)=(0.225,0.16354,0.45096); rgb(9pt)=(0.22875,0.17481,0.47578); rgb(10pt)=(0.23236,0.18603,0.50004); rgb(11pt)=(0.23582,0.1972,0.52373); rgb(12pt)=(0.23915,0.20833,0.54686); rgb(13pt)=(0.24234,0.21941,0.56942); rgb(14pt)=(0.24539,0.23044,0.59142); rgb(15pt)=(0.2483,0.24143,0.61286); rgb(16pt)=(0.25107,0.25237,0.63374); rgb(17pt)=(0.25369,0.26327,0.65406); rgb(18pt)=(0.25618,0.27412,0.67381); rgb(19pt)=(0.25853,0.28492,0.693); rgb(20pt)=(0.26074,0.29568,0.71162); rgb(21pt)=(0.2628,0.30639,0.72968); rgb(22pt)=(0.26473,0.31706,0.74718); rgb(23pt)=(0.26652,0.32768,0.76412); rgb(24pt)=(0.26816,0.33825,0.7805); rgb(25pt)=(0.26967,0.34878,0.79631); rgb(26pt)=(0.27103,0.35926,0.81156); rgb(27pt)=(0.27226,0.3697,0.82624); rgb(28pt)=(0.27334,0.38008,0.84037); rgb(29pt)=(0.27429,0.39043,0.85393); rgb(30pt)=(0.27509,0.40072,0.86692); rgb(31pt)=(0.27576,0.41097,0.87936); rgb(32pt)=(0.27628,0.42118,0.89123); rgb(33pt)=(0.27667,0.43134,0.90254); rgb(34pt)=(0.27691,0.44145,0.91328); rgb(35pt)=(0.27701,0.45152,0.92347); rgb(36pt)=(0.27698,0.46153,0.93309); rgb(37pt)=(0.2768,0.47151,0.94214); rgb(38pt)=(0.27648,0.48144,0.95064); rgb(39pt)=(0.27603,0.49132,0.95857); rgb(40pt)=(0.27543,0.50115,0.96594); rgb(41pt)=(0.27469,0.51094,0.97275); rgb(42pt)=(0.27381,0.52069,0.97899); rgb(43pt)=(0.27273,0.5304,0.98461); rgb(44pt)=(0.27106,0.54015,0.9893); rgb(45pt)=(0.26878,0.54995,0.99303); rgb(46pt)=(0.26592,0.55979,0.99583); rgb(47pt)=(0.26252,0.56967,0.99773); rgb(48pt)=(0.25862,0.57958,0.99876); rgb(49pt)=(0.25425,0.5895,0.99896); rgb(50pt)=(0.24946,0.59943,0.99835); rgb(51pt)=(0.24427,0.60937,0.99697); rgb(52pt)=(0.23874,0.61931,0.99485); rgb(53pt)=(0.23288,0.62923,0.99202); rgb(54pt)=(0.22676,0.63913,0.98851); rgb(55pt)=(0.22039,0.64901,0.98436); rgb(56pt)=(0.21382,0.65886,0.97959); rgb(57pt)=(0.20708,0.66866,0.97423); rgb(58pt)=(0.20021,0.67842,0.96833); rgb(59pt)=(0.19326,0.68812,0.9619); rgb(60pt)=(0.18625,0.69775,0.95498); rgb(61pt)=(0.17923,0.70732,0.94761); rgb(62pt)=(0.17223,0.7168,0.93981); rgb(63pt)=(0.16529,0.7262,0.93161); rgb(64pt)=(0.15844,0.73551,0.92305); rgb(65pt)=(0.15173,0.74472,0.91416); rgb(66pt)=(0.14519,0.75381,0.90496); rgb(67pt)=(0.13886,0.76279,0.8955); rgb(68pt)=(0.13278,0.77165,0.8858); rgb(69pt)=(0.12698,0.78037,0.8759); rgb(70pt)=(0.12151,0.78896,0.86581); rgb(71pt)=(0.11639,0.7974,0.85559); rgb(72pt)=(0.11167,0.80569,0.84525); rgb(73pt)=(0.10738,0.81381,0.83484); rgb(74pt)=(0.10357,0.82177,0.82437); rgb(75pt)=(0.10026,0.82955,0.81389); rgb(76pt)=(0.0975,0.83714,0.80342); rgb(77pt)=(0.09532,0.84455,0.79299); rgb(78pt)=(0.09377,0.85175,0.78264); rgb(79pt)=(0.09287,0.85875,0.7724); rgb(80pt)=(0.09267,0.86554,0.7623); rgb(81pt)=(0.0932,0.87211,0.75237); rgb(82pt)=(0.09451,0.87844,0.74265); rgb(83pt)=(0.09662,0.88454,0.73316); rgb(84pt)=(0.09958,0.8904,0.72393); rgb(85pt)=(0.10342,0.896,0.715); rgb(86pt)=(0.10815,0.90142,0.70599); rgb(87pt)=(0.11374,0.90673,0.69651); rgb(88pt)=(0.12014,0.91193,0.6866); rgb(89pt)=(0.12733,0.91701,0.67627); rgb(90pt)=(0.13526,0.92197,0.66556); rgb(91pt)=(0.14391,0.9268,0.65448); rgb(92pt)=(0.15323,0.93151,0.64308); rgb(93pt)=(0.16319,0.93609,0.63137); rgb(94pt)=(0.17377,0.94053,0.61938); rgb(95pt)=(0.18491,0.94484,0.60713); rgb(96pt)=(0.19659,0.94901,0.59466); rgb(97pt)=(0.20877,0.95304,0.58199); rgb(98pt)=(0.22142,0.95692,0.56914); rgb(99pt)=(0.23449,0.96065,0.55614); rgb(100pt)=(0.24797,0.96423,0.54303); rgb(101pt)=(0.2618,0.96765,0.52981); rgb(102pt)=(0.27597,0.97092,0.51653); rgb(103pt)=(0.29042,0.97403,0.50321); rgb(104pt)=(0.30513,0.97697,0.48987); rgb(105pt)=(0.32006,0.97974,0.47654); rgb(106pt)=(0.33517,0.98234,0.46325); rgb(107pt)=(0.35043,0.98477,0.45002); rgb(108pt)=(0.36581,0.98702,0.43688); rgb(109pt)=(0.38127,0.98909,0.42386); rgb(110pt)=(0.39678,0.99098,0.41098); rgb(111pt)=(0.41229,0.99268,0.39826); rgb(112pt)=(0.42778,0.99419,0.38575); rgb(113pt)=(0.44321,0.99551,0.37345); rgb(114pt)=(0.45854,0.99663,0.3614); rgb(115pt)=(0.47375,0.99755,0.34963); rgb(116pt)=(0.48879,0.99828,0.33816); rgb(117pt)=(0.50362,0.99879,0.32701); rgb(118pt)=(0.51822,0.9991,0.31622); rgb(119pt)=(0.53255,0.99919,0.30581); rgb(120pt)=(0.54658,0.99907,0.29581); rgb(121pt)=(0.56026,0.99873,0.28623); rgb(122pt)=(0.57357,0.99817,0.27712); rgb(123pt)=(0.58646,0.99739,0.26849); rgb(124pt)=(0.59891,0.99638,0.26038); rgb(125pt)=(0.61088,0.99514,0.2528); rgb(126pt)=(0.62233,0.99366,0.24579); rgb(127pt)=(0.63323,0.99195,0.23937); rgb(128pt)=(0.64362,0.98999,0.23356); rgb(129pt)=(0.65394,0.98775,0.22835); rgb(130pt)=(0.66428,0.98524,0.2237); rgb(131pt)=(0.67462,0.98246,0.2196); rgb(132pt)=(0.68494,0.97941,0.21602); rgb(133pt)=(0.69525,0.9761,0.21294); rgb(134pt)=(0.70553,0.97255,0.21032); rgb(135pt)=(0.71577,0.96875,0.20815); rgb(136pt)=(0.72596,0.9647,0.2064); rgb(137pt)=(0.7361,0.96043,0.20504); rgb(138pt)=(0.74617,0.95593,0.20406); rgb(139pt)=(0.75617,0.95121,0.20343); rgb(140pt)=(0.76608,0.94627,0.20311); rgb(141pt)=(0.77591,0.94113,0.2031); rgb(142pt)=(0.78563,0.93579,0.20336); rgb(143pt)=(0.79524,0.93025,0.20386); rgb(144pt)=(0.80473,0.92452,0.20459); rgb(145pt)=(0.8141,0.91861,0.20552); rgb(146pt)=(0.82333,0.91253,0.20663); rgb(147pt)=(0.83241,0.90627,0.20788); rgb(148pt)=(0.84133,0.89986,0.20926); rgb(149pt)=(0.8501,0.89328,0.21074); rgb(150pt)=(0.85868,0.88655,0.2123); rgb(151pt)=(0.86709,0.87968,0.21391); rgb(152pt)=(0.8753,0.87267,0.21555); rgb(153pt)=(0.88331,0.86553,0.21719); rgb(154pt)=(0.89112,0.85826,0.2188); rgb(155pt)=(0.8987,0.85087,0.22038); rgb(156pt)=(0.90605,0.84337,0.22188); rgb(157pt)=(0.91317,0.83576,0.22328); rgb(158pt)=(0.92004,0.82806,0.22456); rgb(159pt)=(0.92666,0.82025,0.2257); rgb(160pt)=(0.93301,0.81236,0.22667); rgb(161pt)=(0.93909,0.80439,0.22744); rgb(162pt)=(0.94489,0.79634,0.228); rgb(163pt)=(0.95039,0.78823,0.22831); rgb(164pt)=(0.9556,0.78005,0.22836); rgb(165pt)=(0.96049,0.77181,0.22811); rgb(166pt)=(0.96507,0.76352,0.22754); rgb(167pt)=(0.96931,0.75519,0.22663); rgb(168pt)=(0.97323,0.74682,0.22536); rgb(169pt)=(0.97679,0.73842,0.22369); rgb(170pt)=(0.98,0.73,0.22161); rgb(171pt)=(0.98289,0.7214,0.21918); rgb(172pt)=(0.98549,0.7125,0.2165); rgb(173pt)=(0.98781,0.7033,0.21358); rgb(174pt)=(0.98986,0.69382,0.21043); rgb(175pt)=(0.99163,0.68408,0.20706); rgb(176pt)=(0.99314,0.67408,0.20348); rgb(177pt)=(0.99438,0.66386,0.19971); rgb(178pt)=(0.99535,0.65341,0.19577); rgb(179pt)=(0.99607,0.64277,0.19165); rgb(180pt)=(0.99654,0.63193,0.18738); rgb(181pt)=(0.99675,0.62093,0.18297); rgb(182pt)=(0.99672,0.60977,0.17842); rgb(183pt)=(0.99644,0.59846,0.17376); rgb(184pt)=(0.99593,0.58703,0.16899); rgb(185pt)=(0.99517,0.57549,0.16412); rgb(186pt)=(0.99419,0.56386,0.15918); rgb(187pt)=(0.99297,0.55214,0.15417); rgb(188pt)=(0.99153,0.54036,0.1491); rgb(189pt)=(0.98987,0.52854,0.14398); rgb(190pt)=(0.98799,0.51667,0.13883); rgb(191pt)=(0.9859,0.50479,0.13367); rgb(192pt)=(0.9836,0.49291,0.12849); rgb(193pt)=(0.98108,0.48104,0.12332); rgb(194pt)=(0.97837,0.4692,0.11817); rgb(195pt)=(0.97545,0.4574,0.11305); rgb(196pt)=(0.97234,0.44565,0.10797); rgb(197pt)=(0.96904,0.43399,0.10294); rgb(198pt)=(0.96555,0.42241,0.09798); rgb(199pt)=(0.96187,0.41093,0.0931); rgb(200pt)=(0.95801,0.39958,0.08831); rgb(201pt)=(0.95398,0.38836,0.08362); rgb(202pt)=(0.94977,0.37729,0.07905); rgb(203pt)=(0.94538,0.36638,0.07461); rgb(204pt)=(0.94084,0.35566,0.07031); rgb(205pt)=(0.93612,0.34513,0.06616); rgb(206pt)=(0.93125,0.33482,0.06218); rgb(207pt)=(0.92623,0.32473,0.05837); rgb(208pt)=(0.92105,0.31489,0.05475); rgb(209pt)=(0.91572,0.3053,0.05134); rgb(210pt)=(0.91024,0.29599,0.04814); rgb(211pt)=(0.90463,0.28696,0.04516); rgb(212pt)=(0.89888,0.27824,0.04243); rgb(213pt)=(0.89298,0.26981,0.03993); rgb(214pt)=(0.88691,0.26152,0.03753); rgb(215pt)=(0.88066,0.25334,0.03521); rgb(216pt)=(0.87422,0.24526,0.03297); rgb(217pt)=(0.8676,0.2373,0.03082); rgb(218pt)=(0.86079,0.22945,0.02875); rgb(219pt)=(0.8538,0.2217,0.02677); rgb(220pt)=(0.84662,0.21407,0.02487); rgb(221pt)=(0.83926,0.20654,0.02305); rgb(222pt)=(0.83172,0.19912,0.02131); rgb(223pt)=(0.82399,0.19182,0.01966); rgb(224pt)=(0.81608,0.18462,0.01809); rgb(225pt)=(0.80799,0.17753,0.0166); rgb(226pt)=(0.79971,0.17055,0.0152); rgb(227pt)=(0.79125,0.16368,0.01387); rgb(228pt)=(0.7826,0.15693,0.01264); rgb(229pt)=(0.77377,0.15028,0.01148); rgb(230pt)=(0.76476,0.14374,0.01041); rgb(231pt)=(0.75556,0.13731,0.00942); rgb(232pt)=(0.74617,0.13098,0.00851); rgb(233pt)=(0.73661,0.12477,0.00769); rgb(234pt)=(0.72686,0.11867,0.00695); rgb(235pt)=(0.71692,0.11268,0.00629); rgb(236pt)=(0.7068,0.1068,0.00571); rgb(237pt)=(0.6965,0.10102,0.00522); rgb(238pt)=(0.68602,0.09536,0.00481); rgb(239pt)=(0.67535,0.0898,0.00449); rgb(240pt)=(0.66449,0.08436,0.00424); rgb(241pt)=(0.65345,0.07902,0.00408); rgb(242pt)=(0.64223,0.0738,0.00401); rgb(243pt)=(0.63082,0.06868,0.00401); rgb(244pt)=(0.61923,0.06367,0.0041); rgb(245pt)=(0.60746,0.05878,0.00427); rgb(246pt)=(0.5955,0.05399,0.00453); rgb(247pt)=(0.58336,0.04931,0.00486); rgb(248pt)=(0.57103,0.04474,0.00529); rgb(249pt)=(0.55852,0.04028,0.00579); rgb(250pt)=(0.54583,0.03593,0.00638); rgb(251pt)=(0.53295,0.03169,0.00705); rgb(252pt)=(0.51989,0.02756,0.0078); rgb(253pt)=(0.50664,0.02354,0.00863); rgb(254pt)=(0.49321,0.01963,0.00955); rgb(255pt)=(0.4796,0.01583,0.01055)}, mesh/rows=8]
table[row sep=crcr, point meta=\thisrow{c}] {%
x	y	c\\
5	6.66233535703154	0.316885056645923\\
5	6.66233535703154	0.316885056645923\\
6	-0.48705565367959	0.381091730319001\\
6	-0.48705565367959	0.381091730319001\\
7	-4.35957677381055	0.419809012586647\\
7	-4.35957677381055	0.419809012586647\\
8	-7.03922992947446	0.448442192057085\\
8	-7.03922992947446	0.448442192057085\\
9	-9.0751772366516	0.471353685405731\\
9	-9.0751772366516	0.471353685405731\\
10	-10.7083990858149	0.490565789183854\\
10	-10.7083990858149	0.490565789183854\\
11	-12.0674624559064	0.50720309524711\\
11	-12.0674624559064	0.50720309524711\\
12	-13.2290170263057	0.521957484490431\\
12	-13.2290170263057	0.521957484490431\\
};
\addplot[
    only marks,
    mark=*,
    mark size=5pt,
    color=red, 
    mark options={
        fill=red,
        fill opacity=0.5, 
        draw opacity=1     
    },
]
coordinates {(12,-13.22)};
\node[color=black, fill=none] at (axis cs: 11.75,-9) {\textcolor{red}{Case 3b}};
\draw[line width=0.5mm,draw=black!80,->] (axis cs:12,-12.7)--(axis cs:12,-10);

\addplot[
    only marks,
    mark=*,
    mark size=5pt,
    color=blue, 
    mark options={
        fill=blue,
        fill opacity=0.5, 
        draw opacity=1     
    },
]
coordinates {(5,6.66)};
\node[color=black, fill=none] at (axis cs: 7.5,5) {\textcolor{blue}{Case 3a}};
\draw[line width=0.5mm,draw=black!80,->] (axis cs:5.2,6.3)--(axis cs:6.5,5);
\end{axis}
\end{tikzpicture}%